\documentclass[fleqn,a4paper]{report} 
\usepackage{epsfig,graphics,graphicx}
\usepackage{pifont}
\usepackage{ifthen}
\usepackage{clshan-math}
\usepackage{color,clshan-color,clshan-font}
\newcounter{mytemplate}
\usecounter{mytemplate}
\newenvironment{templateenumerate}
{\begin{list}
 {\bf Template \arabic{mytemplate}:}
 {\settowidth{\labelwidth}{~~~~~}
  \setlength{\leftmargin}{\labelwidth}
  \renewcommand{\makelabel}{~~~}}}
{\end{list}}
\newcommand{\myitem}               {\refstepcounter{mytemplate}
                                    \item}
\newcounter{myfigure}
\usecounter{myfigure}
\renewcommand{\thefigure}{\arabic{myfigure}}
\newcommand{\picin}            [3] {\refstepcounter{myfigure}
                                    \begin{figure}[h!]
                                    \begin{center}
                                    \begin{picture}(15.5,10)
                                    \blue
                                    {\linethickness{2pt}
                                     \put(0,0){\framebox(15.5,10)
                                               {\includegraphics[width=15cm]{#1}}}}
                                    \end{picture}
                                    \vspace{-0.25cm}
                                    \end{center}
                                    \caption{
                                     #3
                                    }
                                    \label{fig:#1}
                                    \end{figure}}
\newcommand{\picdin}           [4] {\renewcommand{\thefigure}{\arabic{myfigure}a}
                                    \picin{#1-1}{#2-1}{#3}
                                    \vspace{0.75cm}
                                    \addtocounter{myfigure}{-1}%
                                    \renewcommand{\thefigure}{\arabic{myfigure}b}
                                    \picin{#1-2}{#2-2}{#4}
                                    \renewcommand{\thefigure}{\arabic{myfigure}}}
\newcommand{\picqin}           [6] {\renewcommand{\thefigure}{\arabic{myfigure}a}
                                    \picin{#1-1}{#2-1}{#3}
                                    \vspace{0.75cm}
                                    \addtocounter{myfigure}{-1}%
                                    \renewcommand{\thefigure}{\arabic{myfigure}b}
                                    \picin{#1-2}{#2-2}{#4}
                                    \newpage
                                    \addtocounter{myfigure}{-1}%
                                    \renewcommand{\thefigure}{\arabic{myfigure}c}
                                    \picin{#1-3}{#2-3}{#5}
                                    \vspace{0.75cm}
                                    \addtocounter{myfigure}{-1}%
                                    \renewcommand{\thefigure}{\arabic{myfigure}d}
                                    \picin{#1-4}{#2-4}{#6}
                                    \renewcommand{\thefigure}{\arabic{myfigure}}}
\newcommand{\picsin}           [9] {\renewcommand{\thefigure}{\arabic{myfigure}a}
                                    \picin{#1-1}{#2-1}{#3}
                                    \vspace{0.75cm}
                                    \addtocounter{myfigure}{-1}%
                                    \renewcommand{\thefigure}{\arabic{myfigure}b}
                                    \picin{#1-2}{#2-2}{#4}
                                    \newpage
                                    \addtocounter{myfigure}{-1}%
                                    \renewcommand{\thefigure}{\arabic{myfigure}c}
                                    \picin{#1-3}{#2-3}{#5}
                                    \vspace{0.75cm}
                                    \addtocounter{myfigure}{-1}%
                                    \renewcommand{\thefigure}{\arabic{myfigure}d}
                                    \picin{#1-4}{#2-4}{#6}
                                    \newpage
                                    \addtocounter{myfigure}{-1}%
                                    \renewcommand{\thefigure}{\arabic{myfigure}e}
                                    \picin{#1-5}{#2-5}{#7}
                                    \vspace{0.75cm}
                                    \addtocounter{myfigure}{-1}%
                                    \renewcommand{\thefigure}{\arabic{myfigure}f}
                                    \picin{#1-6}{#2-6}{#8}
                                    \newpage
                                    \addtocounter{myfigure}{-1}%
                                    \renewcommand{\thefigure}{\arabic{myfigure}g}
                                    \picin{#1-7}{#2-7}{#9}}
\newcommand{\tabt}             [3] {\begin{flushleft}
                                    \footnotesize
                                    {\bf #1} \\
                                    \vspace{0.2cm}
                                    \renewcommand{\arraystretch}{#2}
                                    \begin{tabular}{#3}
                                    \hline}
\newcommand{\tabc}             [2] {\hline
                                    \end{tabular}
                                    \vspace{0.1cm}
                                    \renewcommand{\arraystretch}{#1}
                                    \begin{tabular}{#2}
                                    \hline}
\newcommand{\tabb}                 {\hline
                                    \end{tabular}
                                    \vspace{0.2cm}
                                    \end{flushleft}}
\newcommand{\deft}                 {\tabt{Script}       {1.15} {|p{15cm}|} }
\newcommand{\defst}                {\tabt{Definitions}  {1.15} {|p{15cm}|} }
\newcommand{\defc}                 {\tabc               {1.15} {|p{15cm}|} }
\newcommand{\modificationdeft}     {\tabt{Modification} {1.15} {|p{15cm}|} }
\newcommand{\Xdeft}            [1] {\tabt{#1}           {1.15} {|p{15cm}|} }
\newcommand{\colordefst}           {\tabt{Definitions} {0.75} {|p{2.5cm} p{2.5cm} p{2.81cm} p{2.5cm} p{3cm}|} }

\newcommand{\colordeftitle}        {\makebox[2   cm][c]{\tt Name}                  &
                                    \makebox[2   cm][c]{\tt In text}               &
                                    \makebox[3.81cm][c]{\tt In background}         &
                                    \makebox[2.5 cm][c]{\tt HEX value}             &
                                    \makebox[3   cm][c]{\tt rgb value \{r, g, b\}} \\}
\newcommand{\colordef}         [3] {#1                 &
                                    \textcolor{#1}{#1} &
                                    \makebox[3.81cm][c]{\colorbox{#1}{\textcolor{#1}{booooooooooooy}}} & 
                                    \makebox[2.5 cm][c]{\tt   #3}   &
                                    \makebox[3   cm][c]{\tt \{#2\}} \\
                                    & & & & \\ }
\newcommand{\redtt}            [1] {\red{\tt #1}}

\newcommand{\bluett}           [1] {\blue{\tt #1}}
\renewcommand{\_}                  {\ttsym{95}}
\newcommand{\mynote}           [2] [1.5cm]{\hspace{#1} {\normalsize \fbox{~#2~}}}
\def \ignore#1 {}
\newcommand{\addemptypage}         {\ifthenelse{\isodd{\thepage}}
                                               {\newpage
                                                ~~
                                               }
                                               {}}
\usepackage{hyperref}
\hypersetup{pdfstartview = FitH,
            pdfborder = {0 0 0.5},
            colorlinks = false,
            citebordercolor = {0  0  1},
            linkbordercolor = {1  0  0},
            urlbordercolor  = {0  1  1},
}
\begin{document}
\begin{titlepage}
\vspace*{3cm}
\begin{flushleft}
{\huge\bf
 Script Handbook for}
\end{flushleft}
\begin{flushright}
{\huge\bf
 Interactive Scientific Website Building}
\end{flushright}
\vspace{2cm}
\begin{center}
{\Large
 Version: 1.73205            \\ \vspace{0.5cm}
 Released: March 25, 2014}  \\
\vspace{14cm}
{\LARGE\bf Chung-Lin Shan}
\\
\vspace{0.5cm}
\href{http://creativecommons.org/licenses/by-sa/3.0/}
     {\includegraphics[width=1.5cm]{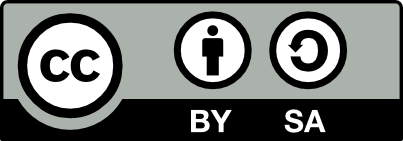}}
\vspace*{-3cm}
\end{center}
\end{titlepage}
\newpage
\thispagestyle{empty}
 ~~
\newpage
\renewcommand{\thepage}{\roman{page}}
\setcounter{page}{1}
\tableofcontents
\addtocontents{toc}{}
\newpage
\renewcommand{\thepage}{\arabic{page}}
\setcounter{page}{0}
\thispagestyle{empty}
\begin{center}
{\LARGE\bf Abstract}
\vspace{0.75cm}
\end{center}
\vspace{-0.5cm}
 In this script handbook,
 we collect the basic (and partially upgraded) PHP scripts
 used for building the {\tt AMIDAS} website
 (http://pisrv0.pit.physik.uni-tuebingen.de/darkmatter/amidas/),
 an online interactive simulation/data analysis system
 for direct Dark Matter detection experiments and phenomenology.
 In this (1.73205) version,
 we add more materials and improve the scripts
 for offering a more convenient, comfortable and user--friendly environment
 on interactive scientific computing websites.
 Some basic, often used commands of
 (X)HTML, CSS, JavaScript, HTML DOM, and PHP
 are also given in Appendix.
 Online demonstrations and downloadable template scripts
 are given on {\tt \url{http://www.tir.tw/iswb/ss2012/templates.php}}.
\newpage
\thispagestyle{empty}
 ~~
\newpage
\renewcommand{\thepage}{\arabic{page}}
\setcounter{page}{1}
\chapter{Basic Structures}
\label{chap:basic_structures}
%
%
%
\newpage
\section{Preparation}
\label{sec:preparation}
%
%
\begin{templateenumerate}
\myitem
\label{item:head_headings}
 {\bf {\cmbfsfx head} and headings}
\deft
\verb+<!DOCTYPE html+                                                   \\
\verb+PUBLIC "-//W3C//DTD XHTML 1.0 +%
  \bluett{Transitional}%
  \verb+//EN"+                                                          \\
\verb+"http://www.w3.org/TR/xhtml1/DTD/xhtml1-+%
  \bluett{transitional}%
  \verb+.dtd">+                                                         \\
\verb++                                                                 \\
\verb+<html xmlns="http://www.w3.org/1999/xhtml">+                      \\
\verb++                                                                 \\
\verb+<head>+                                                           \\
\verb++                                                                 \\
\verb+<title> +%
  \redtt{Demo for Interactive Scientific Websites}%
  \verb+ </title>+                                                      \\
\verb++                                                                 \\
\verb+<meta id="author" content="+%
  \bluett{Chung-Lin Shan}%
  \verb+" />+                                                           \\
\verb+<meta http-equiv="Content-Type" content="text/html; charset=+%
  \bluett{utf-8}%
  \verb+" />+                                                           \\
\verb++                                                                 \\
\verb+<link rel="stylesheet" type="text/css" href="+%
  \bluett{main/main}%
  \verb+.css" />+                                                       \\
\verb++                                                                 \\
\verb+</head>+                                                          \\
\verb++                                                                 \\
\verb+<body +%
  \bluett{class="en"}%
  \verb+>+                                                              \\
\verb++                                                                 \\
\verb+<!-- ******************************************************* -->+ \\
\verb+<!-- Title / authors / upgraded date *********************** -->+ \\
\verb++                                                                 \\
\verb+<h2 id="+%
  \bluett{main}%
  \verb+"> +%
  \redtt{Demo for Interactive Scientific Websites}%
  \verb+ </h2>+                                                         \\
\verb++                                                                 \\
\verb+<h3 class="center"> +%
  \bluett{Chung-Lin Shan}%
  \verb+ </h3>+                                                         \\
\verb++                                                                 \\
\verb+<h6>+                                                             \\
\verb+ Version: +%
  \redtt{3.1623}%
  \verb+ <br />+                                                        \\
\verb+ Last upgraded: +%
  \redtt{January 6, 2013}%
  \verb++                                                               \\
\verb+</h6>+                                                            \\
\verb++                                                                 \\
\verb+<hr />+                                                           \\
\verb++                                                                 \\
\verb+      .+                                                          \\
\verb+      .+ \mynote{Template \ref{item:starting_or_modifying}}       \\
\verb+      .+                                                          \\
\verb++                                                                 \\
\verb+      .+                                                          \\
\verb+      .+ \mynote{Template \ref{item:form}}                        \\
\verb+      .+                                                          \\
\verb++                                                                 \\
\verb+</body>+                                                          \\
\verb++                                                                 \\
\verb+</html>+                                                          \\
\tabb
%
%
%
\modificationdeft
\verb+<!DOCTYPE html+                                                   \\
\verb+PUBLIC "-//W3C//DTD XHTML 1.0 +%
  \bluett{Strict}%
  \verb+//EN"+                                                          \\
\verb+"http://www.w3.org/TR/xhtml1/DTD/xhtml1-+%
  \bluett{strict}%
  \verb+.dtd">+                                                         \\
\tabb
\modificationdeft
\verb+<meta http-equiv="Content-Type" content="text/html; charset=+%
  \bluett{big5}%
  \verb+" />+                                                           \\
\tabb
\newpage
\picin{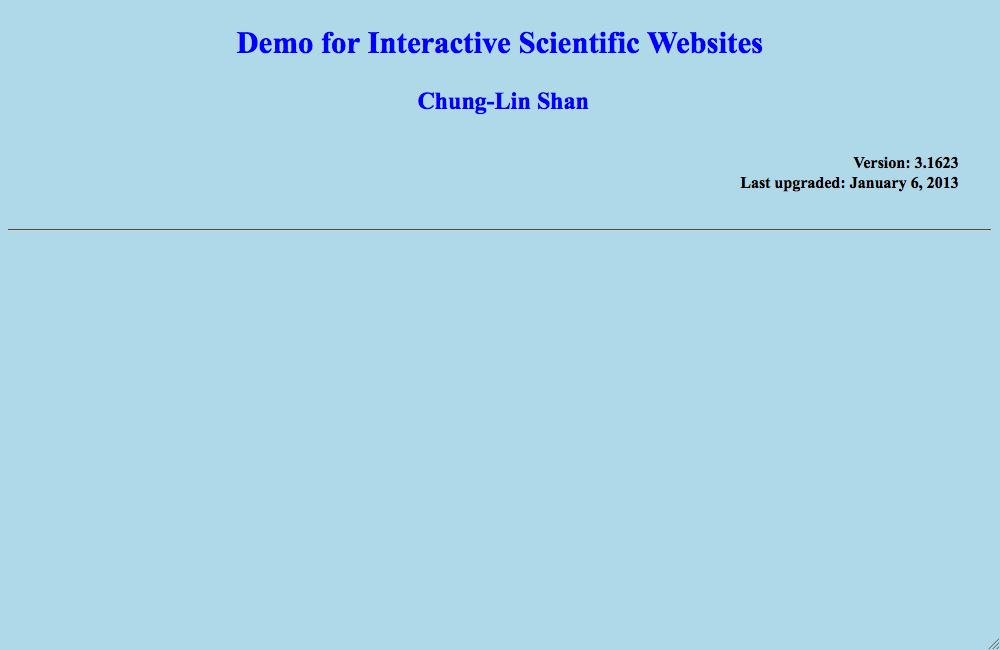}
{head}
{{\sf head} and the headings.}
\end{templateenumerate}
\newpage
\section{\sf form}
\label{sec:form}
%
%
\begin{templateenumerate}
\myitem
\label{item:form}
 {\cmbfsfx form}
\deft
\verb+<form action="+%
  \bluett{main}%
  \verb+.php" method="post" enctype="multipart/form-data">+             \\
\verb++                                                                 \\
\verb+<!-- ******************************************************* -->+ \\
\verb+<!-- +%
  \redtt{Choose one of the main functions}%
  \verb+ ********************** -->+                                    \\
\verb++                                                                 \\
\verb+<h3> +%
  \redtt{List of the main functions}%
  \verb+ </h3>+                                                         \\
\verb++                                                                 \\
\verb+<h4> +%
  \redtt{Choose one of the main functions}%
  \verb+ </h4>+                                                         \\
\verb++                                                                 \\
\verb+<?php+                                                            \\
\verb++                                                                 \\
\verb+      .+                                                          \\
\verb+      .+ \mynote{Template \ref{item:main_function_choosing}}      \\
\verb+      .+                                                          \\
\verb++                                                                 \\
\verb+?>+                                                               \\
\verb++                                                                 \\
\verb+<h5> <a href="#+%
  \bluett{main}%
  \verb+">Top</a> </h5>+ \mynote{{\tt <h2 id="main">}, see Template \ref{item:head_headings}} \\
\verb++                                                                 \\
\verb+<hr />+                                                           \\
\verb++                                                                 \\
\verb+<!-- ******************************************************* -->+ \\
\verb+<!-- ******************************************************* -->+ \\
\verb++                                                                 \\
\verb+<?php+                                                            \\
\verb++                                                                 \\
\verb+      .+                                                          \\
\verb+      .+ \mynote{Template \ref{item:switch_further_step}}         \\
\verb+      .+                                                          \\
\verb++                                                                 \\
\verb+?>+                                                               \\
\verb++                                                                 \\
\verb+<!-- ******************************************************* -->+ \\
\verb+<!-- ******************************************************* -->+ \\
\verb++                                                                 \\
\verb+<input type="submit" class="+%
  \bluett{submit}%
  \verb+" value="+%
  \bluett{Submit}%
  \verb+" />+                                                           \\
\verb+<br />+                                                           \\
\verb++                                                                 \\
\verb+</form>+                                                          \\
\tabb
\newpage
\picin{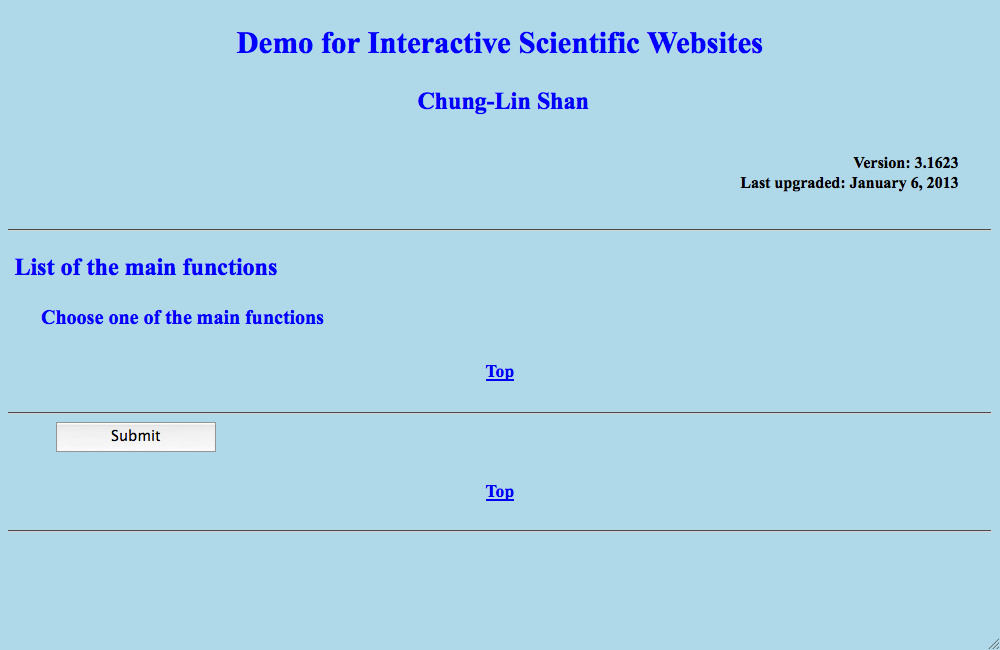}
{form}
{{\sf form}.}
\newpage
\myitem
\label{item:main_function_choosing}
 {\bf Choose one of the main functions}
\deft
\verb+<?php+                                                      \\
\verb+  include("+%
  \bluett{default\_choices/default\_choosing}%
  \verb+.php");+ \mynote{Template \ref{item:default_choosing.php}} \\
\verb++                                                          \\
\verb+  if ($+%
  \bluett{main\_function\_choosing}%
  \verb+ == NULL)+                                                \\
\verb+  {+                                                        \\
\verb+    $+%
  \bluett{main\_function\_choosing}%
  \verb+ = $_POST["+%
  \bluett{main\_function\_choosing}%
  \verb+"];+                                                      \\
\verb+  }+                                                        \\
\verb++                                                           \\
\verb+  include("+%
  \bluett{forms/form-main\_functions}%
  \verb+.php");+ \mynote{Template \ref{item:form-main_functions.php}} \\
\verb++                                                           \\
\verb+  if ($+%
  \bluett{main\_function\_choosing}%
  \verb+ == NULL)+                                                \\
\verb+  {+                                                        \\
\verb+    include("+%
  \bluett{default\_choices/default-main\_function\_choosing}%
  \verb+.php");+ \mynote{Template \ref{item:default-main_function_choosing.php}} \\
\verb+  }+                                                        \\
\verb++                                                           \\
\verb+  else+                                                     \\
\verb+  {+                                                        \\
\verb+    echo '<br />';+                                         \\
\verb+    echo '+%
  \bluett{main\_function\_choosing}%
  \verb+ = ' . $+%
  \bluett{main\_function\_choosing}%
  \verb+ . '<br />';+                                             \\
\verb+    echo '+%
  \bluett{main\_function}%
  \verb+          = ' . $+%
  \bluett{main\_function}%
  \verb+          . '<br />';+                                    \\
\verb+  }+                                                        \\
\verb+?>+                                                         \\
\tabb
\newpage
\myitem
\label{item:form-main_functions.php}
 {\cmbfsfx form-main\_functions.php}
\deft
\verb+<fieldset id="+%
  \bluett{main\_function\_choosing}%
  \verb+" style="width: +%
  \redtt{210}%
  \verb+pt">+                                   \\
\verb++                                         \\
\verb+  <label>+                                \\
\verb+  <input type="radio" name="+%
  \bluett{main\_function\_choosing}%
  \verb+" value="+%
  \redtt{1}%
  \verb+"+                                      \\
\verb+         <?php+                           \\
\verb+           if ($+%
  \bluett{main\_function\_choosing}%
  \verb+ == "+%
  \redtt{1}%
  \verb+")+                                     \\
\verb+           {+                             \\
\verb+             echo 'checked = "checked"';+ \\
\verb++                                         \\
\verb+             $+%
  \bluett{main\_function}%
  \verb+ = "+%
  \redtt{main\_function\_1}%
  \verb+";+                                     \\
\verb+           }+                             \\
\verb+         ?>+                              \\
\verb+         />+                              \\
\verb++                                         \\
\verb+         +%
  \redtt{Description for the main function 1}%
  \verb++                                       \\
\verb+         </label>+                        \\
\verb+         <br />+                          \\
\verb++                                         \\
\verb+            .+                            \\
\verb+            .+                            \\
\verb+            .+                            \\
\verb++                                         \\
\verb+</fieldset>+                              \\
\tabb
\newpage
\myitem
\label{item:default_choosing.php}
 {\cmbfsfx default\_choosing.php}
 (cf.~Template \ref{item:default-main_function_choosing.php})
\deft
\verb+<script type="text/javascript">+                             \\
\verb+  function +%
  \bluett{default\_choosing}%
  \verb+(+%
  \bluett{form\_id}%
  \verb+, +%
  \bluett{n\_IE}%
  \verb+, +%
  \bluett{n\_FF}%
  \verb+)+                                                         \\
\verb+  {+                                                         \\
\verb+    if (navigator.appName == "Microsoft Internet Explorer")+ \\
\verb+    {+                                                       \\
\verb+        document.getElementById(+%
  \bluett{form\_id}%
  \verb+)+                                                         \\
\verb+                .childNodes[+%
  \bluett{n\_IE}%
  \verb+].childNodes[+%
  \bluett{0}%
  \verb+]+                                                         \\
\verb+                .checked+                                    \\
\verb+      = true;+                                               \\
\verb+    }+                                                       \\
\verb++                                                            \\
\verb+    else+                                                    \\
\verb+//  if (navigator.userAgent.search("Chrome")    != -1 ||+    \\
\verb+//      navigator.userAgent.search("Firefox")   != -1 ||+    \\
\verb+//      navigator.userAgent.search("Opera")     != -1 ||+    \\
\verb+//      navigator.userAgent.search("Safari")    != -1 ||+    \\
\verb+//      navigator.userAgent.search("Navigator") != -1   )+   \\
\verb+    {+                                                       \\
\verb+         document.getElementById(+%
  \bluett{form\_id}%
  \verb+)+                                                         \\
\verb+                .childNodes[+%
  \bluett{n\_FF}%
  \verb+].childNodes[+%
  \bluett{1}%
  \verb+]+ \mynote[0.5cm]{{\tt n\_FF} is usually equal to {\tt 2 * n\_IE + 1}} \\
\verb+                .checked+                                    \\
\verb+      = true;+                                               \\
\verb+    }+                                                       \\
\verb++                                                            \\
\verb+      document.getElementById(+%
  \bluett{form\_id}%
  \verb+)+                                                         \\
\verb+              .style.borderColor+                            \\
\verb+    = "+%
  \bluett{red}%
  \verb+";+                                                        \\
\verb+  }+                                                         \\
\verb+</script>+                                                   \\
\tabb
\newpage
\myitem
\label{item:default-main_function_choosing.php}
 {\cmbfsfx default-main\_function\_choosing.php}
 (cf.~Template \ref{item:default_choosing.php})
\deft
\verb+<script type="text/javascript">+                           \\
\verb+<?php+                                                     \\
\verb+/*+                                                        \\
\verb+  if (navigator.appName == "Microsoft Internet Explorer")+ \\
\verb+  {+                                                       \\
\verb+      document.getElementById("+%
  \bluett{main\_function\_choosing}%
  \verb+")+                                                      \\
\verb+              .childNodes[+%
  \redtt{0}%
  \verb+].childNodes[+%
  \bluett{0}%
  \verb+]+                                                       \\
\verb+              .checked+                                    \\
\verb+    = true;+                                               \\
\verb+  }+                                                       \\
\verb++                                                          \\
\verb+  else+                                                    \\
\verb+//if (navigator.userAgent.search("Chrome")    != -1 ||+    \\
\verb+//    navigator.userAgent.search("Firefox")   != -1 ||+    \\
\verb+//    navigator.userAgent.search("Opera")     != -1 ||+    \\
\verb+//    navigator.userAgent.search("Safari")    != -1 ||+    \\
\verb+//    navigator.userAgent.search("Navigator") != -1   )+   \\
\verb+  {+                                                       \\
\verb+      document.getElementById("+%
  \bluett{main\_function\_choosing}%
  \verb+")+                                                      \\
\verb+              .childNodes[+%
  \redtt{1}%
  \verb+].childNodes[+%
  \bluett{1}%
  \verb+]+                                                       \\
\verb+              .checked+                                    \\
\verb+    = true;+                                               \\
\verb+  }+                                                       \\
\verb++                                                          \\
\verb+    document.getElementById("+%
  \bluett{main\_function\_choosing}%
  \verb+")+                                                      \\
\verb+            .style.borderColor+                            \\
\verb+  = "+%
  \bluett{red}%
  \verb+";+                                                      \\
\verb+*/+                                                        \\
\verb+?>+                                                        \\
\verb+  +%
  \bluett{default\_choosing}%
  \verb+("+%
  \bluett{main\_function\_choosing}%
  \verb+", +%
  \redtt{0}%
  \verb+, +%
  \redtt{1}%
  \verb+);+                                                      \\
\verb+</script>+                                                 \\
\tabb
\newpage
\picdin{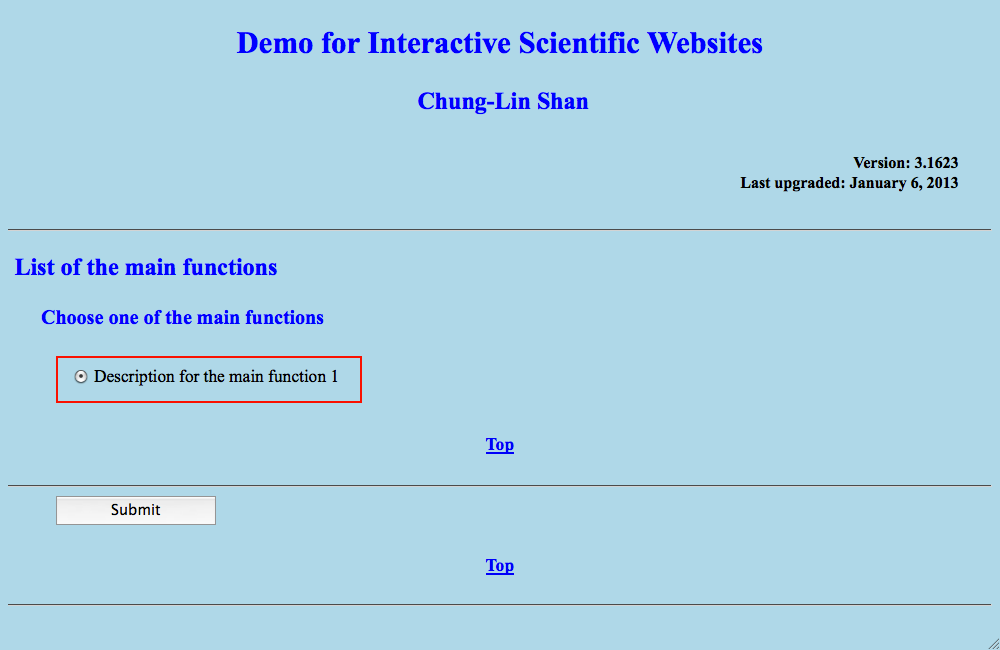}
{main\_function\_choosing}
{Choose one of the main function (before submission).}
{Choose one of the main function (after submission).}
\end{templateenumerate}
\newpage
\section{{\sf switch} for the further step}
\label{sec:switch_further_step}
%
%
\begin{templateenumerate}
\myitem
\label{item:switch_further_step}
 {\bf {\cmbfsfx switch} for the further step}
\deft
\verb+<?php+                                                                             \\
\verb+  if ($+%
  \bluett{main\_function}%
  \verb+ != NULL)+                                                                       \\
\verb+  {+                                                                               \\
\verb+    echo '<script type="text/javascript">'                              . "\r\n";+ \\
\verb+    echo '  if (navigator.appName == "Microsoft Internet Explorer")'    . "\r\n";+ \\
\verb+    echo '  {'                                                          . "\r\n";+ \\
\verb+    echo '    window.location = "#+%
  \bluett{mode-main\_function}%
  \verb+";'                 . "\r\n";+                                                   \\
\verb+    echo '  }'                                                          . "\r\n";+ \\
\verb+    echo '</script>'                                                    . "\r\n";+ \\
\verb++                                                                                  \\
\verb+    include("+%
  \bluett{main/title-main\_functions}%
  \verb+.php");+ \mynote{Template \ref{item:title-main_functions.php}}                   \\
\verb++                                                                                  \\
\verb+    echo '<script type="text/javascript">'                              . "\r\n";+ \\
\verb+    echo '  if (navigator.userAgent.search("Chrome")     != -1     ||'  . "\r\n";+ \\
\verb+    echo '      (navigator.userAgent.search("Firefox")   != -1 &&'      . "\r\n";+ \\
\verb+    echo '       navigator.userAgent.search("Navigator") == -1   ) ||'  . "\r\n";+ \\
\verb+    echo '      navigator.userAgent.search("Opera")      != -1     ||'  . "\r\n";+ \\
\verb+    echo '      (navigator.userAgent.search("Chrome")    == -1 &&'      . "\r\n";+ \\
\verb+    echo '       navigator.userAgent.search("Safari")    != -1   )   )' . "\r\n";+ \\
\verb+    echo '  {'                                                          . "\r\n";+ \\
\verb+    echo '    window.location = "#+%
  \bluett{mode-main\_function}%
  \verb+";'                 . "\r\n";+                                                   \\
\verb+    echo '  }'                                                          . "\r\n";+ \\
\verb+    echo '</script>'                                                    . "\r\n";+ \\
\verb++                                                                                  \\
\verb+    switch ($+%
  \bluett{main\_function}%
  \verb+)+                                                                               \\
\verb+    {+                                                                             \\
\verb+        .+                                                                         \\
\verb+        .+ \mynote{Template \ref{item:mode_choosing}}                              \\
\verb+        .+                                                                         \\
\verb+    }+                                                                             \\
\verb++                                                                                  \\
\verb+    echo '<script type="text/javascript">'                              . "\r\n";+ \\
\verb+    echo '  if (navigator.userAgent.search("Navigator") != -1)'         . "\r\n";+ \\
\verb+    echo '  {'                                                          . "\r\n";+ \\
\verb+    echo '    window.location = "#+%
  \bluett{mode-main\_function}%
  \verb+";'                 . "\r\n";+                                                   \\
\verb+    echo '  }'                                                          . "\r\n";+ \\
\verb+    echo '</script>'                                                    . "\r\n";+ \\
\verb+  }+                                                                               \\
\verb+?>+                                                                                \\
\verb+      .+                                                                           \\
\verb+      .+                                                                           \\
\verb+      .+                                                                           \\
\verb++                                                                                  \\
\verb+<?php+                                                                             \\
\verb+  if ($+%
  \bluett{main\_function}%
  \verb+ != NULL)+                                                                       \\
\verb+  {+                                                                               \\
\verb+    echo '<h5> <a href="#+%
  \bluett{main}%
  \verb+">Top</a> </h5>' . "\r\n";+ \mynote{cf.~Templates \ref{item:form}
                                                      and \ref{item:starting_or_modifying}} \\
\verb+    echo '<hr />'                             . "\r\n";+                           \\
\verb+  }+                                                                               \\
\verb+?>+                                                                                \\
\defc
\verb++                                                                                  \\
\verb+<!-- ******************************************************* -->+                  \\
\verb+<!-- ******************************************************* -->+                  \\
\verb++                                                                                  \\
\verb+<?php+                                                                             \\
\verb++                                                                                  \\
\verb+      .+                                                                           \\
\verb+      .+ \mynote{Template \ref{item:keeping_name}}                                 \\
\verb+      .+                                                                           \\
\verb++                                                                                  \\
\verb+?>+                                                                                \\
\verb++                                                                                  \\
\verb+<?php+                                                                             \\
\verb++                                                                                  \\
\verb+      .+                                                                           \\
\verb+      .+ \mynote{Template \ref{item:modification_switch_further_step}}             \\
\verb+      .+                                                                           \\
\verb++                                                                                  \\
\verb+?>+                                                                                \\
\verb++                                                                                  \\
\verb+<!-- ******************************************************* -->+                  \\
\verb+<!-- ******************************************************* -->+                  \\
\verb++                                                                                  \\
\verb+<?php+                                                                             \\
\verb++                                                                                  \\
\verb+      .+                                                                           \\
\verb+      .+ \mynote{Template \ref{item:changing_name}}                                \\
\verb+      .+                                                                           \\
\verb+?>+                                                                                \\
\verb++                                                                                  \\
\tabb
\newpage
\myitem
\label{item:title-main_functions.php}
 {\cmbfsfx title-main\_functions.php}
\deft
\verb+<?php+                                                                                      \\
\verb+  switch ($+%
  \bluett{main\_function}%
  \verb+)+                                                                                        \\
\verb+  {+                                                                                        \\
\verb+    case "+%
  \redtt{main\_function\_1}%
  \verb+":+                                                                                       \\
\verb++                                                                                           \\
\verb+      echo '<h3 id="+%
  \bluett{mode-main\_function}%
  \verb+">'                                        . "\r\n";+                                     \\
\verb+      echo '  +%
  \redtt{Description for the main function 1}%
  \verb+'                               . "\r\n";+                                                \\
\verb+      echo '</h3>'                                                               . "\r\n";+ \\
\verb++                                                                                           \\
\verb+      echo '<h4> +%
  \redtt{Choose one of the following mode for the main function 1}%
  \verb+ </h4>' . "\r\n";+                                                                        \\
\verb++                                                                                           \\
\verb+      break;+                                                                               \\
\verb++                                                                                           \\
\verb+        .+                                                                                  \\
\verb+        .+                                                                                  \\
\verb+        .+                                                                                  \\
\verb++                                                                                           \\
\verb+  }+                                                                                        \\
\verb+?>+                                                                                         \\
\tabb
\newpage
\picin{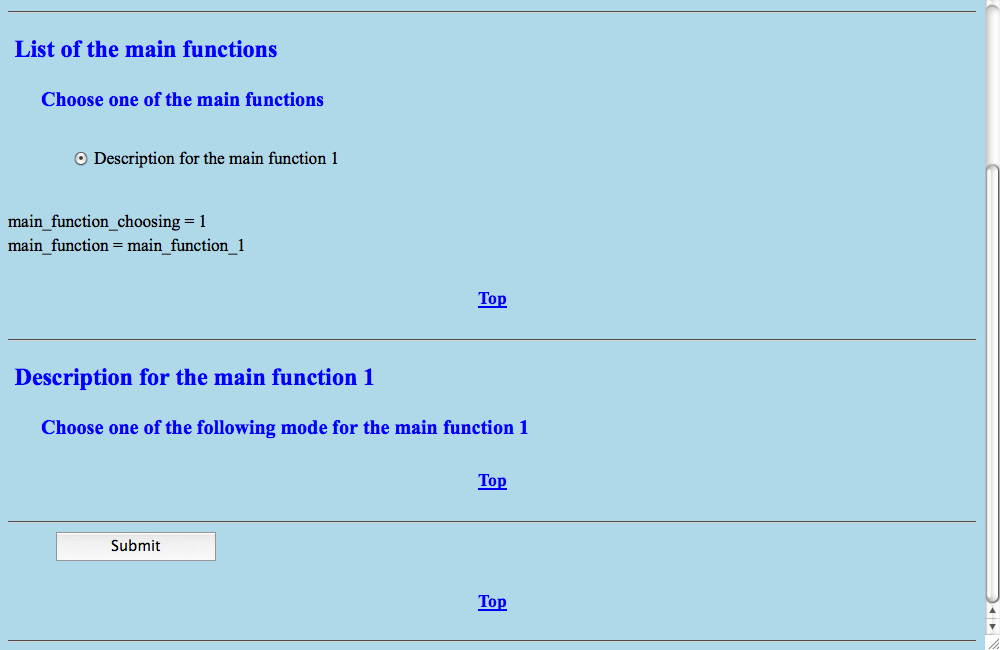}
{switch\_further\_step}
{{\sf switch} for the further step
 (cf.~Figure \ref{fig:form}).}
\newpage
\myitem
\label{item:mode_choosing}
 {\bf Choose the mode for the chosen function}
 (cf.~Template \ref{item:main_function_choosing})
\deft
\verb+<?php+                                                            \\
\verb+  switch ($+%
  \bluett{main\_function}%
  \verb+)+                                                              \\
\verb+  {+                                                              \\
\verb+    /**********************************************************/+ \\
\verb+    /* +%
  \redtt{Mode for the main function 1}%
  \verb+ ***************************/+                                  \\
\verb++                                                                 \\
\verb+    case "+%
  \redtt{main\_function\_1}%
  \verb+":+                                                             \\
\verb++                                                                 \\
\verb+      if ($+%
  \redtt{mode\_main\_function\_1\_choosing}%
  \verb+ == NULL)+                                                      \\
\verb+      {+                                                          \\
\verb+        $+%
  \redtt{mode\_main\_function\_1\_choosing}%
  \verb+ = $_POST["+%
  \redtt{mode\_main\_function\_1\_choosing}%
  \verb+"];+                                                            \\
\verb+      }+                                                          \\
\verb++                                                                 \\
\verb+      include("+%
  \bluett{forms/form-}%
  \verb++%
  \redtt{modes\_main\_function\_1}%
  \verb+.php");+ \mynote{Modification of Template \ref{item:form-main_functions.php}} \\
\verb++                                                                 \\
\verb+      if ($+%
  \redtt{mode\_main\_function\_1\_choosing}%
  \verb+ == NULL)+                                                      \\
\verb+      {+                                                          \\
\verb+        include("+%
  \bluett{default\_choices/default-}%
  \verb++%
  \redtt{mode\_main\_function\_1\_choosing}%
  \verb+.php");+                                                            \\
\verb+      }+                                                          \\
\verb++ \mynote[4cm]{Modification of Template \ref{item:default-main_function_choosing.php}} \\
\verb+      else+                                                       \\
\verb+      {+                                                          \\
\verb+        echo '<br />';+                                           \\
\verb+        echo '+%
  \redtt{mode\_main\_function\_1\_choosing}%
  \verb+ = ' . $+%
  \redtt{mode\_main\_function\_1\_choosing}%
  \verb+ . '<br />';+                                                   \\
\verb+        echo '+%
  \redtt{mode\_main\_function\_1\_choice}%
  \verb+   = ' . $+%
  \redtt{mode\_main\_function\_1\_choice}%
  \verb+   . '<br />';+                                                 \\
\verb+        echo '<br />';+                                           \\
\verb++                                                                 \\
\verb+        $+%
  \redtt{mode\_checked}%
  \verb+ = "yes";+                                                      \\
\verb+      }+                                                          \\
\verb++                                                                 \\
\verb+      break;+                                                     \\
\verb++                                                                 \\
\verb+        .+                                                        \\
\verb+        .+                                                        \\
\verb+        .+                                                        \\
\verb++                                                                 \\
\verb+  }+                                                              \\
\verb+?>+                                                               \\
\tabb
\newpage
\picdin{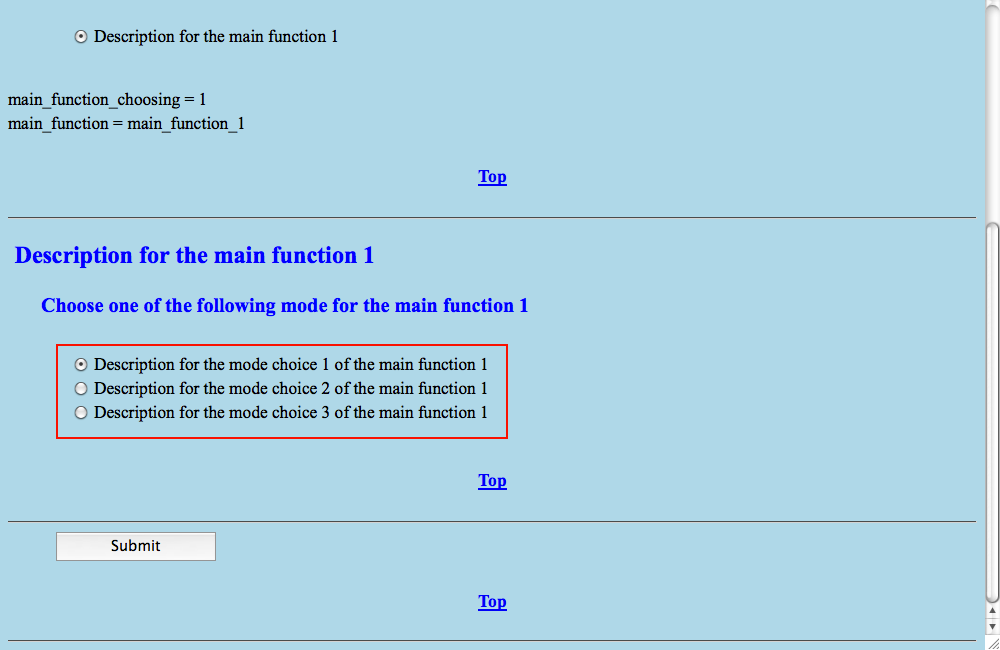}
{mode\_choosing}
{Choose the mode for the chosen function
 (before submission,
  cf.~Figure \ref{fig:main_function_choosing-1}).}
{Choose the mode for the chosen function
 (after submission,
  cf.~Figure \ref{fig:main_function_choosing-2}).}
\end{templateenumerate}
\newpage
\section{\sf input}
\label{sec:input}
%
%
\begin{templateenumerate}
\myitem
\label{item:modification_switch_further_step}
 {\bf Modification of `{\cmbfsfx switch} for the further step'}
 (cf.~Templates \ref{item:switch_further_step}
  and \ref{item:title-main_functions.php})
\deft
\verb+<?php+                                                                             \\
\verb++                                                                                  \\
\verb+      .+                                                                           \\
\verb+      .+ \mynote{Template \ref{item:keeping_name}}                                 \\
\verb+      .+                                                                           \\
\verb++                                                                                  \\
\verb+?>+                                                                                \\
\verb++                                                                                  \\
\verb+<?php+                                                                             \\
\verb+  if ($+%
  \bluett{main\_function}%
  \verb+ != NULL  &&+                                                                    \\
\verb+      $+%
  \redtt{mode\_checked}%
  \verb+  == "yes"   )+                                                                  \\
\verb+  {+                                                                               \\
\verb+    echo '<script type="text/javascript">'                              . "\r\n";+ \\
\verb+    echo '  if (navigator.appName == "Microsoft Internet Explorer")'    . "\r\n";+ \\
\verb+    echo '  {'                                                          . "\r\n";+ \\
\verb+    echo '    window.location = "#+%
  \bluett{setup-main\_function}%
  \verb+";'                . "\r\n";+                                                    \\
\verb+    echo '  }'                                                          . "\r\n";+ \\
\verb+    echo '</script>'                                                    . "\r\n";+ \\
\verb++                                                                                  \\
\verb+    echo '<h3 id="+%
  \bluett{setup-main\_function}%
  \verb+">'                                . "\r\n";+                                    \\
\verb+    echo '  +%
  \redtt{Setup for the main function}%
  \verb+'                                . "\r\n";+                                      \\
\verb+    echo '</h3>'                                                        . "\r\n";+ \\
\verb++                                                                                  \\
\verb+    echo '<script type="text/javascript">'                              . "\r\n";+ \\
\verb+    echo '  if (navigator.userAgent.search("Chrome")     != -1     ||'  . "\r\n";+ \\
\verb+    echo '      (navigator.userAgent.search("Firefox")   != -1 &&'      . "\r\n";+ \\
\verb+    echo '       navigator.userAgent.search("Navigator") == -1   ) ||'  . "\r\n";+ \\
\verb+    echo '      navigator.userAgent.search("Opera")      != -1     ||'  . "\r\n";+ \\
\verb+    echo '      (navigator.userAgent.search("Chrome")    == -1 &&'      . "\r\n";+ \\
\verb+    echo '       navigator.userAgent.search("Safari")    != -1   )   )' . "\r\n";+ \\
\verb+    echo '  {'                                                          . "\r\n";+ \\
\verb+    echo '    window.location = "#+%
  \bluett{setup-main\_function}%
  \verb+";'                . "\r\n";+                                                    \\
\verb+    echo '  }'                                                          . "\r\n";+ \\
\verb+    echo '</script>'                                                    . "\r\n";+ \\
\verb++                                                                                  \\
\verb+        .+                                                                         \\
\verb+        .+ \mynote{Template \ref{item:setup_setting}}                              \\
\verb+        .+                                                                         \\
\verb++                                                                                  \\
\verb+        .+                                                                         \\
\verb+        .+ \mynote{Templates \ref{item:table-setup_item.php}
                                 + \ref{item:default-setup_item.php}~ or
                         Template  \ref{item:table-setup_item_entry}}                    \\
\verb+        .+                                                                         \\
\verb++                                                                                  \\
\verb+    echo '<script type="text/javascript">'                              . "\r\n";+ \\
\verb+    echo '  if (navigator.userAgent.search("Navigator") != -1)'         . "\r\n";+ \\
\verb+    echo '  {'                                                          . "\r\n";+ \\
\verb+    echo '    window.location = "#+%
  \bluett{setup-main\_function}%
  \verb+";'                . "\r\n";+                                                    \\
\verb+    echo '  }'                                                          . "\r\n";+ \\
\verb+    echo '</script>'                                                    . "\r\n";+ \\
\verb+  }+                                                                               \\
\verb+?>+                                                                                \\
\defc
\verb++                                                                                  \\
\verb+<?php+                                                                             \\
\verb+  if ($+%
  \bluett{main\_function}%
  \verb+ != NULL  &&+                                                                    \\
\verb+      $+%
  \redtt{mode\_checked}%
  \verb+  == "yes"   )+                                                                  \\
\verb+  {+                                                                               \\
\verb+    echo '<h5> <a href="#+%
  \bluett{main}%
  \verb+">Top</a> </h5>' . "\r\n";+                                                      \\
\verb+    echo '<hr />'                             . "\r\n";+                           \\
\verb+  }+                                                                               \\
\verb+?>+                                                                                \\
\verb++                                                                                  \\
\verb+<!-- ******************************************************* -->+                  \\
\verb+<!-- ******************************************************* -->+                  \\
\verb++                                                                                  \\
\verb+<?php+                                                                             \\
\verb++                                                                                  \\
\verb+      .+                                                                           \\
\verb+      .+ \mynote{Template \ref{item:changing_name}}                                \\
\verb+      .+                                                                           \\
\verb+?>+                                                                                \\
\verb++                                                                                  \\
\tabb
\newpage
\myitem
\label{item:setup_setting}
 {\bf Setup for the chosen function}
\deft
\verb+<?php+                                                   \\
\verb+  if ($+%
  \bluett{main\_function}%
  \verb+ == "+%
  \redtt{main\_function\_1}%
  \verb+"  &&+                                                 \\
\verb+      $+%
  \redtt{mode\_main\_function\_1\_choice}%
  \verb+ != NULL   )+                                          \\
\verb+  {+                                                     \\
\verb+    echo '<h4> +%
  \redtt{Setup choice for the main function 1}%
  \verb+ </h4>' . "\r\n";+                                     \\
\verb++                                                        \\
\verb+    if ($+%
  \redtt{setup\_main\_function\_1\_choosing}%
  \verb+ == NULL)+                                             \\
\verb+    {+                                                   \\
\verb+      $+%
  \redtt{setup\_main\_function\_1\_choosing}%
  \verb+ = $_POST["+%
  \redtt{setup\_main\_function\_1\_choosing}%
  \verb+"];+                                                   \\
\verb+    }+                                                   \\
\verb++                                                        \\
\verb+    if ($+%
  \redtt{integer\_1}%
  \verb+ == NULL)+                                             \\
\verb+    {+                                                   \\
\verb+      $+%
  \redtt{integer\_1}%
  \verb+ = $_POST["+%
  \redtt{integer\_1}%
  \verb+"];+                                                   \\
\verb+    }+                                                   \\
\verb++                                                        \\
\verb+    if ($+%
  \redtt{integer\_2}%
  \verb+ == NULL)+                                             \\
\verb+    {+                                                   \\
\verb+      $+%
  \redtt{integer\_2}%
  \verb+ = $_POST["+%
  \redtt{integer\_2}%
  \verb+"];+                                                   \\
\verb+    }+                                                   \\
\verb++                                                        \\
\verb+    if ($+%
  \redtt{float\_1\_1}%
  \verb+ == NULL)+                                             \\
\verb+    {+                                                   \\
\verb+      $+%
  \redtt{float\_1\_1}%
  \verb+ = $_POST["+%
  \redtt{float\_1\_1}%
  \verb+"];+                                                   \\
\verb+    }+                                                   \\
\verb++                                                        \\
\verb+    if ($+%
  \redtt{float\_1\_2}%
  \verb+ == NULL)+                                             \\
\verb+    {+                                                   \\
\verb+      $+%
  \redtt{float\_1\_2}%
  \verb+ = $_POST["+%
  \redtt{float\_1\_2}%
  \verb+"];+                                                   \\
\verb+    }+                                                   \\
\verb++                                                        \\
\verb+    if ($+%
  \redtt{float\_2\_1}%
  \verb+ == NULL)+                                             \\
\verb+    {+                                                   \\
\verb+      $+%
  \redtt{float\_2\_1}%
  \verb+ = $_POST["+%
  \redtt{float\_2\_1}%
  \verb+"];+                                                   \\
\verb+    }+                                                   \\
\verb++                                                        \\
\verb+    if ($+%
  \redtt{float\_2\_2}%
  \verb+ == NULL)+                                             \\
\verb+    {+                                                   \\
\verb+      $+%
  \redtt{float\_2\_2}%
  \verb+ = $_POST["+%
  \redtt{float\_2\_2}%
  \verb+"];+                                                   \\
\verb+    }+                                                   \\
\verb++                                                        \\
\verb+    if ($+%
  \redtt{integer\_3}%
  \verb+ == NULL)+                                             \\
\verb+    {+                                                   \\
\verb+      $+%
  \redtt{integer\_3}%
  \verb+ = $_POST["+%
  \redtt{integer\_3}%
  \verb+"];+                                                   \\
\verb+    }+                                                   \\
\verb++                                                        \\
\verb+    if ($+%
  \redtt{float\_3}%
  \verb+ == NULL)+                                             \\
\verb+    {+                                                   \\
\verb+      $+%
  \redtt{float\_3}%
  \verb+ = $_POST["+%
  \redtt{float\_3}%
  \verb+"];+                                                   \\
\verb+    }+                                                   \\
\defc
\verb++                                                        \\
\verb+    include("+%
  \bluett{forms/form-}%
  \verb++%
  \redtt{setup\_main\_function\_1\_choosing}%
  \verb+.php");+ \mynote[0.5cm]{Template \ref{item:form-setup_main_function_1_choosing.php}
                            or \ref{item:modification_form-setup_main_function_1_choosing.php}} \\
\verb++                                                        \\
\verb+    if ($+%
  \redtt{setup\_main\_function\_1\_choosing}%
  \verb+ == NULL)+                                             \\
\verb+    {+                                                   \\
\verb+      include("+%
  \bluett{default\_choices/default-}%
  \verb++%
  \redtt{setup\_main\_function\_1\_choosing}%
  \verb+.php");+                                               \\
\verb+    }+ \mynote[6cm]{Modification of Template \ref{item:default-main_function_choosing.php}} \\
\verb++                                                        \\
\verb+    elseif ($+%
  \redtt{setup\_main\_function\_1\_choosing}%
  \verb+ == "+%
  \redtt{5}%
  \verb+"               &&+                                    \\
\verb+            ($+%
  \redtt{integer\_1}%
  \verb+ == NULL                              ||+              \\
\verb+             $+%
  \redtt{integer\_2}%
  \verb+ == NULL                              ||+              \\
\verb+             ($+%
  \redtt{mode\_main\_function\_1\_choosing}%
  \verb+ == "+%
  \redtt{2}%
  \verb+"      &&+                                             \\
\verb+              ($+%
  \redtt{float\_1\_1}%
  \verb+ == NULL ||+                                           \\
\verb+               $+%
  \redtt{float\_1\_2}%
  \verb+ == NULL   )                      ) ||+ \mynote[1cm]{cf.~Template \ref{item:table-setup_item_entry}} \\
\verb+             ($+%
  \redtt{mode\_main\_function\_1\_choosing}%
  \verb+ == "+%
  \redtt{3}%
  \verb+"      &&+                                             \\
\verb+              ($+%
  \redtt{float\_2\_1}%
  \verb+ == NULL ||+                                           \\
\verb+               $+%
  \redtt{float\_2\_2}%
  \verb+ == NULL   )                      ) ||+                \\
\verb+             (($+%
  \redtt{mode\_main\_function\_1\_choosing}%
  \verb+ == "+%
  \redtt{2}%
  \verb+" ||+                                                  \\
\verb+               $+%
  \redtt{mode\_main\_function\_1\_choosing}%
  \verb+ == "+%
  \redtt{3}%
  \verb+"   ) &&+                                              \\
\verb+              ($+%
  \redtt{integer\_3}%
  \verb+ == NULL ||+                                           \\
\verb+               $+%
  \redtt{float\_3}%
  \verb+   == NULL   )                      )   )   )+         \\
\verb+    {+                                                   \\
\verb+      echo '<script type="text/javascript">' . "\r\n";+  \\
\verb+      echo '  document.getElementById("';+               \\
\verb+      echo '+%
  \redtt{setup\_main\_function\_1\_choosing}%
  \verb+';+                                                    \\
\verb+      echo '").style.borderColor = "+%
  \bluett{red}%
  \verb+";'   . "\r\n";+                                       \\
\verb+      echo '</script>'                       . "\r\n"; + \\
\verb+    }+                                                   \\
\verb++                                                        \\
\verb+    else+                                                \\
\verb+    {+                                                   \\
\verb+      echo '<br />';+                                    \\
\verb+      echo '+%
  \redtt{setup\_main\_function\_1\_choosing}%
  \verb+ = ' . $+%
  \redtt{setup\_main\_function\_1\_choosing}%
  \verb+ . '<br />';+                                          \\
\verb+      echo '<br />';+                                    \\
\verb++                                                        \\
\verb+      if ($+%
  \redtt{setup\_main\_function\_1\_choosing}%
  \verb+ == "+%
  \redtt{5}%
  \verb+")+                                                    \\
\verb+      {+                                                 \\
\verb+        echo '+%
  \redtt{integer\_1}%
  \verb+ = ' . $+%
  \redtt{integer\_1}%
  \verb+ . '<br />';+                                          \\
\verb+        echo '+%
  \redtt{integer\_2}%
  \verb+ = ' . $+%
  \redtt{integer\_2}%
  \verb+ . '<br />';+                                          \\
\verb++                                                        \\
\verb+        if ($+%
  \redtt{mode\_main\_function\_1\_choosing}%
  \verb+ == "+%
  \redtt{2}%
  \verb+")+                                                    \\
\verb+        {+                                               \\
\verb+          echo '+%
  \redtt{float\_1\_1}%
  \verb+ = ' . $+%
  \redtt{float\_1\_1}%
  \verb+ . '<br />';+                                          \\
\verb+          echo '+%
  \redtt{float\_1\_2}%
  \verb+ = ' . $+%
  \redtt{float\_1\_2}%
  \verb+ . '<br />';+                                          \\
\verb+        }+                                               \\
\verb++                                                        \\
\verb+        if ($+%
  \redtt{mode\_main\_function\_1\_choosing}%
  \verb+ == "+%
  \redtt{3}%
  \verb+")+                                                    \\
\verb+        {+                                               \\
\verb+          echo '+%
  \redtt{float\_2\_1}%
  \verb+ = ' . $+%
  \redtt{float\_2\_1}%
  \verb+ . '<br />';+                                          \\
\verb+          echo '+%
  \redtt{float\_2\_2}%
  \verb+ = ' . $+%
  \redtt{float\_2\_2}%
  \verb+ . '<br />';+                                          \\
\verb+        }+                                               \\
\verb++                                                        \\
\verb+        if ($+%
  \redtt{mode\_main\_function\_1\_choosing}%
  \verb+ == "+%
  \redtt{2}%
  \verb+" ||+                                                  \\
\verb+            $+%
  \redtt{mode\_main\_function\_1\_choosing}%
  \verb+ == "+%
  \redtt{3}%
  \verb+"   )+                                                 \\
\verb+        {+                                               \\
\verb+          echo '+%
  \redtt{integer\_3}%
  \verb+ = ' . $+%
  \redtt{integer\_3}%
  \verb+ . '<br />';+                                          \\
\verb+          echo '+%
  \redtt{float\_3}%
  \verb+   = ' . $+%
  \redtt{float\_3}%
  \verb+   . '<br />';+                                        \\
\verb+        }+                                               \\
\verb+      }+                                                 \\
\verb++                                                        \\
\verb+      $+%
  \redtt{setup\_main\_function\_1\_checked}%
  \verb+ = "yes";+                                             \\
\verb+    }+                                                   \\
\verb+  }+                                                     \\
\verb+?>+                                                      \\
\tabb
\newpage
\myitem
\label{item:form-setup_main_function_1_choosing.php}
 {\cmbfsfx form-setup\_main\_function\_1\_choosing.php}
 (cf.~Templates \ref{item:default-main_function_choosing.php}
            and \ref{item:form-main_functions.php})
\deft
\verb+<script type="text/javascript">+                             \\
\verb+  function +%
  \redtt{personal\_setup\_main\_function\_1\_setting}%
  \verb+()+                                                        \\
\verb+  {+                                                         \\
\verb+<?php+                                                       \\
\verb+/*+                                                          \\
\verb+    if (navigator.appName == "Microsoft Internet Explorer")+ \\
\verb+    {+                                                       \\
\verb+        document.getElementById("+%
  \redtt{setup\_main\_function\_1\_choosing}%
  \verb+")+                                                        \\
\verb+                .childNodes[+%
  \redtt{8}%
  \verb+].childNodes[+%
  \bluett{0}%
  \verb+]+                                                         \\
\verb+                .checked+                                    \\
\verb+      = true;+                                               \\
\verb+    }+                                                       \\
\verb++                                                            \\
\verb+    else+                                                    \\
\verb+//  if (navigator.userAgent.search("Chrome")    != -1 ||+    \\
\verb+//      navigator.userAgent.search("Firefox")   != -1 ||+    \\
\verb+//      navigator.userAgent.search("Opera")     != -1 ||+    \\
\verb+//      navigator.userAgent.search("Safari")    != -1 ||+    \\
\verb+//      navigator.userAgent.search("Navigator") != -1   )+   \\
\verb+    {+                                                       \\
\verb+        document.getElementById("+%
  \redtt{setup\_main\_function\_1\_choosing}%
  \verb+")+                                                        \\
\verb+                .childNodes[+%
  \redtt{17}%
  \verb+].childNodes[+%
  \bluett{1}%
  \verb+]+                                                         \\
\verb+                .checked+                                    \\
\verb+      = true;+                                               \\
\verb+    }+                                                       \\
\verb+*/+                                                          \\
\verb+?>+                                                          \\
\verb+    +%
  \bluett{default\_choosing}%
  \verb+("+%
  \redtt{setup\_main\_function\_1\_choosing}%
  \verb+", +%
  \redtt{8}%
  \verb+, +%
  \redtt{17}%
  \verb+);+                                                        \\
\verb+  }+                                                         \\
\verb+</script>+                                                   \\
\defc
\verb++                                         \\
\verb+<fieldset id="+%
  \redtt{setup\_main\_function\_1\_choosing}%
  \verb+" style="width: +%
  \redtt{320}%
  \verb+pt">+                                   \\
\verb++                                         \\
\verb+  <label>+                                \\
\verb+  <input type="radio" name="+%
  \redtt{setup\_main\_function\_1\_choosing}%
  \verb+" value="+%
  \redtt{1}%
  \verb+"+                                      \\
\verb+         <?php+                           \\
\verb+           if ($+%
  \redtt{setup\_main\_function\_1\_choosing}%
  \verb+ == "+%
  \redtt{1}%
  \verb+")+                                     \\
\verb+           {+                             \\
\verb+             echo 'checked = "checked"';+ \\
\verb++                                         \\
\verb+             $+%
  \redtt{setup\_main\_function\_1\_choice}%
  \verb+ = "+%
  \redtt{setup\_main\_function\_1\_choice\_1}%
  \verb+";+                                     \\
\verb+           }+                             \\
\verb+         ?>+                              \\
\verb+         />+                              \\
\verb++                                         \\
\verb+         +%
  \redtt{Description for the setup choice 1 of the main function 1}%
  \verb++                                       \\
\verb+         </label>+                        \\
\verb+         <br />+                          \\
\verb++                                         \\
\verb+            .+                            \\
\verb+            .+                            \\
\verb+            .+                            \\
\verb++                                         \\
\verb+  <label>+                                \\
\verb+  <input type="radio" name="+%
  \redtt{setup\_main\_function\_1\_choosing}%
  \verb+" value="+%
  \redtt{5}%
  \verb+"+                                      \\
\verb+                        id="+%
  \redtt{personal\_setup\_main\_function\_1}%
  \verb+"+                                      \\
\verb+         <?php+                           \\
\verb+           if ($+%
  \redtt{setup\_main\_function\_1\_choosing}%
  \verb+ == "+%
  \redtt{5}%
  \verb+")+                                     \\
\verb+           {+                             \\
\verb+             echo 'checked = "checked"';+ \\
\verb++                                         \\
\verb+             $+%
  \redtt{setup\_main\_function\_1\_choice}%
  \verb+ = "+%
  \redtt{personal\_setup\_main\_function\_1}%
  \verb+";+                                     \\
\verb+           }+                             \\
\verb+         ?>+                              \\
\verb+         />+                              \\
\verb++                                         \\
\verb+         +%
  \redtt{Personal setup for the main function 1:}%
  \verb++                                       \\
\verb+         </label>+                        \\
\verb+         <br />+                          \\
\defc
\verb++                                         \\
\verb+  <span style="margin-left: +%
  \bluett{18}%
  \verb+pt">+                                   \\
\verb+    <label for="+%
  \redtt{personal\_setup\_main\_function\_1}%
  \verb+">+                                     \\
\verb+      +%
  \redtt{integer\_1}%
  \verb+ =+                                     \\
\verb+    </label>+                             \\
\verb+  </span>+                                \\
\verb+  <input class="+%
  \bluett{table}%
  \verb+" type="text" name="+%
  \redtt{integer\_1}%
  \verb+"+                                      \\
\verb+         style="width: +%
  \bluett{60}%
  \verb+pt"+                                    \\
\verb+         onclick="+%
  \redtt{personal\_setup\_main\_function\_1\_setting}%
  \verb+()"+ \mynote[0.5cm]{$\lGetsto$~~{\tt <label for="...">}} \\
\verb+         <?php+                           \\
\verb+           if ($+%
  \redtt{setup\_main\_function\_1\_choosing}%
  \verb+ == "+%
  \redtt{5}%
  \verb+")+                                     \\
\verb+           {+                             \\
\verb+             echo 'value="' . $+%
  \redtt{integer\_1}%
  \verb+ . '"';+                                \\
\verb+           }+                             \\
\verb+         ?>+                              \\
\verb+         /><label for="+%
  \redtt{personal\_setup\_main\_function\_1}%
  \verb+">,+                                    \\
\verb+      +%
  \redtt{integer\_2}%
  \verb+ =+                                     \\
\verb+    </label>+                             \\
\verb+  <input class="+%
  \bluett{table}%
  \verb+" type="text" name="+%
  \redtt{integer\_2}%
  \verb+"+                                      \\
\verb+         style="width: +%
  \bluett{60}%
  \verb+pt"+                                    \\
\verb+         onclick="+%
  \redtt{personal\_setup\_main\_function\_1\_setting}%
  \verb+()"+                                    \\
\verb+         <?php+                           \\
\verb+           if ($+%
  \redtt{setup\_main\_function\_1\_choosing}%
  \verb+ == "+%
  \redtt{5}%
  \verb+")+                                     \\
\verb+           {+                             \\
\verb+             echo 'value="' . $+%
  \redtt{integer\_2}%
  \verb+ . '"';+                                \\
\verb+           }+                             \\
\verb+         ?>+                              \\
\verb+         />+                              \\
\verb+         <br />+                          \\
%
%
\verb++                                                                                 \\
\verb+<?php+                                                                            \\
\verb+  if ($+%
  \redtt{mode\_main\_function\_1\_choosing}%
  \verb+ == "+%
  \redtt{2}%
  \verb+")+                                                                             \\
\verb+  {+                                                                              \\
\verb+    echo '  <span style="margin-left: +%
  \bluett{18}%
  \verb+pt">'                          . "\r\n";+                                       \\
\verb+    echo '    <label for="+%
  \redtt{personal\_setup\_main\_function\_1}%
  \verb+">'            . "\r\n";+                                                       \\
\verb+    echo '      +%
  \redtt{float\_1\_1}%
  \verb+ ='                                           . "\r\n";+                        \\
\verb+    echo '    </label>'                                                . "\r\n";+ \\
\verb+    echo '  </span>'                                                   . "\r\n";+ \\
\verb+    echo '  <input class="+%
  \bluett{table}%
  \verb+" type="text" name="+%
  \redtt{float\_1\_1}%
  \verb+"'         . "\r\n";+                                                           \\
\verb+    echo '         style="width: +%
  \bluett{60}%
  \verb+pt"'                                . "\r\n";+                                  \\
\verb+    echo '         onclick="+%
  \redtt{personal\_setup\_main\_function\_1\_setting}%
  \verb+()"' . "\r\n";+                                                                 \\
\verb+    if ($+%
  \redtt{setup\_main\_function\_1\_choosing}%
  \verb+ == "+%
  \redtt{5}%
  \verb+")+                                                                             \\
\verb+    {+                                                                            \\
\verb+      echo '         value="' . $+%
  \redtt{float\_1\_1}%
  \verb+ . '"'                       . "\r\n";+                                         \\
\verb+    }+                                                                            \\
\verb+    echo '         /><label for="+%
  \redtt{personal\_setup\_main\_function\_1}%
  \verb+">,'    . "\r\n";+                                                              \\
\verb+    echo '      +%
  \redtt{float\_1\_2}%
  \verb+ ='                                           . "\r\n";+                        \\
\verb+    echo '    </label>'                                                . "\r\n";+ \\
\verb+    echo '  <input class="+%
  \bluett{table}%
  \verb+" type="text" name="+%
  \redtt{float\_1\_2}%
  \verb+"'         . "\r\n";+                                                           \\
\verb+    echo '         style="width: +%
  \bluett{60}%
  \verb+pt"'                                . "\r\n";+                                  \\
\verb+    echo '         onclick="+%
  \redtt{personal\_setup\_main\_function\_1\_setting}%
  \verb+()"' . "\r\n";+                                                                 \\
\verb+    if ($+%
  \redtt{setup\_main\_function\_1\_choosing}%
  \verb+ == "+%
  \redtt{5}%
  \verb+")+                                                                             \\
\verb+    {+                                                                            \\
\verb+      echo '         value="' . $+%
  \redtt{float\_1\_2}%
  \verb+ . '"'                       . "\r\n";+                                         \\
\verb+    }+                                                                            \\
\verb+    echo '         />'                                                 . "\r\n".+ \\
\verb+                                                                         "\r\n";+ \\
\verb+  }+                                                                              \\
\verb+?>+                                                                               \\
\defc
\verb++                                                                                 \\
\verb+<?php+                                                                            \\
\verb+  if ($+%
  \redtt{mode\_main\_function\_1\_choosing}%
  \verb+ == "+%
  \redtt{3}%
  \verb+")+                                                                             \\
\verb+  {+                                                                              \\
\verb+    echo '  <span style="margin-left: +%
  \bluett{18}%
  \verb+pt">'                          . "\r\n";+                                       \\
\verb+    echo '    <label for="+%
  \redtt{personal\_setup\_main\_function\_1}%
  \verb+">'            . "\r\n";+                                                       \\
\verb+    echo '      +%
  \redtt{float\_2\_1}%
  \verb+ ='                                           . "\r\n";+                        \\
\verb+    echo '    </label>'                                                . "\r\n";+ \\
\verb+    echo '  </span>'                                                   . "\r\n";+ \\
\verb+    echo '  <input class="+%
  \bluett{table}%
  \verb+" type="text" name="+%
  \redtt{float\_2\_1}%
  \verb+"'         . "\r\n";+                                                           \\
\verb+    echo '         style="width: +%
  \bluett{60}%
  \verb+pt"'                                . "\r\n";+                                  \\
\verb+    echo '         onclick="+%
  \redtt{personal\_setup\_main\_function\_1\_setting}%
  \verb+()"' . "\r\n";+                                                                 \\
\verb+    if ($+%
  \redtt{setup\_main\_function\_1\_choosing}%
  \verb+ == "+%
  \redtt{5}%
  \verb+")+                                                                             \\
\verb+    {+                                                                            \\
\verb+      echo '         value="' . $+%
  \redtt{float\_2\_1}%
  \verb+ . '"'                       . "\r\n";+                                         \\
\verb+    }+                                                                            \\
\verb+    echo '         /><label for="+%
  \redtt{personal\_setup\_main\_function\_1}%
  \verb+">,'    . "\r\n";+                                                              \\
\verb+    echo '      +%
  \redtt{float\_2\_2}%
  \verb+ ='                                           . "\r\n";+                        \\
\verb+    echo '    </label>'                                                . "\r\n";+ \\
\verb+    echo '  <input class="+%
  \bluett{table}%
  \verb+" type="text" name="+%
  \redtt{float\_2\_2}%
  \verb+"'         . "\r\n";+                                                           \\
\verb+    echo '         style="width: +%
  \bluett{60}%
  \verb+pt"'                                . "\r\n";+                                  \\
\verb+    echo '         onclick="+%
  \redtt{personal\_setup\_main\_function\_1\_setting}%
  \verb+()"' . "\r\n";+                                                                 \\
\verb+    if ($+%
  \redtt{setup\_main\_function\_1\_choosing}%
  \verb+ == "+%
  \redtt{5}%
  \verb+")+                                                                             \\
\verb+    {+                                                                            \\
\verb+      echo '         value="' . $+%
  \redtt{float\_2\_2}%
  \verb+ . '"'                       . "\r\n";+                                         \\
\verb+    }+                                                                            \\
\verb+    echo '         />'                                                 . "\r\n".+ \\
\verb+                                                                         "\r\n";+ \\
\verb+  }+                                                                              \\
\verb+?>+                                                                               \\
%
%
\verb++                                                                                 \\
\verb+<?php+                                                                            \\
\verb+  if ($+%
  \redtt{mode\_main\_function\_1\_choosing}%
  \verb+ == "+%
  \redtt{2}%
  \verb+" ||+                                                                           \\
\verb+      $+%
  \redtt{mode\_main\_function\_1\_choosing}%
  \verb+ == "+%
  \redtt{3}%
  \verb+"   )+                                                                          \\
\verb+  {+                                                                              \\
\verb+    echo '         <br />'                                             . "\r\n".+ \\
\verb+                                                                         "\r\n";+ \\
\verb+    echo '  <span style="margin-left: +%
  \bluett{18}%
  \verb+pt">'                          . "\r\n";+                                       \\
\verb+    echo '    <label for="+%
  \redtt{personal\_setup\_main\_function\_1}%
  \verb+">'            . "\r\n";+                                                       \\
\verb+    echo '      +%
  \redtt{integer\_3}%
  \verb+ ='                                           . "\r\n";+                        \\
\verb+    echo '    </label>'                                                . "\r\n";+ \\
\verb+    echo '  </span>'                                                   . "\r\n";+ \\
\verb+    echo '  <input class="+%
  \bluett{table}%
  \verb+" type="text" name="+%
  \redtt{integer\_3}%
  \verb+"'         . "\r\n";+                                                           \\
\verb+    echo '         style="width: +%
  \bluett{60}%
  \verb+pt"'                                . "\r\n";+                                  \\
\verb+    echo '         onclick="+%
  \redtt{personal\_setup\_main\_function\_1\_setting}%
  \verb+()"' . "\r\n";+                                                                 \\
\verb+    if ($+%
  \redtt{setup\_main\_function\_1\_choosing}%
  \verb+ == "+%
  \redtt{5}%
  \verb+")+                                                                             \\
\verb+    {+                                                                            \\
\verb+      echo '         value="' . $+%
  \redtt{integer\_3}%
  \verb+ . '"'                       . "\r\n";+                                         \\
\verb+    }+                                                                            \\
\verb+    echo '         /><label for="+%
  \redtt{personal\_setup\_main\_function\_1}%
  \verb+">,'    . "\r\n";+                                                              \\
\verb+    echo '      +%
  \redtt{float\_3}%
  \verb+ ='                                             . "\r\n";+                      \\
\verb+    echo '    </label>'                                                . "\r\n";+ \\

\verb+    echo '  <input class="+%
  \bluett{table}%
  \verb+" type="text" name="+%
  \redtt{float\_3}%
  \verb+"'           . "\r\n";+                                                         \\
\verb+    echo '         style="width: +%
  \bluett{60}%
  \verb+pt"'                                . "\r\n";+                                  \\
\verb+    echo '         onclick="+%
  \redtt{personal\_setup\_main\_function\_1\_setting}%
  \verb+()"' . "\r\n";+                                                                 \\
\verb+    if ($+%
  \redtt{setup\_main\_function\_1\_choosing}%
  \verb+ == "+%
  \redtt{5}%
  \verb+")+                                                                             \\
\verb+    {+                                                                            \\
\verb+      echo '         value="' . $+%
  \redtt{float\_3}%
  \verb+ . '"'                         . "\r\n";+                                       \\
\verb+    }+                                                                            \\
\verb+    echo '         />'                                                 . "\r\n";+ \\
\verb+    echo '         <br />';+                                                      \\
\verb+  }+                                                                              \\
\verb+?>+                                                                               \\
\verb++                                        \\
\verb+</fieldset>+                             \\
\tabb
\newpage
\picqin{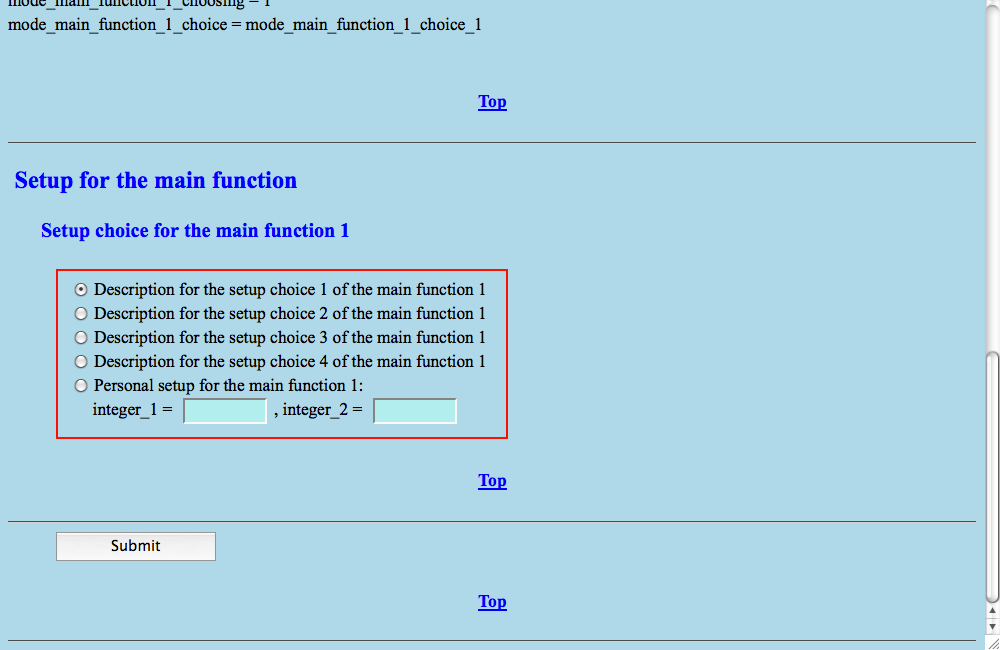}
{setup\_setting}
{Setup for the chosen function
 (before submission,
  cf.~Figure \ref{fig:mode_choosing-1}).}
{Setup for the chosen function
 (after submission,
  cf.~Figure \ref{fig:mode_choosing-2}).}
{Setup for the chosen function with
 {\tt \$mode\_main\_function\_1\_choosing = "2"}
 (before submission,
  cf.~Figure \ref{fig:setup_setting-1}).}
{Setup for the chosen function with
 {\tt \$mode\_main\_function\_1\_choosing = "2"}
 (after submission,
  cf.~Figure \ref{fig:setup_setting-2}).}
\newpage
\myitem
\label{item:modification_form-setup_main_function_1_choosing.php}
 {\bf Modification of {\cmbfsfx form-setup\_main\_function\_1\_choosing.php}}
 (cf.~Template \ref{item:form-setup_main_function_1_choosing.php})
\deft
\verb+<script type="text/javascript">+                             \\
\verb+  function +%
  \redtt{personal\_setup\_main\_function\_1\_setting}%
  \verb+()+                                                        \\
\verb+  {+                                                         \\
\verb+       .+                                                    \\
\verb+       .+                                                    \\
\verb+       .+                                                    \\
\verb+  }+                                                         \\
\verb+</script>+                                                   \\
%
%
\verb++                                              \\
\verb+<fieldset id="+%
  \redtt{setup\_main\_function\_1\_choosing}%
  \verb+" style="width: +%
  \redtt{320}%
  \verb+pt">+                                        \\
\verb++                                              \\
\verb+       .+                                      \\
\verb+       .+                                      \\
\verb+       .+                                      \\
\verb++                                              \\

\verb+  <label>+                                \\
\verb+  <input type="radio" name="+%
  \redtt{setup\_main\_function\_1\_choosing}%
  \verb+" value="+%
  \redtt{5}%
  \verb+"+                                      \\
\verb+                        id="+%
  \redtt{personal\_setup\_main\_function\_1}%
  \verb+"+                                      \\
\verb+         <?php+                           \\
\verb+           if ($+%
  \redtt{setup\_main\_function\_1\_choosing}%
  \verb+ == "+%
  \redtt{5}%
  \verb+")+                                     \\
\verb+           {+                             \\
\verb+             echo 'checked = "checked"';+ \\
\verb++                                         \\
\verb+             $+%
  \redtt{setup\_main\_function\_1\_choice}%
  \verb+ = "+%
  \redtt{personal\_setup\_main\_function\_1}%
  \verb+";+                                     \\
\verb+           }+                             \\
\verb+         ?>+                              \\
\verb+         />+                              \\
\verb++                                         \\
\verb+         +%
  \redtt{Personal setup for the main function 1:}%
  \verb++                                       \\
\verb+         </label>+                        \\
\verb+         <br />+                          \\
\defc
\verb++                                         \\
\verb+  <span style="margin-left: +%
  \bluett{18}%
  \verb+pt">+                                   \\
\verb+    <label for="+%
  \redtt{personal\_setup\_main\_function\_1}%
  \verb+">+                                     \\
\verb+      +%
  \redtt{integer\_1}%
  \verb+ =+                                     \\
\verb+    </label>+                             \\
\verb+  </span>+                                \\
\verb+  <select name="+%
  \redtt{integer\_1}%
  \verb+"+                                           \\
\verb+          onclick="+%
  \redtt{personal\_setup\_main\_function\_1\_setting}%
  \verb+()">+                                        \\
\verb++                                              \\
\verb+    <option value="+%
  \redtt{nn}%
  \verb+"+                                           \\
\verb+            <?php+                             \\
\verb+              if ($+%
  \redtt{setup\_main\_function\_1\_choosing}%
  \verb+ == "+%
  \redtt{5}%
  \verb+" &&+                                        \\
\verb+                  $+%
  \redtt{integer\_1}%
  \verb+ == +%
  \redtt{nn}%
  \verb+                         )+                  \\
\verb+              {+                               \\
\verb+                echo 'selected = "selected"';+ \mynote[0.5cm]{$\lGetsto$~~{\tt checked = "checked"}} \\
\verb+              }+                               \\
\verb+            ?>+                                \\
\verb+            > &nbsp; &nbsp; +%
  \redtt{nn}%
  \verb++                                            \\
\verb+    </option>+                                 \\
\verb++                                              \\
\verb+    <option value="+%
  \redtt{nn}%
  \verb+"+                                           \\
\verb+            <?php+                             \\
\verb+              if ($+%
  \redtt{setup\_main\_function\_1\_choosing}%
  \verb+ == "+%
  \redtt{5}%
  \verb+" &&+                                        \\
\verb+                  $+%
  \redtt{integer\_1}%
  \verb+ == +%
  \redtt{nn}%
  \verb+                         )+                  \\
\verb+              {+                               \\
\verb+                echo 'selected = "selected"';+ \\
\verb+              }+                               \\
\verb+            ?>+                                \\
\verb+            > &nbsp; &nbsp; +%
  \redtt{nn}%
  \verb++                                            \\
\verb+    </option>+                                 \\
\verb++                                              \\
\verb+        .+                                     \\
\verb+        .+                                     \\
\verb+        .+                                     \\
\verb++                                              \\
\verb+  </select><label for="+%
  \redtt{personal\_setup\_main\_function\_1}%
  \verb+">,+                                         \\
\verb+      +%
  \redtt{integer\_2}%
  \verb+ =+                                          \\
\verb+    </label>+                                  \\
\verb+  <select name="+%
  \redtt{integer\_2}%
  \verb+"+                                           \\
\verb+          onclick="+%
  \redtt{personal\_setup\_main\_function\_1\_setting}%
  \verb+()">+                                        \\
\verb++                                              \\
\verb+    <option value="+%
  \redtt{nn}%
  \verb+"+                                           \\
\verb+            <?php+                             \\
\verb+              if ($+%
  \redtt{setup\_main\_function\_1\_choosing}%
  \verb+ == "+%
  \redtt{5}%
  \verb+" &&+                                        \\
\verb+                  $+%
  \redtt{integer\_2}%
  \verb+ == +%
  \redtt{nn}%
  \verb+                         )+                  \\
\verb+              {+                               \\
\verb+                echo 'selected = "selected"';+ \\
\verb+              }+                               \\
\verb+            ?>+                                \\
\verb+            > &nbsp; &nbsp; +%
  \redtt{nn}%
  \verb++                                            \\
\verb+    </option>+                                 \\
\verb++                                              \\
\verb+        .+                                     \\
\verb+        .+                                     \\
\verb+        .+                                     \\
\verb++                                              \\
\verb+  </select>+                                   \\
\verb++                                              \\
\verb+      .+                                       \\
\verb+      .+                                       \\
\verb+      .+                                       \\
\defc
\verb++                                              \\
\verb+<?php+                                         \\
\verb+  if ($+%
  \redtt{mode\_main\_function\_1\_choosing}%
  \verb+ == "+%
  \redtt{2}%
  \verb+")+                                          \\
\verb+  {+                                           \\
\verb+      .+                                       \\
\verb+      .+                                       \\
\verb+      .+                                       \\
\verb+  }+                                           \\
%
%
\verb++                                              \\
\verb+  if ($+%
  \redtt{mode\_main\_function\_1\_choosing}%
  \verb+ == "+%
  \redtt{3}%
  \verb+")+                                          \\
\verb+  {+                                           \\
\verb+      .+                                       \\
\verb+      .+                                       \\
\verb+      .+                                       \\
\verb+  }+                                           \\
%
%
\verb++                                              \\
\verb+  if ($+%
  \redtt{mode\_main\_function\_1\_choosing}%
  \verb+ == "+%
  \redtt{2}%
  \verb+" ||+                                        \\
\verb+      $+%
  \redtt{mode\_main\_function\_1\_choosing}%
  \verb+ == "+%
  \redtt{3}%
  \verb+"   )+                                       \\
\verb+  {+                                           \\
\verb+      .+                                       \\
\verb+      .+                                       \\
\verb+      .+                                       \\
\verb+  }+                                           \\
\verb+?>+                                            \\
\verb++                                              \\
\verb+</fieldset>+                                   \\
\tabb
\newpage
\picin{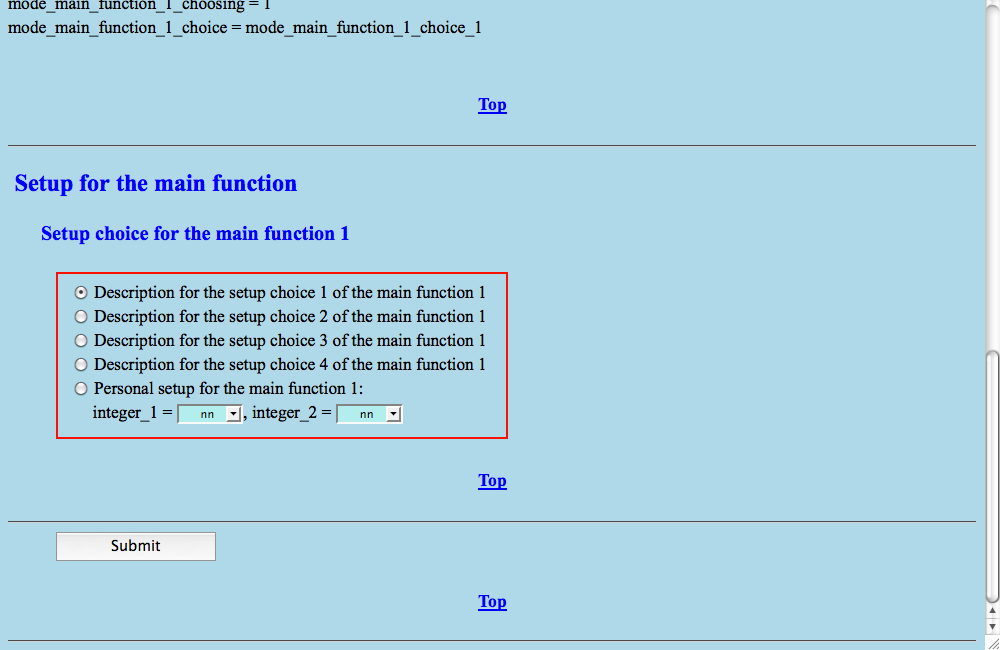}
{modification\_setup\_setting}
{Modification of {\sf form-setup\_main\_function\_1\_choosing.php}
 (cf.~Figures \ref{fig:setup_setting-1} and \ref{fig:setup_setting-3}).}
%
%
\end{templateenumerate}
\newpage
\section{\sf table}
\label{sec:table}
%
%
\begin{templateenumerate}
\myitem
\label{item:table-setup_item.php}
 {\cmbfsfx table-setup\_item.php}
 (see Template \ref{item:table-setup_item_entry})
\deft
\verb+<table id="+%
  \redtt{table-setup\_item}%
  \verb+" border="1">+                            \\
\verb++                                           \\
\verb+  <tr id="+%
  \redtt{setup\_item}%
  \verb+">+                                       \\
\verb++                                           \\
\verb+    <th> +%
  \redtt{setup\_item\_1}%
  \verb+ </th>+                                   \\
\verb+    <th> +%
  \redtt{setup\_item\_2}%
  \verb+ </th>+                                   \\
\verb+    <th> +%
  \redtt{setup\_item\_3}%
  \verb+ </th>+                                   \\
\verb+    <th> +%
  \redtt{setup\_item\_4}%
  \verb+ </th>+                                   \\
\verb+    <th> +%
  \redtt{setup\_item\_5}%
  \verb+ </th>+                                   \\
\verb++                                           \\
\verb+  </tr>+                                    \\
\verb++                                           \\
\verb+  <tr>+                                     \\
\verb++                                           \\
\verb+    <td>+                                   \\
\verb+      <input class="+%
  \bluett{table}%
  \verb+" type="text" name="+%
  \redtt{setup\_item\_1}%
  \verb+"+                                        \\
\verb+             onclick="this.value=''"+       \\
\verb+             <?php+                         \\
\verb+               if ($+%
  \redtt{setup\_item\_1}%
  \verb+ != NULL)+                                \\
\verb+               {+                           \\
\verb+                 echo 'value="' . $+%
  \redtt{setup\_item\_1}%
  \verb+ . '"';+                                  \\
\verb+               }+                           \\
\verb+             ?>+                            \\
\verb+             />+                            \\
\verb+    </td>+                                  \\
\verb++                                           \\
\verb+    <td>+                                   \\
\verb+      <input class="+%
  \bluett{table}%
  \verb+" type="text" name="+%
  \redtt{setup\_item\_2}%
  \verb+"+                                        \\
\verb+             onclick="this.value=''"+       \\
\verb+             <?php+                         \\
\verb+               if ($+%
  \redtt{setup\_item\_2}%
  \verb+ != NULL)+                                \\
\verb+               {+                           \\
\verb+                 if ($+%
  \redtt{setup\_item\_2}%
  \verb+ <= +%
  \redtt{setup\_item\_2\_max}%
  \verb+)+                                        \\
\verb+                 {+                         \\
\verb+                   echo 'value="' . $+%
  \redtt{setup\_item\_2}%
  \verb+ . '"';+                                  \\
\verb+                 }+                         \\
\verb++                                           \\
\verb+                 else+                      \\
\verb+                 {+                         \\
\verb+                   echo 'value="+%
  \redtt{setup\_item\_2\_max}%
  \verb+"';+                                      \\
\verb+                 }+                         \\
\verb+               }+                           \\
\verb+             ?>+                            \\
\verb+             />+                            \\
\verb+    </td>+                                  \\
\defc
\verb++                                           \\
\verb+    <td>+                                   \\
\verb+      <input class="+%
  \bluett{table}%
  \verb+" type="text" name="+%
  \redtt{setup\_item\_3}%
  \verb+"+                                        \\
\verb+             onclick="this.value=''"+       \\
\verb+             <?php+                         \\
\verb+               if ($+%
  \redtt{setup\_item\_3}%
  \verb+ == +%
  \bluett{0.00001}%
  \verb+)+                                        \\
\verb+               {+                           \\
\verb+                 echo 'value="0"';+         \\
\verb+               }+                           \\
\verb++                                           \\
\verb+               elseif ($+%
  \redtt{setup\_item\_3}%
  \verb+ != NULL)+                                \\
\verb+               {+                           \\
\verb+                 echo 'value="' . $+%
  \redtt{setup\_item\_3}%
  \verb+ . '"';+                                  \\
\verb+               }+                           \\
\verb+               ?>+                          \\
\verb+             />+                            \\
\verb+    </td>+ \mynote[4cm]{See Templates \ref{item:sample-input_setup_web_reload.txt}
                              and \ref{item:setup_output_reload.php}} \\
\verb++                                           \\
\verb+       .+                                   \\
\verb+       .+                                   \\
\verb+       .+                                   \\
\verb++                                           \\
\verb+  </tr>+                                    \\
\verb++                                           \\
\verb+</table>+                                   \\
\tabb
\newpage
\picin{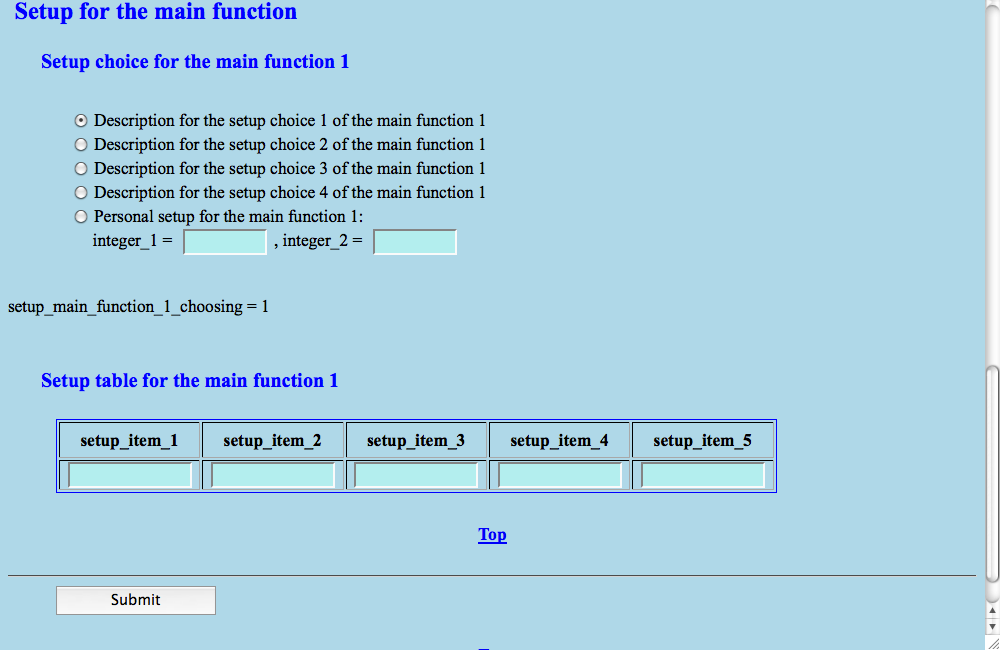}
{table-setup\_item}
{A setup table.}
\newpage
\myitem
\label{item:default-setup_item.php}
 {\cmbfsfx default-setup\_item.php}
\deft
\verb+<script type="text/javascript">+                           \\
\verb+    document.getElementById("+%
  \redtt{table-setup\_item}%
  \verb+")+                                                      \\
\verb+            .style.borderColor+                            \\
\verb+  = "+%
  \bluett{red}%
  \verb+";+                                                      \\
\verb++                                                          \\
\verb+    document.getElementById("+%
  \redtt{table-setup\_item}%
  \verb+")+                                                      \\
\verb+            .border+                                       \\
\verb+  = "+%
  \bluett{2}%
  \verb+";+                                                      \\
\verb++                                                          \\
\verb+  if (navigator.appName == "Microsoft Internet Explorer")+ \\
\verb+  {+                                                       \\
\verb+      var +%
  \redtt{setup\_item\_row}%
  \verb++                                                        \\
\verb+    = document.getElementById("+%
  \redtt{setup\_item}%
  \verb+")+                                                      \\
\verb+              .nextSibling;+                               \\
\verb++                                                          \\
\verb+      +%
  \redtt{setup\_item\_row}%
  \verb+.childNodes[0].childNodes[0]+                            \\
\verb+                    .value+                                \\
\verb+    = "+%
  \redtt{nn}%
  \verb+";+                                                      \\
\verb++                                                          \\
\verb+      +%
  \redtt{setup\_item\_row}%
  \verb+.childNodes[1].childNodes[0]+                            \\
\verb+                    .value+                                \\
\verb+    = "+%
  \redtt{nn}%
  \verb+";+                                                      \\
\verb++                                                          \\
\verb+      +%
  \redtt{setup\_item\_row}%
  \verb+.childNodes[2].childNodes[0]+                            \\
\verb+                    .value+                                \\
\verb+    = "+%
  \redtt{ff.ffff}%
  \verb+";+                                                      \\
\verb++                                                          \\
\verb+      +%
  \redtt{setup\_item\_row}%
  \verb+.childNodes[3].childNodes[0]+                            \\
\verb+                    .value+                                \\
\verb+    = "+%
  \redtt{ff.ffff}%
  \verb+";+                                                      \\
\verb++                                                          \\
\verb+      +%
  \redtt{setup\_item\_row}%
  \verb+.childNodes[4].childNodes[0]+                            \\
\verb+                    .value+                                \\
\verb+    = "+%
  \redtt{ff.ffff}%
  \verb+";+                                                      \\
\verb+  }+                                                       \\
\defc
\verb++                                                          \\
\verb+  else+                                                    \\
\verb+//if (navigator.userAgent.search("Chrome")    != -1 ||+    \\
\verb+//    navigator.userAgent.search("Firefox")   != -1 ||+    \\
\verb+//    navigator.userAgent.search("Opera")     != -1 ||+    \\
\verb+//    navigator.userAgent.search("Safari")    != -1 ||+    \\
\verb+//    navigator.userAgent.search("Navigator") != -1   )+   \\
\verb+  {+                                                       \\
\verb+      var +%
  \redtt{setup\_item\_row}%
  \verb++                                                        \\
\verb+    = document.getElementById("+%
  \redtt{setup\_item}%
  \verb+")+                                                      \\
\verb+              .nextSibling.nextSibling;+                   \\
\verb++                                                          \\
\verb+      +%
  \redtt{setup\_item\_row}%
  \verb+.childNodes[1].childNodes[1]+                            \\
\verb+                    .value+                                \\
\verb+    = "+%
  \redtt{nn}%
  \verb+";+                                                      \\
\verb++                                                          \\
\verb+      +%
  \redtt{setup\_item\_row}%
  \verb+.childNodes[3].childNodes[1]+                            \\
\verb+                    .value+                                \\
\verb+    = "+%
  \redtt{nn}%
  \verb+";+                                                      \\
\verb++                                                          \\
\verb+      +%
  \redtt{setup\_item\_row}%
  \verb+.childNodes[5].childNodes[1]+                            \\
\verb+                    .value+                                \\
\verb+    = "+%
  \redtt{ff.ffff}%
  \verb+";+                                                      \\
\verb++                                                          \\
\verb+      +%
  \redtt{setup\_item\_row}%
  \verb+.childNodes[7].childNodes[1]+                            \\
\verb+                    .value+                                \\
\verb+    = "+%
  \redtt{ff.ffff}%
  \verb+";+                                                      \\
\verb++                                                          \\
\verb+      +%
  \redtt{setup\_item\_row}%
  \verb+.childNodes[9].childNodes[1]+                            \\
\verb+                    .value+                                \\
\verb+    = "+%
  \redtt{ff.ffff}%
  \verb+";+                                                      \\
\verb+  }+                                                       \\
\verb+</script>+                                                 \\
\tabb
\newpage
\picin{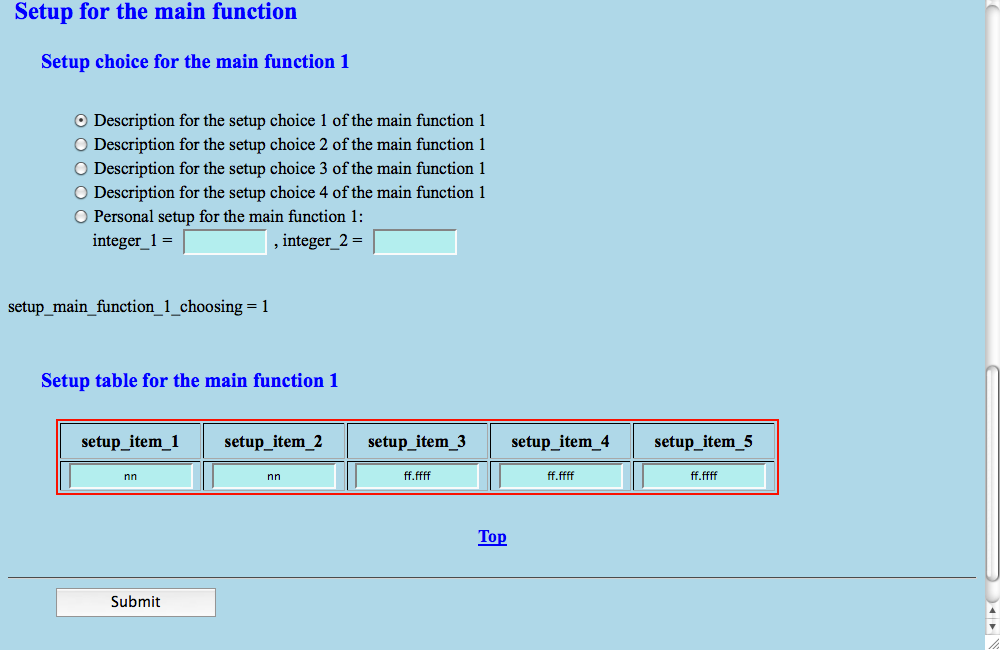}
{default-setup\_item}
{A setup table with default values
 (cf.~Figure \ref{fig:table-setup_item}).}
\newpage
\myitem
\label{item:table-setup_item_entry}
 {\bf Setup table 
      for a group of entry}
  (cf.~Template \ref{item:setup_setting})
\deft
\verb+<?php+                                   \\
\verb+  if ($+%
  \redtt{setup\_main\_function\_1\_checked}%
  \verb+ == "yes")+                            \\
\verb+  {+                                     \\
\verb+    echo '<h4> +%
  \redtt{Setup table for the main function 1}%
  \verb+ </h4>' . "\r\n";+                     \\
\verb++                                        \\
\verb+    $+%
  \redtt{entry\_No}%
  \verb+ = +%
  \redtt{3}%
  \verb+;+                                     \\
\verb++                                        \\
\verb+    for ($+%
  \redtt{n\_entry}%
  \verb+ = 1; $+%
  \redtt{n\_entry}%
  \verb+ <= $+%
  \redtt{entry\_No}%
  \verb+; $+%
  \redtt{n\_entry}%
  \verb- ++)-                                  \\
\verb+    {+                                   \\
\verb+      if ($+%
  \redtt{setup\_item\_1}%
  \verb+[$+%
  \redtt{n\_entry}%
  \verb+] == NULL)+                            \\
\verb+      {+                                 \\
\verb+        $+%
  \redtt{setup\_item\_1}%
  \verb+[$+%
  \redtt{n\_entry}%
  \verb+] = $_POST["+%
  \redtt{setup\_item\_1}%
  \verb+"][$+%
  \redtt{n\_entry}%
  \verb+];+ \mynote{See Sec.~\ref{sec:basic_php_arrays}} \\
\verb+      }+                                 \\
\verb++                                        \\
\verb+      if ($+%
  \redtt{setup\_item\_2}%
  \verb+[$+%
  \redtt{n\_entry}%
  \verb+] == NULL)+                            \\
\verb+      {+                                 \\
\verb+        $+%
  \redtt{setup\_item\_2}%
  \verb+[$+%
  \redtt{n\_entry}%
  \verb+] = $_POST["+%
  \redtt{setup\_item\_2}%
  \verb+"][$+%
  \redtt{n\_entry}%
  \verb+];+                                    \\
\verb+      }+                                 \\
\verb++                                        \\
\verb+      if ($+%
  \redtt{setup\_item\_3}%
  \verb+[$+%
  \redtt{n\_entry}%
  \verb+] == NULL)+                            \\
\verb+      {+                                 \\
\verb+        $+%
  \redtt{setup\_item\_3}%
  \verb+[$+%
  \redtt{n\_entry}%
  \verb+] = $_POST["+%
  \redtt{setup\_item\_3}%
  \verb+"][$+%
  \redtt{n\_entry}%
  \verb+];+                                    \\
\verb+      }+                                 \\
\verb++                                        \\
\verb+      if ($+%
  \redtt{setup\_item\_4}%
  \verb+[$+%
  \redtt{n\_entry}%
  \verb+] == NULL)+                            \\
\verb+      {+                                 \\
\verb+        $+%
  \redtt{setup\_item\_4}%
  \verb+[$+%
  \redtt{n\_entry}%
  \verb+] = $_POST["+%
  \redtt{setup\_item\_4}%
  \verb+"][$+%
  \redtt{n\_entry}%
  \verb+];+                                    \\
\verb+      }+                                 \\
\verb+    }+                                   \\
\verb++                                        \\
\verb+    if ($+%
  \redtt{setup\_entry\_common}%
  \verb+ == NULL)+                             \\
\verb+    {+                                   \\
\verb+      $+%
  \redtt{setup\_entry\_common}%
  \verb+ = $_POST["+%
  \redtt{setup\_entry\_common}%
  \verb+"];+                                   \\
\verb+    }+                                   \\
\verb++                                        \\
\verb+//  include("+%
  \bluett{tables/table-}%
  \verb++%
  \redtt{setup\_item}%
  \verb+.php");+ \mynote[1cm]{Template \ref{item:table-setup_item.php}} \\
\verb++                                        \\
\verb+//  include("+%
  \bluett{default\_choices/default-}%
  \verb++%
  \redtt{setup\_item}%
  \verb+.php");+ \mynote[1cm]{Template \ref{item:default-setup_item.php}} \\
\verb++                                        \\
\verb+            .+                           \\
\verb+            .+ \mynote{Template \ref{item:table-delete_setup_item_4.php}} \\
\verb+            .+                           \\
%
%
\verb++                                        \\
\verb+    include("+%
  \bluett{tables/table-}%
  \verb++%
  \redtt{setup\_item\_head}%
  \verb+.php");+ \mynote{Template \ref{item:table-setup_item_head.php}} \\
\verb+    for ($+%
  \redtt{n\_entry}%
  \verb+ = 1; $+%
  \redtt{n\_entry}%
  \verb+ <= $+%
  \redtt{entry\_No}%
  \verb+; $+%
  \redtt{n\_entry}%
  \verb- ++)-                                  \\
\verb+    {+                                   \\
\verb+      include("+%
  \bluett{tables/table-}%
  \verb++%
  \redtt{setup\_item\_entry}%
  \verb+.php");+ \mynote[1cm]{Template \ref{item:table-setup_item_entry.php}} \\
\verb+    }+                                   \\
\verb+    include("+%
  \bluett{tables/table-}%
  \verb++%
  \redtt{setup\_item\_entry\_common}%
  \verb+.php");+ \mynote[1cm]{Template \ref{item:table-setup_item_entry_common.php}} \\
\verb+    echo '</table>' . "\r\n";+           \\
\defc
\verb++                                        \\
\verb+    for ($+%
  \redtt{n\_entry}%
  \verb+ = 1; $+%
  \redtt{n\_entry}%
  \verb+ <= $+%
  \redtt{entry\_No}%
  \verb+; $+%
  \redtt{n\_entry}%
  \verb- ++)-                                  \\
\verb+    {+                                   \\
\verb+      if ($+%
  \redtt{setup\_item\_1}%
  \verb+[$+%
  \redtt{n\_entry}%
  \verb+] == NULL ||+                          \\
\verb+          $+%
  \redtt{setup\_item\_2}%
  \verb+[$+%
  \redtt{n\_entry}%
  \verb+] == NULL ||+                          \\
\verb+          $+%
  \redtt{setup\_item\_3}%
  \verb+[$+%
  \redtt{n\_entry}%
  \verb+] == NULL ||+                          \\
\verb+          $+%
  \redtt{setup\_item\_4}%
  \verb+[$+%
  \redtt{n\_entry}%
  \verb+] == NULL ||+                          \\
\verb+          $+%
  \redtt{setup\_entry\_common}%
  \verb+     == NULL   )+                      \\
\verb+      {+                                 \\
\verb+        include("+%
  \bluett{default\_choices/default-}%
  \verb++%
  \redtt{setup\_item\_head}%
  \verb+.php");+ \mynote[1cm]{Template \ref{item:default-setup_item_head.php}} \\
\verb+        for ($+%
  \redtt{n\_entry}%
  \verb+ = 1; $+%
  \redtt{n\_entry}%
  \verb+ <= $+%
  \redtt{entry\_No}%
  \verb+; $+%
  \redtt{n\_entry}%
  \verb- ++)-                                  \\
\verb+        {+                               \\
\verb+          include("+%
  \bluett{default\_choices/default-}%
  \verb++%
  \redtt{setup\_item\_entry}%
  \verb+.php");+ \mynote[1cm]{Template \ref{item:default-setup_item_entry.php}} \\
\verb+        }+                               \\
\verb+        include("+%
  \bluett{default\_choices/default-}%
  \verb++%
  \redtt{setup\_item\_entry\_common}%
  \verb+.php");+ \mynote[1cm]{Template \ref{item:default-setup_item_entry_common.php}} \\
\verb++                                        \\
\verb+        break;+                          \\
\verb+      }+                                 \\
\verb++ \mynote[5cm]{cf.~Template \ref{item:setup_setting}} \\
\verb+      else+                              \\
\verb+      {+                                 \\
\verb+        echo '<br />';+                  \\
\verb+        echo '+%
  \redtt{setup\_item\_1}%
  \verb+[' . $+%
  \redtt{n\_entry}%
  \verb+ . '] = ' . $+%
  \redtt{setup\_item\_1}%
  \verb+[$+%
  \redtt{n\_entry}%
  \verb+] . '<br />';+                         \\
\verb+        echo '+%
  \redtt{setup\_item\_2}%
  \verb+[' . $+%
  \redtt{n\_entry}%
  \verb+ . '] = ' . $+%
  \redtt{setup\_item\_2}%
  \verb+[$+%
  \redtt{n\_entry}%
  \verb+] . '<br />';+                         \\
\verb+        echo '+%
  \redtt{setup\_item\_3}%
  \verb+[' . $+%
  \redtt{n\_entry}%
  \verb+ . '] = ' . $+%
  \redtt{setup\_item\_3}%
  \verb+[$+%
  \redtt{n\_entry}%
  \verb+] . '<br />';+                         \\
\verb+        echo '+%
  \redtt{setup\_item\_4}%
  \verb+[' . $+%
  \redtt{n\_entry}%
  \verb+ . '] = ' . $+%
  \redtt{setup\_item\_4}%
  \verb+[$+%
  \redtt{n\_entry}%
  \verb+] . '<br />';+                         \\
\verb+      }+                                 \\
\verb++                                        \\
\verb+      if ($+%
  \redtt{n\_entry}%
  \verb+ == $+%
  \redtt{entry\_No}%
  \verb+)+                                     \\
\verb+      {+                                 \\
\verb+        echo '<br />';+                  \\
\verb+        echo '+%
  \redtt{setup\_entry\_common}%
  \verb+ =             ' . $+%
  \redtt{setup\_entry\_common}%
  \verb+     . '<br />';+                      \\
\verb+        echo '<br />';+                  \\
\verb++                                        \\
\verb+        $+%
  \redtt{setup\_item\_checked}%
  \verb+ = "yes";+                             \\
\verb+      }+                                 \\
\verb+    }+                                   \\
\verb++                                                   \\
\verb+      .+                                            \\
\verb+      .+ \mynote{Templates \ref{item:table-delete_setup_item_3_head.php}
                               + \ref{item:table-delete_setup_item_3_entry.php}} \\
\verb+      .+                                            \\
\verb++                                                   \\
\verb+  }+                                                \\
\verb+?>+                                                 \\
\verb++                                                   \\
\verb+<?php+                                              \\
\verb++                                                   \\
\verb+      .+                                            \\
\verb+      .+ \mynote{Templates \ref{item:modification_table-setup_item_entry}
                          and/or \ref{item:uploading_typing}} \\
\verb+      .+                                            \\
\verb++                                                   \\
\verb+?>+                                                 \\

\tabb
\newpage
\myitem
\label{item:table-setup_item_head.php}
 {\cmbfsfx table-setup\_item\_head.php}
 (cf.~Template \ref{item:table-setup_item.php})
\deft
\verb+<table id="+%
  \redtt{table-setup\_item}%
  \verb+" border="1">+                           \\
\verb++                                          \\
\verb+  <tr id="+%
  \redtt{setup\_item}%
  \verb+">+                                      \\
\verb++                                          \\
\verb+    <th style="width: +%
  \redtt{60}%
  \verb+pt"> +%
  \redtt{entry}%
  \verb+ </th>+                                  \\
\verb++                                          \\
\verb+    <th> +%
  \redtt{setup\_item\_1}%
  \verb+ </th>+                                  \\
\verb+    <th> +%
  \redtt{setup\_item\_2}%
  \verb+ </th>+                                  \\
\verb+    <th> +%
  \redtt{setup\_item\_3}%
  \verb+ </th>+                                  \\
\verb+    <th> +%
  \redtt{setup\_item\_4}%
  \verb+ </th>+                                  \\
\verb++                                          \\
\verb+  </tr>+                                   \\
\tabb
\newpage
\myitem
\label{item:table-setup_item_entry.php}
 {\cmbfsfx table-setup\_item\_entry.php}
 (cf.~Template \ref{item:table-setup_item.php})
\deft
\verb+  <tr>+                                                                               \\
\verb++                                                                                     \\
\verb+    <th style="width: +%
  \redtt{60}%
  \verb+pt">+                                                                               \\
\verb+      <?php echo '+%
  \redtt{entry}%
  \verb+[' . $+%
  \redtt{n\_entry}%
  \verb+ . ']'; ?>+                                                                         \\
\verb+    </th>+                                                                            \\
\verb++                                                                                     \\
\verb+    <td>+                                                                             \\
\verb+      <input class="+%
  \bluett{table}%
  \verb+" type="text" name="+%
  \redtt{setup\_item\_1}%
  \verb+[<?php echo $+%
  \redtt{n\_entry}%
  \verb+; ?>]"+                                                                             \\
\verb+             onclick="this.value=''"+                                                 \\
\verb+             <?php+                                                                   \\
\verb+               if ($+%
  \redtt{setup\_item\_1}%
  \verb+[$+%
  \redtt{n\_entry}%
  \verb+] != NULL)+                                                                         \\
\verb+               {+                                                                     \\
\verb+                 echo 'value="' . $+%
  \redtt{setup\_item\_1}%
  \verb+[$+%
  \redtt{n\_entry}%
  \verb+] . '"';+                                                                           \\
\verb+               }+                                                                     \\
\verb+             ?>+                                                                      \\
\verb+             />+                                                                      \\
\verb+    </td>+                                                                            \\
\verb++                                                                                     \\
\verb+    <td>+                                                                             \\
\verb+      <input class="+%
  \bluett{table}%
  \verb+" type="text" name="+%
  \redtt{setup\_item\_2}%
  \verb+[<?php echo $+%
  \redtt{n\_entry}%
  \verb+; ?>]"+                                                                             \\
\verb+             onclick="this.value=''"+                                                 \\
\verb+             <?php+                                                                   \\
\verb+               if ($+%
  \redtt{setup\_item\_2}%
  \verb+[$+%
  \redtt{n\_entry}%
  \verb+] != NULL)+                                                                         \\
\verb+               {+                                                                     \\
\verb+                 if ($+%
  \redtt{setup\_item\_2}%
  \verb+[$+%
  \redtt{n\_entry}%
  \verb+] <= $+%
  \redtt{setup\_item\_2\_max}%
  \verb+)+                                                                                  \\
\verb+                 {+                                                                   \\
\verb+                   echo 'value="' . $+%
  \redtt{setup\_item\_2}%
  \verb+[$+%
  \redtt{n\_entry}%
  \verb+] . '"';+                                                                           \\
\verb+                 }+                                                                   \\
\verb++                                                                                     \\
\verb+                 else+                                                                \\
\verb+                 {+                                                                   \\
\verb+                   echo 'value="' . $+%
  \redtt{setup\_item\_2\_max}%
  \verb+ . '"';+                                                                            \\
\verb+                 }+                                                                   \\
\verb+               }+                                                                     \\
\verb+             ?>+                                                                      \\
\verb+             />+                                                                      \\
\verb+    </td>+                                                                            \\
\defc
\verb++                                                                                     \\
\verb+    <td>+                                                                             \\
\verb+      <input class="+%
  \bluett{table}%
  \verb+" type="text" name="+%
  \redtt{setup\_item\_3}%
  \verb+[<?php echo $+%
  \redtt{n\_entry}%
  \verb+; ?>]"+                                                                             \\
\verb+             onclick="this.value=''"+                                                 \\
\verb+             <?php+                                                                   \\
\verb+               if ($+%
  \redtt{setup\_item\_3}%
  \verb+[$+%
  \redtt{n\_entry}%
  \verb+] == +%
  \bluett{0.00001}%
  \verb+)+                                                                                  \\
\verb+               {+                                                                     \\
\verb+                 echo 'value="0"';+                                                   \\
\verb+               }+                                                                     \\
\verb++                                                                                     \\
\verb+               elseif ($+%
  \redtt{setup\_item\_3}%
  \verb+[$+%
  \redtt{n\_entry}%
  \verb+] != NULL)+                                                                         \\
\verb+               {+                                                                     \\
\verb+                 echo 'value="' . $+%
  \redtt{setup\_item\_3}%
  \verb+[$+%
  \redtt{n\_entry}%
  \verb+] . '"';+                                                                           \\
\verb+               }+                                                                     \\
\verb+             ?>+                                                                      \\
\verb+             />+                                                                      \\
\verb+    </td>+                                                                            \\
\verb++                                                                                     \\
\verb+    <td>+                                                                             \\
\verb+      <input class="+%
  \bluett{table}%
  \verb+" type="text" name="+%
  \redtt{setup\_item\_4}%
  \verb+[<?php echo $+%
  \redtt{n\_entry}%
  \verb+; ?>]"+                                                                             \\
\verb+             .+                                                                       \\
\verb+             .+                                                                       \\
\verb+             .+                                                                       \\
\verb+    </td>+                                                                            \\
\verb++                                                                                     \\
\verb+  </tr>+                                                                              \\
\tabb
\newpage
\myitem
\label{item:table-setup_item_entry_common.php}
 {\cmbfsfx table-setup\_item\_entry\_common.php}
 (cf.~Template \ref{item:table-setup_item.php})
\deft
\verb+  <tr>+                                                 \\
\verb++                                                       \\
\verb+    <td style="width: +%
  \redtt{60}%
  \verb+pt; border-right-style: none">+                       \\
\verb+      <input class="+%
  \bluett{table}%
  \verb+" type="text" name="+%
  \redtt{setup\_entry\_common}%
  \verb+" style="width: +%
  \redtt{50}%
  \verb+pt"+                                                  \\
\verb+             onclick="this.value=''"+                   \\
\verb+             <?php+                                     \\
\verb+               if ($+%
  \redtt{setup\_entry\_common}%
  \verb+ != NULL)+                                            \\
\verb+               {+                                       \\
\verb+                 if ($+%
  \redtt{setup\_entry\_common}%
  \verb+ <= $+%
  \redtt{setup\_entry\_common\_max}%
  \verb+)+                                                    \\
\verb+                 {+                                     \\
\verb+                   echo 'value="' . $+%
  \redtt{setup\_entry\_common}%
  \verb+     . '"';+                                          \\
\verb+                 }+                                     \\
\verb++                                                       \\
\verb+                 else+                                  \\
\verb+                 {+                                     \\
\verb+                   echo 'value="' . $+%
  \redtt{setup\_entry\_common\_max}%
  \verb+ . '"';+                                              \\
\verb+                 }+                                     \\
\verb+               }+                                       \\
\verb+             ?>+                                        \\
\verb+             />+                                        \\
\verb+    </td>+                                              \\
\verb++                                                       \\
\verb+    <th colspan="+%
  \redtt{4}%
  \verb+" style="text-align: left; border-left-style: none">+ \\
\verb+        +%
  \redtt{unit of setup\_entry\_common}%
  \verb+ (max. <?php echo $+%
  \redtt{setup\_entry\_common\_max}%
  \verb+; ?>)+                                                \\
\verb+    </th>+                                              \\
\verb++                                                       \\
\verb+  </tr>+                                                \\
\tabb
\newpage
\picin{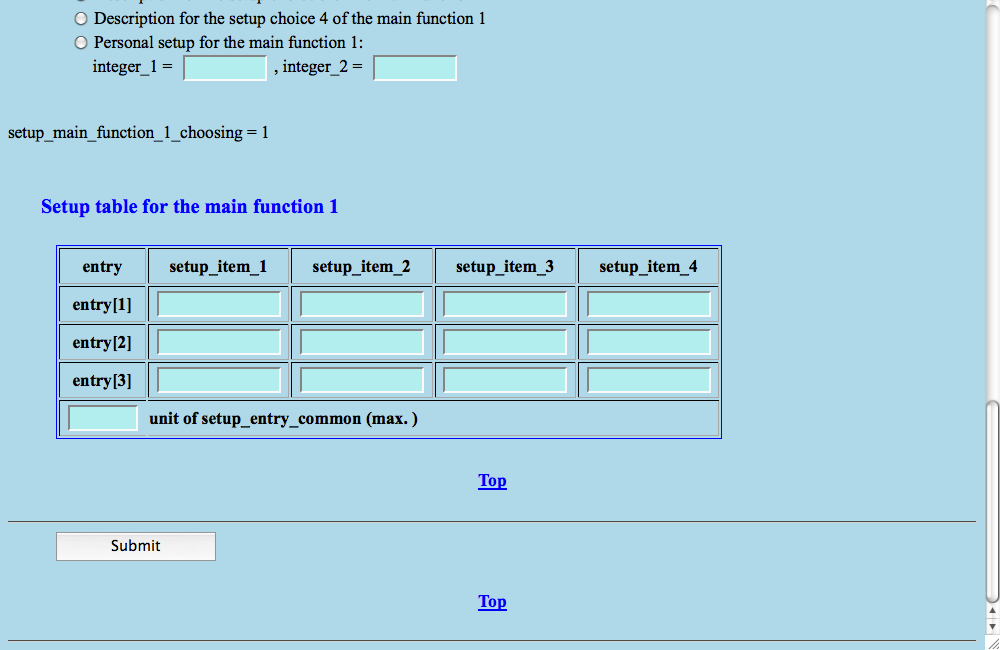}
{table-setup\_item\_entry}
{A setup table for a group of entry
 (cf.~Figure \ref{fig:table-setup_item}).}
\newpage
\myitem
\label{item:default-setup_item_head.php}
 {\cmbfsfx default-setup\_item\_head.php}
 (cf.~Template \ref{item:default-setup_item.php})
\deft
\verb+<script type="text/javascript">+                           \\
\verb+    document.getElementById("+%
  \redtt{table-setup\_item}%
  \verb+")+                                                      \\
\verb+            .style.borderColor+                            \\
\verb+  = "+%
  \bluett{red}%
  \verb+";+                                                      \\
\verb++                                                          \\
\verb+    document.getElementById("+%
  \redtt{table-setup\_item}%
  \verb+")+                                                      \\
\verb+            .border+                                       \\
\verb+  = "+%
  \bluett{2}%
  \verb+";+                                                      \\
\verb++                                                          \\
\verb+  if (navigator.appName == "Microsoft Internet Explorer")+ \\
\verb+  {+                                                       \\
\verb+      var +%
  \redtt{n\_Entry}%
  \verb++                                                        \\
\verb+    = document.getElementById("+%
  \redtt{setup\_item}%
  \verb+")+                                                      \\
\verb+              .nextSibling;+                               \\
\verb+  }+                                                       \\
\verb++                                                          \\
\verb+  else+                                                    \\
\verb+//if (navigator.userAgent.search("Chrome")    != -1 ||+    \\
\verb+//    navigator.userAgent.search("Firefox")   != -1 ||+    \\
\verb+//    navigator.userAgent.search("Opera")     != -1 ||+    \\
\verb+//    navigator.userAgent.search("Safari")    != -1 ||+    \\
\verb+//    navigator.userAgent.search("Navigator") != -1   )+   \\
\verb+  {+                                                       \\
\verb+      var +%
  \redtt{n\_Entry}%
  \verb++                                                        \\
\verb+    = document.getElementById("+%
  \redtt{setup\_item}%
  \verb+")+                                                      \\
\verb+              .nextSibling.nextSibling;+                   \\
\verb+  }+                                                       \\
\verb+</script>+                                                 \\
\tabb
\newpage
\myitem
\label{item:default-setup_item_entry.php}
 {\cmbfsfx default-setup\_item\_entry.php}
 (cf.~Template \ref{item:default-setup_item.php})
\deft
\verb+<script type="text/javascript">+                           \\
\verb+  if (navigator.appName == "Microsoft Internet Explorer")+ \\
\verb+  {+                                                       \\
\verb+      +%
  \redtt{n\_Entry}%
  \verb+.childNodes[1].childNodes[0]+                            \\
\verb+             .value+                                       \\
\verb+    =+                                                     \\
\verb+    <?php+                                                 \\
\verb+      if ($+%
  \redtt{main\_function}%
  \verb+ == "+%
  \redtt{main\_function\_1}%
  \verb+")+                                                      \\
\verb+      {+                                                   \\
\verb+        echo '"+%
  \redtt{nn}%
  \verb+"';+                                                     \\
\verb+      }+                                                   \\
\verb++                                                          \\
\verb+      else+                                                \\
\verb+      {+                                                   \\
\verb+        echo '"+%
  \redtt{nn}%
  \verb+"';+                                                     \\
\verb+      }+                                                   \\
\verb+    ?>;+                                                   \\
\verb++                                                          \\
\verb+      +%
  \redtt{n\_Entry}%
  \verb+.childNodes[2].childNodes[0]+                            \\
\verb+             .value+                                       \\
\verb+    =+                                                     \\
\verb+    <?php+                                                 \\
\verb+      if ($+%
  \redtt{n\_entry}%
  \verb+ == +%
  \redtt{1}%
  \verb+)+                                                       \\
\verb+      {+                                                   \\
\verb+        if ($+%
  \redtt{main\_function}%
  \verb+ == "+%
  \redtt{main\_function\_1}%
  \verb+"   &&+                                                  \\
\verb+            $+%
  \redtt{mode\_main\_function\_1\_choosing}%
  \verb+ == "+%
  \redtt{1}%
  \verb+"   )+                                                   \\
\verb+        {+                                                 \\
\verb+          echo '"+%
  \redtt{nn}%
  \verb+"';+                                                     \\
\verb+        }+                                                 \\
\verb++                                                          \\
\verb+        else+                                              \\
\verb+        {+                                                 \\
\verb+          echo '"+%
  \redtt{nn}%
  \verb+"';+                                                     \\
\verb+        }+                                                 \\
\verb+      }+                                                   \\
\verb++                                                          \\
\verb+      elseif ($+%
  \redtt{n\_entry}%
  \verb+ == +%
  \redtt{2}%
  \verb+ ||+                                                     \\
\verb+              $+%
  \redtt{n\_entry}%
  \verb+ == +%
  \redtt{3}%
  \verb+   )+                                                    \\
\verb+      {+                                                   \\
\verb+        if ($+%
  \redtt{main\_function}%
  \verb+ == "+%
  \redtt{main\_function\_1}%
  \verb+"        &&+                                             \\
\verb+            ($+%
  \redtt{mode\_main\_function\_1\_choosing}%
  \verb+ == "+%
  \redtt{2}%
  \verb+" ||+                                                    \\
\verb+             $+%
  \redtt{mode\_main\_function\_1\_choosing}%
  \verb+ == "+%
  \redtt{3}%
  \verb+"   )   )+                                               \\
\verb+        {+                                                 \\
\verb+          echo '"+%
  \redtt{nn}%
  \verb+"';+                                                     \\
\verb+        }+                                                 \\
\verb++                                                          \\
\verb+        else+                                              \\
\verb+        {+                                                 \\
\verb+          echo '"+%
  \redtt{nn}%
  \verb+"';+                                                     \\
\verb+        }+                                                 \\
\verb+      }+                                                   \\
\verb++                                                          \\
\verb+      else+                                                \\
\verb+      {+                                                   \\
\verb+        echo '"+%
  \redtt{nn}%
  \verb+"';+                                                     \\
\verb+      }+                                                   \\
\verb+    ?>;+                                                   \\
\defc
\verb++                                                          \\
\verb+      +%
  \redtt{n\_Entry}%
  \verb+.childNodes[3].childNodes[0]+                            \\
\verb+             .value+                                       \\
\verb+    = "+%
  \redtt{ff.ffff}%
  \verb+";+                                                      \\
\verb++                                                          \\
\verb+      +%
  \redtt{n\_Entry}%
  \verb+.childNodes[4].childNodes[0]+                            \\
\verb+             .value+                                       \\
\verb+    = "+%
  \redtt{ff.ffff}%
  \verb+";+                                                      \\
\verb++                                                          \\
\verb+      +%
  \redtt{n\_Entry}%
  \verb++                                                        \\
\verb+    = +%
  \redtt{n\_Entry}%
  \verb+.nextSibling;+                                           \\
\verb+  }+                                                       \\
\defc
\verb++                                                          \\
\verb+  else+                                                    \\
\verb+//if (navigator.userAgent.search("Chrome")    != -1 ||+    \\
\verb+//    navigator.userAgent.search("Firefox")   != -1 ||+    \\
\verb+//    navigator.userAgent.search("Opera")     != -1 ||+    \\
\verb+//    navigator.userAgent.search("Safari")    != -1 ||+    \\
\verb+//    navigator.userAgent.search("Navigator") != -1   )+   \\
\verb+  {+                                                       \\
\verb+      +%
  \redtt{n\_Entry}%
  \verb+.childNodes[3].childNodes[1]+                            \\
\verb+             .value+                                       \\
\verb+    =+                                                     \\
\verb+    <?php+                                                 \\
\verb+      if ($+%
  \redtt{main\_function}%
  \verb+ == "+%
  \redtt{main\_function\_1}%
  \verb+")+                                                      \\
\verb+      {+                                                   \\
\verb+        echo '"+%
  \redtt{nn}%
  \verb+"';+                                                     \\
\verb+      }+                                                   \\
\verb++                                                          \\
\verb+      else+                                                \\
\verb+      {+                                                   \\
\verb+        echo '"+%
  \redtt{nn}%
  \verb+"';+                                                     \\
\verb+      }+                                                   \\
\verb+    ?>;+                                                   \\
\verb++                                                          \\
\verb+      +%
  \redtt{n\_Entry}%
  \verb+.childNodes[5].childNodes[1]+                            \\
\verb+             .value+                                       \\
\verb+    =+                                                     \\
\verb+    <?php+                                                 \\
\verb+      if ($+%
  \redtt{n\_entry}%
  \verb+ == +%
  \redtt{1}%
  \verb+)+                                                       \\
\verb+      {+                                                   \\
\verb+        if ($+%
  \redtt{main\_function}%
  \verb+ == "+%
  \redtt{main\_function\_1}%
  \verb+"   &&+                                                  \\
\verb+            $+%
  \redtt{mode\_main\_function\_1\_choosing}%
  \verb+ == "+%
  \redtt{1}%
  \verb+"   )+                                                   \\
\verb+        {+                                                 \\
\verb+          echo '"+%
  \redtt{nn}%
  \verb+"';+                                                     \\
\verb+        }+                                                 \\
\verb++                                                          \\
\verb+        else+                                              \\
\verb+        {+                                                 \\
\verb+          echo '"+%
  \redtt{nn}%
  \verb+"';+                                                     \\
\verb+        }+                                                 \\
\verb+      }+                                                   \\
\verb++                                                          \\
\verb+      elseif ($+%
  \redtt{n\_entry}%
  \verb+ == +%
  \redtt{2}%
  \verb+ ||+                                                     \\
\verb+              $+%
  \redtt{n\_entry}%
  \verb+ == +%
  \redtt{3}%
  \verb+   )+                                                    \\
\verb+      {+                                                   \\
\verb+        if ($+%
  \redtt{main\_function}%
  \verb+ == "+%
  \redtt{main\_function\_1}%
  \verb+"        &&+                                             \\
\verb+            ($+%
  \redtt{mode\_main\_function\_1\_choosing}%
  \verb+ == "+%
  \redtt{2}%
  \verb+" ||+                                                    \\
\verb+             $+%
  \redtt{mode\_main\_function\_1\_choosing}%
  \verb+ == "+%
  \redtt{3}%
  \verb+"   )   )+                                               \\
\verb+        {+                                                 \\
\verb+          echo '"+%
  \redtt{nn}%
  \verb+"';+                                                     \\
\verb+        }+                                                 \\
\verb++                                                          \\
\verb+        else+                                              \\
\verb+        {+                                                 \\
\verb+          echo '"+%
  \redtt{nn}%
  \verb+"';+                                                     \\
\verb+        }+                                                 \\
\verb+      }+                                                   \\
\verb++                                                          \\
\verb+      else+                                                \\
\verb+      {+                                                   \\
\verb+        echo '"+%
  \redtt{nn}%
  \verb+"';+                                                     \\
\verb+      }+                                                   \\
\verb+    ?>;+                                                   \\
\defc
\verb++                                                          \\
\verb+      +%
  \redtt{n\_Entry}%
  \verb+.childNodes[7].childNodes[1]+                            \\
\verb+             .value+                                       \\
\verb+    = "+%
  \redtt{ff.ffff}%
  \verb+";+                                                      \\
\verb++                                                          \\
\verb+      +%
  \redtt{n\_Entry}%
  \verb+.childNodes[9].childNodes[1]+                            \\
\verb+             .value+                                       \\
\verb+    = "+%
  \redtt{ff.ffff}%
  \verb+";+                                                      \\
\verb++                                                          \\
\verb+      +%
  \redtt{n\_Entry}%
  \verb++                                                        \\
\verb+    = +%
  \redtt{n\_Entry}%
  \verb+.nextSibling.nextSibling;+                               \\
\verb+  }+                                                       \\
\verb+</script>+                                                 \\
\tabb
\newpage
\myitem
\label{item:default-setup_item_entry_common.php}
 {\cmbfsfx default-setup\_item\_entry\_common.php}
 (cf.~Template \ref{item:default-setup_item.php})
\deft
\verb+<script type="text/javascript">+                           \\
\verb+  if (navigator.appName == "Microsoft Internet Explorer")+ \\
\verb+  {+                                                       \\
\verb+      +%
  \redtt{n\_Entry}%
  \verb+.childNodes[0].childNodes[0]+                            \\
\verb+             .value+                                       \\
\verb+    = "+%
  \redtt{nn}%
  \verb+";+                                                      \\
\verb++                                                          \\
\verb+      +%
  \redtt{n\_Entry}%
  \verb++                                                        \\
\verb+    = +%
  \redtt{n\_Entry}%
  \verb+.nextSibling;+                                           \\
\verb+  }+                                                       \\
\verb++                                                          \\
\verb+  else+                                                    \\
\verb+//if (navigator.userAgent.search("Chrome")    != -1 ||+    \\
\verb+//    navigator.userAgent.search("Firefox")   != -1 ||+    \\
\verb+//    navigator.userAgent.search("Opera")     != -1 ||+    \\
\verb+//    navigator.userAgent.search("Safari")    != -1 ||+    \\
\verb+//    navigator.userAgent.search("Navigator") != -1   )+   \\
\verb+  {+                                                       \\
\verb+      +%
  \redtt{n\_Entry}%
  \verb+.childNodes[1].childNodes[1]+                            \\
\verb+             .value+                                       \\
\verb+    = "+%
  \redtt{nn}%
  \verb+";+                                                      \\
\verb++                                                          \\
\verb+      +%
  \redtt{n\_Entry}%
  \verb++                                                        \\
\verb+    = +%
  \redtt{n\_Entry}%
  \verb+.nextSibling.nextSibling;+                               \\
\verb+  }+                                                       \\
\verb+</script>+                                                 \\
\tabb
\newpage
\picdin{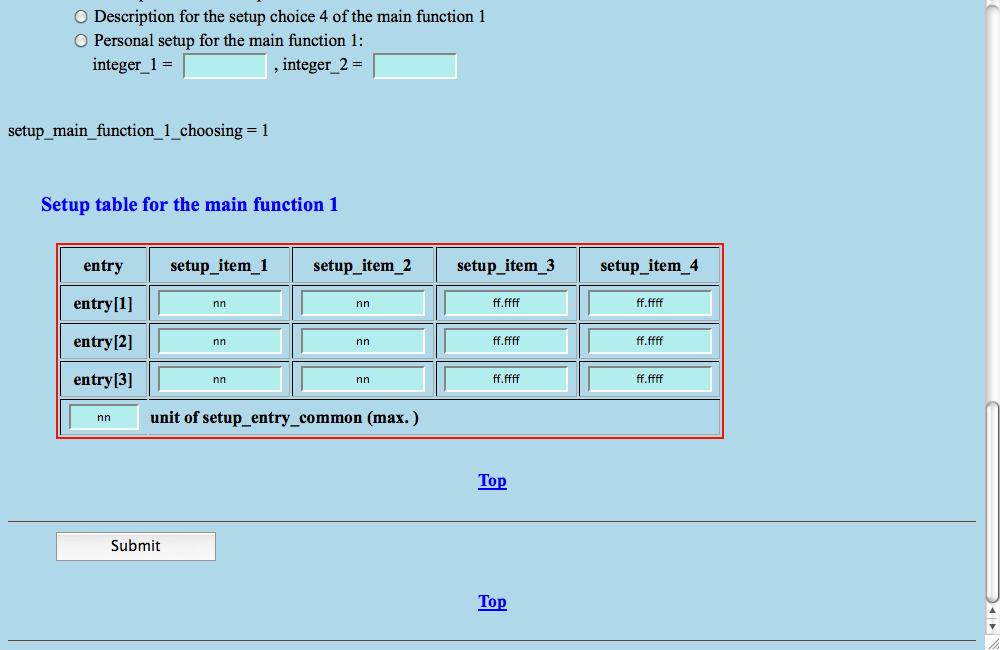}
{default-setup\_item\_entry}
{A setup table for a group of entry with default values
 (before submission,
  cf.~Figures \ref{fig:default-setup_item}
  and \ref{fig:table-setup_item_entry}).}
{A setup table for a group of entry with default values
 (after submission,
  cf.~Figures \ref{fig:default-setup_item}
  and \ref{fig:table-setup_item_entry}).}
\end{templateenumerate}

\addemptypage
\chapter{File Handling}
\label{chap:file_handling}
%
%
%
\newpage
\section{Unique file name generating}
\label{sec:unique_file_name}
%
%
\begin{templateenumerate}
\myitem
\label{item:keeping_name}
 {\bf Keeping the generated name when reloading the current page}
 (cf.~Template \ref{item:changing_name})
\deft
\verb+<?php+                                                                \\
\verb+  if ($+%
  \bluett{main\_function}%
  \verb+ != NULL  &&+                                                       \\
\verb+      $+%
  \redtt{mode\_checked}%
  \verb+  == "yes"   )+                                                     \\
\verb+  {+                                                                  \\
\verb+    /* Generate an unique folder name for uploaded files ******/+     \\
\verb+    /* Folder name will be kept                          ******/+     \\
\verb+    /* when the user "re-submits"                        ******/+     \\
\verb+    /* as well as reloads the page                       ******/+     \\
\verb++                                                                     \\
\verb+    $+%
  \bluett{personal\_file\_folder}%
  \verb+ = "+%
  \redtt{tmp/uploaded\_files}%
  \verb+";+                                                                 \\
\verb++                                                                     \\
\verb+    if ($+%
  \bluett{personal\_file\_date}%
  \verb+ == NULL ||+                                                        \\
\verb+        $+%
  \bluett{personal\_file\_time}%
  \verb+ == NULL   )+                                                       \\
\verb+    {+                                                                \\
\verb+      $+%
  \bluett{personal\_file\_date}%
  \verb+ = $_POST["+%
  \bluett{personal\_file\_date}%
  \verb+"];+                                                                \\
\verb+      $+%
  \bluett{personal\_file\_time}%
  \verb+ = $_POST["+%
  \bluett{personal\_file\_time}%
  \verb+"];+                                                                \\
\verb+    }+                                                                \\
\verb++                                                                     \\
\verb+    if ($_POST["+%
  \bluett{personal\_file\_date}%
  \verb+"] == NULL ||+                                                      \\
\verb+        $_POST["+%
  \bluett{personal\_file\_time}%
  \verb+"] == NULL   )+ \mynote[1cm]{The first time}                         \\
\verb+    {+                                                                \\
\verb+      if ($+%
  \bluett{personal\_file\_date}%
  \verb+ == NULL ||+                                                        \\
\verb+          $+%
  \bluett{personal\_file\_time}%
  \verb+ == NULL   )+                                                       \\
\verb+      {+                                                              \\
\verb+        $+%
  \bluett{personal\_file\_date}%
  \verb+ = gmdate(y) . gmdate(m) . gmdate(d);+                              \\
\verb+        $+%
  \bluett{personal\_file\_time}%
  \verb+ = gmdate(H) . gmdate(i) . gmdate(s);+                              \\
\verb+      }+                                                              \\
\verb+    }+                                                                \\
\verb++                                                                     \\
\verb+    echo '<input type="hidden"'                            . "\r\n";+ \\
\verb+    echo '       name="+%
  \bluett{personal\_file\_date}%
  \verb+"'                . "\r\n";+                                        \\
\verb+    echo '       value="' . $+%
  \bluett{personal\_file\_date}%
  \verb+ . '"'      . "\r\n";+                                              \\
\verb+    echo '       />'                                       . "\r\n";+ \\
\verb++ \mynote[10.3cm]{Submit the fixed date/time}                         \\
\verb+    echo '<input type="hidden"'                            . "\r\n";+ \\
\verb+    echo '       name="+%
  \bluett{personal\_file\_time}%
  \verb+"'                . "\r\n";+                                        \\
\verb+    echo '       value="' . $+%
  \bluett{personal\_file\_time}%
  \verb+ . '"'      . "\r\n";+                                              \\
\verb+    echo '       />'                                       . "\r\n";+ \\
\verb++                                                                     \\
\verb+    echo '+%
  \bluett{personal\_file\_date}%
  \verb+   = ' . $+%
  \bluett{personal\_file\_date}%
  \verb+   . '<br />';+                                                     \\
\verb+    echo '+%
  \bluett{personal\_file\_time}%
  \verb+   = ' . $+%
  \bluett{personal\_file\_time}%
  \verb+   . '<br />';+                                                     \\
\verb+    echo '+%
  \bluett{personal\_file\_folder}%
  \verb+ = ' . $+%
  \bluett{personal\_file\_folder}%
  \verb+ . '<br />';+                                                       \\
\verb+  }+                                                                  \\
\verb+?>+                                                                   \\
\tabb
\newpage
\picdin{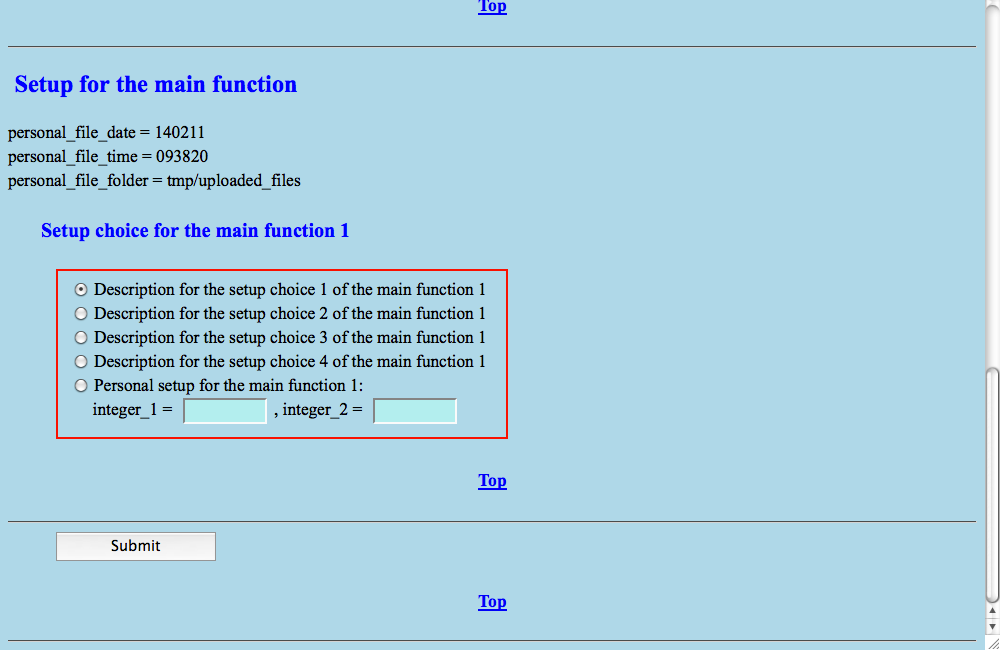}
{keeping\_name}
{Keeping the generated name when reloading the current page (before submission).}
{Keeping the generated name when reloading the current page (after submission).}
\newpage
\myitem
\label{item:changing_name}
 {\bf Changing the generated name when reloading the current page}
 (cf.~Template \ref{item:keeping_name})
\deft
\verb+<?php+                                                                \\
\verb+  if ($+%
  \bluett{main\_function}%
  \verb+ != NULL  &&+                                                       \\
\verb+      $+%
  \redtt{mode\_checked}%
  \verb+  == "yes" &&+                                                      \\
\verb+      $+%
  \redtt{setup\_checked}%
  \verb+ == "yes"   )+                                                      \\
\verb+  {+                                                                  \\
\verb+    /* Generate an unique file name for output plots **********/+     \\
\verb+    /* File name                                       ********/+     \\
\verb+    /* will be changed when the user "re-submits",       ******/+     \\
\verb+    /* while will be kept when the user "reloads" the page ****/+     \\
\verb++                                                                     \\
\verb+    $+%
  \bluett{plot\_output\_folder}%
  \verb+ = "+%
  \redtt{tmp/output\_plots}%
  \verb+";+                                                                 \\
\verb++                                                                     \\
\verb+    if ($+%
  \bluett{plot\_output\_date}%
  \verb+ == NULL ||+                                                        \\
\verb+        $+%
  \bluett{plot\_output\_time}%
  \verb+ == NULL   )+                                                       \\
\verb+    {+                                                                \\
\verb+      $+%
  \bluett{plot\_output\_date}%
  \verb+ = $_POST["+%
  \bluett{plot\_output\_date}%
  \verb+"];+                                                                \\
\verb+      $+%
  \bluett{plot\_output\_time}%
  \verb+ = $_POST["+%
  \bluett{plot\_output\_time}%
  \verb+"];+                                                                \\
\verb+    }+                                                                \\
\verb++                                                                     \\
\verb+      $+%
  \bluett{plot\_output\_file}%
  \verb++                                                                   \\
\verb+    = +%
  \bluett{"plot-" . \$plot\_output\_date . "-" . \$plot\_output\_time}%
  \verb+;+                                                                  \\
\verb++                                                                     \\
\verb+    $+%
  \bluett{plot\_output\_date\_tmp}%
  \verb+ = gmdate(y) . gmdate(m) . gmdate(d);+                              \\
\verb+    $+%
  \bluett{plot\_output\_time\_tmp}%
  \verb+ = gmdate(H) . gmdate(i) . gmdate(s);+ \mynote[1cm]{Generate new date/time} \\
\verb++                                                                     \\
\verb+    echo '<input type="hidden"'                            . "\r\n";+ \\
\verb+    echo '       name="+%
  \bluett{plot\_output\_date}%
  \verb+"'                  . "\r\n";+                                      \\
\verb+    echo '       value="' . $+%
  \bluett{plot\_output\_date\_tmp}%
  \verb+ . '"'    . "\r\n";+                                                \\
\verb+    echo '       />'                                       . "\r\n";+ \\
\verb++ \mynote[10.3cm]{Submit the new date/time}                           \\
\verb+    echo '<input type="hidden"'                            . "\r\n";+ \\
\verb+    echo '       name="+%
  \bluett{plot\_output\_time}%
  \verb+"'                  . "\r\n";+                                      \\
\verb+    echo '       value="' . $+%
  \bluett{plot\_output\_time\_tmp}%
  \verb+ . '"'    . "\r\n";+                                                \\
\verb+    echo '       />'                                       . "\r\n";+ \\
\verb++                                                                     \\
\verb+    echo '+%
  \bluett{plot\_output\_date}%
  \verb+     = ' . $+%
  \bluett{plot\_output\_date}%
  \verb+     . '<br />';+                                                   \\
\verb+    echo '+%
  \bluett{plot\_output\_date\_tmp}%
  \verb+ = ' . $+%
  \bluett{plot\_output\_date\_tmp}%
  \verb+ . '<br />';+                                                       \\
\verb++                                                                     \\
\verb+    echo '+%
  \bluett{plot\_output\_time}%
  \verb+     = ' . $+%
  \bluett{plot\_output\_time}%
  \verb+     . '<br />';+                                                   \\
\verb+    echo '+%
  \bluett{plot\_output\_time\_tmp}%
  \verb+ = ' . $+%
  \bluett{plot\_output\_time\_tmp}%
  \verb+ . '<br />';+                                                       \\
\verb++                                                                     \\
\verb+    echo '<br />';+                                                   \\
\verb+    echo '+%
  \bluett{plot\_output\_folder}%
  \verb+   = ' . $+%
  \bluett{plot\_output\_folder}%
  \verb+   . '<br />';+                                                     \\
\verb+    echo '+%
  \bluett{plot\_output\_file}%
  \verb+     = ' . $+%
  \bluett{plot\_output\_file}%
  \verb+     . '<br />';+                                                   \\
\verb+  }+                                                                  \\
\verb+?>+                                                                   \\
\tabb
\newpage
\picqin{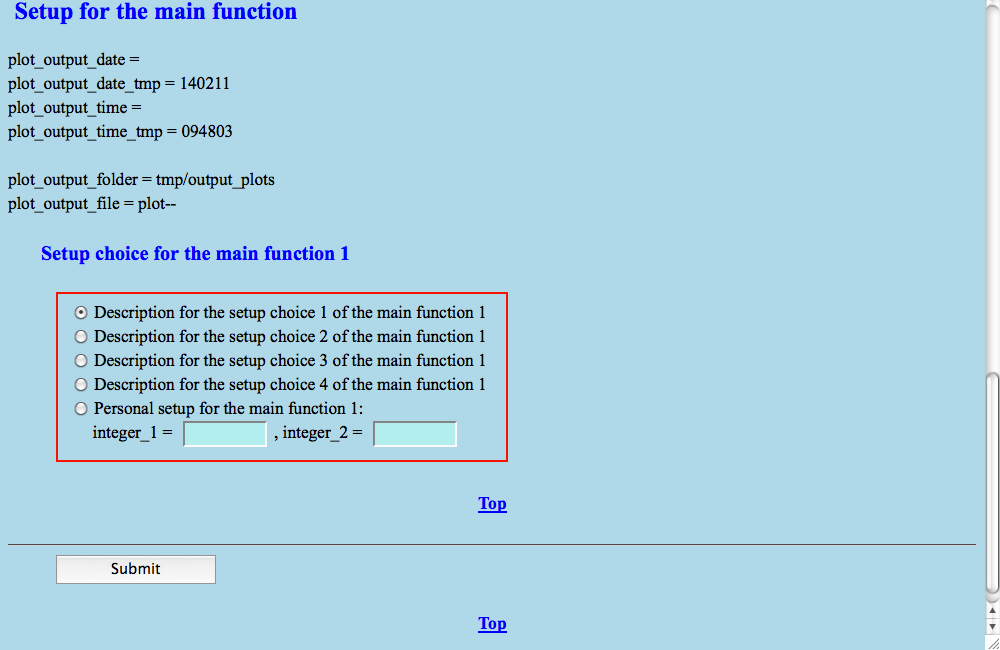}
{changing\_name}
{Changing the generated name when reloading the current page
 (generating the name).}
{Changing the generated name when reloading the current page
 (after submission).}
{Changing the generated name when reloading the current page
 (after reloading).}
{Changing the generated name when reloading the current page
 (after re--submission).}
\end{templateenumerate}
\newpage
\section{File uploading}
\label{sec:file_uploading}
%
%
\begin{templateenumerate}
\myitem
\label{item:modification_table-setup_item_entry}
 {\bf Modification of the `setup table for a group of entry'}
 (cf.~Template \ref{item:table-setup_item_entry})
\deft
\verb+<?php+                                   \\
\verb+  if ($+%
  \redtt{setup\_main\_function\_1\_checked}%
  \verb+ == "yes")+                            \\
\verb+  {+                                     \\
\verb+    echo '<h4> +%
  \redtt{Setup table for the main function 1}%
  \verb+ </h4>' . "\r\n";+                     \\
\verb++                                        \\
\verb+    $+%
  \redtt{entry\_No}%
  \verb+ = +%
  \redtt{3}%
  \verb+;+                                     \\
\verb++                                        \\
\verb+    for ($+%
  \redtt{n\_entry}%
  \verb+ = 1; $+%
  \redtt{n\_entry}%
  \verb+ <= $+%
  \redtt{entry\_No}%
  \verb+; $+%
  \redtt{n\_entry}%
  \verb- ++)-                                  \\
\verb+    {+                                   \\
\verb+      if ($+%
  \redtt{setup\_item\_1}%
  \verb+[$+%
  \redtt{n\_entry}%
  \verb+] == NULL)+                            \\
\verb+      {+                                 \\
\verb+        $+%
  \redtt{setup\_item\_1}%
  \verb+[$+%
  \redtt{n\_entry}%
  \verb+] = $_POST["+%
  \redtt{setup\_item\_1}%
  \verb+"][$+%
  \redtt{n\_entry}%
  \verb+];+                                    \\
\verb+      }+                                 \\
\verb++                                        \\
\verb+          .+                             \\
\verb+          .+                             \\
\verb+          .+                             \\
\verb++                                        \\
\verb+      if ($+%
  \redtt{setup\_item\_4}%
  \verb+[$+%
  \redtt{n\_entry}%
  \verb+] == NULL)+                            \\
\verb+      {+                                 \\
\verb+        $+%
  \redtt{setup\_item\_4}%
  \verb+[$+%
  \redtt{n\_entry}%
  \verb+] = $_POST["+%
  \redtt{setup\_item\_4}%
  \verb+"][$+%
  \redtt{n\_entry}%
  \verb+];+                                    \\
\verb+      }+                                 \\
\verb++                                        \\
\verb+        $+%
  \redtt{personal\_file\_copy}%
  \verb+[$+%
  \redtt{n\_entry}%
  \verb+]+                                     \\
\verb+      = +%
  \bluett{"pf-" . \$personal\_file\_date . "-" . \$personal\_file\_time . "-" . \$}%
  \verb++%
  \redtt{n\_entry}%
  \verb++%
  \bluett{ . ".txt";}%
  \verb++                                      \\
\verb++                                        \\
\verb+      if ($_FILES["+%
  \redtt{personal\_file}%
  \verb+"]["name"][$+%
  \redtt{n\_entry}%
  \verb+] != NULL)+                            \\
\verb+      {+                                 \\
\verb+        $+%
  \redtt{personal\_file}%
  \verb+[$+%
  \redtt{n\_entry}%
  \verb+] = $_FILES["+%
  \redtt{personal\_file}%
  \verb+"]["name"][$+%
  \redtt{n\_entry}%
  \verb+];+ \mynote[0.5cm]{See Sec.~\ref{sec:basic_php_arrays}} \\
\verb++                                        \\
\verb+        move_uploaded_file($_FILES["+%
  \redtt{personal\_file}%
  \verb+"]["tmp_name"][$+%
  \redtt{n\_entry}%
  \verb+],+                                    \\
\verb+                           "$+%
  \bluett{personal\_file\_folder}%
  \verb+/" . $+%
  \redtt{personal\_file\_copy}%
  \verb+[$+%
  \redtt{n\_entry}%
  \verb+]);+                                   \\
\verb+      }+                                 \\
\verb++                                        \\
\verb+      elseif ($_POST["+%
  \redtt{personal\_file}%
  \verb+"][$+%
  \redtt{n\_entry}%
  \verb+] != NULL)+                            \\
\verb+      {+                                 \\
\verb+        $+%
  \redtt{personal\_file}%
  \verb+[$+%
  \redtt{n\_entry}%
  \verb+] = $_POST["+%
  \redtt{personal\_file}%
  \verb+"][$+%
  \redtt{n\_entry}%
  \verb+];+                                    \\
\verb+      }+                                 \\
\verb++                                        \\
\verb+      if (file_exists("$+%
  \bluett{personal\_file\_folder}%
  \verb+/$+%
  \redtt{personal\_file\_copy}%
  \verb+[$+%
  \redtt{n\_entry}%
  \verb+]"))+                                  \\
\verb+      {+                                 \\
\verb+        echo '<input type="hidden"'                            . "\r\n";+ \\
\verb+        echo '       name="+%
  \redtt{personal\_file}%
  \verb+[' . $+%
  \redtt{n\_entry}%
  \verb+ . ']"'   . "\r\n";+                   \\
\verb+        echo '       value="' . $+%
  \redtt{personal\_file}%
  \verb+[$+%
  \redtt{n\_entry}%
  \verb+] . '"' . "\r\n";+                     \\
\verb+        echo '       />'                                       . "\r\n";+ \\
\verb+      }+                                 \\
\verb+    }+                                   \\
\verb++                                        \\
\verb+    if ($+%
  \redtt{setup\_entry\_common}%
  \verb+ == NULL)+                             \\
\verb+    {+                                   \\
\verb+      $+%
  \redtt{setup\_entry\_common}%
  \verb+ = $_POST["+%
  \redtt{setup\_entry\_common}%
  \verb+"];+                                   \\
\verb+    }+                                   \\
\defc
\verb++                                        \\
\verb+    for ($+%
  \redtt{n\_entry}%
  \verb+ = 1; $+%
  \redtt{n\_entry}%
  \verb+ <= $+%
  \redtt{entry\_No}%
  \verb+; $+%
  \redtt{n\_entry}%
  \verb- ++)-                                  \\
\verb+    {+                                   \\
\verb+      if ($+%
  \redtt{personal\_file}%
  \verb+[$+%
  \redtt{n\_entry}%
  \verb+] != NULL)+                            \\
\verb+      {+                                 \\
\verb+        +%
  \redtt{\$personal\_files = "yes"}%
  \verb+;+                                     \\
\verb++                                        \\
\verb+        break;+                          \\
\verb+      }+                                 \\
\verb+    }+                                   \\
\verb++                                        \\
\verb+    include("+%
  \bluett{tables/table-}%
  \verb++%
  \redtt{setup\_item\_head}%
  \verb+.php");+ \mynote{Template \ref{item:modification_table-setup_item_head.php}} \\
\verb+    for ($+%
  \redtt{n\_entry}%
  \verb+ = 1; $+%
  \redtt{n\_entry}%
  \verb+ <= $+%
  \redtt{entry\_No}%
  \verb+; $+%
  \redtt{n\_entry}%
  \verb- ++)-                                  \\
\verb+    {+                                   \\
\verb+      include("+%
  \bluett{tables/table-}%
  \verb++%
  \redtt{setup\_item\_entry}%
  \verb+.php");+ \mynote[1cm]{Template \ref{item:modification_table-setup_item_entry.php}} \\
\verb+    }+                                   \\
\verb+    include("+%
  \bluett{tables/table-}%
  \verb++%
  \redtt{setup\_item\_entry\_common}%
  \verb+.php");+ \mynote[1cm]{Template \ref{item:modification_table-setup_item_entry_common.php}} \\
\verb+    echo '</table>' . "\r\n";+           \\
\verb++                                        \\
\verb+    for ($+%
  \redtt{n\_entry}%
  \verb+ = 1; $+%
  \redtt{n\_entry}%
  \verb+ <= $+%
  \redtt{entry\_No}%
  \verb+; $+%
  \redtt{n\_entry}%
  \verb- ++)-                                  \\
\verb+    {+                                   \\
\verb+      if ($+%
  \redtt{setup\_item\_1}%
  \verb+[$+%
  \redtt{n\_entry}%
  \verb+] == NULL                                            ||+    \\
\verb+          $+%
  \redtt{setup\_item\_2}%
  \verb+[$+%
  \redtt{n\_entry}%
  \verb+] == NULL                                            ||+    \\
\verb+          $+%
  \redtt{setup\_item\_3}%
  \verb+[$+%
  \redtt{n\_entry}%
  \verb+] == NULL                                            ||+    \\
\verb+          $+%
  \redtt{setup\_item\_4}%
  \verb+[$+%
  \redtt{n\_entry}%
  \verb+] == NULL                                            ||+    \\
\verb+          $+%
  \redtt{setup\_entry\_common}%
  \verb+     == NULL                                            ||+ \\
\verb+          (+%
  \redtt{\$personal\_files        == "yes"}%
  \verb+                                              &&+           \\
\verb+           !(file_exists("$+%
  \bluett{personal\_file\_folder}%
  \verb+/$+%
  \redtt{personal\_file\_copy}%
  \verb+[$+%
  \redtt{n\_entry}%
  \verb+]"))   )   )+                                               \\
\verb+      {+ \mynote[10.5cm]{Check uploaded file}                 \\
\verb+        include("+%
  \bluett{default\_choices/default-}%
  \verb++%
  \redtt{setup\_item\_head}%
  \verb+.php");+ \mynote[1cm]{Template \ref{item:default-setup_item_head.php}} \\
\verb+        for ($+%
  \redtt{n\_entry}%
  \verb+ = 1; $+%
  \redtt{n\_entry}%
  \verb+ <= $+%
  \redtt{entry\_No}%
  \verb+; $+%
  \redtt{n\_entry}%
  \verb- ++)-                                  \\
\verb+        {+                               \\
\verb+          include("+%
  \bluett{default\_choices/default-}%
  \verb++%
  \redtt{setup\_item\_entry}%
  \verb+.php");+ \mynote[1cm]{Template \ref{item:default-setup_item_entry.php}} \\
\verb+        }+                               \\
\verb+        include("+%
  \bluett{default\_choices/default-}%
  \verb++%
  \redtt{setup\_item\_entry\_common}%
  \verb+.php");+ \mynote[1cm]{Template \ref{item:default-setup_item_entry_common.php}} \\
\verb++                                        \\
\verb+        break;+                          \\
\verb+      }+                                 \\
\verb++ \mynote[5cm]{cf.~Template \ref{item:setup_setting}} \\
\verb+      else+                              \\
\verb+      {+                                 \\
\verb+        echo '<br />';+                  \\
\verb+        echo '+%
  \redtt{setup\_item\_1}%
  \verb+[' . $+%
  \redtt{n\_entry}%
  \verb+ . '] = '  . $+%
  \redtt{setup\_item\_1}%
  \verb+[$+%
  \redtt{n\_entry}%
  \verb+]  . '<br />';+                        \\
\verb+        echo '+%
  \redtt{setup\_item\_2}%
  \verb+[' . $+%
  \redtt{n\_entry}%
  \verb+ . '] = '  . $+%
  \redtt{setup\_item\_2}%
  \verb+[$+%
  \redtt{n\_entry}%
  \verb+]  . '<br />';+                        \\
\verb+        echo '+%
  \redtt{setup\_item\_3}%
  \verb+[' . $+%
  \redtt{n\_entry}%
  \verb+ . '] = '  . $+%
  \redtt{setup\_item\_3}%
  \verb+[$+%
  \redtt{n\_entry}%
  \verb+]  . '<br />';+                        \\
\verb+        echo '+%
  \redtt{setup\_item\_4}%
  \verb+[' . $+%
  \redtt{n\_entry}%
  \verb+ . '] = '  . $+%
  \redtt{setup\_item\_4}%
  \verb+[$+%
  \redtt{n\_entry}%
  \verb+]  . '<br />';+                        \\
\verb++                                        \\
\verb+        echo '+%
  \redtt{personal\_file}%
  \verb+[' . $+%
  \redtt{n\_entry}%
  \verb+ . '] = ' . $+%
  \redtt{personal\_file}%
  \verb+[$+%
  \redtt{n\_entry}%
  \verb+] . '<br />';+                         \\
\verb+      }+                                 \\
\verb++                                        \\
\verb+      if ($+%
  \redtt{n\_entry}%
  \verb+ == $+%
  \redtt{entry\_No}%
  \verb+)+                                     \\
\verb+      {+                                 \\
\verb+        echo '<br />';+                  \\
\verb+        echo '+%
  \redtt{setup\_entry\_common}%
  \verb+ = '              . $+%
  \redtt{setup\_entry\_common}%
  \verb+      . '<br />';+                     \\
\verb+        echo '<br />';+                  \\
\verb++                                        \\
\verb+        $+%
  \redtt{setup\_item\_checked}%
  \verb+ = "yes";+                             \\
\verb+      }+                                 \\
\verb+    }+                                   \\
\verb+  }+                                     \\
\verb+?>+                                      \\
\tabb
\newpage
\myitem
\label{item:modification_table-setup_item_head.php}
 {\bf Modification of {\cmbfsfx table-setup\_item\_head.php}}
 (cf.~Template \ref{item:table-setup_item_head.php})
\deft
\verb+<table id="+%
  \redtt{table-setup\_item}%
  \verb+" border="1">+                           \\
\verb++                                          \\
\verb+  <tr id="+%
  \redtt{setup\_item}%
  \verb+">+                                      \\
\verb++                                          \\
\verb+    <th style="width: +%
  \redtt{60}%
  \verb+pt"> +%
  \redtt{entry}%
  \verb+ </th>+                                  \\
\verb++                                          \\
\verb+    <th> +%
  \redtt{setup\_item\_1}%
  \verb+ </th>+                                  \\
\verb+    <th> +%
  \redtt{setup\_item\_2}%
  \verb+ </th>+                                  \\
\verb+    <th> +%
  \redtt{setup\_item\_3}%
  \verb+ </th>+                                  \\
\verb+    <th> +%
  \redtt{setup\_item\_4}%
  \verb+ </th>+                                  \\
\verb++                                          \\
\verb+    <th style="width: +%
  \redtt{150}%
  \verb+pt"> +%
  \bluett{Personal files (max.~2 MB)}%
  \verb+ </th>+                                  \\
\verb++                                          \\
\verb+    <?php+                                 \\
\verb+      if (+%
  \redtt{\$personal\_files == "yes"}%
  \verb+)+ \mynote[0.5cm]{See Template \ref{item:modification_table-setup_item_entry}} \\
\verb+      {+                                   \\
\verb+//      for ($+%
  \redtt{n\_entry}%
  \verb+ = 1; $+%
  \redtt{n\_entry}%
  \verb+ <= $+%
  \redtt{entry\_No}%
  \verb+; $+%
  \redtt{n\_entry}%
  \verb- ++)-                                    \\
\verb+//      {+                                 \\
\verb+//        if ($+%
  \redtt{personal\_file}%
  \verb+[$+%
  \redtt{n\_entry}%
  \verb+] != NULL)+                              \\
\verb+//        {+                               \\
\verb+            echo '    <th style="width: +%
  \redtt{150}%
  \verb+pt"> +%
  \bluett{Uploaded files}%
  \verb+ </th>';+                                \\
\verb++                                          \\
\verb+//          break;+                        \\
\verb+//        }+                               \\
\verb+//      }+                                 \\
\verb+      }+                                   \\
\verb+    ?>+                                    \\
\verb++                                          \\
\verb+  </tr>+                                   \\
\tabb
\newpage
\myitem
\label{item:modification_table-setup_item_entry.php}
 {\bf Modification of {\cmbfsfx table-setup\_item\_entry.php}}
 (cf.~Template \ref{item:table-setup_item_entry.php})
\deft
\verb+  <tr>+                                                                             \\
\verb++                                                                                   \\
\verb+    <th style="width: +%
  \redtt{60}%
  \verb+pt">+                                                                             \\
\verb+      <?php echo '+%
  \redtt{entry}%
  \verb+[' . $+%
  \redtt{n\_entry}%
  \verb+ . ']'; ?>+                                                                       \\
\verb+    </th>+                                                                          \\
\verb++                                                                                   \\
\verb+       .+                                                                           \\
\verb+       .+                                                                           \\
\verb+       .+                                                                           \\
\verb++                                                                                   \\
\verb+    <td>+                                                                           \\
\verb+      <input class="+%
  \bluett{table}%
  \verb+" type="text" name="+%
  \redtt{setup\_item\_4}%
  \verb+[<?php echo $+%
  \redtt{n\_entry}%
  \verb+; ?>]"+                                                                           \\
\verb+             .+                                                                     \\
\verb+             .+                                                                     \\
\verb+             .+                                                                     \\
\verb++                                                                                   \\
\verb+    </td>+                                                                          \\
%
%
\verb++                                                                                   \\
\verb+    <td>+                                                                           \\
\verb+      <input class="+%
  \bluett{table}%
  \verb+" type="file" name="+%
  \redtt{personal\_file}%
  \verb+[<?php echo $+%
  \redtt{n\_entry}%
  \verb+; ?>]"+                                                                           \\
\verb+             style="height: +%
  \bluett{19.2}%
  \verb+pt; width: +%
  \bluett{150}%
  \verb+pt; text-align: left"+                                                            \\
\verb+             />+                                                                    \\
\verb+    </td>+                                                                          \\
\verb++                                                                                   \\
\verb+    <?php+                                                                          \\
\verb+      if (+%
  \redtt{\$personal\_files == "yes"}%
  \verb+)+                                                                                \\
\verb+      {+                                                                            \\
\verb+        echo '    <td style="text-align: left; width: +%
  \bluett{140}%
  \verb+pt;'            . "\r\n";+                                                        \\
\verb+        echo '               padding-left: +%
  \bluett{3}%
  \verb+pt; padding-right: +%
  \bluett{3}%
  \verb+pt">'    . "\r\n";+                                                               \\
\verb++                                                                                   \\
\verb+        if ($+%
  \redtt{personal\_file}%
  \verb+[$+%
  \redtt{n\_entry}%
  \verb+] != NULL)+                                                                       \\
\verb+        {+                                                                          \\

\verb+          if (file_exists("$+%
  \bluett{personal\_file\_folder}%
  \verb+/$+%
  \redtt{personal\_file\_copy}%
  \verb+[$+%
  \redtt{n\_entry}%
  \verb+]"))+                                                                             \\
\verb+          {+                                                                        \\
\verb+            echo $+%
  \redtt{personal\_file}%
  \verb+[$+%
  \redtt{n\_entry}%
  \verb+];+                                                                               \\
\verb++                                                                                   \\
\verb+//          echo '      <input type="hidden"'                            . "\r\n";+ \\
\verb+//          echo '             name="+%
  \redtt{personal\_file}%
  \verb+[' . $+%
  \redtt{n\_entry}%
  \verb+ . ']"'   . "\r\n";+                                                              \\
\verb+//          echo '             value="' . $+%
  \redtt{personal\_file}%
  \verb+[$+%
  \redtt{n\_entry}%
  \verb+] . '"' . "\r\n";+                                                                \\
\verb+//          echo '             />'                                       . "\r\n";+ \\
\verb+          }+ \mynote[5cm]{Moved to Template \ref{item:modification_table-setup_item_entry}} \\
\verb++                                                                                   \\
\verb+          else+                                                                     \\
\verb+          {+                                                                        \\
\verb+            echo '<span class="+%
  \bluett{red}%
  \verb+">The file size is too large!</span>';+                                           \\
\verb+          }+                                                                        \\
\verb+        }+                                                                          \\
\verb++                                                                                   \\
\verb+        else+                                                                       \\
\verb+        {+                                                                          \\
\verb+          echo '&nbsp;';+                                                           \\
\verb+        }+                                                                          \\
\verb++                                                                                   \\
\verb+        echo '</td>';+                                                              \\
\verb+      }+                                                                            \\
\verb+    ?>+                                                                             \\
\verb++                                                                                   \\
\verb+  </tr>+                                                                            \\
\tabb
\newpage
\myitem
\label{item:modification_table-setup_item_entry_common.php}
 {\bf Modification of {\cmbfsfx table-setup\_item\_entry\_common.php}}
 (cf.~Template \ref{item:table-setup_item_entry_common.php})
\deft
\verb+  <tr>+                                                     \\
\verb++                                                           \\
\verb+    <td style="width: +%
  \redtt{60}%
  \verb+pt; border-right-style: none">+                           \\
\verb+      <input class="+%
  \bluett{table}%
  \verb+" type="text" name="+%
  \redtt{setup\_entry\_common}%
  \verb+" style="width: +%
  \redtt{50}%
  \verb+pt"+                                                      \\
\verb+             onclick="this.value=''"+                       \\
\verb+             <?php+                                         \\
\verb+               if ($+%
  \redtt{setup\_entry\_common}%
  \verb+ != NULL)+                                                \\
\verb+               {+                                           \\
\verb+                 if ($+%
  \redtt{setup\_entry\_common}%
  \verb+ <= $+%
  \redtt{setup\_entry\_common\_max}%
  \verb+)+                                                        \\
\verb+                 {+                                         \\
\verb+                   echo 'value="' . $+%
  \redtt{setup\_entry\_common}%
  \verb+     . '"';+                                              \\
\verb+                 }+                                         \\
\verb++                                                           \\
\verb+                 else+                                      \\
\verb+                 {+                                         \\
\verb+                   echo 'value="' . $+%
  \redtt{setup\_entry\_common\_max}%
  \verb+ . '"';+                                                  \\
\verb+                 }+                                         \\
\verb+               }+                                           \\
\verb+             ?>+                                            \\
\verb+             />+                                            \\
\verb+    </td>+                                                  \\
\verb++                                                           \\
\verb+    <th colspan=+                                           \\
\verb+        <?php+                                              \\
\verb+          if (+%
  \redtt{\$personal\_files == "yes"}%
  \verb+)+                                                        \\
\verb+          {+                                                \\
\verb+            echo '"+%
  \redtt{6}%
  \verb+"';+ \mynote[0.5cm]{Two columns more}                     \\
\verb+          }+                                                \\
\verb++                                                           \\
\verb+          else+                                             \\
\verb+          {+                                                \\
\verb+            echo '"+%
  \redtt{5}%
  \verb+"';+ \mynote[0.5cm]{One column more}                      \\
\verb+          }+                                                \\
\verb+        ?>+                                                 \\
\verb+        style="text-align: left; border-left-style: none">+ \\
\verb+        +%
  \redtt{unit of setup\_entry\_common}%
  \verb+ (max. +%
  \redtt{setup\_entry\_common\_max}%
  \verb+)+                                                        \\
\verb+    </th>+                                                  \\
\verb++                                                           \\
\verb+  </tr>+                                                    \\
\tabb
\newpage
\picdin{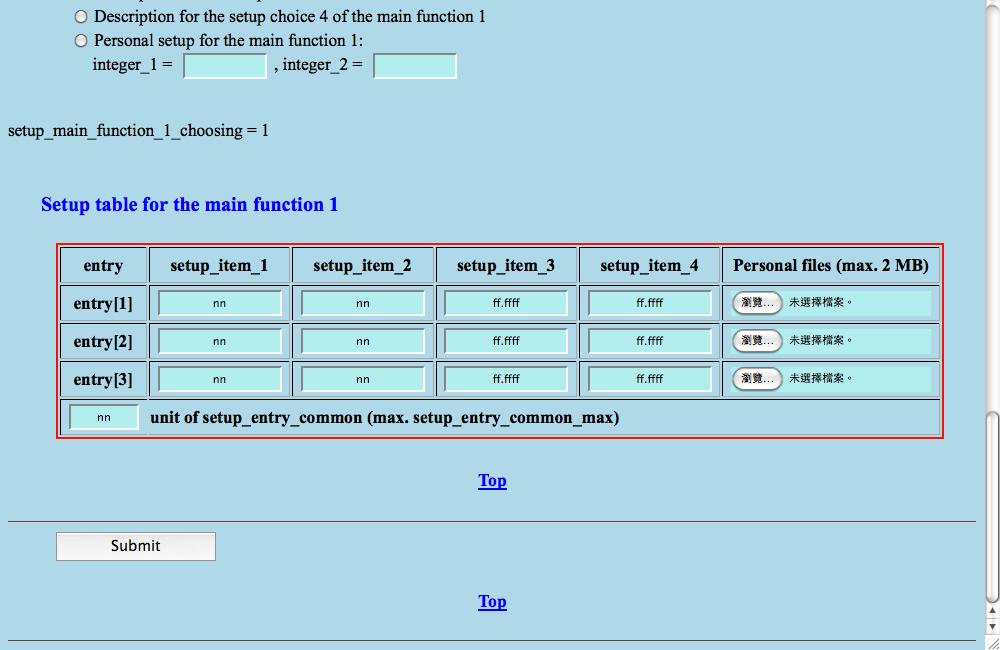}
{modification\_default-setup\_item\_entry}
{Modification of the setup table for uploading files
 (before submission, 
  cf.~Figure \ref{fig:default-setup_item_entry-1}).}
{Modification of setup table for uploading files
 (after submission, 
  cf.~Figure \ref{fig:default-setup_item_entry-2}).}
\newpage
\myitem
\label{item:uploading_typing}
 {\bf Uploading two files or typing texts}
 (cf.~Template \ref{item:modification_table-setup_item_entry})
\deft
\verb+<?php+                                                                                     \\
\verb+  if ($+%
  \bluett{main\_function}%
  \verb+ == "+%
  \redtt{main\_function\_1}%
  \verb+"  &&+                                                                                   \\
\verb+      $+%
  \redtt{mode\_main\_function\_1\_choice}%
  \verb+ != NULL   )+                                                                            \\
\verb+  {+                                                                                       \\
\verb+    echo '<h4> Function choice for the +%
  \redtt{main function 1}%
  \verb+ </h4>' . "\r\n";+                                                                       \\
\verb++                                                                                          \\
\verb+    if ($+%
  \redtt{function\_choosing}%
  \verb+ == NULL)+                                                                               \\
\verb+    {+                                                                                     \\
\verb+      $+%
  \redtt{function\_choosing}%
  \verb+ = $_POST["+%
  \redtt{function\_choosing}%
  \verb+"];+                                                                                     \\
\verb+    }+                                                                                     \\
\verb++                                                                                          \\
\verb+    if ($+%
  \redtt{personal\_function\_text}%
  \verb+ == NULL)+                                                                               \\
\verb+    {+                                                                                     \\
\verb+      $+%
  \redtt{personal\_function\_text}%
  \verb+ = $_POST["+%
  \redtt{personal\_function\_text}%
  \verb+"];+                                                                                     \\
\verb+    }+                                                                                     \\
\verb++                                                                                          \\
\verb+      $+%
  \redtt{personal\_function\_file\_copy}%
  \verb++                                                                                        \\
\verb+    = +%
  \bluett{"pf-" . \$personal\_file\_date . "-" . \$personal\_file\_time . "-copy.txt"}%
  \verb+;+                                                                                       \\
\verb++                                                                                          \\
\verb+    if ($_FILES["+%
  \redtt{personal\_function\_file}%
  \verb+"]["name"] != NULL)+ \mynote[0.5cm]{Just after file 1 uploading}                         \\
\verb+    {+                                                                                     \\
\verb+      $+%
  \redtt{personal\_function\_file}%
  \verb+ = $_FILES["+%
  \redtt{personal\_function\_file}%
  \verb+"]["name"];+                                                                             \\
\verb++                                                                                          \\
\verb+      move_uploaded_file($_FILES["+%
  \redtt{personal\_function\_file}%
  \verb+"]["tmp_name"],+                                                                         \\
\verb+                         "$+%
  \bluett{personal\_file\_folder}%
  \verb+/" . $+%
  \redtt{personal\_function\_file\_copy}%
  \verb+);+                                                                                      \\
\verb+    }+                                                                                     \\
\verb++                                                                                          \\
\verb+    elseif ($_POST["+%
  \redtt{personal\_function\_file}%
  \verb+"] != NULL)+ \mynote[0.5cm]{File 1 not uploaded or already uploaded}                     \\
\verb+    {+                                                                                     \\
\verb+      $+%
  \redtt{personal\_function\_file}%
  \verb+ = $_POST["+%
  \redtt{personal\_function\_file}%
  \verb+"];+                                                                                     \\
\verb+    }+                                                                                     \\
\verb++                                                                                          \\
\verb+    elseif ($+%
  \redtt{function\_choice}%
  \verb+ == "+%
  \redtt{function\_user}%
  \verb+"      &&+                                                                               \\
\verb+            $_POST["+%
  \redtt{personal\_function\_text}%
  \verb+"] != NULL   )+ \mynote[0.5cm]{Copy the typed text 1 into a file}                        \\
\verb+    {+                                                                                     \\
\verb+        $+%
  \redtt{personal\_function\_text\_output}%
  \verb++                                                                                        \\
\verb+      = fopen("$+%
  \bluett{personal\_file\_folder}%
  \verb+/" . $+%
  \redtt{personal\_function\_file\_copy}%
  \verb+, "w");+                                                                                 \\
\verb++                                                                                          \\
\verb+      fputs($+%
  \redtt{personal\_function\_text\_output}%
  \verb+,+                                                                                       \\
\verb+            sprintf("\n"));+                                                               \\
\verb+      fputs($+%
  \redtt{personal\_function\_text\_output}%
  \verb+,+                                                                                       \\
\verb+            sprintf("%s\n", $_POST["+%
  \redtt{personal\_function\_text}%
  \verb+"]));+                                                                                   \\
\verb+      fputs($+%
  \redtt{personal\_function\_text\_output}%
  \verb+,+                                                                                       \\
\verb+            sprintf("\n"));+                                                               \\
\verb++                                                                                          \\
\verb+      fclose($+%
  \redtt{personal\_function\_text\_output}%
  \verb+);+                                                                                      \\
\verb++                                                                                          \\
\verb+      $+%
  \redtt{personal\_function\_file}%
  \verb+ = $+%
  \redtt{personal\_function\_file\_copy}%
  \verb+;+                                                                                       \\
\verb+    }+                                                                                     \\
\verb++                                                                                          \\
\verb+    if (file_exists("$+%
  \bluett{personal\_file\_folder}%
  \verb+/$+%
  \redtt{personal\_function\_file\_copy}%
  \verb+") &&+                                                                                   \\
\verb+        $+%
  \redtt{personal\_function\_file}%
  \verb+ != $+%
  \redtt{personal\_function\_file\_copy}%
  \verb+             )+                                                                          \\
\verb+    {+ \mynote[9.5cm]{File 1 uploaded, not typed}                                          \\
\verb+      echo '<input type="hidden"'                           . "\r\n";+                     \\
\verb+      echo '       name="+%
  \redtt{personal\_function\_file}%
  \verb+"'           . "\r\n";+                                                                  \\
\verb+      echo '       value="' . $+%
  \redtt{personal\_function\_file}%
  \verb+ . '"' . "\r\n";+                                                                        \\
\verb+      echo '       />'                                      . "\r\n";+                     \\
\verb+    }+                                                                                     \\
\defc
\verb++                                                                                          \\
\verb+    if ($+%
  \redtt{personal\_function2\_text}%
  \verb+ == NULL)+                                                                               \\
\verb+    {+                                                                                     \\
\verb+      $+%
  \redtt{personal\_function2\_text}%
  \verb+ = $_POST["+%
  \redtt{personal\_function2\_text}%
  \verb+"];+                                                                                     \\
\verb+    }+                                                                                     \\
\verb++                                                                                          \\
\verb+      $+%
  \redtt{personal\_function2\_file\_copy}%
  \verb++                                                                                        \\
\verb+    = +%
  \bluett{"pf-" . \$personal\_file\_date . "-" . \$personal\_file\_time . "-copy2.txt"}%
  \verb+;+                                                                                       \\
\verb++                                                                                          \\
\verb+       .+                                                                                  \\
\verb+       .+ \mynote{For file/text 2}                                                         \\
\verb+       .+                                                                                  \\
\verb++                                                                                          \\
\verb+    if (file_exists("$+%
  \bluett{personal\_file\_folder}%
  \verb+/$+%
  \redtt{personal\_function2\_file\_copy}%
  \verb+") &&+                                                                                   \\
\verb+        $+%
  \redtt{personal\_function2\_file}%
  \verb+ != $+%
  \redtt{personal\_function2\_file\_copy}%
  \verb+            )+                                                                           \\
\verb+    {+                                                                                     \\
\verb+      echo '<input type="hidden"'                            . "\r\n";+                    \\
\verb+      echo '       name="+%
  \redtt{personal\_function2\_file}%
  \verb+"'           . "\r\n";+                                                                  \\
\verb+      echo '       value="' . $+%
  \redtt{personal\_function2\_file}%
  \verb+ . '"' . "\r\n";+                                                                        \\
\verb+      echo '       />'                                       . "\r\n";+                    \\
\verb+    }+                                                                                     \\
\verb++                                                                                          \\
\verb+    include("+%
  \bluett{forms/form-}%
  \verb++%
  \redtt{function\_defining}%
  \verb+.php");+ \mynote[1cm]{Template \ref{item:form-function_defining.php}}                        \\
\verb++                                                                                          \\
\verb+    if ($+%
  \redtt{function\_choosing}%
  \verb+ == NULL                                                  ||+                            \\
\verb+        ($+%
  \redtt{function\_choice}%
  \verb+  == "+%
  \redtt{function\_user}%
  \verb+"                                  &&+                                                   \\
\verb+         !(file_exists("$+%
  \bluett{personal\_file\_folder}%
  \verb+/$+%
  \redtt{personal\_function\_file\_copy}%
  \verb+"))    )  ||+                                                                            \\
\verb+        ($+%
  \redtt{function\_choice}%
  \verb+  == "+%
  \redtt{function\_user}%
  \verb+"                                  &&+                                                   \\
\verb+         +%
  \redtt{\$function2\_needed == "yes"}%
  \verb+                                            &&+                                          \\
\verb+         !(file_exists("$+%
  \bluett{personal\_file\_folder}%
  \verb+/$+%
  \redtt{personal\_function2\_file\_copy}%
  \verb+"))   )   )+                                                                             \\
\verb+    {+                                                                                     \\
\verb+      include("+%
  \bluett{default\_choices/default-}%
  \verb++%
  \redtt{function\_choosing}%
  \verb+.php");+ \mynote[0.75cm]{Modification of Template \ref{item:default-main_function_choosing.php}} \\
\verb+    }+                                                                                     \\
\verb++                                                                                          \\
\verb+    else+                                                                                  \\
\verb+    {+                                                                                     \\
\verb+      echo '<br />';+                                                                      \\
\verb+      echo '+%
  \redtt{function\_choosing}%
  \verb+ = ' . $+%
  \redtt{function\_choosing}%
  \verb+ . '<br />';+                                                                            \\
\verb+      echo '+%
  \redtt{function\_choice}%
  \verb+   = ' . $+%
  \redtt{function\_choice}%
  \verb+   . '<br />';+                                                                          \\
\verb++                                                                                          \\
\verb+      $+%
  \redtt{function\_choice\_setting}%
  \verb+ = "yes";+                                                                               \\
\verb+    }+                                                                                     \\
\verb+  }+                                                                                       \\
\verb+?>+                                                                                        \\
\tabb
\newpage
\myitem
\label{item:form-function_defining.php}
 {\cmbfsfx form-function\_defining.php}
\deft
\verb+<script type="text/javascript">+                                                         \\
\verb+<?php+                                                                                   \\
\verb+/*+                                                                                      \\
\verb+  function +%
  \redtt{personal\_function\_file\_uploading}%
  \verb+()+                                                                                    \\
\verb+  {+                                                                                     \\
\verb+    if (navigator.appName == "Microsoft Internet Explorer")+                             \\
\verb+    {+                                                                                   \\
\verb+        document.getElementById("+%
  \redtt{setup\_function}%
  \verb+")+                                                                                    \\
\verb+                .childNodes[+%
  \redtt{6}%
  \verb+].childNodes[+%
  \bluett{0}%
  \verb+]+                                                                                     \\
\verb+                .checked+                                                                \\
\verb+       = true;+                                                                          \\
\verb+    }+                                                                                   \\
\verb++                                                                                        \\
\verb+    else+                                                                                \\
\verb+//  if (navigator.userAgent.search("Chrome")    != -1 ||+                                \\
\verb+//      navigator.userAgent.search("Firefox")   != -1 ||+                                \\
\verb+//      navigator.userAgent.search("Opera")     != -1 ||+                                \\
\verb+//      navigator.userAgent.search("Safari")    != -1 ||+                                \\
\verb+//      navigator.userAgent.search("Navigator") != -1   )+                               \\
\verb+    {+                                                                                   \\
\verb+        document.getElementById("+%
  \redtt{setup\_function}%
  \verb+")+                                                                                    \\
\verb+                .childNodes[+%
  \redtt{13}%
  \verb+].childNodes[+%
  \bluett{1}%
  \verb+]+                                                                                     \\
\verb+                .checked+                                                                \\
\verb+       = true;+                                                                          \\
\verb+    }+                                                                                   \\
\verb+  }+                                                                                     \\
\verb+*/+                                                                                      \\
\verb+?>+                                                                                      \\
\verb+  function +%
  \bluett{personal\_function\_file\_uploading}%
  \verb+(+%
  \bluett{form\_id}%
  \verb+, +%
  \bluett{n\_IE}%
  \verb+, +%
  \bluett{n\_FF}%
  \verb+)+                                                         \\
\verb+  {+                                                         \\
\verb+    if (navigator.appName == "Microsoft Internet Explorer")+ \\
\verb+    {+                                                       \\
\verb+        document.getElementById(+%
  \bluett{form\_id}%
  \verb+)+                                                         \\
\verb+                .childNodes[+%
  \bluett{n\_IE}%
  \verb+].childNodes[+%
  \bluett{0}%
  \verb+]+                                                         \\
\verb+                .checked+                                    \\
\verb+      = true;+                                               \\
\verb+    }+                                                       \\
\verb++                                                            \\
\verb+    else+                                                    \\
\verb+//  if (navigator.userAgent.search("Chrome")    != -1 ||+    \\
\verb+//      navigator.userAgent.search("Firefox")   != -1 ||+    \\
\verb+//      navigator.userAgent.search("Opera")     != -1 ||+    \\
\verb+//      navigator.userAgent.search("Safari")    != -1 ||+    \\
\verb+//      navigator.userAgent.search("Navigator") != -1   )+   \\
\verb+    {+                                                       \\
\verb+        document.getElementById(+%
  \bluett{form\_id}%
  \verb+)+                                                         \\
\verb+                .childNodes[+%
  \bluett{n\_FF}%
  \verb+].childNodes[+%
  \bluett{1}%
  \verb+]+ \mynote[0.5cm]{{\tt n\_FF} is usually equal to {\tt 2 * n\_IE + 1}} \\
\verb+                .checked+                                    \\
\verb+      = true;+                                               \\
\verb+    }+                                                       \\
\verb+  }+                                                         \\
\defc
\verb++                                                                                        \\
\verb+  function +%
  \redtt{personal\_function\_text\_expanding}%
  \verb+(+%
  \bluett{text\_id}%
  \verb+, +%
  \bluett{text\_height}%
  \verb+)+                                                                                     \\
\verb+  {+                                                                                     \\
\verb+      document.getElementById(+%
  \bluett{text\_id}%
  \verb+)+                                                                                     \\
\verb+              .style.height+                                                             \\
\verb+    = +%
  \bluett{text\_height}%
  \verb+;+                                                                                     \\
\verb+  }+                                                                                     \\
\verb++                                                                                        \\
\verb+  function +%
  \redtt{personal\_function\_text\_shrinking}%
  \verb+(+%
  \bluett{text\_id}%
  \verb+)+                                                                                     \\
\verb+  {+                                                                                     \\
\verb+      document.getElementById(+%
  \bluett{text\_id}%
  \verb+)+                                                                                     \\
\verb+              .style.height+                                                             \\
\verb+    = "+%
  \bluett{50}%
  \verb+pt";+                                                                                  \\
\verb+  }+                                                                                     \\
\verb+</script>+                                                                               \\
\defc
\verb++                                                                                        \\
\verb+<fieldset id="+%
  \redtt{setup\_function}%
  \verb+" style="width:+                                                                       \\
\verb+  <?php+                                                                                 \\
\verb+    if (+%
  \redtt{\$function2\_needed == "yes"}%
  \verb+)+                                                                                     \\
\verb+    {+                                                                                   \\
\verb+      echo '+%
  \redtt{500}%
  \verb+';+                                                                                    \\
\verb+    }+                                                                                   \\
\verb+    else+                                                                                \\
\verb+    {+                                                                                   \\
\verb+      echo '+%
  \redtt{425}%
  \verb+';+                                                                                    \\
\verb+    }+                                                                                   \\
\verb+  ?>pt">+                                                                                \\
\verb++                                                                                        \\
\verb+  <?php+                                                                                 \\
\verb+    include("+%
  \bluett{forms/form-}%
  \verb++%
  \redtt{function\_choosing}%
  \verb+.php");+ \mynote[1cm]{Template \ref{item:form-function_choosing.php}}                  \\
\verb+  ?>+                                                                                    \\
%
%
\verb++                                                                                        \\
\verb+  <?php+                                                                                 \\
\verb+    echo '<label for="+%
  \redtt{function\_user}%
  \verb+">'                                        . "\r\n";+                                  \\
\verb+    echo '<span style="margin-left: +%
  \redtt{19}%
  \verb+pt">Upload file'                                ;+                                     \\
\verb+    if (+%
  \redtt{\$function2\_needed == "yes"}%
  \verb+)+ \mynote[3cm]{For uploading files}                                                   \\
\verb+    {+                                                                                   \\
\verb+      echo 's';+                                                                         \\
\verb+    }+                                                                                   \\
\verb+    echo ':</span>'                                                           . "\r\n";+ \\
\verb+    echo '</label>'                                                           . "\r\n";+ \\
\verb+  ?>+                                                                                    \\
\verb++                                                                                        \\
\verb+  <input class="upload" type="file" name="+%
  \redtt{personal\_function\_file}%
  \verb+"+                                                                                     \\
\verb+         onclick="+%
  \redtt{personal\_function\_file\_uploading}%
  \verb+('+%
  \redtt{setup\_function}%
  \verb+', +%
  \redtt{6}%
  \verb+, +%
  \redtt{13}%
  \verb+)"+ \mynote[0.75cm]{For file 1}                                                        \\
\verb+         />+                                                                             \\
\verb+         <br />+                                                                         \\
%
%
\verb++                                                                                        \\
\verb+  <?php+                                                                                 \\
\verb+    if (+%
  \redtt{\$function2\_needed == "yes"}%
  \verb+)+                                                                                     \\
\verb+    {+                                                                                   \\
\verb+      echo '<span style="margin-left: +%
  \redtt{19}%
  \verb+pt; color: #+%
  \redtt{ADD8E6}%
  \verb+">'                 . "\r\n";+                                                         \\
\verb+      echo '  Upload files:'                                                  . "\r\n";+ \\
\verb+      echo '</span>'                                                          . "\r\n";+ \\
\verb++                                                                                        \\
\verb+      echo '<input class="upload" type="file" name="+%
  \redtt{personal\_function2\_file}%
  \verb+"' . "\r\n";+                                                                          \\
\verb+//    echo '       onclick="+%
  \redtt{personal\_function\_file\_uploading}%
  \verb+()"'              . "\r\n";+                                                           \\
\verb+      echo '       onclick="+%
  \redtt{personal\_function\_file\_uploading}%
  \verb+('                        ;+ \mynote[0cm]{For file 2}                                  \\
\verb+      echo "'+%
  \redtt{setup\_function}%
  \verb+'" . ', +%
  \redtt{6}%
  \verb+, +%
  \redtt{13}%
  \verb+)"'                                   . "\r\n";+                                       \\
\verb+      echo '       />'                                                        . "\r\n";+ \\
\verb+      echo '       (+%
  \redtt{for ......}%
  \verb+)'                                              . "\r\n";+                             \\
\verb+      echo '       <br />'                                                    . "\r\n";+ \\
\verb+    }+                                                                                   \\
\verb+  ?>+                                                                                    \\
\defc
\verb++                                                                                        \\
\verb+  <?php+                                                                                 \\
\verb+    echo '<label for="+%
  \redtt{function\_user}%
  \verb+">'                                        . "\r\n";+                                  \\
\verb+    echo '<span style="margin-left: +%
  \redtt{19}%
  \verb+pt">Type definition'                            ;+                                     \\
\verb+    if (+%
  \redtt{\$function2\_needed == "yes"}%
  \verb+)+ \mynote[5cm]{For uploading files}                                                   \\
\verb+    {+                                                                                   \\
\verb+      echo 's';+                                                                         \\
\verb+    }+                                                                                   \\
\verb+    echo ':</span>'                                                           . "\r\n";+ \\
\verb+    echo '</label>'                                                           . "\r\n";+ \\
\verb+  ?>+                                                                                    \\
%
%
\verb++                                                                                        \\
\verb+  <?php+                                                                                 \\
\verb+    echo '<textarea name="+%
  \redtt{personal\_function\_text}%
  \verb+"'                            . "\r\n";+                                               \\
\verb+    echo '            id="+%
  \redtt{personal\_function\_text}%
  \verb+"'                            . "\r\n";+                                               \\
\verb+    echo '          style="vertical-align: middle;'                           . "\r\n";+ \\
\verb+    echo '                 font-size: +%
  \redtt{8}%
  \verb+pt; height: +%
  \bluett{50}%
  \verb+pt; width: +%
  \redtt{300}%
  \verb+pt"'       . "\r\n";+ \mynote[-0.15cm]{For text 1}                                     \\
\verb+//  echo '          onclick="+%
  \redtt{personal\_function\_file\_uploading}%
  \verb+();'             . "\r\n";+                                                            \\
\verb+    echo '          onclick="+%
  \redtt{personal\_function\_file\_uploading}%
  \verb+('                       ;+                                                            \\
\verb+    echo "'+%
  \redtt{setup\_function}%
  \verb+'" . ', +%
  \redtt{6}%
  \verb+, +%
  \redtt{13}%
  \verb+)"'                                     . "\r\n";+                                     \\
\verb+    echo '                   +%
  \redtt{personal\_function\_text\_expanding}%
  \verb+('                       ;+                                                            \\
\verb+    echo "'+%
  \redtt{personal\_function\_text}%
  \verb+', '+%
  \redtt{100}%
  \verb+pt'" . ')">'                          . "\r\n";+                                       \\
\verb++                                                                                        \\
\verb+//  if ($+%
  \redtt{personal\_function\_text}%
  \verb+                    == NULL     &&+                                                    \\
\verb+//      ($_FILES["+%
  \redtt{personal\_function\_file}%
  \verb+"]["name"] == NULL &&+ \mynote[2.5cm]{The first time}                                  \\
\verb+//       $_POST["+%
  \redtt{personal\_function\_file}%
  \verb+"]          == NULL   )   )+                                                           \\
\verb+    if ($+%
  \redtt{function\_choice}%
  \verb+ != "+%
  \redtt{function\_user}%
  \verb+")+                                                                                    \\
\verb+    {+                                                                                   \\
\verb+      echo ' +%
  \redtt{double function\_user(double x)}%
  \verb+'                                  . "\r\n";+                                          \\
\verb+      echo ' +%
  \redtt{\{}%
  \verb+'                                                               . "\r\n";+             \\
\verb+      echo '   +%
  \redtt{return ......;}%
  \verb+'                                                . "\r\n";+                            \\
\verb+      echo ' +%
  \redtt{\}}%
  \verb+'                                                               . "\r\n";+             \\
\verb+    }+                                                                                   \\
\verb++                                                                                        \\
\verb+    elseif ($+%
  \redtt{personal\_function\_text}%
  \verb+                    != NULL     &&+                                                    \\
\verb+            ($_FILES["+%
  \redtt{personal\_function\_file}%
  \verb+"]["name"] == NULL &&+ \mynote[0.75cm]{File typed, not uploaded}                       \\
\verb+             $_POST["+%
  \redtt{personal\_function\_file}%
  \verb+"]          == NULL   )   )+                                                           \\
\verb+    {+                                                                                   \\
\verb+      echo $+%
  \redtt{personal\_function\_text}%
  \verb+                                            . "\r\n";+                                 \\
\verb+    }+                                                                                   \\
\verb++                                                                                        \\
\verb+    elseif ($+%
  \redtt{personal\_function\_text}%
  \verb+                    != NULL     &&+                                                    \\
\verb+            ($_FILES["+%
  \redtt{personal\_function\_file}%
  \verb+"]["name"] != NULL ||+ \mynote[0.75cm]{File uploaded, not typed}                       \\
\verb+             $_POST["+%
  \redtt{personal\_function\_file}%
  \verb+"]          != NULL   )   )+                                                           \\
\verb+    {+                                                                                   \\
\verb+      echo                                                                      "\r\n";+ \\
\verb+    }+                                                                                   \\
\verb++                                                                                        \\
\verb+    echo '</textarea>'                                                        . "\r\n";+ \\
\verb+    echo '       <br />'                                                      . "\r\n";+ \\
\verb+  ?>+                                                                                    \\
\defc
\verb++                                                                                        \\
\verb+  <?php+                                                                                 \\
\verb+    if (+%
  \redtt{\$function2\_needed == "yes"}%
  \verb+)+                                                                                     \\
\verb+    {+                                                                                   \\
\verb+      echo '<span style="margin-left: +%
  \redtt{19}%
  \verb+pt; color: #+%
  \redtt{ADD8E6}%
  \verb+">'                 . "\r\n";+                                                         \\
\verb+      echo '  Type definitions:'                                              . "\r\n";+ \\
\verb+      echo '</span>'                                                          . "\r\n";+ \\
\verb++                                                                                        \\

\verb+      echo '<textarea name="+%
  \redtt{personal\_function2\_text}%
  \verb+"'                         . "\r\n";+                                                  \\
\verb+      echo '            id="+%
  \redtt{personal\_function2\_text}%
  \verb+"'                         . "\r\n";+                                                  \\
\verb+      echo '          style="vertical-align: middle;'                         . "\r\n";+ \\
\verb+      echo '                 font-size: +%
  \redtt{8}%
  \verb+pt; height: +%
  \bluett{50}%
  \verb+pt; width: +%
  \redtt{300}%
  \verb+pt"'     . "\r\n";+ \mynote[-0.15cm]{For text 2}                                       \\
\verb+//    echo '          onclick="+%
  \redtt{personal\_function\_file\_uploading}%
  \verb+();'           . "\r\n";+                                                              \\
\verb+      echo '          onclick="+%
  \redtt{personal\_function\_file\_uploading}%
  \verb+('                     ;+                                                              \\
\verb+      echo "'+%
  \redtt{setup\_function}%
  \verb+'" . ', +%
  \redtt{6}%
  \verb+, +%
  \redtt{13}%
  \verb+)"'                                   . "\r\n";+                                       \\
\verb+      echo '                   +%
  \redtt{personal\_function\_text\_expanding}%
  \verb+('                     ;+                                                              \\
\verb+      echo "'+%
  \redtt{personal\_function2\_text}%
  \verb+', '+%
  \redtt{150}%
  \verb+pt'" . ')">'                       . "\r\n";+                                          \\
\verb++                                                                                        \\
\verb+//    if ($+%
  \redtt{personal\_function2\_text}%
  \verb+                    == NULL     &&+                                                    \\
\verb+//        ($_FILES["+%
  \redtt{personal\_function2\_file}%
  \verb+"]["name"] == NULL &&+ \mynote[2cm]{The first time}                                    \\
\verb+//         $_POST["+%
  \redtt{personal\_function2\_file}%
  \verb+"]          == NULL   )   )+                                                           \\
\verb+      if ($+%
  \redtt{function\_choice}%
  \verb+ != "+%
  \redtt{function\_user}%
  \verb+")+                                                                                    \\
\verb+      {+                                                                                 \\
\verb+        echo '    +%
  \redtt{......}%
  \verb+'                                                     . "\r\n";+                       \\
\verb+      }+                                                                                 \\
\verb++                                                                                        \\
\verb+      elseif ($+%
  \redtt{personal\_function2\_text}%
  \verb+                    != NULL     &&+                                                    \\
\verb+              ($_FILES["+%
  \redtt{personal\_function2\_file}%
  \verb+"]["name"] == NULL &&+ \mynote[0.5cm]{File typed, not uploaded}                        \\
\verb+               $_POST["+%
  \redtt{personal\_function2\_file}%
  \verb+"]          == NULL   )   )+                                                           \\
\verb+      {+                                                                                 \\
\verb+        echo $+%
  \redtt{personal\_function2\_text}%
  \verb+                                         . "\r\n";+                                    \\
\verb+      }+                                                                                 \\
\verb++                                                                                        \\
\verb+      elseif ($+%
  \redtt{personal\_function2\_text}%
  \verb+                    != NULL     &&+                                                    \\
\verb+              ($_FILES["+%
  \redtt{personal\_function2\_file}%
  \verb+"]["name"] != NULL ||+ \mynote[0.5cm]{File uploaded, not typed}                        \\
\verb+               $_POST["+%
  \redtt{personal\_function2\_file}%
  \verb+"]          != NULL   )   )+                                                           \\
\verb+      {+                                                                                 \\
\verb+        echo                                                                    "\r\n";+ \\
\verb+      }+                                                                                 \\
\verb++                                                                                        \\
\verb+      echo '</textarea>'                                                      . "\r\n";+ \\
\verb+      echo '       (+%
  \redtt{for ......}%
  \verb+)'                                              . "\r\n";+                             \\
\verb+      echo '       <br />'                                                    . "\r\n";+ \\
\verb+    }+                                                                                   \\
\verb+  ?>+                                                                                    \\
\defc
\verb++                                                                                        \\
\verb+  <?php+                                                                                 \\
\verb+    if ($+%
  \redtt{function\_choice}%
  \verb+ == "+%
  \redtt{function\_user}%
  \verb+")+                                                                                    \\
\verb+    {+                                                                                   \\
\verb+      if ($+%
  \redtt{personal\_function\_file}%
  \verb+ != NULL                         &&+                                                   \\
\verb+          $+%
  \redtt{personal\_function\_file}%
  \verb+ != $+%
  \redtt{personal\_function\_file\_copy}%
  \verb+   )+ \mynote[0.45cm]{File uploaded, not typed}                                        \\
\verb+      {+                                                                                 \\
\verb+        if (file_exists("$+%
  \bluett{personal\_file\_folder}%
  \verb+/$+%
  \redtt{personal\_function\_file\_copy}%
  \verb+"))+                                                                                   \\
\verb+        {+                                                                               \\
\verb+          echo '<b>Uploaded file';+                                                      \\
\verb+          if (+%
  \redtt{\$function2\_needed == "yes"}%
  \verb+                                         &&+                                           \\
\verb+              file_exists("$+%
  \bluett{personal\_file\_folder}%
  \verb+/$+%
  \redtt{personal\_function2\_file\_copy}%
  \verb+") &&+                                                                                 \\
\verb+              $+%
  \redtt{personal\_function2\_file}%
  \verb+ != NULL                                   &&+                                         \\
\verb+              $+%
  \redtt{personal\_function2\_file}%
  \verb+ != $+%
  \redtt{personal\_function2\_file\_copy}%
  \verb+            )+                                                                         \\
\verb+          {+ \mynote[8.5cm]{File uploaded, not typed}                                    \\
\verb+            echo 's';+                                                                   \\
\verb+          }+                                                                             \\
\verb+          echo ':</b> ' . $+%
  \redtt{personal\_function\_file}%
  \verb+;+                                                                                     \\
\verb+        }+                                                                               \\
\verb++                                                                                        \\
\verb+        else+                                                                            \\
\verb+        {+                                                                               \\
\verb+          echo '<span class="+%
  \bluett{red}%
  \verb+">Your file can not be uploaded!</span>'      . "\r\n";+ \mynote[0cm]{For file 1}      \\
\verb+        }+                                                                               \\
\verb+      }+                                                                                 \\
\defc
\verb++                                                                                        \\
\verb+      if (+%
  \redtt{\$function2\_needed == "yes"}%
  \verb+)+                                                                                     \\
\verb+      {+                                                                                 \\
\verb+        if ($+%
  \redtt{personal\_function2\_file}%
  \verb+ != NULL                          &&+                                                  \\
\verb+            $+%
  \redtt{personal\_function2\_file}%
  \verb+ != $+%
  \redtt{personal\_function2\_file\_copy}%
  \verb+   )+                                                                                  \\
\verb+        {+ \mynote[9cm]{File uploaded, not typed}                                        \\
\verb+          if (file_exists("$+%
  \bluett{personal\_file\_folder}%
  \verb+/$+%
  \redtt{personal\_function2\_file\_copy}%
  \verb+"))+                                                                                   \\
\verb+          {+                                                                             \\
\verb+            if (file_exists("+%
  \bluett{personal\_file\_folder}%
  \verb+/$+%
  \redtt{personal\_function\_file\_copy}%
  \verb+") &&+                                                                                 \\
\verb+                $+%
  \redtt{personal\_function\_file}%
  \verb+ != NULL                                  &&+                                          \\
\verb+                $+%
  \redtt{personal\_function\_file}%
  \verb+ != $+%
  \redtt{personal\_function\_file\_copy}%
  \verb+             )+                                                                        \\
\verb+            {+ \mynote[8cm]{File uploaded, not typed}                                    \\
\verb+              echo ', ';+                                                                \\
\verb+            }+                                                                           \\
\verb+            else+                                                                        \\
\verb+            {+                                                                           \\
\verb+              echo '<br />'                                                   . "\r\n";+ \\
\verb+              echo '<b>Uploaded file: </b>'                                   . "\r\n";+ \\
\verb+            }+                                                                           \\
\verb+            echo $+%
  \redtt{personal\_function2\_file}%
  \verb+;+                                                                                     \\
\verb+          }+                                                                             \\
\verb++                                                                                        \\
\verb+          else+                                                                          \\
\verb+          {+                                                                             \\
\verb+            echo '<span class="+%
  \bluett{red}%
  \verb+">';+                                                                                  \\
\verb+            if (file_exists("$+%
  \bluett{personal\_file\_folder}%
  \verb+/$+%
  \redtt{personal\_function\_file\_copy}%
  \verb+"))+                                                                                   \\
\verb+            {+                                                                           \\
\verb+              echo 'your';+                                                              \\
\verb+            }+                                                                           \\
\verb+            else+                                                                        \\
\verb+            {+                                                                           \\
\verb+              echo 'Your';+                                                              \\
\verb+            }+                                                                           \\
\verb+            echo ' file +%
  \redtt{for ......}%
  \verb+ can not be uploaded!</span>'               . "\r\n";+ \mynote[0cm]{For file 2}        \\
\verb+          }+                                                                             \\
\verb+        }+                                                                               \\
\verb+      }+                                                                                 \\
\verb++                                                                                        \\
\verb+      if (($+%
  \redtt{personal\_function\_file}%
  \verb+ != NULL                         &&+                                                   \\
\verb+           $+%
  \redtt{personal\_function\_file}%
  \verb+ != $+%
  \redtt{personal\_function\_file\_copy}%
  \verb+   )        ||+                                                                        \\
\verb+          (+%
  \redtt{\$function2\_needed == "yes"}%
  \verb+                                     &&+                                               \\
\verb+           ($+%
  \redtt{personal\_function2\_file}%
  \verb+ != NULL                          &&+                                                  \\
\verb+            $+%
  \redtt{personal\_function2\_file}%
  \verb+ != $+%
  \redtt{personal\_function2\_file\_copy}%
  \verb+   )   )   )+                                                                          \\
\verb+      {+ \mynote[9cm]{File(s) uploaded, not typed}                                       \\
\verb+        echo '<br />'                                                         . "\r\n";+ \\
\verb+      }+                                                                                 \\
\verb+    }+                                                                                   \\
\verb+  ?>+                                                                                    \\
\defc
\verb++                                                                                        \\
\verb+  <?php+                                                                                 \\
\verb+    if ($+%
  \redtt{function\_choice}%
  \verb+ == "+%
  \redtt{function\_user}%
  \verb+")+                                                                                    \\
\verb+    {+                                                                                   \\
\verb+      if ($+%
  \redtt{personal\_function\_file}%
  \verb+ != NULL                         &&+                                                   \\
\verb+          $+%
  \redtt{personal\_function\_file}%
  \verb+ == $+%
  \redtt{personal\_function\_file\_copy}%
  \verb+   )+ \mynote[0.4cm]{File typed, not uploaded}                                         \\
\verb+      {+                                                                                 \\
\verb+        if (file_exists("$+%
  \bluett{personal\_file\_folder}%
  \verb+/$+%
  \redtt{personal\_function\_file\_copy}%
  \verb+"))+                                                                                   \\
\verb+        {+                                                                               \\
\verb+          echo '<b>Saved into file';+                                                    \\
\verb+          if (+%
  \redtt{\$function2\_needed == "yes"}%
  \verb+                                        &&+                                            \\
\verb+              file_exists("+%
  \bluett{personal\_file\_folder}%
  \verb+/$+%
  \redtt{personal\_function2\_file\_copy}%
  \verb+") &&+                                                                                 \\
\verb+              $+%
  \redtt{personal\_function2\_file}%
  \verb+ != NULL                                  &&+                                          \\
\verb+              $+%
  \redtt{personal\_function2\_file}%
  \verb+ == $+%
  \redtt{personal\_function2\_file\_copy}%
  \verb+            )+                                                                         \\
\verb+          {+  \mynote[8cm]{File typed, not uploaded}                                     \\
\verb+            echo 's';+                                                                   \\
\verb+          }+                                                                             \\
\verb+          echo ':</b> ' . $+%
  \redtt{personal\_function\_file}%
  \verb+                    . "\r\n";+                                                         \\
\verb+        }+                                                                               \\
\verb++                                                                                        \\
\verb+        else+                                                                            \\
\verb+        {+                                                                               \\
\verb+          echo '<span class="+%
  \bluett{red}%
  \verb+">The text can not be saved!</span>' . "\r\n";+ \mynote[0.5cm]{For text 1}             \\
\verb+        }+                                                                               \\
\verb+      }+                                                                                 \\
\defc
\verb++                                                                                        \\
\verb+      if (+%
  \redtt{\$function2\_needed == "yes"}%
  \verb+)+                                                                                     \\
\verb+      {+                                                                                 \\
\verb+        if ($+%
  \redtt{personal\_function2\_file}%
  \verb+ != NULL                          &&+                                                  \\
\verb+            $+%
  \redtt{personal\_function2\_file}%
  \verb+ == $+%
  \redtt{personal\_function2\_file\_copy}%
  \verb+   )+                                                                                  \\
\verb+        {+ \mynote[9cm]{File typed, not uploaded}                                        \\
\verb+          if (file_exists("$+%
  \bluett{personal\_file\_folder}%
  \verb+/$+%
  \redtt{personal\_function2\_file\_copy}%
  \verb+"))+                                                                                   \\
\verb+          {+                                                                             \\
\verb+            if (file_exists("+%
  \bluett{personal\_file\_folder}%
  \verb+/$+%
  \redtt{personal\_function\_file\_copy}%
  \verb+") &&+                                                                                 \\
\verb+                $+%
  \redtt{personal\_function\_file}%
  \verb+ != NULL                                  &&+                                          \\
\verb+                $+%
  \redtt{personal\_function\_file}%
  \verb+ == $+%
  \redtt{personal\_function\_file\_copy}%
  \verb+             )+                                                                        \\
\verb+            {+ \mynote[8cm]{File typed, not uploaded}                                    \\
\verb+              echo ', '                                              . "\r\n";+          \\
\verb+            }+                                                                           \\
\verb+            else+                                                                        \\
\verb+            {+                                                                           \\
\verb+              echo '<b>Saved into file: </b>'                        . "\r\n";+          \\
\verb+            }+                                                                           \\
\verb+            echo $+%
  \redtt{personal\_function2\_file}%
  \verb+                            . "\r\n";+                                                 \\
\verb+          }+                                                                             \\
\verb++                                                                                        \\
\verb+          else+                                                                          \\
\verb+          {+                                                                             \\
\verb+            echo '<span class="+%
  \bluett{red}%
  \verb+">';+                                                                                  \\
\verb+            if (file_exists("$+%
  \bluett{personal\_file\_folder}%
  \verb+/$+%
  \redtt{personal\_function\_file\_copy}%
  \verb+"))+                                                                                   \\
\verb+            {+                                                                           \\
\verb+              echo 'the';+                                                               \\
\verb+            }+                                                                           \\
\verb+            else+                                                                        \\
\verb+            {+                                                                           \\
\verb+              echo 'The';+                                                               \\
\verb+            }+                                                                           \\
\verb+            echo ' text +%
  \redtt{for ......}%
  \verb+ can not be saved!</span>'         . "\r\n";+ \mynote[0.5cm]{For text 2}               \\
\verb+          }+                                                                             \\
\verb+        }+                                                                               \\
\verb+      }+                                                                                 \\
\verb++                                                                                        \\
\verb+      if (($+%
  \redtt{personal\_function\_file}%
  \verb+ != NULL                         &&+                                                   \\
\verb+           $+%
  \redtt{personal\_function\_file}%
  \verb+ == $+%
  \redtt{personal\_function\_file\_copy}%
  \verb+   )        ||+                                                                        \\
\verb+          (+%
  \redtt{\$function2\_needed == "yes"}%
  \verb+                                     &&+                                               \\
\verb+           ($+%
  \redtt{personal\_function2\_file}%
  \verb+ != NULL                          &&+                                                  \\
\verb+            $+%
  \redtt{personal\_function2\_file}%
  \verb+ == $+%
  \redtt{personal\_function2\_file\_copy}%
  \verb+   )   )   )+                                                                          \\
\verb+      {+ \mynote[9cm]{File(s) typed, not uploaded}                                       \\
\verb+        echo '<br />'                                                . "\r\n";+          \\
\verb+      }+                                                                                 \\
\verb+    }+                                                                                   \\
\verb+  ?>+                                                                                    \\
\defc
\verb++                                                                                        \\
\verb+  <b> Download: </b>+                                                                    \\
\verb+  <?php+                                                                                 \\
\verb+    echo '<a href="+%
  \bluett{samples/sample-}%
  \verb++%
  \redtt{function\_user}%
  \verb+.txt"      target="_blank">';+                                                         \\
\verb+    echo ' +%
  \bluett{sample for a user-defined function}%
  \verb+';+ \mynote[4.75cm]{Sample 1}                                                          \\
\verb+    echo '</a>';+                                                                        \\
\verb++                                                                                        \\
\verb+    if (+%
  \redtt{\$function2\_needed == "yes"}%
  \verb+)+                                                                                     \\
\verb+    {+                                                                                   \\
\verb+      echo ', <a href="+%
  \bluett{samples/sample-}%
  \verb++%
  \redtt{function2\_user}%
  \verb+.txt" target="_blank">';+ \mynote[0.5cm]{Sample 2}                                     \\
\verb+      echo '+%
  \redtt{for ......}%
  \verb+</a>'                                        . "\r\n";+                                \\
\verb+    }+                                                                                   \\
\verb+  ?>+                                                                                    \\
\verb+  <br />+                                                                                \\
\verb++                                                                                        \\
\verb+</fieldset>+                                                                             \\
\tabb
\newpage
\myitem
\label{item:form-function_choosing.php}
 {\cmbfsfx form-function\_choosing.php}
 (cf.~Templates \ref{item:form-main_functions.php}
            and \ref{item:form-setup_main_function_1_choosing.php})
\deft
\verb+  <label>+                                \\
\verb+  <input type="radio" name="+%
  \redtt{function\_choosing}%
  \verb+" value="+%
  \redtt{1}%
  \verb+"+                                      \\
\verb+         <?php+                           \\
\verb+           if ($+%
  \redtt{function\_choosing}%
  \verb+ == "+%
  \redtt{1}%
  \verb+")+                                     \\
\verb+           {+                             \\
\verb+             echo 'checked = "checked"';+ \\
\verb++                                         \\
\verb+             $+%
  \redtt{function\_choice}%
  \verb+ = "+%
  \redtt{function\_1}%
  \verb+";+                                     \\
\verb+           }+                             \\
\verb+         ?>+                              \\
\verb+         onclick="+%
  \redtt{personal\_function\_text\_shrinking}%
  \verb+('+%
  \redtt{personal\_function\_text}%
  \verb+');+                                    \\
\verb+                  <?php+                  \\
\verb+                    if (+%
  \redtt{\$function2\_needed == "yes"}%
  \verb+)+                                      \\
\verb+                    {+                    \\
\verb+                      echo '+%
  \redtt{personal\_function\_text\_shrinking}%
  \verb+(' . "'+%
  \redtt{personal\_function2\_text}%
  \verb+'" . ');';+                             \\
\verb+                    }+                    \\
\verb+                  ?>"+                    \\
\verb+         />+                              \\
\verb++                                         \\
\verb+         +%
  \redtt{Description for the function 1}%
  \verb++                                       \\
\verb+         </label>+                        \\
\verb+         <br />+                          \\
\verb++                                         \\
\verb+            .+                            \\
\verb+            .+                            \\
\verb+            .+                            \\
\verb++                                         \\
\verb+  <label>+                                \\
\verb+  <input type="radio" name="+%
  \redtt{function\_choosing}%
  \verb+" value="+%
  \redtt{4}%
  \verb+"+                                      \\
\verb+                        id="+%
  \redtt{function\_user}%
  \verb+"+                                      \\
\verb+         <?php+                           \\
\verb+           if ($+%
  \redtt{function\_choosing}%
  \verb+ == "+%
  \redtt{4}%
  \verb+")+                                     \\
\verb+           {+                             \\
\verb+             echo 'checked = "checked"';+ \\
\verb++                                         \\
\verb+             $+%
  \redtt{function\_choice}%
  \verb+ = "+%
  \redtt{function\_user}%
  \verb+";+                                     \\
\verb+           }+                             \\
\verb+         ?>+                              \\
\verb+         />+                              \\
\verb++                                         \\
\verb+         +%
  \redtt{Personal function}%
  \verb++                                       \\
\verb+         </label>+                        \\
\verb+         <br />+                          \\
\tabb
\newpage
\picqin{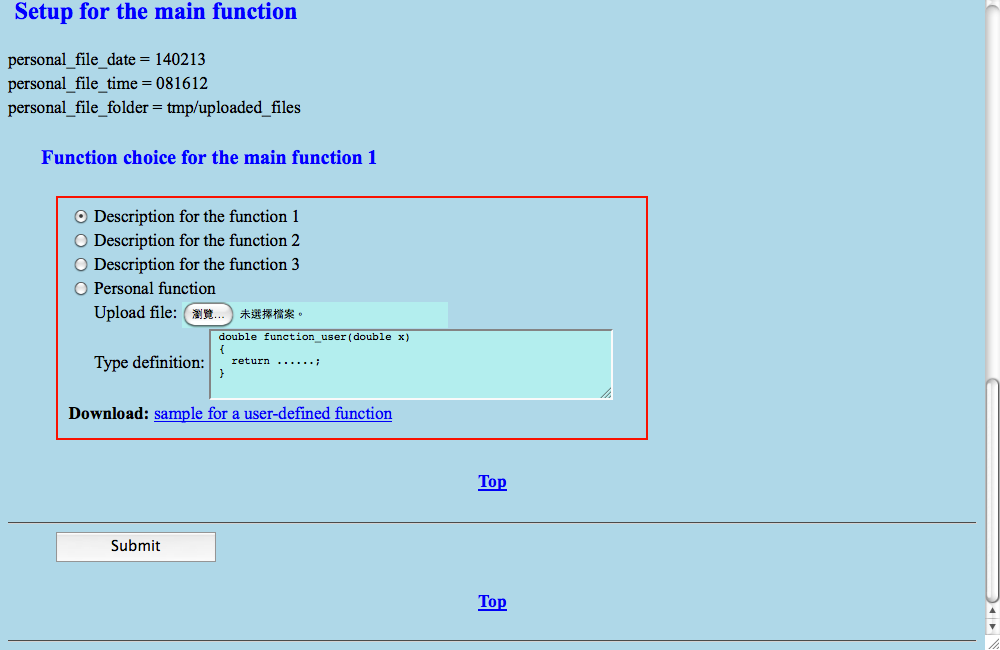}
{uploading\_typing}
{Uploading only one file or typing one text
 (i.e., {\tt \$function2\_needed != "yes"}).}
{Uploading two files or typing texts
 (i.e., {\tt \$function2\_needed == "yes"}).}
{Two files have been uploaded.}
{Only one file has been uploaded,
  the ``second'' required function has been typed and saved.}
\end{templateenumerate}

\addemptypage
\chapter{Special Effects}
\label{chap:special_effects}
%
%
%
\newpage
\section{Option locking}
\label{sec:choice_locking}
%
%
\begin{templateenumerate}
\myitem
\label{item:modification_form-main_functions.php}
 {\bf Modification of {\cmbfsfx form-main\_functions.php}}
 (cf.~Template \ref{item:form-main_functions.php})
\deft
\verb+<fieldset id="+%
  \bluett{main\_function\_choosing}%
  \verb+" style="width: +%
  \redtt{330}%
  \verb+pt">+                                   \\
\verb++                                         \\
\verb+  <label>+                                \\
\verb+  <input type="radio" name="+%
  \bluett{main\_function\_choosing}%
  \verb+" value="+%
  \redtt{1}%
  \verb+"+                                      \\
\verb+         <?php+                           \\
\verb+           if ($+%
  \bluett{main\_function\_choosing}%
  \verb+ == "+%
  \redtt{1}%
  \verb+")+                                     \\
\verb+           {+                             \\
\verb+             echo 'checked = "checked"';+ \\
\verb++                                         \\
\verb+             $+%
  \bluett{main\_function}%
  \verb+ = "+%
  \redtt{main\_function\_1}%
  \verb+";+                                     \\
\verb+           }+                             \\
\verb+         ?>+                              \\
\verb+         />+                              \\
\verb++                                         \\
\verb+         +%
  \redtt{Description for the main function 1}%
  \verb++                                       \\
\verb+         </label>+                        \\
\verb+         <br />+                          \\
\verb++                                         \\
\verb+            .+                            \\
\verb+            .+                            \\
\verb+            .+                            \\
\verb++                                         \\
\verb+  <label>+                                \\
\verb+  <input type="radio" name="+%
  \bluett{main\_function\_choosing}%
  \verb+" value="+%
  \redtt{5}%
  \verb+"+                                      \\
\verb+         onclick="this.checked=false"+    \\
\verb+         onmouseover="this.nextSibling.nextSibling.style.color='+%
  \bluett{red}%
  \verb+'"+                                     \\
\verb+         onmouseout="this.nextSibling.nextSibling.style.color='+%
  \bluett{black}%
  \verb+'"+                                     \\
\verb+         <?php+                           \\
\verb+           if ($+%
  \bluett{main\_function\_choosing}%
  \verb+ == "+%
  \redtt{5}%
  \verb+")+                                     \\
\verb+           {+                             \\
\verb+             echo 'checked = "checked"';+ \\
\verb++                                         \\
\verb+             $+%
  \bluett{main\_function}%
  \verb+ = "+%
  \redtt{main\_function\_5}%
  \verb+";+                                     \\
\verb+           }+                             \\
\verb+         ?>+                              \\
\verb+         />+                              \\
\verb++                                         \\
\verb+         +%
  \redtt{Description for the main function 5}%
  \verb++                                       \\
\verb+         <span>(+%
  \bluett{will be released soon}%
  \verb+)</span> +                              \\
\verb+         </label>+                        \\
\verb+         <br />+                          \\
\verb++                                         \\
\verb+</fieldset>+                              \\
\tabb
\newpage
\picdin{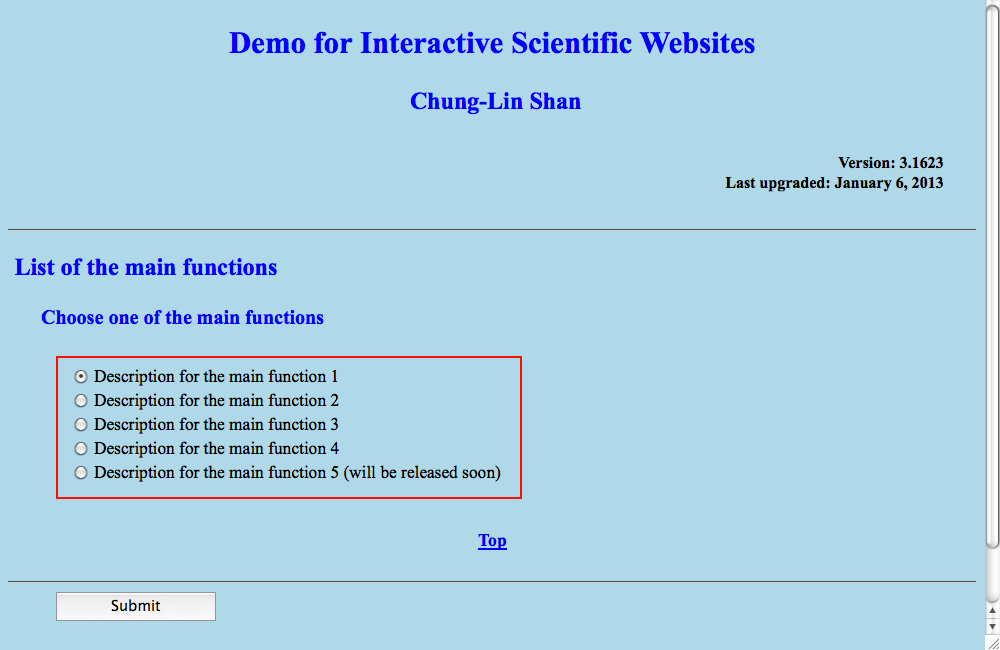}
{modification\_main\_function\_choosing}
{Locking the option 5 for the main function
 (before clicking,
  cf.~Figure \ref{fig:main_function_choosing-1}).}
{Locking the option 5 for the main function
 (after clicking,
  cf.~Figure \ref{fig:main_function_choosing-1}).}
\end{templateenumerate}
\newpage
\section{Equation showing by request}
\label{sec:equation_showing}
%
%
\begin{templateenumerate}
\myitem
\label{item:modification_form-function_defining.php}
 {\bf Modification of {\cmbfsfx form-function\_defining.php}}
 (cf.~Template \ref{item:form-function_defining.php})
\deft
\verb+<script type="text/javascript">+                            \\
\verb+<?php+                                                      \\
\verb+/*+                                                         \\
\verb+  function +%
  \bluett{show\_function}%
  \verb+(+%
  \bluett{function\_name}%
  \verb+)+                                                        \\
\verb+  {+                                                        \\
\verb+      document.getElementById("+%
  \redtt{setup\_function}%
  \verb+")+                                                       \\
\verb+              .nextSibling.nextSibling+                     \\
\verb+              .src+                                         \\
\verb+    = +%
  \bluett{function\_name}%
  \verb+;+                                                        \\
\verb++                                                           \\
\verb+      document.getElementById("+%
  \redtt{setup\_function}%
  \verb+")+                                                       \\
\verb+              .nextSibling.nextSibling+                     \\
\verb+              .style.visibility+                            \\
\verb+    = "visible";+                                           \\
\verb+  }+                                                        \\
\verb++                                                           \\
\verb+  function +%
  \bluett{erase\_function}%
  \verb+()+                                                       \\
\verb+  {+                                                        \\
\verb+      document.getElementById("+%
  \redtt{setup\_function}%
  \verb+")+                                                       \\
\verb+              .nextSibling.nextSibling+                     \\
\verb+              .style.visibility+                            \\
\verb+    = "hidden";+                                            \\
\verb+  }+                                                        \\
\verb+*/+                                                         \\
\verb+?>+                                                         \\

\verb+  function +%
  \bluett{show\_function}%
  \verb+(+%
  \bluett{form\_id}%
  \verb+, +%
  \bluett{function\_name}%
  \verb+)+                                                         \\
\verb+  {+                                                         \\
\verb+      document.getElementById(+%
  \bluett{form\_id}%
  \verb+)+                                                         \\
\verb+              .nextSibling.nextSibling+                      \\
\verb+              .src+                                          \\
\verb+    = +%
  \bluett{function\_name}%
  \verb+;+                                                         \\
\verb++                                                            \\
\verb+      document.getElementById(+%
  \bluett{form\_id}%
  \verb+)+                                                         \\
\verb+              .nextSibling.nextSibling+                      \\
\verb+              .style.visibility+                             \\
\verb+    = "visible";+                                            \\
\verb+  }+                                                         \\
\verb++                                                            \\
\verb+  function +%
  \bluett{erase\_function}%
  \verb+(+%
  \bluett{form\_id}%
  \verb+)+                                                         \\
\verb+  {+                                                         \\
\verb+      document.getElementById(+%
  \bluett{form\_id}%
  \verb+)+                                                         \\
\verb+              .nextSibling.nextSibling+                      \\
\verb+              .style.visibility+                             \\
\verb+    = "hidden";+                                             \\
\verb+  }+                                                         \\

\defc
\verb++                                                           \\
\verb+  function +%
  \bluett{personal\_function\_file\_uploading}%
  \verb+(+%
  \bluett{form\_id}%
  \verb+, +%
  \bluett{n\_IE}%
  \verb+, +%
  \bluett{n\_FF}%
  \verb+)+                                                        \\
\verb+  {+                                                        \\
\verb+      .+                                                    \\
\verb+      .+                                                    \\
\verb+      .+                                                    \\
\verb+  }+                                                        \\
\verb++                                                           \\
\verb+  function +%
  \bluett{personal\_function\_text\_expanding}%
  \verb+(+%
  \bluett{text\_id}%
  \verb+, +%
  \bluett{text\_height}%
  \verb+)+                                                        \\
\verb+  {+                                                        \\
\verb+      .+                                                    \\
\verb+      .+                                                    \\
\verb+      .+                                                    \\
\verb+  }+                                                        \\
\verb++                                                           \\
\verb+  function +%
  \bluett{personal\_function\_text\_shrinking}%
  \verb+(+%
  \bluett{text\_id}%
  \verb+)+                                                        \\
\verb+  {+                                                        \\
\verb+      .+                                                    \\
\verb+      .+                                                    \\
\verb+      .+                                                    \\
\verb+  }+                                                        \\
\verb+</script>+                                                  \\
%
%
\verb++                                                           \\
\verb+<fieldset id="+%
  \redtt{setup\_function}%
  \verb+" style="width: +%
  \redtt{245}%
  \verb+pt">+                                                     \\
\verb++                                                           \\
\verb+  <?php+                                                    \\
\verb+    include("+%
  \bluett{forms/}%
  \verb++%
  \redtt{modification\_}%
  \verb++%
  \bluett{form-}%
  \verb++%
  \redtt{function\_choosing}%
  \verb+.php");+ \mynote[1cm]{Template \ref{item:modification_form-function_choosing.php}} \\
\verb+  ?>+                                                       \\
\verb++                                                           \\
\verb+</fieldset>+                                                \\
\verb++                                                           \\
\verb+<img class="+%
  \bluett{eq-s}%
  \verb+" src="+%
  \bluett{equations/fig-}%
  \verb++%
  \redtt{function\_1}%
  \verb++%
  \bluett{-s.png}%
  \verb+" style="visibility: hidden"/>+                           \\
\tabb
\newpage
\myitem
\label{item:modification_form-function_choosing.php}
 {\bf Modification of {\cmbfsfx form-function\_choosing.php}}
 (cf.~Templates \ref{item:form-main_functions.php}
            and \ref{item:form-function_choosing.php})
\deft
\verb+  <label>+                                  \\
\verb+  <input type="radio" name="+%
  \redtt{function\_choosing}%
  \verb+" value="+%
  \redtt{1}%
  \verb+"+                                        \\
\verb+         <?php+                             \\
\verb+           if ($+%
  \redtt{function\_choosing}%
  \verb+ == "+%
  \redtt{1}%
  \verb+")+                                       \\
\verb+           {+                               \\
\verb+             echo 'checked = "checked"';+   \\
\verb++                                           \\
\verb+             $+%
  \redtt{function\_choice}%
  \verb+ = "+%
  \redtt{function\_1}%
  \verb+";+                                       \\
\verb+           }+                               \\
\verb+         ?>+                                \\
\verb+         />+                                \\
\verb++                                           \\
\verb+         +%
  \redtt{Description for the function 1}%
  \verb++                                         \\
\verb+         </label>+                          \\
\verb+         <a class="+%
  \bluett{eq}%
  \verb+" href="+%
  \bluett{equations/eq-}%
  \verb++%
  \redtt{function\_1}%
  \verb+.html" target="_blank"+ \mynote[0.5cm]{Template \ref{item:eq-function_1.html}} \\
\verb+            onmouseover="+%
  \redtt{show\_function}%
  \verb+('+%
  \redtt{setup\_function}%
  \verb+',+                                      \\
\verb+                                       '+%
  \bluett{equations/fig-}%
  \verb++%
  \redtt{function\_1}%
  \verb++%
  \bluett{-s.png}%
  \verb+')"+                                      \\
\verb+            onmouseout="+%
  \redtt{erase\_function}%
  \verb+('+%
  \redtt{setup\_function}%
  \verb+')"+                                      \\
\verb+            ><sub>analytic form</sub></a>+  \\
\verb+         <br />+                            \\
\verb++                                           \\
\verb+            .+                              \\
\verb+            .+                              \\
\verb+            .+                              \\
\tabb
\newpage
\picqin{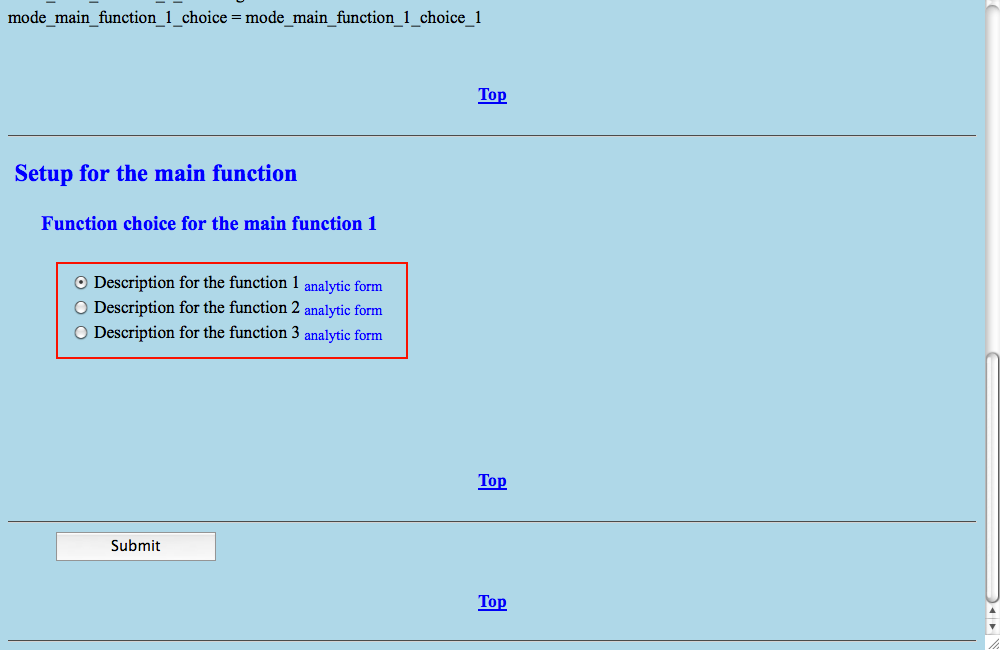}
{modification\_form-function\_defining}
{Function option for the main function 1
 (cf.~Figure \ref{fig:uploading_typing-1}).}
{Showing the equations by moving the cursor onto the ``first'' ``analytic form''
 (cf.~Figure \ref{fig:modification_form-function_defining-1}).}
{Showing the equations by moving the cursor onto the ``second'' ``analytic form''
 (cf.~Figure \ref{fig:modification_form-function_defining-1}).}
{Showing the equations by moving the cursor onto the ``third'' ``analytic form''
 (cf.~Figure \ref{fig:modification_form-function_defining-1}).}
\newpage
\myitem
\label{item:eq-function_1.html}
 {\cmbfsfx eq-function\_1.html}
\deft
\verb+<!DOCTYPE html+                                                \\
\verb+PUBLIC "-//W3C//DTD XHTML 1.0 +%
  \bluett{Transitional}%
  \verb+//EN"+ \mynote[2.25cm]{$\lGetsto$ ``Strict''}                \\
\verb+"http://www.w3.org/TR/xhtml1/DTD/xhtml1-+%
  \bluett{transitional}%
  \verb+.dtd">+ \mynote[1cm]{$\lGetsto$ ``-strict''}                 \\
\verb+<html xmlns="http://www.w3.org/1999/xhtml">+                   \\
\verb++                                                              \\
\verb+<head>+                                                        \\
\verb++                                                              \\
\verb+<title> +%
  \redtt{Function 1}%
  \verb+ </title>+                                                   \\
\verb++                                                              \\
\verb+<meta id="author" content="+%
  \bluett{Chung-Lin Shan}%
  \verb+" />+                                                        \\
\verb+<meta http-equiv="Content-Type" content="text/html; charset=+%
  \bluett{utf-8}%
  \verb+" />+                                                        \\
\verb++                                                              \\
\verb+<link rel="stylesheet" type="text/css" href="+%
  \bluett{../main/main}%
  \verb+.css" />+                                                    \\
\verb++                                                              \\
\verb+</head>+                                                       \\
\verb++                                                              \\
\verb+<body +%
  \bluett{class="en"}%
  \verb+>+                                                           \\
\verb++                                                              \\
\verb+<h4 class="+%
  \bluett{black}%
  \verb+"> +%
  \redtt{Function 1}%
  \verb+ </h4>+                                                      \\
\verb++                                                              \\
\verb+<p class="+%
  \bluett{EqCard}%
  \verb+">+                                                          \\
\verb++                                                              \\
\verb+<img class="+%
  \bluett{eq}%
  \verb+" src="+%
  \redtt{fig-function\_1}%
  \verb+.png" />+                                                    \\
\verb+<br />+                                                        \\
\verb++                                                              \\
\verb++%
  \redtt{ Here ...}%
  \verb++                                                            \\
\verb+<br />+                                                        \\
\verb++                                                              \\
\verb+<img class="+%
  \bluett{eq}%
  \verb+" src="+%
  \redtt{fig-function\_1\_1}%
  \verb+.png" />+                                                    \\
\verb+<br />+                                                        \\
\verb++                                                              \\
\verb++%
  \redtt{ is ..., and}%
  \verb++                                                            \\
\verb+<br />+                                                        \\
\verb++                                                              \\
\verb+<img class="+%
  \bluett{eq}%
  \verb+" src="+%
  \redtt{fig-function\_1\_2}%
  \verb+.png" />+                                                    \\
\verb+<br />+                                                        \\
\verb++                                                              \\
\verb++%
  \redtt{ is ...}%
  \verb++                                                            \\
\verb+<br />+                                                        \\
\verb++                                                              \\
\verb+<br />+                                                        \\
\verb+</p>+                                                          \\
\verb++                                                              \\
\verb+<hr />+                                                        \\
\defc
\verb++                                                              \\
\verb+<h4 class="+%
  \bluett{black}%
  \verb+"> References </h4>+                                         \\
\verb++                                                              \\
\verb+<ol>+                                                          \\
\verb+ <li> +%
  \redtt{...}%
  \verb++                                                            \\
\verb+      </li>+                                                   \\
\verb++                                                              \\
\verb+ <li> +%
  \bluett{G. Jungman, M. Kamionkowski, and K. Griest}%
  \verb+,+                                                           \\
\verb+      ''+%
  \redtt{Supersymmetric Dark Matter}%
  \verb+'',+                                                         \\
\verb+      <i>+%
  \redtt{Phys. Rep.}%
  \verb+</i> <b>+%
  \redtt{267}%
  \verb+</b>, +%
  \redtt{195}%
  \verb+ (+%
  \redtt{1996}%
  \verb+)+                                                           \\
\verb+      (<a href="http://+%
  \redtt{www.sciencedirect.com/science/article/pii/0370157395000585}%
  \verb+"+                                                           \\
\verb+          target="_blank">abstract</a>);+                      \\
\verb+      <b>arXiv:+%
  \redtt{hep-ph/9506380}%
  \verb+</b>+                                                        \\
\verb+      (<a href="http://de.arxiv.org/abs/+%
  \redtt{hep-ph/9506380}%
  \verb+" target="_blank">abstract</a>,+                             \\
\verb+       <a href="http://de.arxiv.org/pdf/+%
  \redtt{hep-ph/9506380}%
  \verb+" target="_blank">pdf</a>).+                                 \\
\verb+      </li>+                                                   \\
\verb++                                                              \\
\verb+ <li> +%
  \redtt{...}%
  \verb++                                                            \\
\verb+      </li>+                                                   \\
\verb+</ol>+                                                         \\
\verb++                                                              \\
\verb+</body>+                                                       \\
\verb++                                                              \\
\verb+</html>+                                                       \\
\tabb
\newpage
\picin{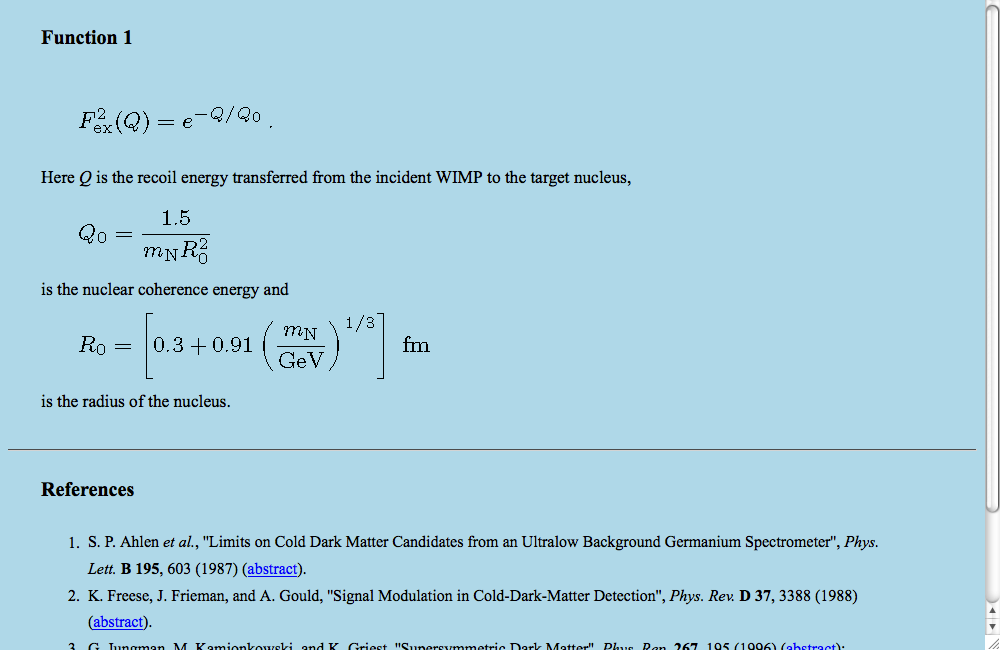}
{eq-function\_1}
{{\sf eq-function\_1.html}.}
\end{templateenumerate}
\newpage
\section{Notation definition showing by request}
\label{sec:notation_definition_showing}
%
%
\begin{templateenumerate}
\myitem
\label{item:show_definition.php}
 {\cmbfsfx show\_definition.php}
\deft
\verb+<script type="text/javascript">+                                                 \\
\verb+    var +%
  \bluett{show\_definition}%
  \verb++                                                                              \\
\verb+  = "<sup>* Move the cursor onto a notation for checking its definition.</sup>"+ \\
\verb+</script>+                                                                       \\
\verb++                                                            \\
\verb+<script type="text/javascript">+                             \\
\verb+  function +%
  \bluett{show\_def}%
  \verb+(+%
  \bluett{table\_id}%
  \verb+, +%
  \bluett{item\_def}%
  \verb+)+                                                         \\
\verb+  {+                                                         \\
\verb+    if (navigator.appName == "Microsoft Internet Explorer")+ \\
\verb+    {+                                                       \\
\verb+        document.getElementById(+%
  \bluett{table\_id}%
  \verb+)+                                                         \\
\verb+                .nextSibling+                                \\
\verb+                .innerHTML+                                  \\
\verb+      = +%
  \bluett{item\_def}%
  \verb+;+                                                         \\
\verb+    }+                                                       \\
\verb++                                                            \\
\verb+    else+                                                    \\
\verb+//  if (navigator.userAgent.search("Chrome")    != -1 ||+    \\
\verb+//      navigator.userAgent.search("Firefox")   != -1 ||+    \\
\verb+//      navigator.userAgent.search("Opera")     != -1 ||+    \\
\verb+//      navigator.userAgent.search("Safari")    != -1 ||+    \\
\verb+//      navigator.userAgent.search("Navigator") != -1   )+   \\
\verb+    {+                                                       \\
\verb+        document.getElementById(+%
  \bluett{table\_id}%
  \verb+)+                                                         \\
\verb+                .nextSibling.nextSibling+                    \\
\verb+                .innerHTML+                                  \\
\verb+      = +%
  \bluett{def\_name}%
  \verb+;+                                                         \\
\verb+    }+                                                       \\
\verb+  }+                                                         \\
\verb++                                                            \\
\verb+  function +%
  \bluett{erase\_def}%
  \verb+(+%
  \bluett{table\_id}%
  \verb+)+                                                         \\
\verb+  {+                                                         \\
\verb+    if (navigator.appName == "Microsoft Internet Explorer")+ \\
\verb+    {+                                                       \\
\verb+        document.getElementById(+%
  \bluett{table\_id}%
  \verb+)+                                                         \\
\verb+                .nextSibling+                                \\
\verb+                .innerHTML+                                  \\
\verb+      = +%
  \bluett{show\_definition}%
  \verb+;+                                                         \\
\verb+    }+                                                       \\
\verb++                                                            \\
\verb+    else+                                                    \\
\verb+//  if (navigator.userAgent.search("Chrome")    != -1 ||+    \\
\verb+//      navigator.userAgent.search("Firefox")   != -1 ||+    \\
\verb+//      navigator.userAgent.search("Opera")     != -1 ||+    \\
\verb+//      navigator.userAgent.search("Safari")    != -1 ||+    \\
\verb+//      navigator.userAgent.search("Navigator") != -1   )+   \\
\verb+    {+                                                       \\
\verb+        document.getElementById(+%
  \bluett{table\_id}%
  \verb+)+                                                         \\
\verb+                .nextSibling.nextSibling+                    \\
\verb+                .innerHTML+                                  \\
\verb+      = +%
  \bluett{show\_definition}%
  \verb+;+                                                         \\
\verb+    }+                                                       \\
\verb+  }+                                                         \\
\verb+</script>+                                                   \\
\tabb
\newpage
\myitem
\label{item:modification2_table-setup_item.php}
 {\bf Modification 2 of {\cmbfsfx table-setup\_item.php}}
 (cf.~Templates \ref{item:show_definition.php}
            and \ref{item:table-setup_item.php})
\deft
\verb+<script type="text/javascript">+                             \\
\verb+<?php+                                                       \\
\verb+/*+                                                          \\
\verb+  function +%
  \bluett{show\_def\_item}%
  \verb+(+%
  \bluett{item\_def}%
  \verb+)+                                                         \\
\verb+  {+                                                         \\
\verb+    if (navigator.appName == "Microsoft Internet Explorer")+ \\
\verb+    {+                                                       \\
\verb+        document.getElementById("+%
  \redtt{table-setup\_item}%
  \verb+")+                                                        \\
\verb+                .nextSibling+                                \\
\verb+                .innerHTML+                                  \\
\verb+      = +%
  \bluett{item\_def}%
  \verb+;+                                                         \\
\verb+    }+                                                       \\
\verb++                                                            \\
\verb+    else+                                                    \\
\verb+//  if (navigator.userAgent.search("Chrome")    != -1 ||+    \\
\verb+//      navigator.userAgent.search("Firefox")   != -1 ||+    \\
\verb+//      navigator.userAgent.search("Opera")     != -1 ||+    \\
\verb+//      navigator.userAgent.search("Safari")    != -1 ||+    \\
\verb+//      navigator.userAgent.search("Navigator") != -1   )+   \\
\verb+    {+                                                       \\
\verb+        document.getElementById("+%
  \redtt{table-setup\_item}%
  \verb+")+                                                        \\
\verb+                .nextSibling.nextSibling+                    \\
\verb+                .innerHTML+                                  \\
\verb+      = +%
  \bluett{item\_def}%
  \verb+;+                                                         \\
\verb+    }+                                                       \\
\verb+  }+                                                         \\
\verb++                                                            \\
\verb+  function +%
  \bluett{erase\_def\_item}%
  \verb+()+                                                        \\
\verb+  {+                                                         \\
\verb+    if (navigator.appName == "Microsoft Internet Explorer")+ \\
\verb+    {+                                                       \\
\verb+        document.getElementById("+%
  \redtt{table-setup\_item}%
  \verb+")+                                                        \\
\verb+                .nextSibling+                                \\
\verb+                .innerHTML+                                  \\
\verb+      = +%
  \bluett{show\_definition}%
  \verb+;+                                                         \\
\verb+    }+                                                       \\
\verb++                                                            \\
\verb+    else+                                                    \\
\verb+//  if (navigator.userAgent.search("Chrome")    != -1 ||+    \\
\verb+//      navigator.userAgent.search("Firefox")   != -1 ||+    \\
\verb+//      navigator.userAgent.search("Opera")     != -1 ||+    \\
\verb+//      navigator.userAgent.search("Safari")    != -1 ||+    \\
\verb+//      navigator.userAgent.search("Navigator") != -1   )+   \\
\verb+    {+                                                       \\
\verb+        document.getElementById("+%
  \redtt{table-setup\_item}%
  \verb+")+                                                        \\
\verb+                .nextSibling.nextSibling+                    \\
\verb+                .innerHTML+                                  \\
\verb+      = +%
  \bluett{show\_definition}%
  \verb+;+                                                         \\
\verb+    }+                                                       \\
\verb+  }+                                                         \\
\verb+</script>+                                                   \\
\verb+*/+                                                          \\
\verb+?>+                                                          \\
\defc
\verb++                                 \\
\verb+<table id="+%
  \redtt{table-setup\_item}%
  \verb+" border="1">+                  \\
\verb++                                 \\
\verb+  <tr id="+%
  \redtt{setup\_item}%
  \verb+">+                             \\
\verb++                                 \\
\verb+    <th>+                         \\
\verb+<?php+                            \\
\verb+/*+                               \\
\verb+      <span onmouseover="+%
  \bluett{show\_def\_item}%
  \verb+('+%
  \redtt{setup\_item\_1}%
  \verb+: +%
  \redtt{definition of setup\_item\_1}%
  \verb+')"+                            \\
\verb+            onmouseout="+%
  \bluett{erase\_def\_item}%
  \verb+()"+                            \\
\verb+*/+                               \\
\verb+?>+                               \\
\verb+      <span onmouseover="+%
  \bluett{show\_def}%
  \verb+('+%
  \redtt{table-setup\_item}%
  \verb+',+                             \\
\verb+                                  '+%
  \redtt{setup\_item\_1}%
  \verb+: +%
  \redtt{definition of setup\_item\_1}%
  \verb+')"+                            \\
\verb+            onmouseout="+%
  \bluett{erase\_def}%
  \verb+('+%
  \redtt{table-setup\_item}%
  \verb+')"+ \mynote[2.5cm]{Only in `one' line!} \\
\verb+            > +%
  \redtt{setup\_item\_1}%
  \verb+ </span>+                       \\
\verb+    </th>+                        \\
\verb++                                 \\
\verb+    <th>+                         \\
\verb+      <span onmouseover="+%
  \bluett{show\_def}%
  \verb+('+%
  \redtt{table-setup\_item}%
  \verb+',+                             \\
\verb+                                  '+%
  \redtt{setup\_item\_1}%
  \verb+: +%
  \redtt{definition of setup\_item\_1}%
  \verb+')"+                            \\
\verb+            onmouseout="+%
  \bluett{erase\_def}%
  \verb+('+%
  \redtt{table-setup\_item}%
  \verb+')"+ \mynote[2.5cm]{Only in `one' line!} \\
\verb+            > +%
  \redtt{setup\_item\_2}%
  \verb+ </span>+                       \\
\verb+    </th>+                        \\
\verb++                                 \\
\verb+       .+                         \\
\verb+       .+                         \\
\verb+       .+                         \\
\verb++                                 \\
\verb+  </tr>+                          \\
\verb++                                 \\
\verb+  <tr>+                           \\
\verb++                                 \\
\verb+       .+                         \\
\verb+       .+                         \\
\verb+       .+                         \\
\verb++                                 \\
\verb+  </tr>+                          \\
\verb++                                 \\
\verb+</table>+                         \\
\verb++                                 \\
\verb+<span class="+%
  \bluett{def}%
  \verb+">+                             \\
\verb+<script type="text/javascript">+  \\
\verb+  document.write(+%
  \bluett{show\_definition}%
  \verb+);+ \mynote{Template \ref{item:show_definition.php},
                    see Template \ref{item:modification2_table-setup_item_entry}} \\
\verb+</script>+                        \\
\verb+</span>+                          \\
\tabb
\newpage
\picdin{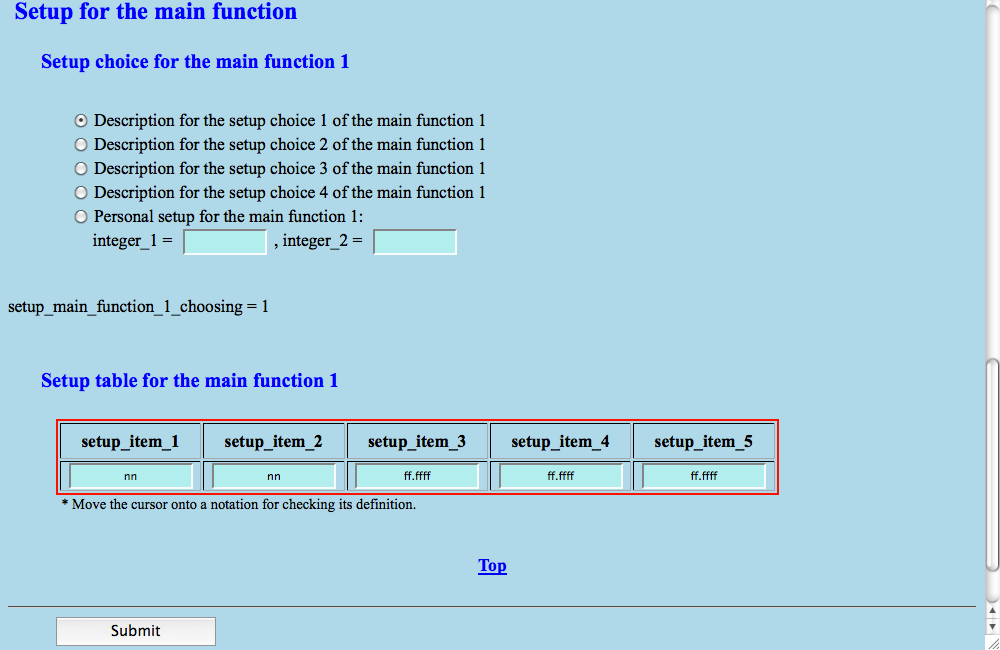}
{modification2\_default-setup\_item}
{A setup table with default values
 (cf.~Figure \ref{fig:default-setup_item}).}
{Showing its definition by moving the cursor onto a notation
 (cf.~Figure \ref{fig:default-setup_item}).}
\end{templateenumerate}
\newpage
\section{Item value calculating by given values of other items}
\label{sec:item_value_calculating}
%
%
\begin{templateenumerate}
\myitem
\label{item:modification2_table-setup_item_entry}
 {\bf Modification 2 of the `setup table for a group of entry'}
 (cf.~Template \ref{item:table-setup_item_entry})
\deft
\verb+<?php+                                   \\
\verb+  include("+%
  \bluett{tables/show\_definition}%
  \verb+.php");+ \mynote{Template \ref{item:show_definition.php}} \\
\verb++                                        \\
\verb+  include("+%
  \bluett{tables/table-delete}%
  \verb+.php");+ \mynote[2cm]{Template \ref{item:table-delete.php}} \\
\verb++                                        \\
\verb+  include("+%
  \bluett{tables/table-delete\_head}%
  \verb+.php");+ \mynote{Template \ref{item:table-delete_head.php}} \\
\verb+  include("+%
  \bluett{tables/table-delete\_entry}%
  \verb+.php");+ \mynote{Template \ref{item:table-delete_entry.php}} \\
\verb++                                        \\
\verb+  if ($+%
  \redtt{setup\_main\_function\_1\_checked}%
  \verb+ == "yes")+                            \\
\verb+  {+                                     \\
\verb+    echo '<h4> +%
  \redtt{Setup table for the main function 1}%
  \verb+ </h4>' . "\r\n";+                     \\
\verb++                                        \\
\verb+    $+%
  \redtt{entry\_No}%
  \verb+ = +%
  \redtt{3}%
  \verb+;+                                     \\
\verb++                                        \\
\verb+    for ($+%
  \redtt{n\_entry}%
  \verb+ = 1; $+%
  \redtt{n\_entry}%
  \verb+ <= $+%
  \redtt{entry\_No}%
  \verb+; $+%
  \redtt{n\_entry}%
  \verb- ++)-                                  \\
\verb+    {+                                   \\
\verb++                                        \\
\verb+        .+                               \\
\verb+        .+                               \\
\verb+        .+                               \\
\verb++                                        \\
\verb+    }+                                   \\
\verb++                                        \\
\verb+    if ($+%
  \redtt{setup\_entry\_common}%
  \verb+ == NULL)+                             \\
\verb+    {+                                   \\
\verb+      $+%
  \redtt{setup\_entry\_common}%
  \verb+ = $_POST["+%
  \redtt{setup\_entry\_common}%
  \verb+"];+                                   \\
\verb+    }+                                   \\
%
%
\verb++                                        \\
\verb+    include("+%
  \bluett{tables/table-}%
  \verb++%
  \redtt{setup\_item\_head}%
  \verb+.php");+ \mynote[2cm]{Template \ref{item:modification2_table-setup_item_head.php}} \\
\verb+    for ($+%
  \redtt{n\_entry}%
  \verb+ = 1; $+%
  \redtt{n\_entry}%
  \verb+ <= $+%
  \redtt{entry\_No}%
  \verb+; $+%
  \redtt{n\_entry}%
  \verb- ++)-                                  \\
\verb+    {+                                   \\
\verb+      include("+%
  \bluett{tables/table-}%
  \verb++%
  \redtt{setup\_item\_entry}%
  \verb+.php");+ \mynote[1.6cm]{Template \ref{item:modification2_table-setup_item_entry.php}} \\
\verb+    }+                                   \\
\verb+    include("+%
  \bluett{tables/table-}%
  \verb++%
  \redtt{setup\_item\_entry\_common}%
  \verb+.php");+ \mynote[1.6cm]{Template \ref{item:table-setup_item_entry_common.php}} \\
\verb+    echo '</table>'                           . "\r\n";+ \\
\verb++                                                        \\
\verb+    echo '<span class="+%
  \bluett{def}%
  \verb+">'                 . "\r\n";+                         \\
\verb+    echo '<script type="text/javascript">'    . "\r\n";+ \\
\verb+    echo '  document.write(+%
  \bluett{show\_definition}%
  \verb+);' . "\r\n";+ \mynote[1cm]{Template \ref{item:show_definition.php},
                                cf.~Template \ref{item:modification2_table-setup_item.php}} \\
\verb+    echo '</script>'                          . "\r\n";+ \\
\verb+    echo '</span>'                            . "\r\n";+ \\
\verb++                                                        \\
\verb+    for ($+%
  \redtt{n\_entry}%
  \verb+ = 1; $+%
  \redtt{n\_entry}%
  \verb+ <= $+%
  \redtt{entry\_No}%
  \verb+; $+%
  \redtt{n\_entry}%
  \verb- ++)-                                  \\
\verb+    {+                                   \\
\verb++                                        \\
\verb+        .+                               \\
\verb+        .+                               \\
\verb+        .+                               \\
\verb++                                        \\
\verb+    }+                                   \\
\verb+  }+                                     \\
\verb+?>+                                      \\
\tabb
\newpage
\myitem
\label{item:modification2_table-setup_item_head.php}
 {\bf Modification 2 of {\cmbfsfx table-setup\_item\_head.php}}
 (cf.~Templates \ref{item:table-setup_item_head.php}
  and \ref{item:modification2_table-setup_item.php})
\deft
\verb+<script type="text/javascript">+                             \\
\verb+  var +%
  \bluett{run}%
  \verb+;+                                                         \\
\verb+  var +%
  \bluett{run\_max}%
  \verb+;+                                                         \\
\verb++                                                            \\
\verb+  function +%
  \redtt{erase\_setup\_item\_4}%
  \verb+(+%
  \bluett{run\_max}%
  \verb+)+                                                         \\
\verb+  {+                                                         \\
\verb+      var +%
  \redtt{n\_Entry}%
  \verb++                                                          \\
\verb+    = document.getElementById("+%
  \redtt{setup\_item}%
  \verb+");+                                                       \\
\verb++                                                            \\
\verb+    <?php+                                                   \\
\verb+      if (+%
  \redtt{\$setup\_item2\_needed == "yes"}%
  \verb+)+ \mynote[1cm]{Check if Table 2 exists!}                  \\
\verb+      {+                                                     \\
\verb+        echo '      var +%
  \redtt{n\_Entry2}%
  \verb+'                                . "\r\n";+                \\
\verb+        echo '    = document.getElementById("+%
  \redtt{setup\_item2}%
  \verb+");'     . "\r\n";+ \mynote[0.5cm]{For table 2}            \\
\verb+      }+                                                     \\
\verb+    ?>+                                                      \\
\verb++                                                            \\
\verb+    if (navigator.appName == "Microsoft Internet Explorer")+ \\
\verb+    {+                                                       \\
\verb+      for (+%
  \bluett{run}%
  \verb+ = 1; +%
  \bluett{run}%
  \verb+ <= +%
  \bluett{run\_max}%
  \verb+; +%
  \bluett{run}%
  \verb- ++)- \mynote[1cm]{Skip run\_max rows/n\_entry's}          \\
\verb+      {+                                                     \\
\verb+          +%
  \redtt{n\_Entry}%
  \verb++                                                          \\
\verb+        = +%
  \redtt{n\_Entry}%
  \verb+.nextSibling;+                                             \\
\verb++                                                            \\
\verb+        <?php+                                               \\
\verb+          if (+%
  \redtt{\$setup\_item2\_needed == "yes"}%
  \verb+)+ \mynote[1cm]{Check if Table 2 exists!}                  \\
\verb+          {+                                                 \\
\verb+            echo '          +%
  \redtt{n\_Entry2}%
  \verb+'                            . "\r\n";+                    \\
\verb+            echo '        = +%
  \redtt{n\_Entry2}%
  \verb+.nextSibling;'               . "\r\n";+ \mynote[0.5cm]{For table 2} \\
\verb+          }+                                                 \\
\verb+        ?>+                                                  \\
\verb+      }+                                                     \\
\verb++                                                            \\
\verb+        +%
  \redtt{n\_Entry}%
  \verb+.childNodes[+%
  \redtt{4}%
  \verb+].childNodes[0]+                                           \\
\verb+               .value+                                       \\
\verb+      = "";+                                                 \\
\verb++                                                            \\
\verb+      <?php+                                                 \\
\verb+        if (+%
  \redtt{\$setup\_item2\_needed == "yes"}%
  \verb+)+ \mynote[1cm]{Check if Table 2 exists!}                  \\
\verb+        {+                                                   \\
\verb+          echo '        +%
  \redtt{n\_Entry2}%
  \verb+.childNodes[+%
  \redtt{4}%
  \verb+].innerHTML = "";'  . "\r\n";+ \mynote[0.5cm]{For table 2} \\
\verb+          echo '        +%
  \redtt{n\_Entry2}%
  \verb+.childNodes[+%
  \redtt{5}%
  \verb+].innerHTML = "";'  . "\r\n";+                             \\
\verb+        }+                                                   \\
\verb+      ?>+                                                    \\
\verb+    }+                                                       \\
\defc
\verb++                                                            \\
\verb+    else+                                                    \\
\verb+//  if (navigator.userAgent.search("Chrome")    != -1 ||+    \\
\verb+//      navigator.userAgent.search("Firefox")   != -1 ||+    \\
\verb+//      navigator.userAgent.search("Opera")     != -1 ||+    \\
\verb+//      navigator.userAgent.search("Safari")    != -1 ||+    \\
\verb+//      navigator.userAgent.search("Navigator") != -1   )+   \\
\verb+    {+                                                       \\
\verb+      for (+%
  \bluett{run}%
  \verb+ = 1; +%
  \bluett{run}%
  \verb+ <= +%
  \bluett{run\_max}%
  \verb+; +%
  \bluett{run}%
  \verb- ++)- \mynote[1cm]{Skip run\_max rows/n\_entry's}          \\
\verb+      {+                                                     \\
\verb+          +%
  \redtt{n\_Entry}%
  \verb++                                                          \\
\verb+        = +%
  \redtt{n\_Entry}%
  \verb+.nextSibling.nextSibling;+                                 \\
\verb++                                                            \\
\verb+        <?php+                                               \\
\verb+          if (+%
  \redtt{\$setup\_item2\_needed == "yes"}%
  \verb+)+ \mynote[1cm]{Check if Table 2 exists!}                  \\
\verb+          {+                                                 \\
\verb+            echo '          +%
  \redtt{n\_Entry2}%
  \verb+'                            . "\r\n";+                    \\
\verb+            echo '        = +%
  \redtt{n\_Entry2}%
  \verb+.nextSibling.nextSibling;'   . "\r\n";+ \mynote[0.5cm]{For table 2} \\
\verb+          }+                                                 \\
\verb+        ?>+                                                  \\
\verb+      }+                                                     \\
\verb++                                                            \\
\verb+        +%
  \redtt{n\_Entry}%
  \verb+.childNodes[+%
  \redtt{9}%
  \verb+].childNodes[0]+                                           \\
\verb+               .value+                                       \\
\verb+      = "";+                                                 \\
\verb++                                                            \\
\verb+      <?php+                                                 \\
\verb+        if (+%
  \redtt{\$setup\_item2\_needed == "yes"}%
  \verb+)+ \mynote[1cm]{Check if Table 2 exists!}                  \\
\verb+        {+                                                   \\
\verb+          echo '        +%
  \redtt{n\_Entry2}%
  \verb+.childNodes[+%
  \redtt{ 9}%
  \verb+].innerHTML = "";' . "\r\n";+ \mynote[0.5cm]{For table 2} \\
\verb+          echo '        +%
  \redtt{n\_Entry2}%
  \verb+.childNodes[+%
  \redtt{11}%
  \verb+].innerHTML = "";' . "\r\n";+                              \\
\verb+        }+                                                   \\
\verb+      ?>+                                                    \\
\verb+    }+                                                       \\
\verb+  }+                                                         \\
\verb+</script>+                                                   \\
\defc
\verb+<?php+                                                       \\
\verb+/*+                                                          \\
\verb+<script type="text/javascript">+                             \\
\verb+  function +%
  \bluett{show\_def\_item}%
  \verb+(+%
  \bluett{item\_def}%
  \verb+)+                                                         \\
\verb+  {+                                                         \\
\verb+      .+                                                     \\
\verb+      .+                                                     \\
\verb+      .+                                                     \\
\verb+  }+                                                         \\
\verb++ \mynote[5cm]{See Templates \ref{item:show_definition.php}
                               and \ref{item:modification2_table-setup_item.php}} \\
\verb+  function +%
  \bluett{erase\_def\_item}%
  \verb+()+                                                        \\
\verb+  {+                                                         \\
\verb+      .+                                                     \\
\verb+      .+                                                     \\
\verb+      .+                                                     \\
\verb+  }+                                                         \\
\verb+</script>+                                                   \\
\verb+*/+                                                          \\
\verb+?>+                                                          \\
%
%
\verb+<table id="+%
  \redtt{table-setup\_item}%
  \verb+" border="1">+                                             \\
\verb++                                                            \\
\verb+  <tr id="+%
  \redtt{setup\_item}%
  \verb+">+                                                        \\
\verb++                                                            \\
\verb+    <th style="width: +%
  \redtt{60}%
  \verb+pt"> +%
  \redtt{entry}%
  \verb+ </th>+                                                    \\
\verb++                                                            \\
\verb+    <th>+                                                    \\
\verb+<?php+                                                       \\
\verb+/*+                                                          \\
\verb+      <span onmouseover="+%
  \bluett{show\_def\_item}%
  \verb+('+%
  \redtt{setup\_item\_1}%
  \verb+: +%
  \redtt{definition of setup\_item\_1}%
  \verb+')"+                                                       \\
\verb+            onmouseout="+%
  \bluett{erase\_def\_item}%
  \verb+()"+                                                       \\
\verb+*/+                                                          \\
\verb+?>+                                                          \\
\verb+      <span onmouseover="+%
  \bluett{show\_def}%
  \verb+('+%
  \redtt{table-setup\_item}%
  \verb+',+                                                        \\
\verb+                                  '+%
  \redtt{setup\_item\_1}%
  \verb+: +%
  \redtt{definition of setup\_item\_1}%
  \verb+')"+                                                       \\
\verb+            onmouseout="+%
  \bluett{erase\_def}%
  \verb+('+%
  \redtt{table-setup\_item}%
  \verb+')"+ \mynote[2.5cm]{Only in `one' line!}                   \\
\verb+            > +%
  \redtt{setup\_item\_1}%
  \verb+ </span>+                                                  \\
\verb+    </th>+                                                   \\
\verb++                                                            \\
\verb+       .+                                                    \\
\verb+       .+ \mynote{See Templates \ref{item:table-setup_item_head.php}
                        and \ref{item:modification2_table-setup_item.php}} \\
\verb+       .+                                                    \\
\verb++                                                            \\
\verb+    <th>+                                                    \\
\verb+      <span onmouseover="+%
  \bluett{show\_def}%
  \verb+('+%
  \redtt{table-setup\_item}%
  \verb+',+                                                        \\
\verb+                                  '+%
  \redtt{setup\_item\_4}%
  \verb+: +%
  \redtt{definition of setup\_item\_4}%
  \verb+')"+                                                       \\
\verb+            onmouseout="+%
  \bluett{erase\_def}%
  \verb+('+%
  \redtt{table-setup\_item}%
  \verb+')"+ \mynote[2.5cm]{Only in `one' line!}                   \\
\verb+            > +%
  \redtt{setup\_item\_4}%
  \verb+ </span>+                                                  \\
\verb+    </th>+                                                   \\
\verb++                                                            \\
\verb+  </tr>+                                                     \\
\tabb
\newpage
\myitem
\label{item:modification2_table-setup_item_entry.php}
 {\bf Modification 2 of {\cmbfsfx table-setup\_item\_entry.php}}
 (cf.~Template \ref{item:table-setup_item_entry.php})
\deft
\verb+  <tr>+                                      \\
\verb++                                            \\
\verb+    <th style="width: +%
  \redtt{60}%
  \verb+pt">+                                      \\
\verb+      <?php echo '+%
  \redtt{entry}%
  \verb+[' . $+%
  \redtt{n\_entry}%
  \verb+ . ']'; ?>+                                \\
\verb+    </th>+                                   \\
\verb++                                            \\
%
%
\verb+       .+                                    \\
\verb+       .+                                    \\
\verb+       .+                                    \\
\verb++                                            \\
\verb+    <td>+                                    \\
\verb+      <input class="+%
  \bluett{table}%
  \verb+" type="text" name="+%
  \redtt{setup\_item\_4}%
  \verb+[<?php echo $+%
  \redtt{n\_entry}%
  \verb+; ?>]"+                                    \\
\verb+             onclick="+%
  \redtt{erase\_setup\_item\_4}%
  \verb+(<?php echo $+%
  \redtt{n\_entry}%
  \verb+; ?>)"+                                    \\
\verb+             <?php+                          \\
\verb+               if (+%
  \redtt{\$calculate\_setup\_item2\_4 == "yes"}%
  \verb+)+ \mynote[1cm]{Check if the calculation is needed!}  \\
\verb+               {+                            \\
\verb+                 echo 'onkeyup="+%
  \redtt{calculate\_setup\_item2\_4}%
  \verb+(' . $+%
  \redtt{n\_entry}%
  \verb+ . ')"';+                                  \\
\verb+               }+                            \\
\verb+             ?>+                             \\
\verb+             />+                             \\
\verb+    </td>+                                   \\
\verb++                                            \\
\verb+       .+                                    \\
\verb+       .+                                    \\
\verb+       .+                                    \\
\verb++                                            \\
\verb+  </tr>+                                     \\
\tabb
\newpage
\picdin{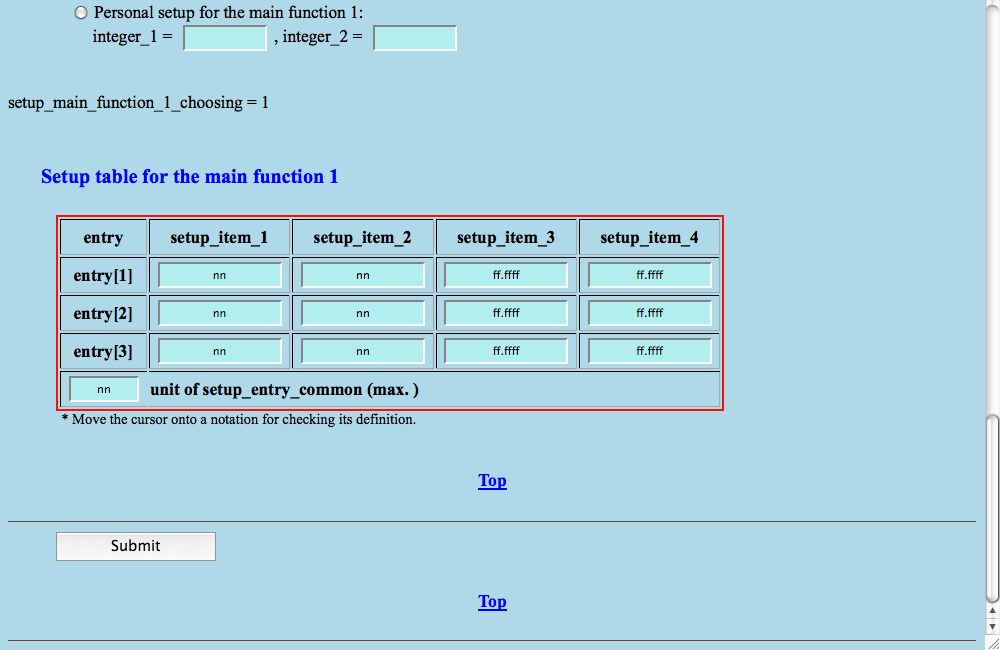}
{modification2\_default-setup\_item\_entry}
{A setup table for a group of entry with default values
 (cf.~Figures \ref{fig:default-setup_item_entry-1}
  and \ref{fig:modification2_default-setup_item-1}).}
{Showing its definition by moving the cursor onto a notation
 (cf.~Figures \ref{fig:default-setup_item_entry-1}
  and \ref{fig:modification2_default-setup_item-2}).}
\newpage
\myitem
\label{item:modification2_table-setup_item2_head.php}
 {\bf Modification 2 of {\cmbfsfx table-setup\_item2\_head.php}}
 (cf.~Templates \ref{item:table-setup_item_head.php}
  and \ref{item:modification2_table-setup_item_head.php})
\deft
\verb+<script type="text/javascript">+                             \\
\verb+  var +%
  \bluett{run}%
  \verb+;+                                                         \\
\verb+  var +%
  \bluett{run\_max}%
  \verb+;+                                                         \\
\verb++                                                            \\
\verb+  function +%
  \redtt{erase\_setup\_item2\_4}%
  \verb+(+%
  \bluett{run\_max}%
  \verb+)+ \mynote[1cm]{Erase both {\tt item2\_4} and {\tt item2\_5}!} \\
\verb+  {+                                                         \\
\verb+      var +%
  \redtt{n\_Entry2}%
  \verb++                                                          \\
\verb+    = document.getElementById("+%
  \redtt{setup\_item2}%
  \verb+");+                                                       \\
\verb++                                                            \\
\verb+    if (navigator.appName == "Microsoft Internet Explorer")+ \\
\verb+    {+                                                       \\
\verb+      for (+%
  \bluett{run}%
  \verb+ = 1; +%
  \bluett{run}%
  \verb+ <= +%
  \bluett{run\_max}%
  \verb+; +%
  \bluett{run}%
  \verb- ++)-                                                      \\
\verb+      {+                                                     \\
\verb+          +%
  \redtt{n\_Entry2}%
  \verb++                                                          \\
\verb+        = +%
  \redtt{n\_Entry2}%
  \verb+.nextSibling;+                                             \\
\verb+      }+                                                     \\
\verb++                                                            \\
\verb+        +%
  \redtt{n\_Entry2}%
  \verb+.childNodes[+%
  \redtt{3}%
  \verb+].childNodes[0]+                                           \\
\verb+                .value+                                      \\
\verb+      = "";+                                                 \\
\verb++                                                            \\
\verb+      +%
  \redtt{n\_Entry2}%
  \verb+.childNodes[+%
  \redtt{4}%
  \verb+].innerHTML = "";+                                         \\
\verb+      +%
  \redtt{n\_Entry2}%
  \verb+.childNodes[+%
  \redtt{5}%
  \verb+].innerHTML = "";+ \mynote[0.75cm]{Use {\tt innerHTML}, not {\tt childNodes[0].value}!} \\
\verb+    }+                                                       \\
%
%
\verb++                                                            \\
\verb+    else+                                                    \\
\verb+//  if (navigator.userAgent.search("Chrome")    != -1 ||+    \\
\verb+//      navigator.userAgent.search("Firefox")   != -1 ||+    \\
\verb+//      navigator.userAgent.search("Opera")     != -1 ||+    \\
\verb+//      navigator.userAgent.search("Safari")    != -1 ||+    \\
\verb+//      navigator.userAgent.search("Navigator") != -1   )+   \\
\verb+    {+                                                       \\
\verb+      for (+%
  \bluett{run}%
  \verb+ = 1; +%
  \bluett{run}%
  \verb+ <= +%
  \bluett{run\_max}%
  \verb+; +%
  \bluett{run}%
  \verb- ++)-                                                      \\
\verb+      {+                                                     \\
\verb+          +%
  \redtt{n\_Entry2}%
  \verb++                                                          \\
\verb+        = +%
  \redtt{n\_Entry2}%
  \verb+.nextSibling.nextSibling;+                                 \\
\verb+      }+                                                     \\
\verb++                                                            \\
\verb+        +%
  \redtt{n\_Entry2}%
  \verb+.childNodes[+%
  \redtt{7}%
  \verb+].childNodes[1]+                                           \\
\verb+                .value = "";+                                \\
\verb+      = "";+                                                 \\
\verb++                                                            \\
\verb+      +%
  \redtt{n\_Entry2}%
  \verb+.childNodes[+%
  \redtt{ 9}%
  \verb+].innerHTML = "";+                                         \\
\verb+      +%
  \redtt{n\_Entry2}%
  \verb+.childNodes[+%
  \redtt{11}%
  \verb+].innerHTML = "";+ \mynote[0.75cm]{Use {\tt innerHTML}, not {\tt childNodes[1].value}!} \\
\verb+    }+                                                       \\
\verb+  }+                                                         \\
\verb+</script>+                                                   \\
\defc
\verb++                                                            \\
\verb+<script type="text/javascript">+                             \\
\verb+  function +%
  \redtt{calculate\_setup\_item2\_4}%
  \verb+(+%
  \bluett{run\_max}%
  \verb+)+                                                         \\
\verb+  {+                                                         \\
\verb+      var +%
  \redtt{n\_Entry}%
  \verb++                                                          \\
\verb+    = document.getElementById("+%
  \redtt{setup\_item}%
  \verb+");+                                                       \\
\verb++                                                            \\
\verb+      var +%
  \redtt{n\_Entry2}%
  \verb++                                                          \\
\verb+    = document.getElementById("+%
  \redtt{setup\_item2}%
  \verb+");+                                                       \\
\verb++                                                            \\
\verb+    if (navigator.appName == "Microsoft Internet Explorer")+ \\
\verb+    {+                                                       \\
\verb+      for (+%
  \bluett{run}%
  \verb+ = 1; +%
  \bluett{run}%
  \verb+ <= +%
  \bluett{run\_max}%
  \verb+; +%
  \bluett{run}%
  \verb- ++)-                                                      \\
\verb+      {+                                                     \\
\verb+          +%
  \redtt{n\_Entry}%
  \verb++                                                          \\
\verb+        = +%
  \redtt{n\_Entry}%
  \verb+.nextSibling;+                                             \\
\verb++                                                            \\
\verb+          +%
  \redtt{n\_Entry2}%
  \verb++                                                          \\
\verb+        = +%
  \redtt{n\_Entry2}%
  \verb+.nextSibling;+                                             \\
\verb+      }+                                                     \\
\verb++                                                            \\
\verb+      if (  +%
  \redtt{n\_Entry2}%
  \verb+.childNodes[+%
  \redtt{3}%
  \verb+].childNodes[0].value+                                     \\
\verb+          > +%
  \redtt{setup\_item2\_3\_max}%
  \verb+                         )+                                \\
\verb+      {+                                                     \\
\verb+          +%
  \redtt{n\_Entry2}%
  \verb+.childNodes[+%
  \redtt{3}%
  \verb+].childNodes[0]+                                           \\
\verb+                  .value+                                    \\
\verb+        = +%
  \redtt{setup\_item2\_3\_max}%
  \verb+;+                                                         \\
\verb+      }+                                                     \\
\verb++                                                            \\
\verb+         +%
  \redtt{n\_Entry2}%
  \verb+.childNodes[+%
  \redtt{4}%
  \verb+].innerHTML+                                               \\
\verb+      =  +%
  \redtt{n\_Entry}%
  \verb+ .childNodes[+%
  \redtt{4}%
  \verb+].childNodes[0].value *+                                   \\
\verb+         +%
  \redtt{n\_Entry2}%
  \verb+.childNodes[+%
  \redtt{3}%
  \verb+].childNodes[0].value;+                                    \\
\verb++                                                            \\
\verb+         +%
  \redtt{n\_Entry2}%
  \verb+.childNodes[+%
  \redtt{5}%
  \verb+].innerHTML+                                               \\
\verb+      =  +%
  \redtt{n\_Entry}%
  \verb+ .childNodes[+%
  \redtt{4}%
  \verb+].childNodes[0].value+                                     \\
\verb+       - +%
  \redtt{n\_Entry}%
  \verb+ .childNodes[+%
  \redtt{4}%
  \verb+].childNodes[0].value *+ \mynote[0.15cm]{Do not use {\tt [5] = [4] * (1.0 - [3])}!} \\
\verb+         +%
  \redtt{n\_Entry2}%
  \verb+.childNodes[+%
  \redtt{3}%
  \verb+].childNodes[0].value;+                                    \\
\verb+    }+                                                       \\
\defc
\verb++                                                            \\
\verb+    else+                                                    \\
\verb+//  if (navigator.userAgent.search("Chrome")    != -1 ||+    \\
\verb+//      navigator.userAgent.search("Firefox")   != -1 ||+    \\
\verb+//      navigator.userAgent.search("Opera")     != -1 ||+    \\
\verb+//      navigator.userAgent.search("Safari")    != -1 ||+    \\
\verb+//      navigator.userAgent.search("Navigator") != -1   )+   \\
\verb+    {+                                                       \\
\verb+      for (+%
  \bluett{run}%
  \verb+ = 1; +%
  \bluett{run}%
  \verb+ <= +%
  \bluett{run\_max}%
  \verb+; +%
  \bluett{run}%
  \verb- ++)-                                                      \\
\verb+      {+                                                     \\
\verb+          +%
  \redtt{n\_Entry}%
  \verb++                                                          \\
\verb+        = +%
  \redtt{n\_Entry}%
  \verb+.nextSibling.nextSibling;+                                 \\
\verb++                                                            \\
\verb+          +%
  \redtt{n\_Entry2}%
  \verb++                                                          \\
\verb+        = +%
  \redtt{n\_Entry2}%
  \verb+.nextSibling.nextSibling;+                                 \\
\verb+      }+                                                     \\
\verb++                                                            \\
\verb+      if (  +%
  \redtt{n\_Entry2}%
  \verb+.childNodes[+%
  \redtt{7}%
  \verb+].childNodes[1].value+                                     \\
\verb+          > +%
  \redtt{setup\_item2\_3\_max}%
  \verb+                         )+                                \\
\verb+      {+                                                     \\
\verb+          +%
  \redtt{n\_Entry2}%
  \verb+.childNodes[+%
  \redtt{7}%
  \verb+].childNodes[1]+                                           \\
\verb+                  .value+                                    \\
\verb+        = +%
  \redtt{setup\_item2\_3\_max}%
  \verb+;+                                                         \\
\verb+      }+                                                     \\
\verb++                                                            \\
\verb+         +%
  \redtt{n\_Entry2}%
  \verb+.childNodes[+%
  \redtt{ 9}%
  \verb+].innerHTML+                                               \\
\verb+      =  +%
  \redtt{n\_Entry}%
  \verb+ .childNodes[+%
  \redtt{ 9}%
  \verb+].childNodes[1].value *+                                   \\
\verb+         +%
  \redtt{n\_Entry2}%
  \verb+.childNodes[+%
  \redtt{ 7}%
  \verb+].childNodes[1].value;+                                    \\
\verb++                                                            \\
\verb+         +%
  \redtt{n\_Entry2}%
  \verb+.childNodes[+%
  \redtt{11}%
  \verb+].innerHTML+                                               \\
\verb+      =  +%
  \redtt{n\_Entry}%
  \verb+ .childNodes[+%
  \redtt{ 9}%
  \verb+].childNodes[1].value+                                     \\
\verb+       - +%
  \redtt{n\_Entry}%
  \verb+ .childNodes[+%
  \redtt{ 9}%
  \verb+].childNodes[1].value *+ \mynote[0.15cm]{Do not use {\tt [11] = [9] * (1.0 - [7])}!} \\
\verb+         +%
  \redtt{n\_Entry2}%
  \verb+.childNodes[+%
  \redtt{ 7}%
  \verb+].childNodes[1].value;+                                    \\
\verb+    }+                                                       \\
\verb+  }+                                                         \\
\verb+</script>+                                                   \\

\defc
\verb++                                                            \\
\verb+<?php+                                                       \\
\verb+/*+                                                          \\
\verb+<script type="text/javascript">+                             \\
\verb+  function +%
  \bluett{show\_def\_item2}%
  \verb+(+%
  \bluett{item2\_def}%
  \verb+)+                                                         \\
\verb+  {+                                                         \\
\verb+      .+                                                     \\
\verb+      .+                                                     \\
\verb+      .+                                                     \\
\verb+  }+                                                         \\
\verb++ \mynote[5cm]{See Templates \ref{item:show_definition.php},
                                   \ref{item:modification2_table-setup_item.php}
                               and \ref{item:modification2_table-setup_item_head.php}} \\
\verb+  function +%
  \bluett{erase\_def\_item2}%
  \verb+()+                                                        \\
\verb+  {+                                                         \\
\verb+      .+                                                     \\
\verb+      .+                                                     \\
\verb+      .+                                                     \\
\verb+  }+                                                         \\
\verb+</script>+                                                   \\
\verb+*/+                                                          \\
\verb+?>+                                                          \\
%
%
\verb+<table id="+%
  \redtt{table-setup\_item2}%
  \verb+" border="1">+                                             \\
\verb++                                                            \\
\verb+  <tr id="+%
  \redtt{setup\_item2}%
  \verb+">+                                                        \\
\verb++                                                            \\
\verb+    <th style="width: +%
  \redtt{60}%
  \verb+pt"> +%
  \redtt{entry}%
  \verb+ </th>+                                                    \\
\verb++                                                            \\
\verb+    <th>+                                                    \\
\verb+<?php+                                                       \\
\verb+/*+                                                          \\
\verb+      <span onmouseover="+%
  \bluett{show\_def\_item2}%
  \verb+('+%
  \redtt{setup\_item2\_1}%
  \verb+: +%
  \redtt{definition of setup\_item2\_1}%
  \verb+')"+                                                       \\
\verb+            onmouseout="+%
  \bluett{erase\_def\_item2}%
  \verb+()"+                                                       \\
\verb+*/+                                                          \\
\verb+?>+                                                          \\
\verb+      <span onmouseover="+%
  \bluett{show\_def}%
  \verb+('+%
  \redtt{table-setup\_item2}%
  \verb+',+                                                        \\
\verb+                                  '+%
  \redtt{setup\_item2\_1}%
  \verb+: +%
  \redtt{definition of setup\_item2\_1}%
  \verb+')"+                                                       \\
\verb+            onmouseout="+%
  \bluett{erase\_def}%
  \verb+('+%
  \redtt{table-setup\_item2}%
  \verb+')"+ \mynote[2.5cm]{Only in `one' line!}                   \\
\verb+            > +%
  \redtt{setup\_item2\_1}%
  \verb+ </span>+                                                  \\
\verb+    </th>+                                                   \\
\verb++                                                            \\
\verb+       .+                                                    \\
\verb+       .+ \mynote{See Templates \ref{item:table-setup_item_head.php}
                        and \ref{item:modification2_table-setup_item.php}} \\
\verb+       .+                                                    \\
\verb++                                                            \\
\verb+    <th>+                                                    \\
\verb+      <span onmouseover="+%
  \bluett{show\_def}%
  \verb+('+%
  \redtt{table-setup\_item2}%
  \verb+',+                                                        \\
\verb+                                  '+%
  \redtt{setup\_item2\_5}%
  \verb+: +%
  \redtt{definition of setup\_item2\_5}%
  \verb+')"+                                                       \\
\verb+            onmouseout="+%
  \bluett{erase\_def}%
  \verb+('+%
  \redtt{table-setup\_item2}%
  \verb+')"+ \mynote[2.5cm]{Only in `one' line!}                   \\
\verb+            > +%
  \redtt{setup\_item2\_5}%
  \verb+ </span>+                                                  \\
\verb+    </th>+                                                   \\
\verb++                                                            \\
\verb+  </tr>+                                                     \\
\tabb
\newpage
\myitem
\label{item:modification2_table-setup_item2_entry.php}
 {\bf Modification 2 of {\cmbfsfx table-setup\_item2\_entry.php}}
 (cf.~Templates \ref{item:table-setup_item_entry.php}
  and \ref{item:modification2_table-setup_item_entry.php})
\deft
\verb+  <tr>+                                      \\
\verb++                                            \\
\verb+    <th style="width: +%
  \bluett{60}%
  \verb+pt">+                                      \\
\verb+      <?php echo '+%
  \redtt{entry}%
  \verb+[' . $+%
  \redtt{n\_entry}%
  \verb+ . ']'; ?>+                                \\
\verb+    </th>+                                   \\
\verb++                                            \\
\verb+    <td>+                                    \\
\verb+      <input class="+%
  \bluett{table}%
  \verb+" type="text" name="+%
  \redtt{setup\_item2\_1}%
  \verb+[<?php echo $+%
  \redtt{n\_entry}%
  \verb+; ?>]"+                                    \\
\verb+             onclick="this.value=''"+        \\
\verb+             <?php+                          \\
\verb+               if ($+%
  \redtt{setup\_item2\_1}%
  \verb+[$+%
  \redtt{n\_entry}%
  \verb+] != NULL)+                                \\
\verb+               {+                            \\
\verb+                 echo 'value="' . $+%
  \redtt{setup\_item2\_1}%
  \verb+[$+%
  \redtt{n\_entry}%
  \verb+] . '"';+                                  \\
\verb+               }+                            \\
\verb+             ?>+                             \\
\verb+             />+                             \\
\verb+    </td>+                                   \\
\verb++                                            \\
\verb+    <td>+                                    \\
\verb+      <input class="+%
  \bluett{table}%
  \verb+" type="text" name="+%
  \redtt{setup\_item2\_2}%
  \verb+[<?php echo $+%
  \redtt{n\_entry}%
  \verb+; ?>]"+                                    \\
\verb+             onclick="this.value=''"+        \\
\verb+             <?php+                          \\
\verb+               if ($+%
  \redtt{setup\_item2\_2}%
  \verb+[$+%
  \redtt{n\_entry}%
  \verb+] == +%
  \bluett{0.00001}%
  \verb+)+                                         \\
\verb+               {+                            \\
\verb+                 echo 'value="0"';+          \\
\verb+               }+                            \\
\verb++                                            \\
\verb+               elseif ($+%
  \redtt{setup\_item2\_2}%
  \verb+[$+%
  \redtt{n\_entry}%
  \verb+] != NULL)+                                \\
\verb+               {+                            \\
\verb+                 echo 'value="' . $+%
  \redtt{setup\_item2\_2}%
  \verb+[$+%
  \redtt{n\_entry}%
  \verb+] . '"';+                                  \\
\verb+               }+                            \\
\verb+             ?>+                             \\
\verb+             />+                             \\
\verb+    </td>+                                   \\
\verb++                                            \\
\verb+    <td>+                                    \\
\verb+      <input class="+%
  \bluett{table}%
  \verb+" type="text" name="+%
  \redtt{setup\_item2\_3}%
  \verb+[<?php echo $+%
  \redtt{n\_entry}%
  \verb+; ?>]"+                                    \\
\verb+             onclick="+%
  \redtt{erase\_setup\_item2\_4}%
  \verb+(<?php echo $+%
  \redtt{n\_entry}%
  \verb+; ?>)"+                                    \\
\verb+             onkeyup="+%
  \redtt{calculate\_setup\_item2\_4}%
  \verb+(<?php echo $+%
  \redtt{n\_entry}%
  \verb+; ?>)"+                                    \\
\verb+             <?php+                          \\
\verb+               if ($+%
  \redtt{setup\_item2\_3}%
  \verb+[$+%
  \redtt{n\_entry}%
  \verb+] != NULL)+                                \\
\verb+               {+                            \\
\verb+                 if ($+%
  \redtt{setup\_item2\_3}%
  \verb+[$+%
  \redtt{n\_entry}%
  \verb+] <= $+%
  \redtt{setup\_item2\_3\_max}%
  \verb+)+                                         \\
\verb+                 {+                          \\
\verb+                   echo 'value="' . $+%
  \redtt{setup\_item2\_3}%
  \verb+[$+%
  \redtt{n\_entry}%
  \verb+] . '"';+                                  \\
\verb+                 }+                          \\
\verb++                                            \\
\verb+                 else+                       \\
\verb+                 {+                          \\
\verb+                   echo 'value="' . $+%
  \redtt{setup\_item2\_3\_max}%
  \verb+ . '"';+                                   \\
\verb+                 }+                          \\
\verb+               }+                            \\
\verb+             ?>+                             \\
\verb+             />+                             \\
\verb+    </td>+                                   \\
\defc
\verb++                                            \\
\verb+    <td>+                                    \\
\verb+      <?php+                                 \\
\verb+        if ($+%
  \redtt{setup\_item\_4}%
  \verb+[$+%
  \redtt{n\_entry}%
  \verb+]  != NULL &&+                             \\
\verb+            $+%
  \redtt{setup\_item2\_3}%
  \verb+[$+%
  \redtt{n\_entry}%
  \verb+] != NULL   )+                             \\
\verb+        {+                                   \\
\verb+          echo $+%
  \redtt{setup\_item\_4}%
  \verb+[$+%
  \redtt{n\_entry}%
  \verb+] * $+%
  \redtt{setup\_item2\_3}%
  \verb+[$+%
  \redtt{n\_entry}%
  \verb+];+                                        \\
\verb+        }+                                   \\
\verb+      ?>+                                    \\
\verb+    </td>+                                   \\
\verb++ \mynote[10cm]{Calculate the values}        \\
\verb+    <td>+                                    \\
\verb+      <?php+                                 \\
\verb+        if ($+%
  \redtt{setup\_item\_4}%
  \verb+[$n+%
  \redtt{\_entry}%
  \verb+]  != NULL &&+                             \\
\verb+            $+%
  \redtt{setup\_item2\_3}%
  \verb+[$+%
  \redtt{n\_entry}%
  \verb+] != NULL   )+                             \\
\verb+        {+                                   \\
\verb+          echo $+%
  \redtt{setup\_item\_4}%
  \verb+[$+%
  \redtt{n\_entry}%
  \verb+] - $+%
  \redtt{setup\_item\_4}%
  \verb+[$+%
  \redtt{n\_entry}%
  \verb+] * $+%
  \redtt{setup\_item2\_3}%
  \verb+[$+%
  \redtt{n\_entry}%
  \verb+];+                                        \\
\verb+        }+                                   \\
\verb+      ?>+ \mynote[3.5cm]{Do not use {\tt 4[] * (1.0 - 2\_3[])}!} \\
\verb+    </td>+                                   \\
\verb++                                            \\
\verb+  </tr>+                                     \\
\tabb
\newpage
\myitem
\label{item:default-setup_item2_head.php}
 {\cmbfsfx default-setup\_item2\_head.php}
 (cf.~Template \ref{item:default-setup_item_head.php})
\deft
\verb+<script type="text/javascript">+                           \\
\verb+    document.getElementById("+%
  \redtt{table-setup\_item2}%
  \verb+")+                                                      \\
\verb+            .style.borderColor+                            \\
\verb+  = "+%
  \bluett{red}%
  \verb+";+                                                      \\
\verb++                                                          \\
\verb+    document.getElementById("+%
  \redtt{table-setup\_item2}%
  \verb+")+                                                      \\
\verb+            .border+                                       \\
\verb+  = "+%
  \bluett{2}%
  \verb+";+                                                      \\
\verb++                                                          \\
\verb+  if (navigator.appName == "Microsoft Internet Explorer")+ \\
\verb+  {+                                                       \\
\verb+      var +%
  \redtt{n\_Entry}%
  \verb++                                                        \\
\verb+    = document.getElementById("+%
  \redtt{setup\_item}%
  \verb+")+                                                      \\
\verb+              .nextSibling;+                               \\
\verb++                                                          \\
\verb+      var +%
  \redtt{n\_Entry2}%
  \verb++                                                        \\
\verb+    = document.getElementById("+%
  \redtt{setup\_item2}%
  \verb+")+                                                      \\
\verb+              .nextSibling;+                               \\
\verb+  }+                                                       \\
\verb++                                                          \\
\verb+  else+                                                    \\
\verb+//if (navigator.userAgent.search("Chrome")    != -1 ||+    \\
\verb+//    navigator.userAgent.search("Firefox")   != -1 ||+    \\
\verb+//    navigator.userAgent.search("Opera")     != -1 ||+    \\
\verb+//    navigator.userAgent.search("Safari")    != -1 ||+    \\
\verb+//    navigator.userAgent.search("Navigator") != -1   )+   \\
\verb+  {+                                                       \\
\verb+      var +%
  \redtt{n\_Entry}%
  \verb++                                                        \\
\verb+    = document.getElementById("+%
  \redtt{setup\_item}%
  \verb+")+                                                      \\
\verb+              .nextSibling.nextSibling;+                   \\
\verb++                                                          \\
\verb+      var +%
  \redtt{n\_Entry2}%
  \verb++                                                        \\
\verb+    = document.getElementById("+%
  \redtt{setup\_item2}%
  \verb+")+                                                      \\
\verb+              .nextSibling.nextSibling;+                   \\
\verb+  }+                                                       \\
\verb+</script>+                                                 \\
\tabb
\newpage
\myitem
\label{item:default-setup_item2_entry.php}
 {\cmbfsfx default-setup\_item2\_entry.php}
 (cf.~Templates \ref{item:default-setup_item_entry.php}
  and \ref{item:modification2_table-setup_item2_head.php})
\deft
\verb+<script type="text/javascript">+                           \\
\verb+  if (navigator.appName == "Microsoft Internet Explorer")+ \\
\verb+  {+                                                       \\
\verb+       +%
  \redtt{n\_Entry2}%
  \verb+.childNodes[1].childNodes[0]+                            \\
\verb+               .value+                                     \\
\verb+    =  "+%
  \redtt{nn}%
  \verb+";+                                                      \\
\verb++                                                          \\
\verb+       +%
  \redtt{n\_Entry2}%
  \verb+.childNodes[2].childNodes[0]+                            \\
\verb+               .value+                                     \\
\verb+    =  "+%
  \redtt{nn}%
  \verb+";+                                                      \\
\verb++                                                          \\
\verb+       +%
  \redtt{n\_Entry2}%
  \verb+.childNodes[3].childNodes[0]+                            \\
\verb+               .value+                                     \\
\verb+    =  "+%
  \redtt{ff.ffff}%
  \verb+";+                                                      \\
\verb++                                                          \\
\verb+       +%
  \redtt{n\_Entry2}%
  \verb+.childNodes[+%
  \redtt{4}%
  \verb+].innerHTML+                                             \\
\verb+    =  +%
  \redtt{n\_Entry}%
  \verb+ .childNodes[+%
  \redtt{4}%
  \verb+].childNodes[0].value *+                                 \\
\verb+       +%
  \redtt{n\_Entry2}%
  \verb+.childNodes[+%
  \redtt{3}%
  \verb+].childNodes[0].value;+                                  \\
\verb++ \mynote[9cm]{See Template \ref{item:modification2_table-setup_item2_head.php}} \\
\verb+       +%
  \redtt{n\_Entry2}%
  \verb+.childNodes[+%
  \redtt{5}%
  \verb+].innerHTML+                                             \\
\verb+    =  +%
  \redtt{n\_Entry}%
  \verb+ .childNodes[+%
  \redtt{4}%
  \verb+].childNodes[0].value+                                   \\
\verb+     - +%
  \redtt{n\_Entry}%
  \verb+ .childNodes[+%
  \redtt{4}%
  \verb+].childNodes[0].value *+                                 \\
\verb+       +%
  \redtt{n\_Entry2}%
  \verb+.childNodes[+%
  \redtt{3}%
  \verb+].childNodes[0].value;+                                  \\
\verb++                                                          \\
\verb+      +%
  \redtt{n\_Entry}%
  \verb++                                                        \\
\verb+    = +%
  \redtt{n\_Entry}%
  \verb+ .nextSibling;+                                          \\
\verb++                                                          \\
\verb+      +%
  \redtt{n\_Entry2}%
  \verb++                                                        \\
\verb+    = +%
  \redtt{n\_Entry2}%
  \verb+.nextSibling;+                                           \\
\verb+  }+                                                       \\
\defc
\verb++                                                          \\
\verb+  else+                                                    \\
\verb+//if (navigator.userAgent.search("Chrome")    != -1 ||+    \\
\verb+//    navigator.userAgent.search("Firefox")   != -1 ||+    \\
\verb+//    navigator.userAgent.search("Opera")     != -1 ||+    \\
\verb+//    navigator.userAgent.search("Safari")    != -1 ||+    \\
\verb+//    navigator.userAgent.search("Navigator") != -1   )+   \\
\verb+  {+                                                       \\
\verb+       +%
  \redtt{n\_Entry2}%
  \verb+.childNodes[3].childNodes[1]+                            \\
\verb+               .value+                                     \\
\verb+    =  "+%
  \redtt{nn}%
  \verb+";+                                                      \\
\verb++                                                          \\
\verb+       +%
  \redtt{n\_Entry2}%
  \verb+.childNodes[5].childNodes[1]+                            \\
\verb+               .value+                                     \\
\verb+    =  "+%
  \redtt{nn}%
  \verb+";+                                                      \\
\verb++                                                          \\
\verb+       +%
  \redtt{n\_Entry2}%
  \verb+.childNodes[7].childNodes[1]+                            \\
\verb+               .value+                                     \\
\verb+    =  "+%
  \redtt{ff.ffff}%
  \verb+";+                                                      \\
\verb++                                                          \\
\verb+       +%
  \redtt{n\_Entry2}%
  \verb+.childNodes[+%
  \redtt{ 9}%
  \verb+].innerHTML+                                             \\
\verb+    =  +%
  \redtt{n\_Entry}%
  \verb+ .childNodes[+%
  \redtt{ 9}%
  \verb+].childNodes[1].value *+                                 \\
\verb+       +%
  \redtt{n\_Entry2}%
  \verb+.childNodes[+%
  \redtt{ 7}%
  \verb+].childNodes[1].value;+                                  \\
\verb++ \mynote[9cm]{See Template \ref{item:modification2_table-setup_item2_head.php}} \\
\verb+       +%
  \redtt{n\_Entry2}%
  \verb+.childNodes[+%
  \redtt{11}%
  \verb+].innerHTML+                                             \\
\verb+    =  +%
  \redtt{n\_Entry}%
  \verb+ .childNodes[+%
  \redtt{ 9}%
  \verb+].childNodes[1].value+                                   \\
\verb+     - +%
  \redtt{n\_Entry}%
  \verb+ .childNodes[+%
  \redtt{ 9}%
  \verb+].childNodes[1].value *+                                 \\
\verb+       +%
  \redtt{n\_Entry2}%
  \verb+.childNodes[+%
  \redtt{ 7}%
  \verb+].childNodes[1].value;+                                  \\
\verb++                                                          \\
\verb+      +%
  \redtt{n\_Entry}%
  \verb++                                                        \\
\verb+    = +%
  \redtt{n\_Entry}%
  \verb+ .nextSibling.nextSibling;+                              \\
\verb++                                                          \\
\verb+      +%
  \redtt{n\_Entry2}%
  \verb++                                                        \\
\verb+    = +%
  \redtt{n\_Entry2}%
  \verb+.nextSibling.nextSibling;+                               \\
\verb+  }+                                                       \\
\verb+</script>+                                                 \\
\tabb
\newpage
\picqin{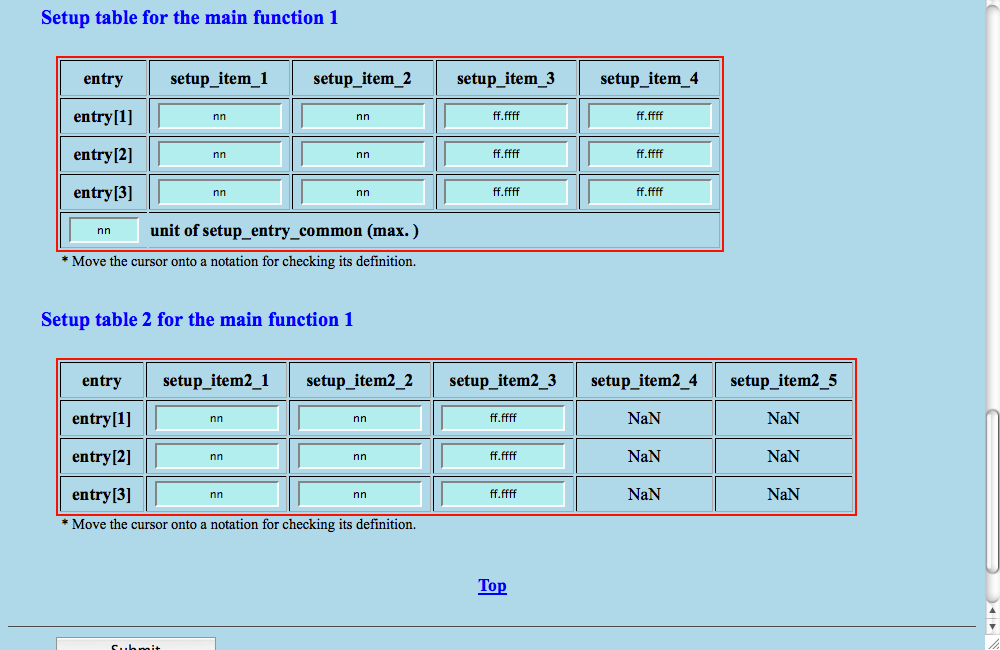}
{modification2\_default-setup\_item2\_entry}
{Two setup tables with default values
 (cf.~Figure \ref{fig:modification2_default-setup_item_entry-1}).}
{Showing its definition by moving the cursor onto a notation
 (cf.~Figure \ref{fig:modification2_default-setup_item_entry-2}).}
{Clicking the ``setup\_item\_4'' of ``entry[1]''.}
{The ``setup\_item2\_4'' and ``setup\_item2\_5'' of ``entry[1]''
 have been calculated automatically.}
\end{templateenumerate}
\newpage
\section{Table column eliminating}
\label{sec:table_column_eliminating}
%
%
\begin{templateenumerate}
\myitem
\label{item:table-delete.php}
 {\cmbfsfx table-delete.php}
 (cf.~Template \ref{item:table-delete_setup_item_4.php})
\deft
\verb+<script type="text/javascript">+                             \\
\verb+  function +%
  \bluett{delete\_cell}%
  \verb+(+%
  \bluett{table\_tr\_id}%
  \verb+, +%
  \bluett{cell\_No}%
  \verb+)+                                                         \\
\verb+  {+                                                         \\
\verb+    document.getElementById(+%
  \bluett{table\_tr\_id}%
  \verb+)+                                                         \\
\verb+            .deleteCell(+%
  \bluett{cell\_No}%
  \verb+);+                                                        \\
\verb++                                                            \\
\verb+    if (navigator.appName == "Microsoft Internet Explorer")+ \\
\verb+    {+                                                       \\
\verb+      document.getElementById(+%
  \bluett{table\_tr\_id}%
  \verb+)+                                                         \\
\verb+              .nextSibling+                                  \\
\verb+              .deleteCell(+%
  \bluett{cell\_No}%
  \verb+);+                                                        \\
\verb+    }+                                                       \\
\verb++                                                            \\
\verb+    else+                                                    \\
\verb+//  if (navigator.userAgent.search("Chrome")    != -1 ||+    \\
\verb+//      navigator.userAgent.search("Firefox")   != -1 ||+    \\
\verb+//      navigator.userAgent.search("Opera")     != -1 ||+    \\
\verb+//      navigator.userAgent.search("Safari")    != -1 ||+    \\
\verb+//      navigator.userAgent.search("Navigator") != -1   )+   \\
\verb+    {+                                                       \\
\verb+      document.getElementById(+%
  \bluett{table\_tr\_id}%
  \verb+)+                                                         \\
\verb+              .nextSibling.nextSibling+                      \\
\verb+              .deleteCell(+%
  \bluett{cell\_No}%
  \verb+);+                                                        \\
\verb+    }+                                                       \\
\verb+  }+                                                         \\
\verb+</script>+                                                   \\
\tabb
\newpage
\myitem
\label{item:table-delete_setup_item_4.php}
 {\cmbfsfx table-delete\_setup\_item\_4.php}
 (cf.~Template \ref{item:table-delete.php},
  see also Template \ref{item:modification3_table-setup_item_entry})
\deft
\verb+<script type="text/javascript">+                             \\
\verb+<?php+                                                       \\
\verb+/*+                                                          \\
\verb+  document.getElementById("+%
  \redtt{setup\_item}%
  \verb+")+                                                        \\
\verb+          .deleteCell(+%
  \redtt{3}%
  \verb+);+ \mynote[2.5cm]{Delete the 4th column ({\tt setup\_item\_4})} \\
\verb++                                                            \\
\verb+  if (navigator.appName == "Microsoft Internet Explorer")+   \\
\verb+  {+                                                         \\
\verb+    document.getElementById("+%
  \redtt{setup\_item}%
  \verb+")+                                                        \\
\verb+            .nextSibling+                                    \\
\verb+            .deleteCell(+%
  \redtt{3}%
  \verb+);+                                                        \\
\verb+  }+                                                         \\
\verb++                                                            \\
\verb+  else+                                                      \\
\verb+//if (navigator.userAgent.search("Chrome")    != -1 ||+      \\
\verb+//    navigator.userAgent.search("Firefox")   != -1 ||+      \\
\verb+//    navigator.userAgent.search("Opera")     != -1 ||+      \\
\verb+//    navigator.userAgent.search("Safari")    != -1 ||+      \\
\verb+//    navigator.userAgent.search("Navigator") != -1   )+     \\
\verb+  {+                                                         \\
\verb+    document.getElementById("+%
  \redtt{setup\_item}%
  \verb+")+                                                        \\
\verb+            .nextSibling.nextSibling+                        \\
\verb+            .deleteCell(+%
  \redtt{3}%
  \verb+);+                                                        \\
\verb+  }+                                                         \\
\verb+*/+                                                          \\
\verb+?>+                                                          \\
\verb+  +%
  \bluett{delete\_cell}%
  \verb+("+%
  \redtt{setup\_item}%
  \verb+", +%
  \redtt{3}%
  \verb+);+ \mynote{Delete the 4th column ({\tt setup\_item\_4})}  \\
\verb+</script>+                                                   \\
\tabb
\newpage
\picin{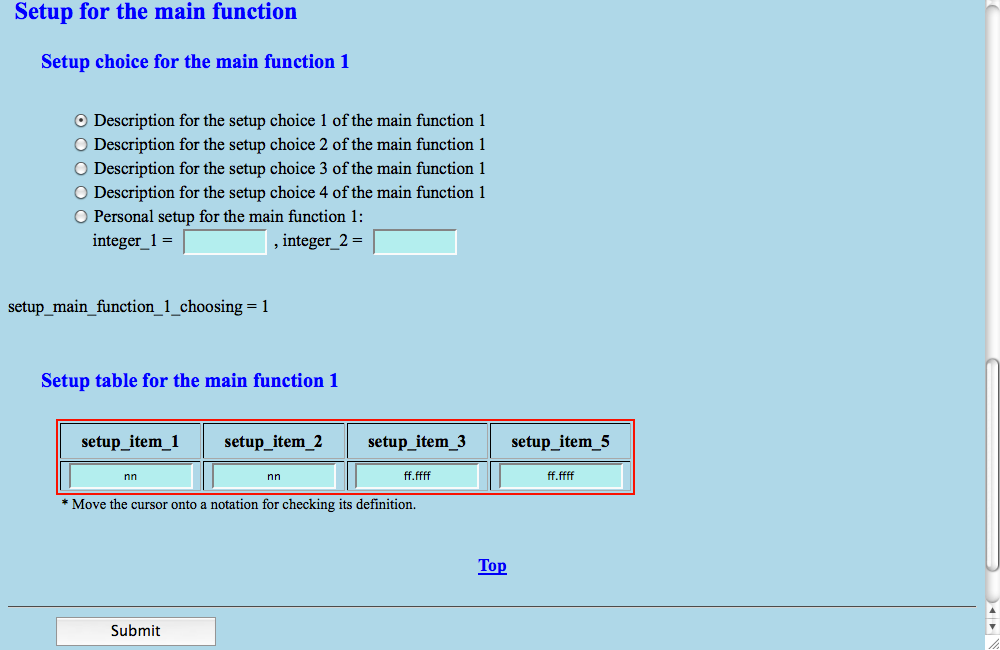}
{table-delete\_setup\_item\_4}
{The 4th column of the setup table has been deleted
 (cf.~Figures \ref{fig:default-setup_item}
  and \ref{fig:modification2_default-setup_item-1}).}
\newpage
\myitem
\label{item:modification3_table-setup_item_entry}
 {\bf Modification 3 of the `setup table for the chosen function'}
 (cf.~Template \ref{item:table-setup_item_entry})
\deft
\verb+<?php+                                   \\
\verb+  include("+%
  \bluett{tables/show\_definition}%
  \verb+.php");+ \mynote{Template \ref{item:show_definition.php}}   \\
\verb++                                        \\
\verb+//include("+%
  \bluett{tables/table-delete}%
  \verb+.php");+ \mynote[2cm]{Template \ref{item:table-delete.php}} \\
\verb++                                        \\
\verb+  include("+%
  \bluett{tables/table-delete\_head}%
  \verb+.php");+ \mynote{Template \ref{item:table-delete_head.php}} \\
\verb+//for ($+%
  \redtt{n\_entry}%
  \verb+ = 1; $+%
  \redtt{n\_entry}%
  \verb+ <= $+%
  \redtt{entry\_No}%
  \verb+; $+%
  \redtt{n\_entry}%
  \verb- ++)-                                  \\
\verb+  {+                                     \\
\verb+    include("+%
  \bluett{tables/table-delete\_entry}%
  \verb+.php");+ \mynote{Template \ref{item:table-delete_entry.php}} \\
\verb+  }+                                     \\
\verb++                                        \\
\verb+  if ($+%
  \redtt{setup\_main\_function\_1\_checked}%
  \verb+ == "yes")+                            \\
\verb+  {+                                     \\
\verb+    echo '<h4> +%
  \redtt{Setup table for the main function 1}%
  \verb+ </h4>' . "\r\n";+                     \\
\verb++                                        \\
\verb+    $+%
  \redtt{entry\_No}%
  \verb+ = +%
  \redtt{3}%
  \verb+;+                                     \\
\verb++                                        \\
\verb+         .+                              \\
\verb+         .+                              \\
\verb+         .+                              \\
\verb++                                        \\
\verb+      if ($+%
  \redtt{n\_entry}%
  \verb+ == $+%
  \redtt{entry\_No}%
  \verb+)+                                     \\
\verb+      {+                                 \\
\verb+        echo '<br />';+                  \\
\verb+        echo '+%
  \redtt{setup\_entry\_common}%
  \verb+ =             ' . $+%
  \redtt{setup\_entry\_common}%
  \verb+     . '<br />';+                      \\
\verb+        echo '<br />';+                  \\
\verb++                                        \\
\verb+        $+%
  \redtt{setup\_item\_checked}%
  \verb+ = "yes";+                             \\
\verb+      }+                                 \\
\verb+    }+                                   \\
\verb++                                        \\
\verb+    include("+%
  \bluett{tables/table-}%
  \verb++%
  \redtt{delete\_setup\_item\_3\_head}%
  \verb+.php");+ \mynote{Template \ref{item:table-delete_setup_item_3_head.php}} \\
\verb+    for ($+%
  \redtt{n\_entry}%
  \verb+ = 1; $+%
  \redtt{n\_entry}%
  \verb+ <= $+%
  \redtt{entry\_No}%
  \verb+; $+%
  \redtt{n\_entry}%
  \verb- ++)-                                  \\
\verb+    {+                                   \\
\verb+      include("+%
  \bluett{tables/table-}%
  \verb++%
  \redtt{delete\_setup\_item\_3\_entry}%
  \verb+.php");+ \mynote{Template \ref{item:table-delete_setup_item_3_entry.php}} \\
\verb+    }+                                   \\
\verb+  }+ \mynote[5cm]{Delete the 4th column ({\tt setup\_item\_3[]}) row by row} \\
\verb+?>+                                      \\
\tabb

\newpage
\myitem
\label{item:table-delete_head.php}
 {\cmbfsfx table-delete\_head.php}
 (cf.~Templates \ref{item:table-delete.php}
            and \ref{item:table-delete_setup_item_3_head.php})
\deft
\verb+<script type="text/javascript">+                             \\
\verb+  function +%
  \bluett{delete\_cell\_head}%
  \verb+(+%
  \bluett{table\_tr\_id}%
  \verb+, +%
  \bluett{cell\_No}%
  \verb+)+                                                         \\
\verb+  {+                                                         \\
\verb+    document.getElementById(+%
  \bluett{table\_tr\_id}%
  \verb+)+                                                         \\
\verb+            .deleteCell(+%
  \bluett{cell\_No}%
  \verb+);+                                                        \\
\verb+<?php+                                                       \\
\verb+/*+                                                          \\
\verb+    if (navigator.appName == "Microsoft Internet Explorer")+ \\
\verb+    {+                                                       \\
\verb+        var +%
  \redtt{n\_Enrty\_D}%
  \verb++                                                          \\
\verb+      = document.getElementById(+%
  \bluett{table\_tr\_id}%
  \verb+)+ \mynote{Moved to Template \ref{item:table-delete_entry.php}} \\
\verb+                .nextSibling;+                               \\
\verb+    }+                                                       \\
\verb++                                                            \\
\verb+    else+                                                    \\
\verb+//  if (navigator.userAgent.search("Chrome")    != -1 ||+    \\
\verb+//      navigator.userAgent.search("Firefox")   != -1 ||+    \\
\verb+//      navigator.userAgent.search("Opera")     != -1 ||+    \\
\verb+//      navigator.userAgent.search("Safari")    != -1 ||+    \\
\verb+//      navigator.userAgent.search("Navigator") != -1   )+   \\
\verb+    {+                                                       \\
\verb+        var +%
  \redtt{n\_Enrty\_D}%
  \verb++                                                          \\
\verb+      = document.getElementById(+%
  \bluett{table\_tr\_id}%
  \verb+)+ \mynote{Moved to Template \ref{item:table-delete_entry.php}} \\
\verb+                .nextSibling.nextSibling;+                   \\
\verb+    }+                                                       \\
\verb+*/+                                                          \\
\verb+?>+                                                          \\
\verb+  }+                                                         \\
\verb+</script>+                                                   \\
\tabb
\newpage
\myitem
\label{item:table-delete_entry.php}
 {\cmbfsfx table-delete\_entry.php}
 (cf.~Templates \ref{item:table-delete_head.php},
                \ref{item:table-delete_setup_item_3_head.php}
            and \ref{item:table-delete_setup_item_3_entry.php})
\deft
\verb+<script type="text/javascript">+                               \\
\verb+  function +%
  \bluett{delete\_cell\_entry}%
  \verb+(+%
  \bluett{table\_tr\_id}%
  \verb+, +%
  \bluett{cell\_No}%
  \verb+, +%
  \bluett{n\_Enrty\_No}%
  \verb+)+                                                           \\
\verb+  {+                                                           \\
\verb+    if (navigator.appName == "Microsoft Internet Explorer")+   \\
\verb+    {+                                                         \\
\verb+        var +%
  \redtt{n\_Enrty\_D}%
  \verb++                                                            \\
\verb+      = document.getElementById(+%
  \bluett{table\_tr\_id}%
  \verb+)+                                                           \\
\verb+                .nextSibling;+                                 \\
\verb+    }+                                                         \\
\verb++                                                              \\
\verb+    else+                                                      \\
\verb+//  if (navigator.userAgent.search("Chrome")    != -1 ||+      \\
\verb+//      navigator.userAgent.search("Firefox")   != -1 ||+      \\
\verb+//      navigator.userAgent.search("Opera")     != -1 ||+      \\
\verb+//      navigator.userAgent.search("Safari")    != -1 ||+      \\
\verb+//      navigator.userAgent.search("Navigator") != -1   )+     \\
\verb+    {+                                                         \\
\verb+        var +%
  \redtt{n\_Enrty\_D}%
  \verb++                                                            \\
\verb+      = document.getElementById(+%
  \bluett{table\_tr\_id}%
  \verb+)+                                                           \\
\verb+                .nextSibling.nextSibling;+                     \\
\verb+    }+                                                         \\
\verb++                                                              \\
\verb+    for (+%
  \redtt{n\_Enrty}%
  \verb+ = 1; +%
  \redtt{n\_Enrty}%
  \verb+ <= +%
  \bluett{n\_Enrty\_No}%
  \verb+; +%
  \redtt{n\_Enrty}%
  \verb- ++)-                                                        \\
\verb+    {+                                                         \\
\verb+      +%
  \redtt{n\_Enrty\_D}%
  \verb+.deleteCell(+%
  \bluett{cell\_No}%
  \verb+);+                                                          \\
\verb++                                                              \\
\verb+      if (navigator.appName == "Microsoft Internet Explorer")+ \\
\verb+      {+                                                       \\
\verb+          +%
  \redtt{n\_Enrty\_D}%
  \verb++                                                            \\
\verb+        = +%
  \redtt{n\_Enrty\_D}%
  \verb+.nextSibling;+                                               \\
\verb+      }+                                                       \\
\verb++                                                              \\
\verb+      else+                                                    \\
\verb+//    if (navigator.userAgent.search("Chrome")    != -1 ||+    \\
\verb+//        navigator.userAgent.search("Firefox")   != -1 ||+    \\
\verb+//        navigator.userAgent.search("Opera")     != -1 ||+    \\
\verb+//        navigator.userAgent.search("Safari")    != -1 ||+    \\
\verb+//        navigator.userAgent.search("Navigator") != -1   )+   \\
\verb+      {+                                                       \\
\verb+          +%
  \redtt{n\_Enrty\_D}%
  \verb++                                                            \\
\verb+        = +%
  \redtt{n\_Enrty\_D}%
  \verb+.nextSibling.nextSibling;+                                   \\
\verb+      }+                                                       \\
\verb+    }+                                                         \\
\verb+  }+                                                           \\
\verb+</script>+                                                     \\
\tabb
\newpage
\myitem
\label{item:table-delete_setup_item_3_head.php}
 {\cmbfsfx table-delete\_setup\_item\_3\_head.php}
 (cf.~Templates \ref{item:table-delete_setup_item_4.php},
                \ref{item:table-delete_head.php}
            and \ref{item:table-delete_entry.php})
\deft
\verb+<script type="text/javascript">+                           \\
\verb+<?php+                                                     \\
\verb+/*+                                                        \\
\verb+  document.getElementById("+%
  \redtt{setup\_item}%
  \verb+")+                                                      \\
\verb+          .deleteCell(+%
  \redtt{3}%
  \verb+);+ \mynote[2.5cm]{Delete the 4th column ({\tt setup\_item\_4[]})} \\
\verb++                                                          \\
\verb+  if (navigator.appName == "Microsoft Internet Explorer")+ \\
\verb+  {+                                                       \\
\verb+      var +%
  \redtt{n\_Enrty\_D}%
  \verb++                                                        \\
\verb+    = document.getElementById("+%
  \redtt{setup\_item}%
  \verb+")+                                                      \\
\verb+              .nextSibling;+                               \\
\verb+  }+                                                       \\
\verb++                                                          \\
\verb+  else+                                                    \\
\verb+//if (navigator.userAgent.search("Chrome")    != -1 ||+    \\
\verb+//    navigator.userAgent.search("Firefox")   != -1 ||+    \\
\verb+//    navigator.userAgent.search("Opera")     != -1 ||+    \\
\verb+//    navigator.userAgent.search("Safari")    != -1 ||+    \\
\verb+//    navigator.userAgent.search("Navigator") != -1   )+   \\
\verb+  {+                                                       \\
\verb+      var +%
  \redtt{n\_Enrty\_D}%
  \verb++                                                        \\
\verb+    = document.getElementById("+%
  \redtt{setup\_item}%
  \verb+")+                                                      \\
\verb+              .nextSibling.nextSibling;+                   \\
\verb+  }+                                                       \\
\verb+*/+                                                        \\
\verb+?>+                                                        \\
\verb+  +%
  \bluett{delete\_cell\_head}%
  \verb+("+%
  \redtt{setup\_item}%
  \verb+", +%
  \redtt{3}%
  \verb+);+ \mynote[1cm]{Delete the 4th column ({\tt setup\_item\_4})} \\
\verb+</script>+                                                 \\
\tabb
\newpage
\myitem
\label{item:table-delete_setup_item_3_entry.php}
 {\cmbfsfx table-delete\_setup\_item\_3\_entry.php}
 (cf.~Templates \ref{item:table-delete_setup_item_4.php}
            and \ref{item:table-delete_entry.php})
\deft
\verb+<script type="text/javascript">+                           \\
\verb+<?php+                                                     \\
\verb+/*+                                                        \\
\verb+  +%
  \redtt{n\_Enrty\_D}%
  \verb+.deleteCell(+%
  \redtt{3}%
  \verb+);+ \mynote[1cm]{Delete the 4th column ({\tt setup\_item\_4[]})} \\
\verb++                                                          \\
\verb+  if (navigator.appName == "Microsoft Internet Explorer")+ \\
\verb+  {+                                                       \\
\verb+      +%
  \redtt{n\_Enrty\_D}%
  \verb++                                                        \\
\verb+    = +%
  \redtt{n\_Enrty\_D}%
  \verb+.nextSibling;+                                           \\
\verb+  }+                                                       \\
\verb++                                                          \\
\verb+  else+                                                    \\
\verb+//if (navigator.userAgent.search("Chrome")    != -1 ||+    \\
\verb+//    navigator.userAgent.search("Firefox")   != -1 ||+    \\
\verb+//    navigator.userAgent.search("Opera")     != -1 ||+    \\
\verb+//    navigator.userAgent.search("Safari")    != -1 ||+    \\
\verb+//    navigator.userAgent.search("Navigator") != -1   )+   \\
\verb+  {+                                                       \\
\verb+      +%
  \redtt{n\_Enrty\_D}%
  \verb++                                                        \\
\verb+    = +%
  \redtt{n\_Enrty\_D}%
  \verb+.nextSibling.nextSibling;+                               \\
\verb+  }+                                                       \\
\verb+*/+                                                        \\
\verb+?>+ \mynote[7.5cm]{Delete the 4th column ({\tt setup\_item\_4})} \\
\verb+  +%
  \bluett{delete\_cell\_entry}%
  \verb+("+%
  \redtt{setup\_item}%
  \verb+", +%
  \redtt{3}%
  \verb+, <?php echo $+%
  \redtt{entry\_No}%
  \verb+; ?>);+                                                  \\
\verb+</script>+                                                 \\
\tabb
\newpage
\picin{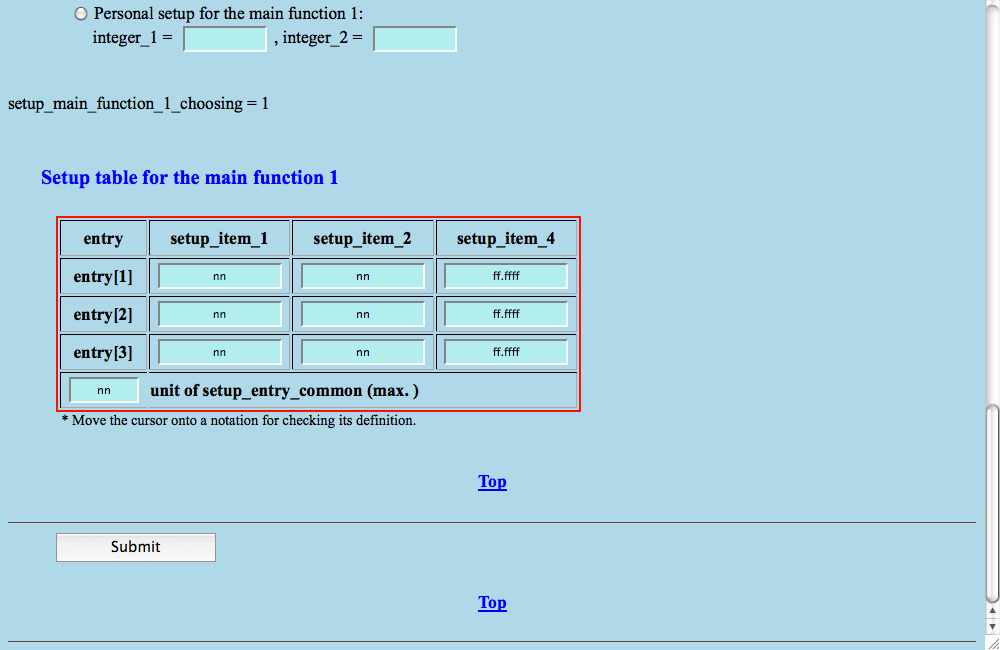}
{table-delete\_setup\_item\_3\_entry}
{The 4th column of the setup table has been deleted row by row
 (cf.~Figures \ref{fig:default-setup_item_entry-1}
  and \ref{fig:modification2_default-setup_item_entry-1}).}
\end{templateenumerate}

\addemptypage
\chapter{Result Presentation}
\label{chap:result_presentation}
%
%
%
\newpage
\section{Program starting}
\label{sec:program_starting}
%
%
\begin{templateenumerate}
\myitem
\label{item:starting_or_modifying}
 {\bf Starting a program or modifying the setup}
\deft
\verb+<?php+                                                                               \\
\verb+  if ($+%
  \bluett{main\_function}%
  \verb+ != NULL  &&+                                                                      \\
\verb+      $+%
  \redtt{mode\_checked}%
  \verb+  == "yes" &&+                                                                     \\
\verb+      $+%
  \redtt{setup\_checked}%
  \verb+ == "yes"   )+                                                                     \\
\verb+  {+                                                                                 \\
\verb+    if ($_POST["+%
  \bluett{start}%
  \verb+"] == NULL && + \mynote{For {\tt main.php}, not {\tt results.php},
                                cf.~Template \ref{item:going_to_results}}                  \\
\verb+        $+%
  \bluett{output\_series}%
  \verb+  == NULL   )+                                                                     \\
\verb+    {+                                                                               \\
\verb+      echo '<script type="text/javascript">'                              . "\r\n";+ \\
\verb+      echo '  if (navigator.appName == "Microsoft Internet Explorer")'    . "\r\n";+ \\
\verb+      echo '  {'                                                          . "\r\n";+ \\
\verb+      echo '    window.location = "#+%
  \bluett{start}%
  \verb+";'                              . "\r\n";+                                        \\
\verb+      echo '  }'                                                          . "\r\n";+ \\
\verb+      echo '</script>'                                                    . "\r\n";+ \\
\verb++                                                                                    \\
\verb+      echo '<h3 id="+%
  \bluett{start}%
  \verb+"> +%
  \bluett{Start Program}%
  \verb+ </h3>'                          . "\r\n";+                                        \\
\verb++                                                                                    \\
\verb+      echo '<script type="text/javascript">'                              . "\r\n";+ \\
\verb+      echo '  if (navigator.userAgent.search("Chrome")     != -1     ||'  . "\r\n";+ \\
\verb+      echo '      (navigator.userAgent.search("Firefox")   != -1 &&'      . "\r\n";+ \\
\verb+      echo '       navigator.userAgent.search("Navigator") == -1   ) ||'  . "\r\n";+ \\
\verb+      echo '      navigator.userAgent.search("Opera")      != -1     ||'  . "\r\n";+ \\
\verb+      echo '      (navigator.userAgent.search("Chrome")    == -1 &&'      . "\r\n";+ \\
\verb+      echo '       navigator.userAgent.search("Safari")    != -1   )   )' . "\r\n";+ \\
\verb+      echo '  {'                                                          . "\r\n";+ \\
\verb+      echo '    window.location = "#+%
  \bluett{start}%
  \verb+";'                              . "\r\n";+                                        \\
\verb+      echo '  }'                                                          . "\r\n";+ \\
\verb+      echo '</script>'                                                    . "\r\n";+ \\
\defc
\verb++                                                                                    \\
\verb+      echo '<h4> +%
  \bluett{Check the above setup}%
  \verb+ </h4>'                             . "\r\n";+                                     \\
\verb++                                                                                    \\
\verb+      echo '<input type="submit" class="+%
  \bluett{submit}%
  \verb+"'                          . "\r\n";+                                             \\
\verb+      echo '       name="+%
  \bluett{start}%
  \verb+" value="+%
  \bluett{Start}%
  \verb+"'                            . "\r\n";+                                           \\
\verb+      if ($_POST["+%
  \bluett{start}%
  \verb+"] != NULL &&+                                                                     \\
\verb+          $+%
  \bluett{output\_series}%
  \verb+  != NULL   )+                                                                     \\
\verb+      {+                                                                             \\
\verb+        echo '       disabled = "disabled"';+                                        \\
\verb+      }+                                                                             \\
\verb+      echo '       />'                                                    . "\r\n";+ \\
\verb+      echo '<br />'                                                       . "\r\n";+ \\
\verb++                                                                                    \\
\verb+      echo '<br />'                                                       . "\r\n";+ \\
\verb+      echo '<input type="submit" class="+%
  \bluett{submit}%
  \verb+"'                          . "\r\n";+                                             \\
\verb+      echo '       name="+%
  \bluett{re\_submit}%
  \verb+" value="+%
  \bluett{Modify and re-submit}%
  \verb+"'         . "\r\n";+                                                              \\
\verb+      if ($_POST["+%
  \bluett{start}%
  \verb+"] != NULL &&+                                                                     \\
\verb+          $+%
  \bluett{output\_series}%
  \verb+  != NULL   )+                                                                     \\
\verb+      {+                                                                             \\
\verb+        echo '       disabled = "disabled"';+                                        \\
\verb+      }+                                                                             \\
\verb+      echo '       />'                                                    . "\r\n";+ \\
\verb+      echo '<br />'                                                       . "\r\n";+ \\
\verb++                                                                                    \\
\verb+      echo '<script type="text/javascript">'                              . "\r\n";+ \\
\verb+      echo '  if (navigator.userAgent.search("Navigator") != -1)'         . "\r\n";+ \\
\verb+      echo '  {'                                                          . "\r\n";+ \\
\verb+      echo '    window.location = "#+%
  \bluett{start}%
  \verb+";'                              . "\r\n";+                                        \\
\verb+      echo '  }'                                                          . "\r\n";+ \\
\verb+      echo '</script>'                                                    . "\r\n";+ \\
\verb++                                                                                    \\
\verb+         .+                                                                          \\
\verb+         .+ \mynote{Template \ref{item:changing_name}}                               \\
\verb+         .+                                                                          \\
\verb+    }+                                                                               \\
\verb+  }+                                                                                 \\
\defc
\verb++                                                                                    \\
\verb+  elseif (!($+%
  \bluett{main\_function}%
  \verb+ != NULL  &&+                                                                      \\
\verb+            $+%
  \redtt{mode\_checked}%
  \verb+  == "yes" &&+                                                                     \\
\verb+            $+%
  \redtt{setup\_checked}%
  \verb+ == "yes"   ) &&+                                                                  \\

\verb+          $+%
  \bluett{output\_series}%
  \verb+   == NULL        )+                                                               \\
\verb+  {+                                                                                 \\
\verb+    echo '<input type="submit" class="+%
  \bluett{submit}%
  \verb+" value="+%
  \bluett{Submit}%
  \verb+" />'          . "\r\n";+                                                          \\
\verb+    echo '<br />'                                                         . "\r\n";+ \\
\verb+  }+ \mynote[7.5cm]{cf.~Template \ref{item:form}}                                    \\
\verb+?>+                                                                                  \\
\verb++                                                                                    \\
\verb+</form>+                                                                             \\
\verb++                                                                                    \\
\verb+<?php+                                                                               \\
\verb+  if ($_POST["+%
  \bluett{start}%
  \verb+"] == NULL &&+                                                                     \\
\verb+      $+%
  \bluett{output\_series}%
  \verb+  == NULL   )+                                                                     \\
\verb+  {+ \mynote[7.5cm]{cf.~Template \ref{item:switch_further_step}}                     \\
\verb+    echo '<h5> <a href="#+%
  \bluett{main}%
  \verb+">Top</a> </h5>'                             . "\r\n";+                            \\
\verb+    echo '<hr />'                                                         . "\r\n";+ \\
\verb+  }+                                                                                 \\
\verb+?>+                                                                                  \\
\verb+         .+                                                                          \\
\verb+         .+ \mynote{Template \ref{item:copying_uploaded_files}}                      \\
\verb+         .+                                                                          \\
\verb++                                                                                    \\
\verb+         .+                                                                          \\
\verb+         .+ \mynote{Template \ref{item:going_to_results}}                            \\
\verb+         .+                                                                          \\
\tabb
\newpage
\picin{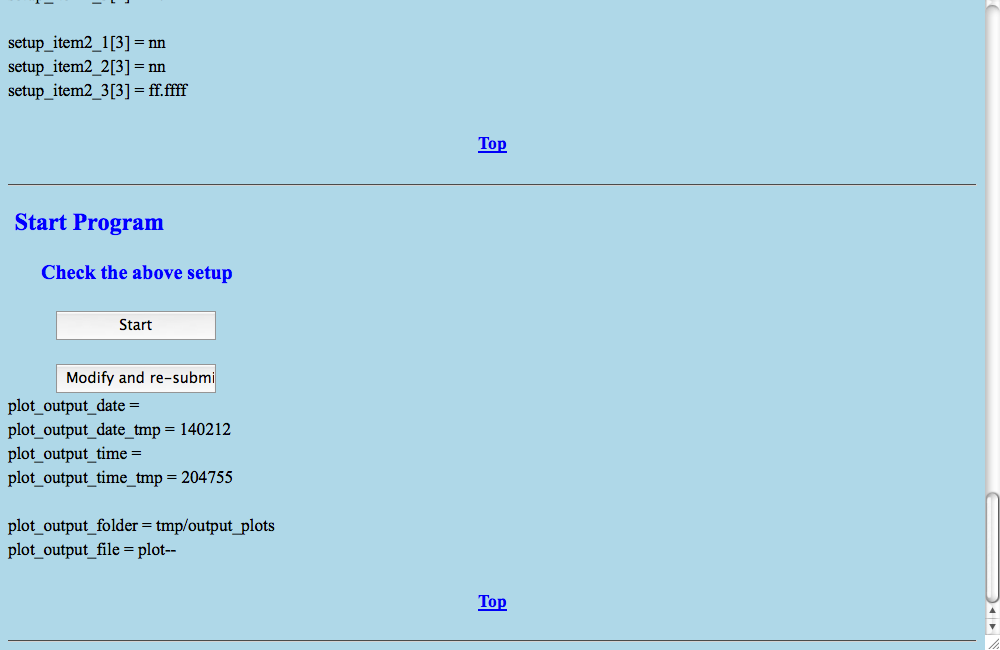}
{start\_program}
{Starting a program or modifying the setup.}
%
%
%
%
\newpage
%
%
%
%
\myitem
\label{item:copying_uploaded_files}
 {\bf Copying uploaded files}
\deft
\verb+<?php+                                                                       \\
\verb+  if ($+%
  \bluett{main\_function}%
  \verb+  != NULL  &&+                                                             \\
\verb+      $+%
  \redtt{mode\_checked}%
  \verb+   == "yes" &&+                                                            \\
\verb+      $+%
  \redtt{setup\_checked}%
  \verb+  == "yes" &&+                                                             \\
\verb+      $_POST["+%
  \bluett{start}%
  \verb+"] != NULL    )+ \mynote[1cm]{After {\tt "Start"} clicking, not {\tt "Modify and re-submit"}} \\
\verb+  {+                                                                         \\
\verb+    include("+%
  \bluett{main/setup\_output\_file}%
  \verb+.php");+ \mynote[1cm]{Template \ref{item:setup_output_file.php}}           \\
\verb++                                                                            \\
\verb+    if ($+%
  \redtt{function\_choice}%
  \verb+ == "+%
  \redtt{function\_user}%
  \verb+")+                                                                        \\
\verb+    {+                                                                       \\
\verb+        $+%
  \bluett{code\_folder}%
  \verb++                                                                          \\
\verb+      = "+%
  \redtt{/home/}%
  \verb++%
  \bluett{clshan}%
  \verb++%
  \redtt{/main}%
  \verb+";+ \mynote{Folder of the main program}                                    \\
\verb++                                                                            \\
\verb+        $+%
  \bluett{personal\_file\_folder\_copy}%
  \verb++                                                                          \\
\verb+      = +%
  \bluett{\$code\_folder .~"/" .~\$personal\_file\_folder}%
  \verb+;+                                                                         \\
\verb++                                                                            \\
%
%
\verb+        $+%
  \bluett{cp\_cmd}%
  \verb++                                                                          \\
\verb+      = "cp " . $+%
  \bluett{personal\_file\_folder}%
  \verb+      . "/" . $+%
  \redtt{personal\_function\_file\_copy}%
  \verb+ . " "+                                                                    \\
\verb+              . $+%
  \bluett{personal\_file\_folder\_copy}%
  \verb+ . "/" . "+%
  \redtt{function\_user.c}%
  \verb+";+                                                                        \\
\verb++                                                                            \\
\verb+      shell_exec("$+%
  \bluett{cp\_cmd}%
  \verb+");+                                                                       \\
\verb++                                                                            \\
\verb+      if (+%
  \redtt{\$function2\_needed == "yes"}%
  \verb+)+                                                                         \\
\verb+      {+                                                                     \\
%
%
\verb+          $+%
  \bluett{cp\_cmd}%
  \verb++                                                                          \\
\verb+        = "cp " . $+%
  \bluett{personal\_file\_folder}%
  \verb+ . "/" . $+%
  \redtt{personal\_function2\_file\_copy}%
  \verb+ . " "+                                                                    \\
\verb+                . $+%
  \bluett{personal\_file\_folder}%
  \verb+ . "/" . "+%
  \redtt{function2\_user}%
  \verb+.txt";+                                                                    \\
\verb++                                                                            \\
\verb++                                                                            \\
\verb+        shell_exec("$+%
  \bluett{cp\_cmd}%
  \verb+");+                                                                       \\
\verb+      }+                                                                     \\
\verb++                                                                            \\
\verb+      shell_exec("gcc -g -o +%
  \bluett{main-web-user}%
  \verb+ +%
  \bluett{\$code\_folder/main-web.c}%
  \verb+ -lm");+                                                                   \\
\verb+      shell_exec("./+%
  \bluett{main-web-user}%
  \verb+");+                                                                       \\
\verb+    }+ \mynote[2.5cm]{Re-compile and execute the main program with user-uploaded file(s)} \\
\verb++                                                                            \\
\verb+    else+                                                                    \\
\verb+    {+                                                                       \\
\verb+      shell_exec("./+%
  \bluett{main-web}%
  \verb+");+ \mynote[1cm]{Execute the default main program}                        \\
\verb+    }+                                                                       \\
\defc
\verb++                                                                            \\
\verb+    if ($+%
  \redtt{function\_choice}%
  \verb+ == "+%
  \redtt{function\_user}%
  \verb+")+                                                                        \\
\verb+    {+                                                                       \\
\verb+        $+%
  \bluett{cp\_cmd}%
  \verb++                                                                          \\
\verb+      = "cp +%
  \bluett{samples/sample-}%
  \verb++%
  \redtt{function\_user}%
  \verb+.txt "+                                                                    \\
\verb+          . $+%
  \bluett{personal\_file\_folder\_copy}%
  \verb+ . "/" . "+%
  \redtt{function\_user.c}%
  \verb+";+                                                                        \\
\verb++                                                                            \\
\verb+      shell_exec("$+%
  \bluett{cp\_cmd}%
  \verb+");+ \mynote{Copy the sample file(s) to replace the user-uploaded files}   \\
\verb++                                                                            \\
\verb+      if (+%
  \redtt{\$function2\_needed == "yes"}%
  \verb+)+                                                                         \\
\verb+      {+                                                                     \\
\verb+          $+%
  \bluett{cp\_cmd}%
  \verb++                                                                          \\
\verb+        = "cp +%
  \bluett{samples/sample-}%
  \verb++%
  \redtt{function2\_user}%
  \verb+.txt "+                                                                    \\
\verb+            . $+%
  \bluett{personal\_file\_folder}%
  \verb+ . "/" . "+%
  \redtt{function2\_user}%
  \verb+.txt";+                                                                    \\
\verb++                                                                            \\
\verb+        shell_exec("$+%
  \bluett{cp\_cmd}%
  \verb+");+                                                                       \\
\verb+      }+                                                                     \\
\verb+    }+                                                                       \\
\verb++                                                                            \\
\verb+    include("+%
  \bluett{main/setup\_output\_reload}%
  \verb+.php");+ \mynote[1cm]{Template \ref{item:setup_output_reload.php}}         \\
\verb++                                                                            \\
\verb+    echo '<script type="text/javascript">'         . "\r\n";+                \\
\verb+    echo '  this.location.replace("+%
  \bluett{results}%
  \verb+.php");' . "\r\n";+ \mynote[1cm]{Redirect to {\tt results.php}}            \\
\verb+    echo '</script>'                               . "\r\n";+                \\
\verb+  }+                                                                         \\
\verb+?>+                                                                          \\
\tabb
\newpage
\myitem
\label{item:sample-input_setup_web.txt}
 {\cmbfsfx sample-input\_setup\_web.txt}
 (cf.~Templates \ref{item:setup_output_file.php},
  \ref{item:sample-input_setup_web_reload.txt}
  and \ref{item:setup-reader.c})
\deft
\verb++%
  \bluett{0}%
  \verb+  +%
  \bluett{Main\_function}%
  \verb++                              \\
\verb++                                \\
\verb+   +%
  \bluett{1}%
  \verb+   +%
  \bluett{1}%
  \verb+,   +%
  \bluett{main\_function\_choosing}%
  \verb+ = +%
  \redtt{1}%
  \verb+ ;      +%
  \bluett{0}%
  \verb+  +%
  \redtt{...\_/\_...\_/\_...\_/\_...}%
  \verb++                              \\
\verb++                                \\
\verb++%
  \bluett{0}%
  \verb+  +%
  \redtt{Main\_function\_1}%
  \verb++                              \\
\verb++                                \\
\verb+   +%
  \redtt{2}%
  \verb+   +%
  \redtt{1}%
  \verb+,   +%
  \redtt{mode\_choosing}%
  \verb+          = +%
  \redtt{1}%
  \verb+ ;      +%
  \bluett{0}%
  \verb+  +%
  \redtt{...\_/\_...\_/\_...\_/\_...}%
  \verb++                              \\
\verb++                                \\
\verb+   +%
  \redtt{2}%
  \verb+   +%
  \redtt{2}%
  \verb+,   +%
  \redtt{integer\_1}%
  \verb+     = +%
  \bluett{nn}%
  \verb+ ;+                            \\
\verb+   +%
  \redtt{2}%
  \verb+   +%
  \redtt{3}%
  \verb+,   +%
  \redtt{integer\_2}%
  \verb+     = +%
  \bluett{nn}%
  \verb+ ;+                            \\
\verb++                                \\
\verb+   +%
  \redtt{2}%
  \verb+   +%
  \redtt{4}%
  \verb+   1,   +%
  \redtt{float\_1\_1}%
  \verb+ = +%
  \bluett{ff.ffff}%
  \verb+ ;+                            \\
\verb+   +%
  \redtt{2}%
  \verb+   +%
  \redtt{4}%
  \verb+   2,   +%
  \redtt{float\_1\_2}%
  \verb+ = +%
  \bluett{ff.ffff}%
  \verb+ ;+                            \\
\verb++                                \\
\verb+   +%
  \redtt{2}%
  \verb+   +%
  \redtt{5}%
  \verb+   1,   +%
  \redtt{float\_2\_1}%
  \verb+ = +%
  \bluett{ff.ffff}%
  \verb+ ;+                            \\
\verb+   +%
  \redtt{2}%
  \verb+   +%
  \redtt{5}%
  \verb+   2,   +%
  \redtt{float\_2\_2}%
  \verb+ = +%
  \bluett{ff.ffff}%
  \verb+ ;+                            \\
\verb++                                \\
\verb+   +%
  \redtt{2}%
  \verb+   +%
  \redtt{6}%
  \verb+   1,   +%
  \redtt{integer\_3}%
  \verb+ = +%
  \bluett{nn}%
  \verb+      ;+                       \\
\verb+   +%
  \redtt{2}%
  \verb+   +%
  \redtt{6}%
  \verb+   2,   +%
  \redtt{float\_3}%
  \verb+   = +%
  \bluett{ff.ffff}%
  \verb+ ;+                            \\
\verb++                                \\
\verb++%
  \bluett{0}%
  \verb+  For_temporary_folders+       \\
\verb++                                \\
\verb+   +%
  \redtt{1}%
  \verb+   +%
  \redtt{5}%
  \verb+   1,   +%
  \bluett{personal\_file\_date}%
  \verb+ = +%
  \bluett{140325}%
  \verb+ ;+                            \\
\verb+   +%
  \redtt{1}%
  \verb+   +%
  \redtt{5}%
  \verb+   2,   +%
  \bluett{personal\_file\_time}%
  \verb+ = +%
  \bluett{060000}%
  \verb+ ;+                            \\
\verb++                                \\
\verb+   +%
  \redtt{1}%
  \verb+   +%
  \redtt{5}%
  \verb+   3,   +%
  \bluett{plot\_output\_date}%
  \verb+   = +%
  \bluett{140325}%
  \verb+ ;+                            \\
\verb+   +%
  \redtt{1}%
  \verb+   +%
  \redtt{5}%
  \verb+   4,   +%
  \bluett{plot\_output\_time}%
  \verb+   = +%
  \bluett{090000}%
  \verb+ ;+                            \\
\verb++                                \\
\verb+   +%
  \redtt{1}%
  \verb+   +%
  \redtt{5}%
  \verb+   5,   +%
  \bluett{random\_gen}%
  \verb+         =   +%
  \bluett{1732}%
  \verb+ ;+                            \\
\tabb
\newpage
\myitem
\label{item:setup_output_file.php}
 {\cmbfsfx setup\_output\_file.php}
 (cf.~Templates \ref{item:sample-input_setup_web.txt},
  \ref{item:setup_output_reload.php}
  and \ref{item:setup-reader.c})
\deft
\verb+<?php+                                                                               \\
\verb+  $+%
  \bluett{setup\_output}%
  \verb+ = fopen("+%
  \bluett{tmp/input\_setup\_web/input\_setup\_web}%
  \verb+.txt", "w");+                                                                      \\
\verb++                                                                                    \\
\verb+  fputs($+%
  \bluett{setup\_output}%
  \verb+, sprintf("\n"));+                                                                 \\
\verb++                                                                                    \\
\verb+  /* +%
  \bluett{Main function}%
  \verb+ ***************************************************************/+                 \\
\verb++                                                                                    \\
\verb+  fputs($+%
  \bluett{setup\_output}%
  \verb+, sprintf("+%
  \bluett{0}%
  \verb+  +%
  \bluett{Main\_function}%
  \verb+\n"));+                                                                            \\
\verb+  fputs($+%
  \bluett{setup\_output}%
  \verb+, sprintf("\n"));+                                                                 \\
\verb++                                                                                    \\
\verb+  fputs($+%
  \bluett{setup\_output}%
  \verb+,+                                                                                 \\
\verb+        sprintf("   +%
  \bluett{1}%
  \verb+   +%
  \bluett{1}%
  \verb+,   +%
  \bluett{main\_function\_choosing}%
  \verb+ = %d ;\n",+                                                                       \\
\verb+                $+%
  \bluett{main\_function\_choosing}%
  \verb+));+                                                                               \\
\verb++                                                                                    \\
\verb+  fputs($+%
  \bluett{setup\_output}%
  \verb+, sprintf("\n"));+                                                                 \\
\verb+  fputs($+%
  \bluett{setup\_output}%
  \verb+, sprintf("\n"));+                                                                 \\
\defc
\verb++                                                                                    \\
\verb+  /* +%
  \redtt{Main\_function\_1}%
  \verb+ *************************************************************/+                   \\
\verb++                                                                                    \\
\verb+  if ($+%
  \bluett{main\_function}%
  \verb+ == "+%
  \redtt{main\_function\_1}%
  \verb+")+                                                                                \\
\verb+  {+                                                                                 \\
\verb+    fputs($+%
  \bluett{setup\_output}%
  \verb+, sprintf("  +%
  \bluett{0}%
  \verb+  +%
  \redtt{Main\_function\_1}%
  \verb+\n"));+                                                                            \\
\verb+    fputs($+%
  \bluett{setup\_output}%
  \verb+, sprintf("\n"));+                                                                 \\
\verb++                                                                                    \\
\verb+    fputs($+%
  \bluett{setup\_output}%
  \verb+,+                                                                                 \\
\verb+          sprintf("   +%
  \redtt{2}%
  \verb+   +%
  \redtt{1}%
  \verb+,   +%
  \redtt{mode\_choosing}%
  \verb+        = %d ;\n",+                                                                \\
\verb+                  $+%
  \redtt{mode\_choosing}%
  \verb+));+                                                                               \\
\verb+    fputs($+%
  \bluett{setup\_output}%
  \verb+, sprintf("\n"));+                                                                 \\
\verb++                                                                                    \\
\verb+    fputs($+%
  \bluett{setup\_output}%
  \verb+,+                                                                                 \\
\verb+          sprintf("   +%
  \redtt{2}%
  \verb+   +%
  \redtt{2}%
  \verb+,   +%
  \redtt{integer\_1}%
  \verb+     = %s ;\n",+ \mynote[1cm]{Not {\tt \%d}!}                                      \\
\verb+                  $+%
  \redtt{integer\_1}%
  \verb+));+                                                                               \\
\verb+    fputs($+%
  \bluett{setup\_output}%
  \verb+,+                                                                                 \\
\verb+          sprintf("   +%
  \redtt{2}%
  \verb+   +%
  \redtt{3}%
  \verb+,   +%
  \redtt{integer\_2}%
  \verb+     = %s ;\n",+                                                                   \\
\verb+                  $+%
  \redtt{integer\_2}%
  \verb+));+                                                                               \\
\verb+    fputs($+%
  \bluett{setup\_output}%
  \verb+, sprintf("\n"));+                                                                 \\
\verb++                                                                                    \\
\verb+    fputs($+%
  \bluett{setup\_output}%
  \verb+,+                                                                                 \\
\verb+          sprintf("   +%
  \redtt{2}%
  \verb+   +%
  \redtt{4}%
  \verb+   1,   +%
  \redtt{float\_1\_1}%
  \verb+ = %s ;\n",+ \mynote[1cm]{Not {\tt \%f}!}                                          \\
\verb+                  $+%
  \redtt{float\_1\_1}%
  \verb+));+                                                                               \\
\verb+    fputs($+%
  \bluett{setup\_output}%
  \verb+,+                                                                                 \\
\verb+          sprintf("   +%
  \redtt{2}%
  \verb+   +%
  \redtt{4}%
  \verb+   2,   +%
  \redtt{float\_1\_2}%
  \verb+ = %s ;\n",+                                                                       \\
\verb+                  $+%
  \redtt{float\_1\_2}%
  \verb+));+                                                                               \\
\verb+    fputs($+%
  \bluett{setup\_output}%
  \verb+, sprintf("\n"));+                                                                 \\
\verb++                                                                                    \\
\verb+    fputs($+%
  \bluett{setup\_output}%
  \verb+,+                                                                                 \\
\verb+          sprintf("   +%
  \redtt{2}%
  \verb+   +%
  \redtt{5}%
  \verb+   1,   +%
  \redtt{float\_2\_1}%
  \verb+ = %s ;\n",+                                                                       \\
\verb+                  $+%
  \redtt{float\_2\_1}%
  \verb+));+                                                                               \\
\verb+    fputs($+%
  \bluett{setup\_output}%
  \verb+,+                                                                                 \\
\verb+          sprintf("   +%
  \redtt{2}%
  \verb+   +%
  \redtt{5}%
  \verb+   2,   +%
  \redtt{float\_2\_2}%
  \verb+ = %s ;\n",+                                                                       \\
\verb+                  $+%
  \redtt{float\_2\_2}%
  \verb+));+                                                                               \\
\verb+    fputs($+%
  \bluett{setup\_output}%
  \verb+, sprintf("\n"));+                                                                 \\
\verb++                                                                                    \\
\verb+    fputs($+%
  \bluett{setup\_output}%
  \verb+,+                                                                                 \\
\verb+          sprintf("   +%
  \redtt{2}%
  \verb+   +%
  \redtt{6}%
  \verb+   1,   +%
  \redtt{integer\_3}%
  \verb+ = %s ;\n",+                                                                       \\
\verb+                  $+%
  \redtt{integer\_3}%
  \verb+));+                                                                               \\
\verb++                                                                                    \\
\verb+    fputs($+%
  \bluett{setup\_output}%
  \verb+,+                                                                                 \\
\verb+          sprintf("   +%
  \redtt{2}%
  \verb+   +%
  \redtt{6}%
  \verb+   2,   +%
  \redtt{float\_3}%
  \verb+   = %s ;\n",+                                                                     \\
\verb+                  $+%
  \redtt{float\_3}%
  \verb+));+                                                                               \\
\verb++                                                                                    \\
\verb+    fputs($+%
  \bluett{setup\_output}%
  \verb+, sprintf("\n"));+                                                                 \\
\verb+  }+                                                                                 \\
\defc
\verb++                                                                                    \\
\verb+  /* For temporary folders *******************************************************/+ \\
\verb++                                                                                    \\
\verb+  fputs($+%
  \bluett{setup\_output}%
  \verb+,+                                                                                 \\
\verb+        sprintf("+%
  \bluett{0}%
  \verb+  For_temporary_folders\n"));+                                                     \\
\verb+  fputs($+%
  \bluett{setup\_output}%
  \verb+, sprintf("\n"));+                                                                 \\
\verb++                                                                                    \\
\verb+  fputs($+%
  \bluett{setup\_output}%
  \verb+,+                                                                                 \\
\verb+        sprintf("   +%
  \redtt{1}%
  \verb+   +%
  \redtt{5}%
  \verb+   1,   +%
  \bluett{personal\_file\_date}%
  \verb+ = %s ;\n",+ \mynote[1cm]{Not {\tt \%d}!}                                          \\
\verb+                $+%
  \bluett{personal\_function\_date}%
  \verb+));+                                                                               \\
\verb+  fputs($+%
  \bluett{setup\_output}%
  \verb+,+                                                                                 \\
\verb+        sprintf("   +%
  \redtt{1}%
  \verb+   +%
  \redtt{5}%
  \verb+   2,   +%
  \bluett{personal\_file\_time}%
  \verb+ = %s ;\n",+                                                                       \\
\verb+                $+%
  \bluett{personal\_function\_time}%
  \verb+));+                                                                               \\
\verb+  fputs($+%
  \bluett{setup\_output}%
  \verb+, sprintf("\n"));+                                                                 \\
\verb++                                                                                    \\
\verb+  fputs($+%
  \bluett{setup\_output}%
  \verb+,+                                                                                 \\
\verb+        sprintf("   +%
  \redtt{1}%
  \verb+   +%
  \redtt{5}%
  \verb+   3,   +%
  \bluett{plot\_output\_date}%
  \verb+   = %s ;\n",+                                                                     \\
\verb+                $+%
  \bluett{plot\_output\_date}%
  \verb+));+                                                                               \\
\verb+  fputs($+%
  \bluett{setup\_output}%
  \verb+,+                                                                                 \\
\verb+        sprintf("   +%
  \redtt{1}%
  \verb+   +%
  \redtt{5}%
  \verb+   4,   +%
  \bluett{plot\_output\_time}%
  \verb+   = %s ;\n",+                                                                     \\
\verb+                $+%
  \bluett{plot\_output\_time}%
  \verb+));+                                                                               \\
\verb+  fputs($+%
  \bluett{setup\_output}%
  \verb+, sprintf("\n"));+                                                                 \\
\verb++                                                                                    \\
\verb+  $+%
  \bluett{random\_gen}%
  \verb+ = ( (gmdate(d) . gmdate(H)) * (gmdate(i) . gmdate(s)) ) % 10000;+                 \\
\verb++                                                                                    \\
\verb+  if ($+%
  \bluett{random\_gen}%
  \verb+ == 0)+                                                                            \\
\verb+  {+                                                                                 \\
\verb+    $+%
  \bluett{random\_gen}%
  \verb+ = 909;+                                                                           \\
\verb+  }+                                                                                 \\
\verb++                                                                                    \\
\verb+  elseif ($+%
  \bluett{random\_gen}%
  \verb+ >=    1 &&+                                                                       \\
\verb+          $+%
  \bluett{random\_gen}%
  \verb+ <    10   )+                                                                      \\
\verb+  {+                                                                                 \\
\verb+    $+%
  \bluett{random\_gen}%
  \verb+ = $+%
  \bluett{random\_gen}%
  \verb+ * 909;+                                                                           \\
\verb+  }+                                                                                 \\
\verb++                                                                                    \\
\verb+  elseif ($+%
  \bluett{random\_gen}%
  \verb+ >=   10 &&+ \mynote{\bluett{909} $<$ \bluett{random\_gen} $\le$ \bluett{9999}}    \\
\verb+          $+%
  \bluett{random\_gen}%
  \verb+ <   100   )+                                                                      \\
\verb+  {+                                                                                 \\
\verb+    $+%
  \bluett{random\_gen}%
  \verb+ = $+%
  \bluett{random\_gen}%
  \verb+ * 101;+                                                                           \\
\verb+  }+                                                                                 \\
\verb++                                                                                    \\
\verb+  elseif ($+%
  \bluett{random\_gen}%
  \verb+ >=  100 &&+                                                                       \\
\verb+          $+%
  \bluett{random\_gen}%
  \verb+ <  1000   )+                                                                      \\
\verb+  {+                                                                                 \\
\verb+    $+%
  \bluett{random\_gen}%
  \verb+ = $+%
  \bluett{random\_gen}%
  \verb- + ($-%
  \bluett{random\_gen}%
  \verb+ / 100) * 1000;+                                                                   \\
\verb+  }+                                                                                 \\
\verb++                                                                                    \\
\verb+  fputs($+%
  \bluett{setup\_output}%
  \verb+,+                                                                                 \\
\verb+        sprintf("   +%
  \redtt{1}%
  \verb+   +%
  \redtt{5}%
  \verb+   5,   +%
  \bluett{random\_gen}%
  \verb+         = %d ;\n",+                                                               \\
\verb+                $+%
  \bluett{random\_gen}%
  \verb+));+                                                                               \\
\verb++                                                                                    \\
\verb+  fputs($+%
  \bluett{setup\_output}%
  \verb+, sprintf("\n"));+                                                                 \\
\verb+  fputs($+%
  \bluett{setup\_output}%
  \verb+, sprintf("\n"));+                                                                 \\
\verb++                                                                                    \\
\verb+  /*******************************************************************************/+ \\
\verb++                                                                                    \\
\verb+  fclose($+%
  \bluett{setup\_output}%
  \verb+);+                                                                                \\
\verb+?>+                                                                                  \\
\tabb
\newpage
\myitem
\label{item:sample-input_setup_web_reload.txt}
 {\cmbfsfx sample-input\_setup\_web\_reload.txt}
 (cf.~Templates \ref{item:sample-input_setup_web.txt}
  and \ref{item:setup_output_reload.php})
\deft
\verb+<?php+                                             \\
\verb+   $+%
  \bluett{output\_series}%
  \verb+ = "+%
  \redtt{output-140325-090000}%
  \verb+" ;+                                             \\
\verb++                                                  \\
\verb+// +%
  \bluett{Main\_function}%
  \verb++                                                \\
\verb++                                                  \\
\verb+   $+%
  \bluett{main\_function\_choosing}%
  \verb+  = +%
  \redtt{1}%
  \verb+ ;      //  +%
  \redtt{...\_/\_...\_/\_...\_/\_...}%
  \verb++                                                \\
\verb++                                                  \\
\verb+// +%
  \redtt{Main\_function\_1}%
  \verb++                                                \\
\verb++                                                  \\
\verb+   $+%
  \redtt{mode\_choosing}%
  \verb+           = +%
  \redtt{1}%
  \verb+ ;      //  +%
  \redtt{...\_/\_...\_/\_...\_/\_...}%
  \verb++                                                \\
\verb++                                                  \\
\verb+   $+%
  \bluett{personal\_function\_file}%
  \verb+  = "+%
  \redtt{tmp/uploaded\_functions/function\_user}%
  \verb+.txt"  ;+ \mynote[0.5cm]{See Template \ref{item:setup_output_reload.php}} \\
\verb+   $+%
  \bluett{personal\_function2\_file}%
  \verb+ = "+%
  \redtt{tmp/uploaded\_functions/function2\_user}%
  \verb+.txt" ;+                                         \\
\verb++                                                  \\
\verb+   $+%
  \redtt{integer\_1}%
  \verb+ = +%
  \bluett{nn}%
  \verb+ ;+                                              \\
\verb+   $+%
  \redtt{integer\_2}%
  \verb+ = +%
  \bluett{nn}%
  \verb+ ;+                                              \\
\verb++                                                  \\
\verb+   $+%
  \redtt{float\_1\_1}%
  \verb+ = +%
  \bluett{0.00001}%
  \verb+ ;              //  Replacement_for_+%
  \redtt{float\_1\_1}%
  \verb+_=_0!!+                                          \\
\verb+   $+%
  \redtt{float\_1\_2}%
  \verb+ = +%
  \bluett{ff.ffff}%
  \verb+ ;+                                              \\
\verb++                                                  \\
\verb+   $+%
  \redtt{float\_2\_1}%
  \verb+ = +%
  \bluett{ff.ffff}%
  \verb+ ;+                                              \\
\verb+   $+%
  \redtt{float\_2\_2}%
  \verb+ = +%
  \bluett{ff.ffff}%
  \verb+ ;+                                              \\
\verb++                                                  \\
\verb+   $+%
  \redtt{integer\_3}%
  \verb+ = +%
  \bluett{nn}%
  \verb+      ;+                                         \\
\verb+   $+%
  \redtt{float\_3}%
  \verb+   = +%
  \bluett{ff.ffff}%
  \verb+ ;+                                              \\
\verb++                                                  \\
\verb+// For_temporary_folders+                          \\
\verb++                                                  \\
\verb+   $+%
  \bluett{personal\_file\_date}%
  \verb+ = "+%
  \bluett{140325}%
  \verb+" ;+                                             \\
\verb+   $+%
  \bluett{personal\_file\_time}%
  \verb+ = "+%
  \bluett{060000}%
  \verb+" ;+                                             \\
\verb++                                                  \\
\verb+   $+%
  \bluett{plot\_output\_date}%
  \verb+   = "+%
  \bluett{140325}%
  \verb+" ;+                                             \\
\verb+   $+%
  \bluett{plot\_output\_time}%
  \verb+   = "+%
  \bluett{090000}%
  \verb+" ;+                                             \\
\verb+?>+                                                \\
\tabb
\newpage
\myitem
\label{item:setup_output_reload.php}
 {\cmbfsfx setup\_output\_reload.php}
 (cf.~Templates \ref{item:setup_output_file.php}
  and \ref{item:sample-input_setup_web_reload.txt})
\deft
\verb+<?php+                                                                               \\
\verb+  $+%
  \bluett{setup\_output}%
  \verb+ = fopen("+%
  \bluett{tmp/input\_setup\_web/input\_setup\_web\_reload}%
  \verb+.txt", "w");+                                                                      \\
\verb++                                                                                    \\
\verb+  fputs($+%
  \bluett{setup\_output}%
  \verb+, sprintf("<?php\n"));+                                                            \\
\verb+  fputs($+%
  \bluett{setup\_output}%
  \verb+, sprintf("\n"));+                                                                 \\
\verb++                                                                                    \\
\verb+  /*******************************************************************************/+ \\
\verb++                                                                                    \\
\verb+  fputs($+%
  \bluett{setup\_output}%
  \verb+,+                                                                                 \\
\verb+        sprintf("   \$+%
  \bluett{output\_series}%
  \verb+ = \"+%
  \bluett{plot-\%s-\%s}%
  \verb+\" ;\n",+                                                                          \\
\verb+                     $+%
  \bluett{plot\_output\_date}%
  \verb+, $+%
  \bluett{plot\_output\_time}%
  \verb+));+                                                                               \\
\verb++                                                                                    \\
\verb+  fputs($+%
  \bluett{setup\_output}%
  \verb+, sprintf("\n"));+                                                                \\
\verb+  fputs($+%
  \bluett{setup\_output}%
  \verb+, sprintf("\n"));+                                                                 \\
\verb++                                                                                    \\
\verb+  /* +%
  \bluett{Main function}%
  \verb+ ***************************************************************/+                 \\
\verb++                                                                                    \\
\verb+  fputs($+%
  \bluett{setup\_output}%
  \verb+, sprintf("//  +%
  \bluett{Main\_function}%
  \verb+\n"));+                                                                            \\
\verb+  fputs($+%
  \bluett{setup\_output}%
  \verb+, sprintf("\n"));+                                                                 \\
\verb++                                                                                    \\
\verb+  fputs($+%
  \bluett{setup\_output}%
  \verb+,+                                                                                 \\
\verb+        sprintf("   \$+%
  \bluett{main\_function\_choosing}%
  \verb+  = %d ;\n",+                                                                      \\
\verb+                     $+%
  \bluett{main\_function\_choosing}%
  \verb+));+                                                                               \\
\verb++                                                                                    \\
\verb+  fputs($+%
  \bluett{setup\_output}%
  \verb+, sprintf("\n"));+                                                                 \\
\verb+  fputs($+%
  \bluett{setup\_output}%
  \verb+, sprintf("\n"));+                                                                 \\
\defc
\verb++                                                                                    \\
\verb+  /* +%
  \redtt{main\_function\_1}%
  \verb+ *************************************************************/+                   \\
\verb++                                                                                    \\
\verb+  if ($+%
  \bluett{main\_function}%
  \verb+ == "+%
  \redtt{main\_function\_1}%
  \verb+")+                                                                                \\
\verb+  {+                                                                                 \\
\verb+    fputs($+%
  \bluett{setup\_output}%
  \verb+, sprintf("//  +%
  \redtt{Main\_function\_1}%
  \verb+\n"));+                                                                            \\
\verb+    fputs($+%
  \bluett{setup\_output}%
  \verb+, sprintf("\n"));+                                                                 \\
\verb++                                                                                    \\
\verb+    fputs($+%
  \bluett{setup\_output}%
  \verb+,+                                                                                 \\
\verb+          sprintf("   \$+%
  \redtt{mode\_choosing}%
  \verb+           = %d ;\n",+                                                             \\
\verb+                       $+%
  \redtt{mode\_choosing}%
  \verb+));+                                                                               \\
\verb++                                                                                    \\
\verb+    fputs($+%
  \bluett{setup\_output}%
  \verb+, sprintf("\n"));+                                                                 \\
\verb++                                                                                    \\
\verb+    if ($+%
  \bluett{function\_choice == "function\_user"}%
  \verb+)+                                                                                 \\
\verb+    {+                                                                               \\
\verb+      if ($+%
  \bluett{personal\_function\_file}%
  \verb+ != NULL                         &&+                                               \\
\verb+          $+%
  \bluett{personal\_function\_file}%
  \verb+ != $+%
  \bluett{personal\_function\_file\_copy}%
  \verb+   )+                                                                              \\
\verb+      {+                                                                             \\
\verb+        fputs($+%
  \bluett{setup\_output}%
  \verb+,+ \mynote[6cm]{File 1 uploaded, not typed}                                        \\
\verb+              sprintf("  \$+%
  \bluett{personal\_function\_file}%
  \verb+  = \"%s\" ;\n",+                                                                  \\
\verb+                          $+%
  \bluett{personal\_function\_file}%
  \verb+));+                                                                               \\
\verb++                                                                                    \\
\verb+          fputs($+%
  \bluett{setup\_output}%
  \verb+, sprintf("\n"));+                                                                 \\
\verb+      }+                                                                             \\
\verb++                                                                                    \\
\verb+      if ($+%
  \redtt{function2\_needed == "yes"}%
  \verb+)+                                                                                 \\
\verb+      {+                                                                             \\
\verb+        if ($+%
  \bluett{personal\_function2\_file}%
  \verb+ != NULL                          &&+                                              \\
\verb+            $+%
  \bluett{personal\_function2\_file}%
  \verb+ != $+%
  \bluett{personal\_function2\_file\_copy}%
  \verb+   )+                                                                              \\
\verb+        {+                                                                           \\
\verb+          fputs($+%
  \bluett{setup\_output}%
  \verb+,+ \mynote[5.8cm]{File 2 uploaded, not typed}                                      \\
\verb+                sprintf("  \$+%
  \bluett{personal\_function2\_file}%
  \verb+ = \"%s\" ;\n",+                                                                   \\
\verb+                            $+%
  \bluett{personal\_function2\_file}%
  \verb+));+                                                                               \\
\verb++                                                                                    \\
\verb+          fputs($+%
  \bluett{setup\_output}%
  \verb+, sprintf("\n"));+                                                                 \\
\verb+        }+                                                                           \\
\verb+      }+                                                                             \\
\verb++                                                                                    \\
\verb+      if ($+%
  \bluett{personal\_function\_file}%
  \verb+ != NULL                         &&+                                               \\
\verb+          $+%
  \bluett{personal\_function\_file}%
  \verb+ == $+%
  \bluett{personal\_function\_file\_copy}%
  \verb+   )+                                                                              \\
\verb+      {+                                                                             \\
\verb+          fputs($+%
  \bluett{setup\_output}%
  \verb+,+ \mynote[5.8cm]{File 1 typed, not uploaded}                                      \\
\verb+              sprintf("  \$+%
  \bluett{personal\_function\_text}%
  \verb+  = \"%s\" ;\n",+                                                                  \\
\verb+                          $+%
  \bluett{personal\_function\_text}%
  \verb+));+                                                                               \\
\verb++                                                                                    \\
\verb+          fputs($+%
  \bluett{setup\_output}%
  \verb+, sprintf("\n"));+                                                                 \\
\verb+      }+                                                                             \\
\verb++                                                                                    \\
\verb+      if ($+%
  \redtt{function2\_needed == "yes"}%
  \verb+)+                                                                                 \\
\verb+      {+                                                                             \\
\verb+        if ($+%
  \bluett{personal\_function2\_file}%
  \verb+ != NULL                          &&+                                              \\
\verb+            $+%
  \bluett{personal\_function2\_file}%
  \verb+ == $+%
  \bluett{personal\_function2\_file\_copy}%
  \verb+   )+                                                                              \\
\verb+        {+                                                                           \\
\verb+          fputs($+%
  \bluett{setup\_output}%
  \verb+,+ \mynote[5.8cm]{File 2 typed, not uploaded}                                      \\
\verb+                sprintf("  \$+%
  \bluett{personal\_function2\_text}%
  \verb+ = \"%s\" ;\n",+                                                                   \\
\verb+                            $+%
  \bluett{personal\_function2\_text}%
  \verb+));+                                                                               \\
\verb++                                                                                    \\
\verb+          fputs($+%
  \bluett{setup\_output}%
  \verb+, sprintf("\n"));+                                                                 \\
\verb+        }+                                                                           \\
\verb+      }+                                                                             \\
\verb+    }+                                                                               \\
\defc
\verb++                                                                                    \\
\verb+    fputs($+%
  \bluett{setup\_output}%
  \verb+,+                                                                                 \\
\verb+        sprintf("   \$+%
  \redtt{integer\_1}%
  \verb+ = %s ;\n", $+%
  \redtt{integer\_1}%
  \verb+));+                                                                               \\
\verb+    fputs($+%
  \bluett{setup\_output}%
  \verb+,+                                                                                 \\
\verb+        sprintf("   \$+%
  \redtt{integer\_2}%
  \verb+ = %s ;\n", $+%
  \redtt{integer\_2}%
  \verb+));+                                                                               \\
\verb+    fputs($+%
  \bluett{setup\_output}%
  \verb+, sprintf("\n"));+                                                                 \\
\verb++ \mynote[6cm]{\redtt{integer\_1} and \redtt{integer\_2} could practically also be 0!} \\
\verb+    if ($+%
  \redtt{float\_1\_1}%
  \verb+ != 0.0)+                                                                          \\
\verb+    {+                                                                               \\
\verb+      fputs($+%
  \bluett{setup\_output}%
  \verb+,+                                                                                 \\
\verb+            sprintf("   \$+%
  \redtt{float\_1\_1}%
  \verb+ = %s ;\n", $+%
  \redtt{float\_1\_1}%
  \verb+));+                                                                               \\
\verb+    }+                                                                               \\
\verb+    else+                                                                            \\
\verb+    {+                                                                               \\
\verb+      fputs($+%
  \bluett{setup\_output}%
  \verb+,+                                                                                 \\
\verb+            sprintf("   \$+%
  \redtt{float\_1\_1}%
  \verb+ = +%
  \bluett{0.00001}%
  \verb+ ;\n"));+                                                                          \\
\verb+    }+                                                                               \\
\verb++                                                                                    \\
\verb+    if ($+%
  \redtt{float\_1\_2}%
  \verb+ != 0.0)+                                                                          \\
\verb+    {+                                                                               \\
\verb+      fputs($+%
  \bluett{setup\_output}%
  \verb+,+                                                                                 \\
\verb+            sprintf("   \$+%
  \redtt{float\_1\_2}%
  \verb+ = %s ;\n", $+%
  \redtt{float\_1\_2}%
  \verb+));+                                                                               \\
\verb+    }+                                                                               \\
\verb+    else+                                                                            \\
\verb+    {+                                                                               \\
\verb+      fputs($+%
  \bluett{setup\_output}%
  \verb+,+                                                                                 \\
\verb+            sprintf("   \$+%
  \redtt{float\_1\_2}%
  \verb+ = +%
  \bluett{0.00001}%
  \verb+ ;\n"));+                                                                          \\
\verb+    }+                                                                               \\
\verb++                                                                                    \\
\verb+    fputs($+%
  \bluett{setup\_output}%
  \verb+, sprintf("\n"));+                                                                 \\
\verb++                                                                                    \\
\verb+    if ($+%
  \redtt{float\_2\_1}%
  \verb+ != 0.0)+                                                                          \\
\verb+    {+                                                                               \\
\verb+      fputs($+%
  \bluett{setup\_output}%
  \verb+,+                                                                                 \\
\verb+            sprintf("   \$+%
  \redtt{float\_2\_1}%
  \verb+ = %s ;\n", $+%
  \redtt{float\_2\_1}%
  \verb+));+                                                                               \\
\verb+    }+                                                                               \\
\verb+    else+                                                                            \\
\verb+    {+                                                                               \\
\verb+      fputs($+%
  \bluett{setup\_output}%
  \verb+,+                                                                                 \\
\verb+            sprintf("   \$+%
  \redtt{float\_2\_1}%
  \verb+ = +%
  \bluett{0.00001}%
  \verb+ ;\n"));+                                                                          \\
\verb+    }+                                                                               \\
\verb++                                                                                    \\
\verb+    if ($+%
  \redtt{float\_2\_2}%
  \verb+ != 0.0)+                                                                          \\
\verb+    {+                                                                               \\
\verb+      fputs($+%
  \bluett{setup\_output}%
  \verb+,+                                                                                 \\
\verb+            sprintf("   \$+%
  \redtt{float\_2\_2}%
  \verb+ = %s ;\n", $+%
  \redtt{float\_2\_2}%
  \verb+));+                                                                               \\
\verb+    }+                                                                               \\
\verb+    else+                                                                            \\
\verb+    {+                                                                               \\
\verb+      fputs($+%
  \bluett{setup\_output}%
  \verb+,+                                                                                 \\
\verb+            sprintf("   \$+%
  \redtt{float\_2\_2}%
  \verb+ = +%
  \bluett{0.00001}%
  \verb+ ;\n"));+                                                                          \\
\verb+    }+                                                                               \\
\verb++                                                                                    \\
\verb+    fputs($+%
  \bluett{setup\_output}%
  \verb+, sprintf("\n"));+                                                                 \\
\defc
\verb++                                                                                    \\
\verb+    fputs($+%
  \bluett{setup\_output}%
  \verb+,+                                                                                 \\
\verb+        sprintf("   \$+%
  \redtt{integer\_3}%
  \verb+ = %s ;\n", $+%
  \redtt{integer\_3}%
  \verb+));+                                                                               \\
\verb++ \mynote[8cm]{\redtt{integer\_3} could practically also be 0!}                      \\
\verb+    if ($+%
  \redtt{float\_3}%
  \verb+ != 0.0)+                                                                          \\
\verb+    {+                                                                               \\
\verb+      fputs($+%
  \bluett{setup\_output}%
  \verb+,+                                                                                 \\
\verb+            sprintf("   \$+%
  \redtt{float\_3}%
  \verb+   = %s ;\n", $+%
  \redtt{float\_3}%
  \verb+));+                                                                               \\
\verb+    }+                                                                               \\
\verb+    else+                                                                            \\
\verb+    {+                                                                               \\
\verb+      fputs($+%
  \bluett{setup\_output}%
  \verb+,+                                                                                 \\
\verb+            sprintf("   \$+%
  \redtt{float\_3}%
  \verb+ = +%
  \bluett{0.00001}%
  \verb+ ;\n"));+                                                                          \\
\verb+    }+                                                                               \\
\verb++                                                                                    \\
\verb+    fputs($+%
  \bluett{setup\_output}%
  \verb+, sprintf("\n"));+                                                                 \\
\verb+    fputs($+%
  \bluett{setup\_output}%
  \verb+, sprintf("\n"));+                                                                 \\
\verb+  }+                                                                                 \\
\verb++                                                                                    \\
\verb+  /* For temporary folders *******************************************************/+ \\
\verb++                                                                                    \\
\verb+  fputs($+%
  \bluett{setup\_output}%
  \verb+,+                                                                                 \\
\verb+        sprintf("//  For_temporary_folders\n"));+                                    \\
\verb+  fputs($+%
  \bluett{setup\_output}%
  \verb+, sprintf("\n"));+                                                                 \\
\verb++                                                                                    \\
\verb+  fputs($+%
  \bluett{setup\_output}%
  \verb+,+                                                                                 \\
\verb+        sprintf("   \$+%
  \bluett{personal\_file\_date}%
  \verb+ = \"%s\" ;\n",+ \mynote[1cm]{Not {\tt \%d}!}                                      \\
\verb+                     $+%
  \bluett{personal\_file\_date}%
  \verb+));+                                                                               \\
\verb+  fputs($+%
  \bluett{setup\_output}%
  \verb+,+                                                                                 \\
\verb+        sprintf("   \$+%
  \bluett{personal\_file\_time}%
  \verb+ = \"%s\" ;\n",+                                                                   \\
\verb+                     $+%
  \bluett{personal\_file\_time}%
  \verb+));+                                                                               \\
\verb++                                                                                    \\
\verb+  fputs($+%
  \bluett{setup\_output}%
  \verb+, sprintf("\n"));+                                                                 \\
\verb++                                                                                    \\
\verb+  fputs($+%
  \bluett{setup\_output}%
  \verb+,+                                                                                 \\
\verb+        sprintf("   \$+%
  \bluett{plot\_output\_date}%
  \verb+   = \"%s\" ;\n",+                                                                 \\
\verb+                     $+%
  \bluett{plot\_output\_date}%
  \verb+));+                                                                               \\
\verb+  fputs($+%
  \bluett{setup\_output}%
  \verb+,+                                                                                 \\
\verb+        sprintf("   \$+%
  \bluett{plot\_output\_time}%
  \verb+   = \"%s\" ;\n",+                                                                 \\
\verb+                     $+%
  \bluett{plot\_output\_time}%
  \verb+));+                                                                               \\
\verb++                                                                                    \\
\verb+  fputs($+%
  \bluett{setup\_output}%
  \verb+, sprintf("\n"));+                                                                 \\
\verb+  fputs($+%
  \bluett{setup\_output}%
  \verb+, sprintf("\n"));+                                                                \\
\verb++                                                                                    \\
\verb+  /*******************************************************************************/+ \\
\verb++                                                                                    \\
\verb+  fputs($+%
  \bluett{setup\_output}%
  \verb+, sprintf("?>\n"));+                                                               \\
\verb+  fputs($+%
  \bluett{setup\_output}%
  \verb+, sprintf("\n"));+                                                                 \\
\verb++                                                                                    \\
\verb+  fclose($+%
  \bluett{setup\_output}%
  \verb+);+                                                                                \\
\verb+?>+                                                                                  \\
\tabb
\end{templateenumerate}
\newpage
\section{Result presenting}
\label{sec:result_presenting}
%
%
\begin{templateenumerate}
\myitem
\label{item:reloading_setup}
 {\bf Reloading the setup}
\Xdeft{Script in {\tt main.php}}
\verb+<?php+                                                            \\
\verb+  /* Only for "+%
  \bluett{results}%
  \verb+.php" ***********************************/+                     \\
\verb++%
  \redtt{/*}%
  \verb++                                                               \\
\verb+  include("+%
  \bluett{tmp/input\_setup\_web/input\_setup\_web\_reload}%
  \verb+.txt");+ \mynote[1cm]{Template \ref{item:sample-input_setup_web_reload.txt}} \\
\verb++%
  \redtt{*/}%
  \verb++                                                               \\
\verb+?>+                                                               \\
\verb++                                                                 \\
\verb+<form action="+%
  \bluett{main}%
  \verb+.php" method="post" enctype="multipart/form-data">+             \\
\verb++                                                                 \\
\verb+      .+                                                          \\
\verb+      .+ \mynote{Template \ref{item:form}}                        \\
\verb+      .+                                                          \\
\tabb
\Xdeft{Script in {\tt results.php}}
\verb+<?php+                                                            \\
\verb+  /* Only for "+%
  \bluett{results}%
  \verb+.php" ***********************************/+                     \\
\verb++                                                                 \\
\verb+  include("+%
  \bluett{tmp/input\_setup\_web/input\_setup\_web\_reload}%
  \verb+.txt");+ \mynote[1cm]{Template \ref{item:sample-input_setup_web_reload.txt}} \\
\verb++                                                                 \\
\verb+?>+                                                               \\
\verb++                                                                 \\
\verb+<form action="+%
  \bluett{main}%
  \verb+.php" method="post" enctype="multipart/form-data">+             \\
\verb++                                                                 \\
\verb+      .+                                                          \\
\verb+      .+ \mynote{Template \ref{item:form}}                        \\
\verb+      .+                                                          \\
\tabb
 This is the unique required modification
 from \verb+main.php+ to \verb+results.php+.
%
%
%
%
\newpage
%
%
%
%
\myitem
\label{item:going_to_results}
 {\bf Going to ``Results''}
 (cf.~Templates \ref{item:switch_further_step}
  and \ref{item:modification_switch_further_step})
\deft
\verb+<?php+                                                                             \\
\verb+  if ($+%
  \bluett{main\_function}%
  \verb+ != NULL  &&+                                                                    \\
\verb+      $+%
  \redtt{mode\_checked}%
  \verb+  == "yes" &&+                                                                   \\
\verb+      $+%
  \redtt{setup\_checked}%
  \verb+ == "yes" &&+                                                                    \\
\verb+      $+%
  \bluett{output\_series}%
  \verb+ != NULL    )+ \mynote[1cm]{For {\tt results.php}, not {\tt main.php},
                                    cf.~Template \ref{item:starting_or_modifying}}       \\
\verb+  {+                                                                               \\
\verb+    echo '<script type="text/javascript">'                              . "\r\n";+ \\
\verb+    echo '  if (navigator.appName == "Microsoft Internet Explorer")'    . "\r\n";+ \\
\verb+    echo '  {'                                                          . "\r\n";+ \\
\verb+    echo '    window.location = "#+%
  \bluett{results}%
  \verb+";'                            . "\r\n";+                                        \\
\verb+    echo '  }'                                                          . "\r\n";+ \\
\verb+    echo '</script>'                                                    . "\r\n";+ \\
\verb++                                                                                  \\
\verb+    echo '<h3 id="+%
  \bluett{results}%
  \verb+"> +%
  \bluett{Results}%
  \verb+ </h3>'                              . "\r\n";+                                  \\
\verb++                                                                                  \\
\verb+    echo '<script type="text/javascript">'                              . "\r\n";+ \\
\verb+    echo '  if (navigator.userAgent.search("Chrome")     != -1     ||'  . "\r\n";+ \\
\verb+    echo '      (navigator.userAgent.search("Firefox")   != -1 &&'      . "\r\n";+ \\
\verb+    echo '       navigator.userAgent.search("Navigator") == -1   ) ||'  . "\r\n";+ \\
\verb+    echo '      navigator.userAgent.search("Opera")      != -1     ||'  . "\r\n";+ \\
\verb+    echo '      (navigator.userAgent.search("Chrome")    == -1 &&'      . "\r\n";+ \\
\verb+    echo '       navigator.userAgent.search("Safari")    != -1   )   )' . "\r\n";+ \\
\verb+    echo '  {'                                                          . "\r\n";+ \\
\verb+    echo '    window.location = "#+%
  \bluett{results}%
  \verb+";'                            . "\r\n";+                                        \\
\verb+    echo '  }'                                                          . "\r\n";+ \\
\verb+    echo '</script>'                                                    . "\r\n";+ \\
\verb++                                                                                  \\
\verb+        .+                                                                         \\
\verb+        .+ \mynote{Template \ref{item:forms_plots_tables}}                         \\
\verb+        .+                                                                         \\
\verb++                                                                                  \\
\verb+        .+                                                                         \\
\verb+        .+ \mynote{Template \ref{item:downloadable_output_data_files}}             \\
\verb+        .+                                                                         \\
\verb++                                                                                  \\
\verb+    echo '<script type="text/javascript">'                              . "\r\n";+ \\
\verb+    echo '  if (navigator.userAgent.search("Navigator") != -1)'         . "\r\n";+ \\
\verb+    echo '  {'                                                          . "\r\n";+ \\
\verb+    echo '    window.location = "#+%
  \bluett{results}%
  \verb+";'                            . "\r\n";+                                        \\
\verb+    echo '  }'                                                          . "\r\n";+ \\
\verb+    echo '</script>'                                                    . "\r\n";+ \\
\verb++                                                                                  \\
\verb+    echo '<h5> <a href="#+%
  \bluett{main}%
  \verb+">Top</a> </h5>'                           . "\r\n";+                            \\
\verb+    echo '<hr />'                                                       . "\r\n";+ \\
\verb+  }+                                                                               \\
\verb+?>+                                                                                \\
\tabb
\newpage
\myitem
\label{item:forms_plots_tables}
 {\bf Results shown in forms of plots and tables}
\deft
\verb+<?php+                         \\
\verb+    if ($+%
  \bluett{main\_function}%
  \verb+ == "+%
  \redtt{main\_function\_1}%
  \verb+" ||+                        \\
\verb+        $+%
  \bluett{main\_function}%
  \verb+ == "+%
  \redtt{main\_function\_2}%
  \verb+"   )+                       \\
\verb+    {+                         \\
\verb+      include("+%
  \bluett{setup/setup-plotNo}%
  \verb+.php");+ \mynote{Template \ref{item:setup-plotNo.php}} \\
\verb++                              \\
\verb+      for ($+%
  \bluett{n\_plot}%
  \verb+ = 1; $+%
  \bluett{n\_plot}%
  \verb+ <= $+%
  \bluett{output\_plot\_No}%
  \verb+; $+%
  \bluett{n\_plot}%
  \verb- ++)-                        \\
\verb+      {+                       \\
\verb+        switch ($+%
  \bluett{main\_function}%
  \verb+)+                           \\
\verb+        {+                     \\
\verb+          case "+%
  \redtt{main\_function\_1}%
  \verb+":+                          \\
\verb+            echo '<h4> +%
  \redtt{......}%
  \verb+ </h4>' . "\r\n";+           \\
\verb++                              \\
\verb+            include("+%
  \bluett{main/plot\_output}%
  \verb+.php");+ \mynote[3cm]{Templates \ref{item:plot_output.php}
                              and \ref{item:plot_output_animation.php}} \\
\verb+            include("+%
  \bluett{captions/caption-}%
  \verb++%
  \redtt{main\_function\_1}%
  \verb+.php");+                     \\
\verb++                              \\
\verb+            break;+            \\
\verb++                              \\
\verb+              .+               \\
\verb+              .+               \\
\verb+              .+               \\
\verb+        }+                     \\
\verb+      }+                       \\
\verb+    }+                         \\
\verb++                              \\
\verb+    elseif ($+%
  \bluett{main\_function}%
  \verb+ == "+%
  \redtt{main\_function\_3}%
  \verb+" ||+                        \\
\verb+            $+%
  \bluett{main\_function}%
  \verb+ == "+%
  \redtt{main\_function\_4}%
  \verb+"   )+                       \\
\verb+    {+                         \\
\verb+      include("+%
  \bluett{setup/setup-tableNo}%
  \verb+.php");+ \mynote{Template \ref{item:setup-tableNo.php}} \\
\verb++                              \\
\verb+      for ($+%
  \bluett{n\_table}%
  \verb+ = 1; $+%
  \bluett{n\_table}%
  \verb+ <= $+%
  \bluett{output\_table\_No}%
  \verb+; $+%
  \bluett{n\_table}%
  \verb- ++)-                        \\
\verb+      {+                       \\
\verb+          $+%
  \bluett{plot\_output\_file\_copy}%
  \verb++ \mynote{Result output by the main program in PHP syntax} \\
\verb+        = $+%
  \bluett{plot\_output\_file}%
  \verb+ . "+%
  \bluett{-0}%
  \verb+.txt";+                      \\
\verb++                              \\
\verb+        include("$+%
  \bluett{plot\_output\_folder}%
  \verb+/" . $+%
  \bluett{plot\_output\_file\_copy}%
  \verb+);+ \mynote[0.5cm]{Read the output results} \\
\verb++                              \\
\verb+        switch ($+%
  \bluett{main\_function}%
  \verb+)+                           \\
\verb+        {+                     \\
\verb+          case "+%
  \redtt{main\_function\_3}%
  \verb+":+                          \\
\verb+            echo '<h4> +%
  \redtt{......}%
  \verb+ </h4>' . "\r\n";+           \\
\verb++                              \\
\verb+            include("+%
  \bluett{tables/table-output\_}%
  \verb++%
  \redtt{main\_function\_3}%
  \verb+.php");+                     \\
\verb+            include("+%
  \bluett{captions/caption-}%
  \verb++%
  \redtt{main\_function\_3}%
  \verb+.php");+                     \\
\verb++                              \\
\verb+            break;+            \\
\verb++                              \\
\verb+              .+               \\
\verb+              .+               \\
\verb+              .+               \\
\verb+        }+                     \\
\verb+      }+                       \\
\verb+    }+                         \\
\verb+?>+                            \\
\tabb
\newpage
\myitem
\label{item:downloadable_output_data_files}
 {\bf Downloadable output data files}
 (cf.~Template \ref{item:plot_output.php})
\deft
\verb+<?php+                                                                \\
\verb+    echo '<h4> Data </h4>'                                 . "\r\n";+ \\
\verb++                                                                     \\
\verb+    echo '<span class="+%
  \bluett{plot\_link}%
  \verb+">'                        . "\r\n";+                               \\
\verb+    echo ' <b> TXT: </b>'                                  . "\r\n";+ \\
\verb+    echo ' <a href="+%
  \bluett{tmp/output\_results}%
  \verb+/';+                                                                \\
\verb+    echo      $+%
  \bluett{plot\_output\_file}%
  \verb+ . '.txt" target="_blank">' . "\r\n";+                              \\
\verb+    echo      $+%
  \bluett{plot\_output\_file}%
  \verb+ . '.txt</a>'               . "\r\n";+                              \\
\verb+    echo '</span>'                                         . "\r\n";+ \\
\verb+    echo '<br />'                                          . "\r\n";+ \\
\verb+    echo '<br />'                                          . "\r\n";+ \\
\verb+?>+                                                                   \\
\tabb
\newpage
\myitem
\label{item:setup-plotNo.php}
 {\cmbfsfx setup-plotNo.php}
 (cf.~Template \ref{item:setup-tableNo.php})
\deft
\verb+<?php+                                               \\
\verb++                                                    \\
\verb+/* --------------------------------------------- */+ \\
\verb+//+                                                  \\
\verb+// Set the number of output plots+                   \\
\verb+//+                                                  \\
\verb+/* --------------------------------------------- */+ \\
\verb++                                                    \\
\verb+  if ($+%
  \bluett{main\_function}%
  \verb+ == "+%
  \redtt{main\_function\_1}%
  \verb+")+                                                \\
\verb+  {+                                                 \\
\verb+    $+%
  \bluett{output\_plot\_No}%
  \verb+ = +%
  \redtt{1}%
  \verb+;+                                                 \\
\verb+  }+                                                 \\
\verb++                                                    \\
\verb+/* --------------------------------------------- */+ \\
\verb++                                                    \\
\verb+  if ($+%
  \bluett{main\_function}%
  \verb+ == "+%
  \redtt{main\_function\_2}%
  \verb+")+                                                \\
\verb+  {+                                                 \\
\verb+    $+%
  \bluett{output\_plot\_No}%
  \verb+ = +%
  \redtt{2}%
  \verb+;+                                                 \\
\verb+  }+                                                 \\
\verb++                                                    \\
\verb+/* --------------------------------------------- */+ \\
\verb++                                                    \\
\verb+  if ($+%
  \bluett{main\_function}%
  \verb+ == "+%
  \redtt{main\_function\_3}%
  \verb+")+                                                \\
\verb+  {+                                                 \\
\verb+    $+%
  \bluett{output\_plot\_No}%
  \verb+ = +%
  \redtt{0}%
  \verb+;+                                                 \\
\verb+  }+                                                 \\
\verb++                                                    \\
\verb+/* --------------------------------------------- */+ \\
\verb++                                                    \\
\verb+  if ($+%
  \bluett{main\_function}%
  \verb+ == "+%
  \redtt{main\_function\_4}%
  \verb+")+                                                \\
\verb+  {+                                                 \\
\verb+    $+%
  \bluett{output\_plot\_No}%
  \verb+ = +%
  \redtt{0}%
  \verb+;+                                                 \\
\verb+  }+                                                 \\
\verb++                                                    \\
\verb+/* --------------------------------------------- */+ \\
\verb++                                                    \\
\verb+?>+                                                  \\
\tabb
\newpage
\myitem
\label{item:setup-tableNo.php}
 {\cmbfsfx setup-tableNo.php}
 (cf.~Template \ref{item:setup-plotNo.php})
\deft
\verb+<?php+                                               \\
\verb++                                                    \\
\verb+/* --------------------------------------------- */+ \\
\verb+//+                                                  \\
\verb+// Set the number of output tables+                  \\
\verb+//+                                                  \\
\verb+/* --------------------------------------------- */+ \\
\verb++                                                    \\
\verb+  if ($+%
  \bluett{main\_function}%
  \verb+ == "+%
  \redtt{main\_function\_1}%
  \verb+")+                                                \\
\verb+  {+                                                 \\
\verb+    $+%
  \bluett{output\_table\_No}%
  \verb+ = +%
  \redtt{0}%
  \verb+;+                                                 \\
\verb+  }+                                                 \\
\verb++                                                    \\
\verb+/* --------------------------------------------- */+ \\
\verb++                                                    \\
\verb+  if ($+%
  \bluett{main\_function}%
  \verb+ == "+%
  \redtt{main\_function\_2}%
  \verb+")+                                                \\
\verb+  {+                                                 \\
\verb+    $+%
  \bluett{output\_table\_No}%
  \verb+ = +%
  \redtt{0}%
  \verb+;+                                                 \\
\verb+  }+                                                 \\
\verb++                                                    \\
\verb+/* --------------------------------------------- */+ \\
\verb++                                                    \\
\verb+  if ($+%
  \bluett{main\_function}%
  \verb+ == "+%
  \redtt{main\_function\_3}%
  \verb+")+                                                \\
\verb+  {+                                                 \\
\verb+    $+%
  \bluett{output\_table\_No}%
  \verb+ = +%
  \redtt{2}%
  \verb+;+                                                 \\
\verb+  }+                                                 \\
\verb++                                                    \\
\verb+/* --------------------------------------------- */+ \\
\verb++                                                    \\
\verb+  if ($+%
  \bluett{main\_function}%
  \verb+ == "+%
  \redtt{main\_function\_4}%
  \verb+")+                                                \\
\verb+  {+                                                 \\
\verb+    $+%
  \bluett{output\_table\_No}%
  \verb+ = +%
  \redtt{1}%
  \verb+;+                                                 \\
\verb+  }+                                                 \\
\verb++                                                    \\
\verb+/* --------------------------------------------- */+ \\
\verb++                                                    \\
\verb+?>+                                                  \\
\tabb
\newpage
\myitem
\label{item:plot_output.php}
 {\cmbfsfx plot\_output.php}
 (cf.~Template \ref{item:downloadable_output_data_files})
\deft
\verb+<span class="+%
  \bluett{plot\_link}%
  \verb+">+                                      \\
\verb+ <b> PS: </b>+                             \\
\verb+ <a href="<?php echo $+%
  \bluett{plot\_output\_folder}%
  \verb+ . '/' .+                                \\
\verb+                     $+%
  \bluett{plot\_output\_file}%
  \verb+   . '-' ;+                              \\
\verb+                if ($+%
  \bluett{n\_plot}%
  \verb+ <= 9)+                                  \\
\verb+                {+                         \\
\verb+                  echo '0';+               \\
\verb+                }+                         \\
\verb+                echo $+%
  \bluett{n\_plot}%
  \verb+; ?>.ps" target="_blank">+               \\
\verb+          <?php echo $+%
  \bluett{plot\_output\_file}%
  \verb+   . '-' ;+                              \\
\verb+                if ($+%
  \bluett{n\_plot}%
  \verb+ <= 9)+                                  \\
\verb+                {+                         \\
\verb+                  echo '0';+               \\
\verb+                }+                         \\
\verb+                echo $+%
  \bluett{n\_plot}%
  \verb+; ?>.ps</a>+                             \\
\verb+</span>+                                   \\
\verb+<br />+                                    \\
\verb++                                          \\
\verb+<span class="+%
  \bluett{plot\_link}%
  \verb+">+                                      \\
\verb+ <b> EPS: </b>+                            \\
\verb+ <a href="<?php echo $+%
  \bluett{plot\_output\_folder}%
  \verb+ . '/' .+                                \\
\verb+                     $+%
  \bluett{plot\_output\_file}%
  \verb+   . '-' ;+                              \\
\verb+                if ($+%
  \bluett{n\_plot}%
  \verb+ <= 9)+                                  \\
\verb+                {+                         \\
\verb+                  echo '0';+               \\
\verb+                }+                         \\
\verb+                echo $+%
  \bluett{n\_plot}%
  \verb+; ?>.eps" target="_blank">+              \\
\verb+          <?php echo $+%
  \bluett{plot\_output\_file}%
  \verb+   . '-' ;+                              \\
\verb+                if ($+%
  \bluett{n\_plot}%
  \verb+ <= 9)+                                  \\
\verb+                {+                         \\
\verb+                  echo '0';+               \\
\verb+                }+                         \\
\verb+                echo $+%
  \bluett{n\_plot}%
  \verb+; ?>.eps</a>+                            \\
\verb+</span>+                                   \\
\verb+<br />+                                    \\
\verb++                                          \\
\verb+<span class="+%
  \bluett{plot\_link}%
  \verb+">+                                      \\
\verb+ <b> PDF: </b>+                            \\
\verb+ <a href="<?php echo $+%
  \bluett{plot\_output\_folder}%
  \verb+ . '/' .+                                \\
\verb+                     $+%
  \bluett{plot\_output\_file}%
  \verb+   . '-' ;+                              \\
\verb+                if ($+%
  \bluett{n\_plot}%
  \verb+ <= 9)+                                  \\
\verb+                {+                         \\
\verb+                  echo '0';+               \\
\verb+                }+                         \\
\verb+                echo $+%
  \bluett{n\_plot}%
  \verb+; ?>.pdf" target="_blank">+              \\
\verb+          <?php echo $+%
  \bluett{plot\_output\_file}%
  \verb+   . '-' ;+                              \\
\verb+                if ($+%
  \bluett{n\_plot}%
  \verb+ <= 9)+                                  \\
\verb+                {+                         \\
\verb+                  echo '0';+               \\
\verb+                }+                         \\
\verb+                echo $+%
  \bluett{n\_plot}%
  \verb+; ?>.pdf</a>+                            \\
\verb+</span>+                                   \\
\verb+<br />+                                    \\
\defc
\verb++                                          \\
\verb+<span class="+%
  \bluett{plot\_link}%
  \verb+">+                                      \\
\verb+ <b> PNG: </b>+                            \\
\verb+ <a href="<?php echo $+%
  \bluett{plot\_output\_folder}%
  \verb+ . '/' .+                                \\
\verb+                     $+%
  \bluett{plot\_output\_file}%
  \verb+   . '-' ;+                              \\
\verb+                if ($+%
  \bluett{n\_plot}%
  \verb+ <= 9)+                                  \\
\verb+                {+                         \\
\verb+                  echo '0';+               \\
\verb+                }+                         \\
\verb+                echo $+%
  \bluett{n\_plot}%
  \verb+; ?>.png" target="_blank">+              \\
\verb+          <?php echo $+%
  \bluett{plot\_output\_file}%
  \verb+   . '-' ;+                              \\
\verb+                if ($+%
  \bluett{n\_plot}%
  \verb+ <= 9)+                                  \\
\verb+                {+                         \\
\verb+                  echo '0';+               \\
\verb+                }+                         \\
\verb+                echo $+%
  \bluett{n\_plot}%
  \verb+; ?>.png</a>+                            \\
\verb+</span>+                                   \\
\verb+<br />+                                    \\
\verb++                                          \\
\verb+<br />+                                    \\
\verb+<span class="+%
  \bluett{plot\_link}%
  \verb+">+                                      \\
\verb+ <b> Plot preview </b>+                    \\
\verb+</span>+                                   \\
\verb+<br />+                                    \\
\verb+ <a href="<?php echo $+%
  \bluett{plot\_output\_folder}%
  \verb+ . '/' .+                                \\
\verb+                     $+%
  \bluett{plot\_output\_file}%
  \verb+   . '-' ;+                              \\
\verb+                if ($+%
  \bluett{n\_plot}%
  \verb+ <= 9)+                                  \\
\verb+                {+                         \\
\verb+                  echo '0';+               \\
\verb+                }+                         \\
\verb+                echo $+%
  \bluett{n\_plot}%
  \verb+; ?>.png"+                               \\
\verb+    target="_blank">+                      \\
\verb+    <img class="+%
  \bluett{output\_plot}%
  \verb+"+                                       \\
\verb+         src="<?php echo $+%
  \bluett{plot\_output\_folder}%
  \verb+ . '/' .+                                \\
\verb+                         $+%
  \bluett{plot\_output\_file}%
  \verb+   . '-' ;+                              \\
\verb+                if ($+%
  \bluett{n\_plot}%
  \verb+ <= 9)+                                  \\
\verb+                {+                         \\
\verb+                  echo '0';+               \\
\verb+                }+                         \\
\verb+                echo $+%
  \bluett{n\_plot}%
  \verb+; ?>.png"+                               \\
\verb+         alt="<?php echo $+%
  \bluett{plot\_output\_file}%
  \verb+   . '-' ;+                              \\
\verb+                if ($+%
  \bluett{n\_plot}%
  \verb+ <= 9)+                                  \\
\verb+                {+                         \\
\verb+                  echo '0';+               \\
\verb+                }+                         \\
\verb+                echo $+%
  \bluett{n\_plot}%
  \verb+; ?>.png" /></a>+                        \\
\verb+<br />+                                    \\
\verb++                                          \\
\verb+<span class="+%
  \bluett{credit}%
  \verb+">+                                      \\
\verb+ <b> Plot credit: </b>+                    \\
\verb+ +%
  \bluett{TiResearch}%
  \verb+ &nbsp; +%
  \bluett{http://www.tir.tw/xxxxxx/}%
  \verb++                                        \\
\verb+</span>+                                   \\
\verb+<br />+                                    \\
\verb++                                          \\
\verb+<br />+                                    \\
\tabb
\newpage
\myitem
\label{item:plot_output_animation.php}
 {\cmbfsfx plot\_output\_animation.php}
 (cf.~Template \ref{item:plot_output.php})
\deft
\verb+<script type="text/javascript">+           \\
\verb+  var +%
  \bluett{n\_plot\_animation}%
  \verb+ = 2;+                                   \\
\verb++                                          \\
\verb+  function +%
  \bluett{plot\_change}%
  \verb+()+                                      \\
\verb+  {+                                       \\
\verb+    document.getElementById("+%
  \bluett{plot\_output\_animation}%
  \verb+").src =+                                \\
\verb+    <?php+                                 \\
\verb+      echo '"' . $+%
  \bluett{plot\_output\_folder}%
  \verb+ . '/' . $+%
  \bluett{plot\_output\_file}%
  \verb+ . '-';+                                 \\
\verb+      if ($+%
  \bluett{n\_plot}%
  \verb+ <= 9)+                                  \\
\verb+      {+                                   \\
\verb+        echo '0';+                         \\
\verb+      }+                                   \\
\verb-    ?>" + -%
  \bluett{n\_plot\_animation}%
  \verb- + ".png";-                              \\
\verb++                                          \\
\verb+    if (+%
  \bluett{n\_plot\_animation}%
  \verb+ == +%
  \redtt{4}%
  \verb+)+ \mynote{Total plot number}            \\
\verb+    {+                                     \\
\verb+      +%
  \bluett{n\_plot\_animation}%
  \verb+ = 1;+                                   \\
\verb+    }+                                     \\
\verb++                                          \\
\verb+    else+                                  \\
\verb+    {+                                     \\
\verb+      +%
  \bluett{n\_plot\_animation}%
  \verb+ = +%
  \bluett{n\_plot\_animation}%
  \verb- + 1;- \mynote{Cycling the plots}        \\
\verb+    }+                                     \\
\verb+  }+                                       \\
\verb++                                          \\
\verb+  function +%
  \bluett{plot\_animate}%
  \verb+()+                                      \\
\verb+  {+                                       \\
\verb+    var +%
  \bluett{t}%
  \verb+ = setTimeout("+%
  \bluett{plot\_change}%
  \verb+()", +%
  \redtt{1000}%
  \verb+);+ \mynote[0.5cm]{1000 ns, time separation between two plots} \\
\verb+  }+                                       \\
\verb+</script>+                                 \\
\verb++                                          \\
\verb+<span class="+%
  \bluett{plot\_link}%
  \verb+">+                                      \\
\verb+ <b> Plot animation </b>+                  \\
\verb+</span>+                                   \\
\verb+<br />+                                    \\
\verb+    <img class="+%
  \bluett{output\_plot}%
  \verb+"+                                       \\
\verb+         id="+%
  \bluett{plot\_outout\_animation}%
  \verb+"+                                       \\
\verb+         src="<?php echo $+%
  \bluett{plot\_output\_folder}%
  \verb+ . '/' .+                                \\
\verb+                         $+%
  \bluett{plot\_output\_file}%
  \verb+   . '-' ;+                              \\
\verb+                if ($+%
  \bluett{n\_plot}%
  \verb+ <= 9)+                                  \\
\verb+                {+                         \\
\verb+                  echo '0';+               \\
\verb+                }+                         \\
\verb+                echo $+%
  \bluett{n\_plot}%
  \verb+; ?>.png"+                               \\
\verb+         alt="<?php echo $+%
  \bluett{plot\_output\_file}%
  \verb+   . '-' ;+                              \\
\verb+                if ($+%
  \bluett{n\_plot}%
  \verb+ <= 9)+                                  \\
\verb+                {+                         \\
\verb+                  echo '0';+               \\
\verb+                }+                         \\
\verb+                echo $+%
  \bluett{n\_plot}%
  \verb+; ?>.png"+                               \\
\verb+         onload="+%
  \bluett{plot\_animate}%
  \verb+()" />+                                  \\
\verb+<br />+                                    \\
\verb++                                          \\
\verb+<span class="+%
  \bluett{credit}%
  \verb+">+                                      \\
\verb+ <b> Plot credit: </b>+                    \\
\verb+ +%
  \bluett{TiResearch}%
  \verb+ &nbsp; +%
  \bluett{http://www.tir.tw/xxxxxx/}%
  \verb++                                        \\
\verb+</span>+                                   \\
\verb+<br />+                                    \\
\verb++                                          \\
\verb+<br /> <br />+                             \\
\tabb
\newpage
\picdin{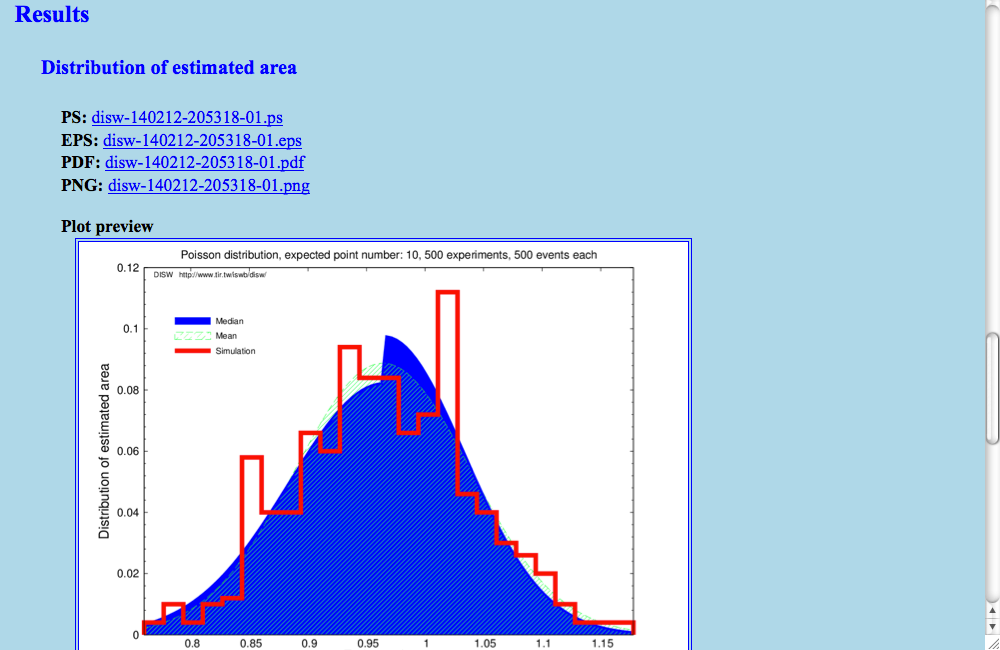}
{results}
{Result in form of plots.}
{Results in form of a table.}
\end{templateenumerate}

\addemptypage
\chapter{Miscellaneous}
\label{chap:misc}
%
%
%
\newpage
\section{File removing}
\label{sec:file_removing}
%
%
\begin{templateenumerate}
\myitem
\label{item:eraser.php}
 {\cmbfsfx eraser.php}
\deft
\verb+<!DOCTYPE html+                                                \\
\verb+PUBLIC "-//W3C//DTD XHTML 1.0 +%
  \bluett{Strict}%
  \verb+//EN"+                                                       \\
\verb+"http://www.w3.org/TR/xhtml1/DTD/xhtml1-+%
  \bluett{strict}%
  \verb+.dtd">+                                                      \\
\verb+<html xmlns="http://www.w3.org/1999/xhtml">+                   \\
\verb++                                                              \\
\verb+<head>+                                                        \\
\verb++                                                              \\
\verb+<title> +%
  \bluett{File Eraser}%
  \verb+ </title>+                                                   \\
\verb++                                                              \\
\verb+<meta id="author" content="+%
  \bluett{Chung-Lin Shan}%
  \verb+" />+                                                        \\
\verb+<meta http-equiv="Content-Type" content="text/html; charset=+%
  \bluett{utf-8}%
  \verb+" />+                                                        \\
\verb++                                                              \\
\verb+<link rel="stylesheet" type="text/css" href="+%
  \bluett{main/main}%
  \verb+.css" />+                                                    \\
\verb++                                                              \\
\verb+</head>+                                                       \\
\verb++                                                              \\
\verb+<body +%
  \bluett{class="en"}%
  \verb+>+                                                           \\
\verb++                                                              \\
\verb+<h3 +%
  \bluett{class="center"}%
  \verb+> +%
  \bluett{File Eraser}%
  \verb+ </h3>+                                                      \\
\verb++                                                              \\
\verb+<?php+                                                         \\
\verb+  shell_exec("rm -r +%
  \bluett{tmp/input\_setup\_web}%
  \verb+/*.txt");+                                                   \\
\verb++                                                              \\
\verb+  shell_exec("rm -r +%
  \bluett{tmp/output\_plots}%
  \verb+/*.eps");+                                                   \\
\verb+  shell_exec("rm -r +%
  \bluett{tmp/output\_plots}%
  \verb+/*.ps");+                                                    \\
\verb+  shell_exec("rm -r +%
  \bluett{tmp/output\_plots}%
  \verb+/*.pdf");+                                                   \\
\verb+  shell_exec("rm -r +%
  \bluett{tmp/output\_plots}%
  \verb+/*.png");+                                                   \\
\verb+  shell_exec("rm -r +%
  \bluett{tmp/output\_plots}%
  \verb+/*.txt");+                                                   \\
\verb++                                                              \\
\verb+  shell_exec("rm -r +%
  \bluett{tmp/output\_results}%
  \verb+/*.txt");+                                                   \\
\verb++                                                              \\
\verb+  shell_exec("rm -r +%
  \bluett{tmp/uploaded\_files}%
  \verb+/*.txt");+                                                   \\
\verb+?>+                                                            \\
\verb++                                                              \\
\verb+</body>+                                                       \\
\verb++                                                              \\
\verb+</html>+                                                       \\
\tabb
\end{templateenumerate}
\newpage
\section{Redirection}
\label{sec:redirection}
%
%
\begin{templateenumerate}
\myitem
\label{item:redirection_parent_directory}
 {\bf Redirection to the parent directory}
\deft
\verb+<!DOCTYPE html+                                                \\
\verb+PUBLIC "-//W3C//DTD XHTML 1.0 +%
  \bluett{Strict}%
  \verb+//EN"+                                                       \\
\verb+"http://www.w3.org/TR/xhtml1/DTD/xhtml1-+%
  \bluett{strict}%
  \verb+.dtd">+                                                      \\
\verb+<html xmlns="http://www.w3.org/1999/xhtml">+                   \\
\verb++                                                              \\
\verb+<head>+                                                        \\
\verb++                                                              \\
\verb+<title> </title>+                                              \\
\verb++                                                              \\
\verb+<meta id="author" content="+%
  \bluett{Chung-Lin Shan}%
  \verb+" />+                                                        \\
\verb+<meta http-equiv="Content-Type" content="text/html; charset=+%
  \bluett{utf-8}%
  \verb+" />+                                                        \\
\verb++                                                              \\
\verb+</head>+                                                       \\
\verb++                                                              \\
\verb+<body onload="this.location.replace('../')">+                  \\
\verb++                                                              \\
\verb+</body>+                                                       \\
\verb++                                                              \\
\verb+</html>+                                                       \\
\tabb
\newpage
\myitem
\label{item:redirection_webpage}
 {\bf Redirection to a web page}
\deft
\verb+<!DOCTYPE html+                                                \\
\verb+PUBLIC "-//W3C//DTD XHTML 1.0 +%
  \bluett{Strict}%
  \verb+//EN"+                                                       \\
\verb+"http://www.w3.org/TR/xhtml1/DTD/xhtml1-+%
  \bluett{strict}%
  \verb+.dtd">+                                                      \\
\verb+<html xmlns="http://www.w3.org/1999/xhtml">+                   \\
\verb++                                                              \\
\verb+<head>+                                                        \\
\verb++                                                              \\
\verb+<title> </title>+                                              \\
\verb++                                                              \\
\verb+<meta id="author" content="+%
  \bluett{Chung-Lin Shan}%
  \verb+" />+                                                        \\
\verb+<meta http-equiv="Content-Type" content="text/html; charset=+%
  \bluett{utf-8}%
  \verb+" />+                                                        \\
\verb++                                                              \\
\verb+</head>+                                                       \\
\verb++                                                              \\
\verb+<body onload="http://+%
  \bluett{www.tir.tw/}%
  \verb+">+                                                          \\
\verb++                                                              \\
\verb+</body>+                                                       \\
\verb++                                                              \\
\verb+</html>+                                                       \\
\tabb
\newpage
\myitem
\label{item:redirection_webpage_countdown}
 {\bf Redirection to a web page with countdown}
\deft
\verb+<!DOCTYPE html+                                                \\
\verb+PUBLIC "-//W3C//DTD XHTML 1.0 +%
  \bluett{Strict}%
  \verb+//EN"+                                                       \\
\verb+"http://www.w3.org/TR/xhtml1/DTD/xhtml1-+%
  \bluett{strict}%
  \verb+.dtd">+                                                      \\
\verb+<html xmlns="http://www.w3.org/1999/xhtml">+                   \\
\verb++                                                              \\
\verb+<head>+                                                        \\
\verb++                                                              \\
\verb+<title> </title>+                                              \\
\verb++                                                              \\
\verb+<meta id="author" content="+%
  \bluett{Chung-Lin Shan}%
  \verb+" />+                                                        \\
\verb+<meta http-equiv="Content-Type" content="text/html; charset=+%
  \bluett{utf-8}%
  \verb+" />+                                                        \\
\verb++                                                              \\
\verb+<meta http-equiv="Refresh" content="+%
  \redtt{15}%
  \verb+; url='+%
  \bluett{www.tir.tw/}%
  \verb+'" />+ \mynote{See Template \ref{item:basic_html_head}}      \\
\verb++                                                              \\
\verb+</head>+                                                       \\
\verb++                                                              \\
\verb+<body onload="+%
  \bluett{countdown}%
  \verb+()" +%
  \redtt{style="background-color:~\symbol{35}AFEEEE"}%
  \verb+>+                                                           \\
\verb++                                                              \\
\verb+<script type="text/javascript">+                               \\
\verb+  var +%
  \bluett{show\_rest\_time}%
  \verb+;+                                                           \\
\verb+  var +%
  \bluett{rest\_time}%
  \verb+ = +%
  \redtt{15}%
  \verb+;+                                                           \\
\verb++                                                              \\
\verb+  function +%
  \bluett{countdown}%
  \verb+()+                                                          \\
\verb+  {+                                                           \\
\verb+    if (+%
  \bluett{rest\_time}%
  \verb+ == 1)+                                                      \\
\verb+    {+                                                         \\
\verb+      document.getElementById("+%
  \bluett{rest\_time}%
  \verb+").innerHTML = +%
  \bluett{rest\_time}%
  \verb- + " second";-                                               \\
\verb+    }+                                                         \\
\verb++                                                              \\
\verb+    else+                                                      \\
\verb+    {+                                                         \\
\verb+      document.getElementById("+%
  \bluett{rest\_time}%
  \verb+").innerHTML = +%
  \bluett{rest\_time}%
  \verb- + " seconds";-                                              \\
\verb+    }+                                                         \\
\verb++                                                              \\
\verb+    if (+%
  \bluett{rest\_time}%
  \verb+ >= 1)+                                                      \\
\verb+    {+                                                         \\
\verb+      +%
  \bluett{rest\_time}%
  \verb+ = +%
  \bluett{rest\_time}%
  \verb+ - 1;+                                                       \\
\verb++                                                              \\
\verb+      +%
  \bluett{show\_rest\_time}%
  \verb+ = setTimeout("+%
  \bluett{countdown}%
  \verb+()", 1000);+                                                 \\
\verb+    }+                                                         \\
\verb+  }+                                                           \\
\verb+</script>+                                                     \\
\verb++                                                              \\
\verb+<p style="font-family: '+%
  \bluett{Times New Roman}%
  \verb+'; font-size: +%
  \bluett{14}%
  \verb+pt;+                                                         \\
\verb+          line-height: +%
  \bluett{18}%
  \verb+pt; text-align: +%
  \bluett{center}%
  \verb+">+                                                          \\
\verb+ <br />+                                                       \\
\verb+ This page will be redirected in <span id="+%
  \bluett{rest\_time}%
  \verb+"></span>,+ \mynote[0.5cm]{No space! Nothing!}               \\
\verb+ or you could click the URL below.+                            \\
\verb+ <br />+                                                       \\
\verb+ You could also bookmark this URL for your next visit:+        \\
\verb+ <a href="http://+%
  \bluett{www.tir.tw/}%
  \verb+">http://+%
  \bluett{www.tir.tw/}%
  \verb+</a>+                                                        \\
\verb+</p>+                                                          \\
\verb++                                                              \\
\verb+</body>+                                                       \\
\verb++                                                              \\
\verb+</html>+                                                       \\
\tabb
\end{templateenumerate}

\addemptypage
\begin{appendix}
\chapter{Some Auxiliary Files}
\label{chap:auxiliary_files}
%
%
%
\newpage
\section{\sf main.css}
\label{sec:main.css}
%
%
\begin{templateenumerate}
\myitem
\label{item:css_headings}
 {\bf Headings}
\defst
\verb+h1                {font-size: 26pt; color: blue;  text-align: center}+                    \\
\verb+h2                {font-size: 23pt; color: blue;  text-align: center}+                    \\
\verb+h3                {font-size: 18pt; color: blue;  text-align: left;  margin-left:   5pt}+ \\
\verb+h4                {font-size: 15pt; color: blue;  text-align: left;  margin-left:  25pt}+ \\
\verb+h5                {font-size: 13pt; color: blue;  text-align: center}+                    \\
\verb+h6                {font-size: 12pt; color: black; text-align: right; margin-right: 25pt}+ \\
\tabb
\myitem
\label{item:css_form}
 {\cmbfsfx form}
\defst
\verb+form              {font-size: 13pt; line-height: 18pt}+             \\
\verb++                                                                   \\
\verb+fieldset          {margin-left: 36pt;+                              \\
\verb+                   border-style: solid; border-color: transparent}+ \\
\tabb
\myitem
\label{item:css_input}
 {\cmbfsfx input}
\defst
\verb+input.table       {line-height: 15pt; text-align: center;+                            \\
\verb+                   margin-left:  5pt; margin-right: 5pt; height: 15pt; width: 90pt;+  \\
\verb+                   background-color: #AFEEEE}+                                        \\
\verb+                   /* PaleTurquoise, RGB={175, 238, 238} */+                          \\
\verb++                                                                                     \\
\verb+input.upload      {line-height: 15pt; text-align: left;+                              \\
\verb+                   height: 19.2pt; width: 200pt;+                                     \\
\verb+                   background-color: #AFEEEE}+                                        \\
\verb++                                                                                     \\
\verb+input.text        {line-height: 15pt; text-align: center;+                            \\
\verb+                   margin-left:  5pt; margin-right: 5pt; height: 15pt; width: 40pt;+  \\
\verb+                   background-color: #AFEEEE}+                                        \\
\verb++                                                                                     \\
\verb+input.submit      {font-size: 11pt;+                                                  \\
\verb+                   margin-left: 36pt; margin-right: 5pt; height: 22pt; width: 120pt}+ \\
\verb++                                                                                     \\
\verb+textarea          {height: 150pt; width: 300pt; background-color: #AFEEEE}+           \\
\verb++                                                                                     \\
\verb+select            {padding-left: 3pt; height: 15pt; width: 50pt;+                     \\
\verb+                   background-color: #AFEEEE}+                                        \\
\tabb
\myitem
\label{item:css_li}
 {\cmbfsfx li}
\defst
\verb+ul                {list-style-type: disc;    margin-left:  0pt; margin-right: 25pt}+ \\
\verb+ol                {list-style-type: decimal; margin-left:  5pt; margin-right: 25pt}+ \\
\verb++                                                                                    \\
\verb+li                {line-height: 20pt; margin-left: 25pt; margin-right: 25pt}+        \\
\tabb
\myitem
\label{item:css_table}
 {\cmbfsfx table}
\defst
\verb+table             {line-height: 24pt; margin-left: 36pt;+    \\
\verb+                   border-style: solid; border-color: blue}+ \\
\verb++                                                            \\
\verb+th                {text-align: center; width: 100pt}+        \\
\verb++                                                            \\
\verb+td                {text-align: center; width: 100pt}+        \\
\verb+td.result         {background-color: #AFEEEE}+               \\
\tabb
\myitem
\label{item:css_img}
 {\cmbfsfx img}
\defst
\verb+img.eq            {margin-left: 25pt;+                                           \\
\verb+                   border-top-width:  5pt; border-bottom-width: 4pt;+            \\
\verb+                   border-left-width: 0pt; border-right-width:  0pt;+            \\
\verb+                   border-style: solid; border-color: transparent}+              \\
\verb++                                                                                \\
\verb+img.eq-s          {border-width: 0px; border-style: solid; border-color: blue}+  \\
\verb++                                                                                \\
\verb+img.output_plot   {margin-left: 50pt; height: 320pt; width: auto;+               \\
\verb+                   border-width: 4px; border-style: double; border-color: blue}+ \\
\tabb
\myitem
\label{item:css_p}
 {\cmbfsfx p}
\defst
\verb+body              {background-color: #ADD8E6}+                             \\
\verb+                  /* lightblue, RGB={173, 216, 230} */+                    \\
\verb++                                                                          \\
\verb+div               {font-size: 13pt; text-indent: 20pt; line-height: 16pt}+ \\
\verb++                                                                          \\
\verb+p                 {color: black; text-align: justify;+                     \\
\verb+                   margin-left: 25pt; margin-right: 25pt}+                 \\
\verb++                                                                          \\
\verb+p.EqCard          {font-size: 13pt}+                                       \\
\verb++                                                                          \\
\verb+p.caption         {font-size: 13pt; line-height: 20pt;+                    \\
\verb+                   margin-left: 36pt; width: 600pt}+                       \\
\verb++                                                                          \\
\verb+p.date            {text-align: right; margin-right: 50pt}+                 \\
\tabb
\myitem
\label{item:css_font}
 {\cmbfsfx font}
\defst
\verb+ .en              {font-family: "Times New Roman"}+ \\
\verb++                                                   \\
\verb+ .bf              {font-weight: bold}+              \\
\verb+ .it              {font-style:  italic}+            \\
\verb+ .sf              {font-family: sans-serif}+        \\
\tabb
\myitem
\label{item:css_color}
 {\cmbfsfx color}
\defst
\verb+ .black           {color: black}+   \\
\verb+ .white           {color: white}+   \\
\verb++                                   \\
\verb+ .red             {color: red}+     \\
\verb++                                   \\
\verb+ .green           {color: green}+   \\
\verb++                                   \\
\verb+ .blue            {color: blue}+    \\
\verb+ .lightblue       {color: #ADD8E6}+ \\
\tabb
\myitem
\label{item:css_text-align}
 {\cmbfsfx text-align}
\defst
\verb+ .center          {text-align: center}+ \\
\verb+ .left            {text-align: left}+   \\
\verb+ .right           {text-align: right}+  \\
\tabb
\myitem
\label{item:css_a}
 {\cmbfsfx a}
\defst
\verb+a:visited         {color: #00008B}                     /* darkblue, RGB={0, 0, 139} */+ \\
\verb+a.eq              {color: blue; text-decoration: none}+                                 \\
\tabb
\myitem
\label{item:css_footnotes}
 {\bf Footnotes}
\defst
\verb+ .def             {line-height: 20pt; margin-left: 40pt}+                  \\
\verb+ .plot_link       {font-size: 13pt; line-height: 17pt; margin-left: 40pt}+ \\
\verb+ .credit          {font-size: 11pt; line-height: 13pt; margin-left: 60pt}+ \\
\tabb
\end{templateenumerate}
\newpage
\section{\sf setup-reader.c}
\label{sec:setup-reader.c}
%
%
\begin{templateenumerate}
\myitem
\label{item:including_setup-reader.c}
 {\bf Including {\cmbfsfx setup-reader.c}}
\deft
\verb++%
  \bluett{personal\_setup}%
  \verb+ = fopen("+%
  \bluett{tmp/input\_setup\_web/input\_setup\_web}%
  \verb+.txt", "r");+                                   \\
\verb++                                                 \\
\verb+  #include "+%
  \bluett{main-sub/setup-reader}%
  \verb+.c"+ \mynote{Template \ref{item:setup-reader.c}} \\
\verb++                                                 \\
\verb+fclose(+%
  \bluett{personal\_setup}%
  \verb+);+                                             \\
\tabb
\newpage
\myitem
\label{item:setup-reader.c}
 {\cmbfsfx setup-reader.c}
 (cf.~Template \ref{item:sample-input_setup_web.txt})
\deft
\verb+/* --------------------------------------------- */+      \\
\verb+//+                                                       \\
\verb+// A subroutine for reading input setup from the website+ \\
\verb+//+                                                       \\
\verb+/* --------------------------------------------- */+      \\
\verb++                                                         \\
\verb+do+                                                       \\
\verb+{+                                                        \\
\verb+  fscanf(+%
  \bluett{personal\_setup}%
  \verb+, "%d", &+%
  \bluett{read\_setup\_category}%
  \verb+);+                                                     \\
\verb++                                                         \\
\verb+  switch (+%
  \bluett{read\_setup\_category}%
  \verb+)+                                                      \\
\verb+  {+                                                      \\
\verb++                                                         \\
\verb+/* --------------------------------------------- */+      \\
\verb+//+                                                       \\
\verb+// For users' comments+                                   \\
\verb+//+                                                       \\
\verb++                                                         \\
\verb+    case +%
  \bluett{0}%
  \verb+:+                                                      \\
\verb+      fscanf(+%
  \bluett{personal\_setup}%
  \verb+, "%*s");+                                              \\
\verb++                                                         \\
\verb+      break;+                                             \\
\defc
\verb++                                                         \\
\verb+/* --------------------------------------------- */+      \\
\verb+//+                                                       \\
\verb+// Setup for main functions+                              \\
\verb+//+                                                       \\
\verb++                                                         \\
\verb+    case +%
  \bluett{1}%
  \verb+:+                                                      \\
\verb+      fscanf(+%
  \bluett{personal\_setup}%
  \verb+, "%d", &+%
  \bluett{read\_setup\_item}%
  \verb+);+                                                     \\
\verb++                                                         \\
\verb+      switch (+%
  \bluett{read\_setup\_item}%
  \verb+)+                                                      \\
\verb+      {+                                                  \\
\verb+        case +%
  \bluett{1}%
  \verb+:+                                                      \\
\verb+          fscanf(+%
  \bluett{personal\_setup}%
  \verb+, "%*s%*s%*s%d%*s", &+%
  \bluett{read\_setup\_data\_int}%
  \verb+);+                                                     \\
\verb++                                                         \\
\verb+          switch (+%
  \bluett{read\_setup\_data\_int}%
  \verb+)+                                                      \\
\verb+          {+                                              \\
\verb+            case 1:+                                      \\
\verb+              +%
  \bluett{main\_function}%
  \verb+ = "+%
  \redtt{main\_function\_1}%
  \verb+";+                                                     \\
\verb+              break;+                                     \\
\verb++                                                         \\
\verb+            case 2:+                                      \\
\verb+              +%
  \bluett{main\_function}%
  \verb+ = "+%
  \redtt{main\_function\_2}%
  \verb+";+                                                     \\
\verb+              break;+                                     \\
\verb++                                                         \\
\verb+            case 3:+                                      \\
\verb+              +%
  \bluett{main\_function}%
  \verb+ = "+%
  \redtt{main\_function\_3}%
  \verb+";+                                                     \\
\verb+              break;+                                     \\
\verb++                                                         \\
\verb+            case 4:+                                      \\
\verb+              +%
  \bluett{main\_function}%
  \verb+ = "+%
  \redtt{main\_function\_4}%
  \verb+";+                                                     \\
\verb+              break;+                                     \\
\verb++                                                         \\
\verb+            default:+                                     \\
\verb+              return 0;+                                  \\
\verb+          }+                                              \\
\verb++                                                         \\
\verb+          break;+                                         \\
\verb++                                                         \\
\verb+            .+                                            \\
\verb+            .+                                            \\
\verb+            .+                                            \\
\defc
\verb++                                                         \\
\verb+        case +%
  \redtt{5}%
  \verb+:+                                                      \\
\verb+          fscanf(+%
  \bluett{personal\_setup}%
  \verb+, "%d", &+%
  \bluett{read\_setup\_entry}%
  \verb+);+                                                     \\
\verb+          fscanf(+%
  \bluett{personal\_setup}%
  \verb+, "%*s%*s%*s%d%*s", &+%
  \bluett{read\_setup\_data\_int}%
  \verb+);+                                                     \\
\verb++                                                         \\
\verb+          switch (+%
  \bluett{read\_setup\_entry}%
  \verb+)+                                                      \\
\verb+          {+                                              \\
\verb+            case 1:+                                      \\
\verb+              sprintf(+%
  \bluett{read\_setup\_data\_date}%
  \verb+, "%d", +%
  \bluett{read\_setup\_data\_int}%
  \verb+);+                                                     \\
\verb++                                                         \\
\verb+//            if (+%
  \bluett{read\_setup\_data\_int}%
  \verb+ < 100000)+                                             \\
\verb+//            {+                                          \\
\verb+//              strcpy(+%
  \bluett{personal\_file\_date}%
  \verb+, "0");+                                                \\
\verb+//              strcat(+%
  \bluett{personal\_file\_date}%
  \verb+, +%
  \bluett{read\_setup\_data\_date}%
  \verb+);+                                                     \\
\verb+//            }+                                          \\
\verb+//            else+ \mynote[8cm]{Keep 6 digitals!}        \\
\verb+              {+                                          \\
\verb+                strcpy(+%
  \bluett{personal\_file\_date}%
  \verb+, +%
  \bluett{read\_setup\_data\_date}%
  \verb+);+                                                     \\
\verb+              }+                                          \\
\verb++                                                         \\
\verb+              break;+                                     \\
\verb++                                                         \\
\verb+            case 2:+                                      \\
\verb+              sprintf(+%
  \bluett{read\_setup\_data\_date}%
  \verb+, "%d", +%
  \bluett{read\_setup\_data\_int}%
  \verb+);+                                                     \\
\verb++                                                         \\
\verb+              if (+%
  \bluett{read\_setup\_data\_int}%
  \verb+ < 10)+                                                 \\
\verb+              {+                                          \\
\verb+                strcpy(+%
  \bluett{personal\_file\_time}%
  \verb+, "00000");+                                            \\
\verb+                strcat(+%
  \bluett{personal\_file\_time}%
  \verb+, +%
  \bluett{read\_setup\_data\_date}%
  \verb+);+                                                     \\
\verb+              }+                                          \\
\verb+              else+                                       \\
\verb+              if (+%
  \bluett{read\_setup\_data\_int}%
  \verb+ < 100)+                                                \\
\verb+              {+                                          \\
\verb+                strcpy(+%
  \bluett{personal\_file\_time}%
  \verb+, "0000");+                                             \\
\verb+                strcat(+%
  \bluett{personal\_file\_time}%
  \verb+, +%
  \bluett{read\_setup\_data\_date}%
  \verb+);+                                                     \\
\verb+              }+                                          \\
\verb+              else+                                       \\
\verb+              if (+%
  \bluett{read\_setup\_data\_int}%
  \verb+ < 1000)+                                               \\
\verb+              {+                                          \\
\verb+                strcpy(+%
  \bluett{personal\_file\_time}%
  \verb+, "000");+                                              \\
\verb+                strcat(+%
  \bluett{personal\_file\_time}%
  \verb+, +%
  \bluett{read\_setup\_data\_date}%
  \verb+);+                                                     \\
\verb+              }+                                          \\
\verb+              else+ \mynote[8cm]{Keep 6 digitals!}        \\
\verb+              if (+%
  \bluett{read\_setup\_data\_int}%
  \verb+ < 10000)+                                              \\
\verb+              {+                                          \\
\verb+                strcpy(+%
  \bluett{personal\_file\_time}%
  \verb+, "00");+                                               \\
\verb+                strcat(+%
  \bluett{personal\_file\_time}%
  \verb+, +%
  \bluett{read\_setup\_data\_date}%
  \verb+);+                                                     \\
\verb+              }+                                          \\
\verb+              else+                                       \\
\verb+              if (+%
  \bluett{read\_setup\_data\_int}%
  \verb+ < 100000)+                                             \\
\verb+              {+                                          \\
\verb+                strcpy(+%
  \bluett{personal\_file\_time}%
  \verb+, "0");+                                                \\
\verb+                strcat(+%
  \bluett{personal\_file\_time}%
  \verb+, +%
  \bluett{read\_setup\_data\_date}%
  \verb+);+                                                     \\
\verb+              }+                                          \\
\verb+              else+                                       \\
\verb+              {+                                          \\
\verb+                strcpy(+%
  \bluett{personal\_file\_time}%
  \verb+, +%
  \bluett{read\_setup\_data\_date}%
  \verb+);+                                                     \\
\verb+              }+                                          \\
\verb++                                                         \\
\verb+              break;+                                     \\
\defc
\verb++                                                         \\
\verb+            case 3:+                                      \\
\verb+              sprintf(+%
  \bluett{read\_setup\_data\_date}%
  \verb+, "%d", +%
  \bluett{read\_setup\_data\_int}%
  \verb+);+                                                     \\
\verb++                                                         \\
\verb+//            if (+%
  \bluett{read\_setup\_data\_int}%
  \verb+ < 100000)+                                             \\
\verb+//            {+                                          \\
\verb+//              strcpy(+%
  \bluett{plot\_output\_date}%
  \verb+, "0");+                                                \\
\verb+//              strcat(+%
  \bluett{plot\_output\_date}%
  \verb+, +%
  \bluett{read\_setup\_data\_date}%
  \verb+);+                                                     \\
\verb+//            }+                                          \\
\verb+//            else+ \mynote[8cm]{Keep 6 digitals!}        \\
\verb+              {+                                          \\
\verb+                strcpy(+%
  \bluett{plot\_output\_date}%
  \verb+, +%
  \bluett{read\_setup\_data\_date}%
  \verb+);+                                                     \\
\verb+              }+                                          \\
\verb++                                                         \\
\verb+              break;+                                     \\
\verb++                                                         \\
\verb+            case 4:+                                      \\
\verb+              sprintf(+%
  \bluett{read\_setup\_data\_date}%
  \verb+, "%d", +%
  \bluett{read\_setup\_data\_int}%
  \verb+);+                                                     \\
\verb++                                                         \\
\verb+              if (+%
  \bluett{read\_setup\_data\_int}%
  \verb+ < 10)+                                                 \\
\verb+              {+                                          \\
\verb+                strcpy(+%
  \bluett{plot\_output\_time}%
  \verb+, "00000");+                                            \\
\verb+                strcat(+%
  \bluett{plot\_output\_time}%
  \verb+, +%
  \bluett{read\_setup\_data\_date}%
  \verb+);+                                                     \\
\verb+              }+                                          \\
\verb+              else+                                       \\
\verb+              if (+%
  \bluett{read\_setup\_data\_int}%
  \verb+ < 100)+                                                \\
\verb+              {+                                          \\
\verb+                strcpy(+%
  \bluett{plot\_output\_time}%
  \verb+, "0000");+                                             \\
\verb+                strcat(+%
  \bluett{plot\_output\_time}%
  \verb+, +%
  \bluett{read\_setup\_data\_date}%
  \verb+);+                                                     \\
\verb+              }+                                          \\
\verb+              else+                                       \\
\verb+              if (+%
  \bluett{read\_setup\_data\_int}%
  \verb+ < 1000)+                                               \\
\verb+              {+                                          \\
\verb+                strcpy(+%
  \bluett{plot\_output\_time}%
  \verb+, "000");+                                              \\
\verb+                strcat(+%
  \bluett{plot\_output\_time}%
  \verb+, +%
  \bluett{read\_setup\_data\_date}%
  \verb+);+                                                     \\
\verb+              }+                                          \\
\verb+              else+ \mynote[8cm]{Keep 6 digitals!}        \\
\verb+              if (+%
  \bluett{read\_setup\_data\_int}%
  \verb+ < 10000)+                                              \\
\verb+              {+                                          \\
\verb+                strcpy(+%
  \bluett{plot\_output\_time}%
  \verb+, "00");+                                               \\
\verb+                strcat(+%
  \bluett{plot\_output\_time}%
  \verb+, +%
  \bluett{read\_setup\_data\_date}%
  \verb+);+                                                     \\
\verb+              }+                                          \\
\verb+              else+                                       \\
\verb+              if (+%
  \bluett{read\_setup\_data\_int}%
  \verb+ < 100000)+                                             \\
\verb+              {+                                          \\
\verb+                strcpy(+%
  \bluett{plot\_output\_time}%
  \verb+, "0");+                                                \\
\verb+                strcat(+%
  \bluett{plot\_output\_time}%
  \verb+, +%
  \bluett{read\_setup\_data\_date}%
  \verb+);+                                                     \\
\verb+              }+                                          \\
\verb+              else+                                       \\
\verb+              {+                                          \\
\verb+                strcpy(+%
  \bluett{plot\_output\_time}%
  \verb+, +%
  \bluett{read\_setup\_data\_date}%
  \verb+);+                                                     \\
\verb+              }+                                          \\
\verb++                                                         \\
\verb+              break;+                                     \\
\defc
\verb++                                                         \\
\verb+            case 5:+                                      \\
\verb+              +%
  \bluett{random\_gen}%
  \verb+ = +%
  \bluett{read\_setup\_data\_int}%
  \verb+;+                                                      \\
\verb++                                                         \\
\verb+              break;+                                     \\
\verb++                                                         \\
\verb+            default:+                                     \\
\verb+              return 0;+                                  \\
\verb+          }+                                              \\
\verb++                                                         \\
\verb+          break;+                                         \\
\verb++                                                         \\
\verb+            .+                                            \\
\verb+            .+                                            \\
\verb+            .+                                            \\
\verb++                                                         \\
\verb+        default:+                                         \\
\verb+          return 0;+                                      \\
\verb+      }+                                                  \\
\verb++                                                         \\
\verb+      break;+                                             \\
\defc
\verb++                                                         \\
\verb+/* --------------------------------------------- */+      \\
\verb+//+                                                       \\
\verb+// Setup for "+%
  \redtt{main\_function\_1}%
  \verb+"+                                                      \\
\verb+//+                                                       \\
\verb++                                                         \\
\verb+    case +%
  \redtt{2}%
  \verb+:+                                                      \\
\verb+      fscanf(+%
  \bluett{personal\_setup}%
  \verb+, "%d", &+%
  \bluett{read\_setup\_item}%
  \verb+);+                                                     \\
\verb++                                                         \\
\verb+      switch (+%
  \bluett{read\_setup\_item}%
  \verb+)+                                                      \\
\verb+      {+                                                  \\
\verb+        case +%
  \redtt{1}%
  \verb+:+                                                      \\
\verb+          fscanf(+%
  \bluett{personal\_setup}%
  \verb+, "%*s%*s%*s%d%*s", &+%
  \bluett{read\_setup\_data\_int}%
  \verb+);+                                                     \\
\verb++                                                         \\
\verb+          switch (+%
  \bluett{read\_setup\_data\_int}%
  \verb+)+                                                      \\
\verb+          {+                                              \\
\verb+            case 1:+                                      \\
\verb+              +%
  \redtt{mode\_choice}%
  \verb+ = "+%
  \redtt{mode\_choice\_1}%
  \verb+";+                                                     \\
\verb+              break;+                                     \\
\verb++                                                         \\
\verb+            case 2:+                                      \\
\verb+              +%
  \redtt{mode\_choice}%
  \verb+ = "+%
  \redtt{mode\_choice\_2}%
  \verb+";+                                                     \\
\verb+              break;+                                     \\
\verb++                                                         \\
\verb+            case 3:+                                      \\
\verb+              +%
  \redtt{mode\_choice}%
  \verb+ = "+%
  \redtt{mode\_choice\_3}%
  \verb+";+                                                     \\
\verb+              break;+                                     \\
\verb++                                                         \\
\verb+            case 4:+                                      \\
\verb+              +%
  \redtt{mode\_choice}%
  \verb+ = "+%
  \redtt{mode\_choice\_4}%
  \verb+";+                                                     \\
\verb+              break;+                                     \\
\verb++                                                         \\
\verb+            default:+                                     \\
\verb+              return 0;+                                  \\
\verb+          }+                                              \\
\verb++                                                         \\
\verb+          break;+                                         \\
\verb++                                                         \\
\verb+        case +%
  \redtt{2}%
  \verb+:+                                                      \\
\verb+          fscanf(+%
  \bluett{personal\_setup}%
  \verb+, "%*s%*s%*s+%
  \bluett{\%d}%
  \verb+%*s", &+%
  \bluett{read\_setup\_data\_int}%
  \verb+);+                                                     \\
\verb+          +%
  \redtt{integer\_1}%
  \verb+ = +%
  \bluett{read\_setup\_data\_int}%
  \verb+;+                                                      \\
\verb++                                                         \\
\verb+          break;+                                         \\
\verb++                                                         \\
\verb+        case +%
  \redtt{3}%
  \verb+:+                                                      \\
\verb+          fscanf(+%
  \bluett{personal\_setup}%
  \verb+, "%*s%*s%*s+%
  \bluett{\%d}%
  \verb+%*s", &+%
  \bluett{read\_setup\_data\_int}%
  \verb+);+                                                     \\
\verb+          +%
  \redtt{integer\_2}%
  \verb+ = +%
  \bluett{read\_setup\_data\_int}%
  \verb+;+                                                      \\
\verb++                                                         \\
\verb+          break;+                                         \\
\defc
\verb++                                                         \\
\verb+        case +%
  \redtt{4}%
  \verb+:+                                                      \\
\verb+          fscanf(+%
  \bluett{personal\_setup}%
  \verb+, "%d", &+%
  \bluett{read\_setup\_entry}%
  \verb+);+                                                     \\
\verb+          fscanf(+%
  \bluett{personal\_setup}%
  \verb+, "%*s%*s%*s+%
  \bluett{\%f}%
  \verb+%*s", &+%
  \bluett{read\_setup\_data\_float}%
  \verb+);+                                                     \\
\verb++                                                         \\
\verb+          switch (+%
  \bluett{read\_setup\_entry}%
  \verb+)+                                                      \\
\verb+          {+                                              \\
\verb+            case 1:+                                      \\
\verb+              +%
  \redtt{float\_1\_1}%
  \verb+ = +%
  \bluett{read\_setup\_data\_float}%
  \verb+;+                                                      \\
\verb+              break;+                                     \\
\verb++                                                         \\
\verb+            case 2:+                                      \\
\verb+              +%
  \redtt{float\_1\_2}%
  \verb+ = +%
  \bluett{read\_setup\_data\_float}%
  \verb+;+                                                      \\
\verb+              break;+                                     \\
\verb++                                                         \\
\verb+            default:+                                     \\
\verb+              return 0;+                                  \\
\verb+          }+                                              \\
\verb++                                                         \\
\verb+          break;+                                         \\
\verb++                                                         \\
\verb+        case +%
  \redtt{5}%
  \verb+:+                                                      \\
\verb+          fscanf(+%
  \bluett{personal\_setup}%
  \verb+, "%d", &+%
  \bluett{read\_setup\_entry}%
  \verb+);+                                                     \\
\verb+          fscanf(+%
  \bluett{personal\_setup}%
  \verb+, "%*s%*s%*s+%
  \bluett{\%f}%
  \verb+%*s", &+%
  \bluett{read\_setup\_data\_float}%
  \verb+);+                                                     \\
\verb++                                                         \\
\verb+          switch (+%
  \bluett{read\_setup\_entry}%
  \verb+)+                                                      \\
\verb+          {+                                              \\
\verb+            case 1:+                                      \\
\verb+              +%
  \redtt{float\_2\_1}%
  \verb+ = +%
  \bluett{read\_setup\_data\_float}%
  \verb+;+                                                      \\
\verb+              break;+                                     \\
\verb++                                                         \\
\verb+            case 2:+                                      \\
\verb+              +%
  \redtt{float\_2\_2}%
  \verb+ = +%
  \bluett{read\_setup\_data\_float}%
  \verb+;+                                                      \\
\verb+              break;+                                     \\
\verb++                                                         \\
\verb+            default:+                                     \\
\verb+              return 0;+                                  \\
\verb+          }+                                              \\
\verb++                                                         \\
\verb+          break;+                                         \\
\verb++                                                         \\
\verb+        case +%
  \redtt{6}%
  \verb+:+                                                      \\
\verb+          fscanf(+%
  \bluett{personal\_setup}%
  \verb+, "%d", &+%
  \bluett{read\_setup\_entry}%
  \verb+);+                                                     \\
\verb++                                                         \\
\verb+          switch (+%
  \bluett{read\_setup\_entry}%
  \verb+)+                                                      \\
\verb+          {+                                              \\
\verb+            case +%
  \redtt{1}%
  \verb+:+                                                      \\
\verb+              fscanf(+%
  \bluett{personal\_setup}%
  \verb+, "%*s%*s%*s+%
  \bluett{\%d}%
  \verb+%*s", &+%
  \bluett{read\_setup\_data\_int}%
  \verb+);+  \\
\verb+              +%
  \redtt{integer\_3}%
  \verb+ = +%
  \bluett{read\_setup\_data\_int}%
  \verb+;+                                                      \\
\verb+              break;+                                     \\
\verb++                                                         \\
\verb+            case +%
  \redtt{2}%
  \verb+:+                                                      \\
\verb+              fscanf(+%
  \bluett{personal\_setup}%
  \verb+, "%*s%*s%*s+%
  \bluett{\%f}%
  \verb+%*s", &+%
  \bluett{read\_setup\_data\_float}%
  \verb+);+ \\
\verb+              +%
  \redtt{float\_3}%
  \verb+   = +%
  \bluett{read\_setup\_data\_float}%
  \verb+;+                                                      \\
\verb+              break;+                                     \\
\verb++                                                         \\
\verb+            default:+                                     \\
\verb+              return 0;+                                  \\
\verb+          }+                                              \\
\verb++                                                         \\
\verb+          break;+                                         \\
\defc
\verb++                                                         \\
\verb+/* --------------------------------------------- */+      \\
\verb++                                                         \\
\verb+    default:+                                             \\
\verb+      return 0;+                                          \\
\verb+  }+                                                      \\
\verb+}+                                                        \\
\verb+while (!feof(+%
  \bluett{personal\_setup}%
  \verb+));+                                                    \\
\tabb
\end{templateenumerate}

\addemptypage
\chapter{Basic Commands for Website Building}
\label{chap:basic_commands_web}
 In this chapter
 I give some basic and frequently used commands
 for building a website.
 Full website building tutorials with
 more detailed explanations and interactive examples
 can be found in Ref.~\cite{W3Schools}.
 Two validator links
 for validating the HTML and CSS syntax of a web page
 are also given in Ref.~\cite{validators}.
\newpage
\section{HTML (Hyper Text Markup Language) and XHTML (EXtensible HTML)}
\label{sec:basic_html}
%
%
\subsection{\sf html}
\label{sec:basic_html_html}
%
%
\begin{templateenumerate}
\myitem
\label{item:basic_html}
 {\bf Basic structure of an {\cmbfsfx .html} file}
 (see Template \ref{item:head_headings})
\deft
\verb+<!DOCTYPE html+                              \\
\verb+PUBLIC "-//W3C//DTD XHTML 1.0 +%
  \bluett{Strict}%
  \verb+//EN"+                                     \\
\verb+"http://www.w3.org/TR/xhtml1/DTD/xhtml1-+%
  \bluett{strict}%
  \verb+.dtd">+                                    \\
\verb++                                            \\
\verb+<html xmlns="http://www.w3.org/1999/xhtml">+ \\
\verb++                                            \\
\verb+  <head>+                                    \\
\verb++                                            \\
\verb+    .+                                       \\
\verb+    .+                                       \\
\verb+    .+                                       \\
\verb++                                            \\
\verb+  </head>+                                   \\
\verb++                                            \\
\verb++                                            \\
\verb+  <body>+                                    \\
\verb++                                            \\
\verb+    .+                                       \\
\verb+    .+                                       \\
\verb+    .+                                       \\
\verb++                                            \\
\verb+  </body>+                                   \\
\verb++                                            \\
\verb+</html>+                                     \\
\tabb
\modificationdeft
\verb+<!DOCTYPE html+                              \\
\verb+PUBLIC "-//W3C//DTD XHTML 1.0 +%
  \bluett{Transitional}%
  \verb+//EN"+                                     \\
\verb+"http://www.w3.org/TR/xhtml1/DTD/xhtml1-+%
  \bluett{transitional}%
  \verb+.dtd">+                                    \\
\tabb
\modificationdeft
\verb+<!DOCTYPE html+                              \\
\verb+PUBLIC "-//W3C//DTD XHTML 1.0 +%
  \bluett{Frameset}%
  \verb+//EN"+                                     \\
\verb+"http://www.w3.org/TR/xhtml1/DTD/xhtml1-+%
  \bluett{frameset}%
  \verb+.dtd">+                                    \\
\tabb
\end{templateenumerate}
\newpage
\subsection{\sf head}
\label{sec:basic_html_head}
%
%
\begin{templateenumerate}
\myitem
\label{item:basic_html_head}
 {\bf Basic structure of the {\cmbfsfx head} of an {\cmbfsfx .html} file}
 (see Template \ref{item:head_headings})
\deft
\verb+<head>+                                                           \\
\verb++                                                                 \\
\verb+  <title> +%
  \redtt{Title of the web page}%
  \verb+ </title>+                                                      \\
\verb++                                                                 \\
\verb+  <meta id="author"               content="+%
  \redtt{name(s) of author(s)}%
  \verb+" />+                                                           \\
\verb+  <meta http-equiv="Content-Type" content="text/html; charset=+%
  \bluett{utf-8}%
  \verb+" />+                                                           \\
\verb+  <meta http-equiv="Refresh"      content="+%
  \redtt{15}%
  \verb+; url='+%
  \redtt{www.xxxxxx.xxx/xxxxxx/}%
  \verb+'" />+                                                          \\
\verb++                                                                 \\
\verb+  <base href="+%
  \redtt{http://www.xxxxxx.xxx/xxxxxx}%
  \verb+/" />+                                                          \\
\verb+  <base target="+%
  \bluett{\_top}%
  \verb+" />+ \mynote[3.5cm]{See Sec.~\ref{sec:basic_html_a} for links} \\
\verb++                                                                 \\
\verb++                                                                 \\
\verb+  <link rel="stylesheet"          type="text/css"+                \\
\verb+        href="+%
  \redtt{folder of the css file}%
  \verb+/+%
  \redtt{name of the css file}%
  \verb+.css" />+ \mynote[0.5cm]{See Sec.~\ref{sec:basic_css} for CSS}  \\
\verb++                                                                 \\
\verb+  <style type="text/css">+                                        \\
\verb+       .+                                                         \\
\verb+       .+ \mynote{See Sec.~\ref{sec:basic_css} for CSS}           \\
\verb+       .+                                                         \\
\verb+  </style>+                                                       \\
\verb++                                                                 \\
\verb++                                                                 \\
\verb+  <script type="text/javascript">+                                \\
\verb+       .+                                                         \\
\verb+       .+  \mynote{See Sec.~\ref{sec:basic_js} for JavaScript}    \\
\verb+       .+                                                         \\
\verb+  </script>+                                                      \\
\verb++                                                                 \\
\verb+</head>+                                                          \\
\tabb
\modificationdeft
\verb+  <meta http-equiv="Content-Type" content="text/html; charset=+%
  \bluett{big5}%
  \verb+" />+                                                           \\
\tabb
\modificationdeft
\verb+  <base target="+%
  \bluett{\_blank}%
  \verb+" />+                                                           \\
\tabb
\Xdeft{Corresponding command for {\tt<base>} in {\tt<body>}}
\verb+  <a href="http:">+%
  \redtt{Description of this link}%
  \verb+</a>+                       \\
\tabb
\end{templateenumerate}
\newpage
\subsection{Comments}
\label{sec:basic_html_comments}
%
%
\deft
\verb+<!--   +%
  \redtt{Comment line}%
  \verb+   -->+              \\
\verb++                      \\
\verb+<!--+                  \\
\verb+  +%
  \redtt{Some comments ...}%
  \verb++                    \\
\verb+      .+               \\
\verb+      .+               \\
\verb+      .+               \\
\verb+      .+               \\
\verb+-->+                   \\
\tabb
 See Secs.~\ref{sec:basic_css_comments},
 \ref{sec:basic_js_comments},
 and \ref{sec:basic_php_comments}
 for comments in CSS, JavaScript, and PHP,
 respectively.
\newpage
\subsection{Tags and elements}
\label{sec:basic_html_tags}
%
%
\Xdeft{Start/opening tags}
\verb+<html>, <head>, <body>,+   \\
\verb+<title>, <div>, <p>, <a>,+ \\
\verb+<span>+                    \\
\tabb
\Xdeft{End/closing tags}
\verb+</html>, </head>, </body>,+    \\
\verb+</title>, </div>, </p>, </a>,+ \\
\verb+</span>+                       \\
\tabb
\Xdeft{Elements}
\\
\(
%
{\rm Element}
\cleft{\begin{array}{l}
         \mbox{\tt<html>}                                            \\
                                                                     \\ \\
%
{\rm Element}
\cleft{\begin{array}{l}
         \mbox{\tt<head>}                                            \\
                                                                     \\ \\
           {\tt~~~~~~}\stackrel{\underbrace{\mbox{\tt<title> ...... </title>}}}{\rm Element} \\
                                                                     \\ \\
         \mbox{\tt</head>}                                           \\
       \end{array}}                                                  \\
                                                                     \\ \\
%
{\rm Element}
\cleft{\begin{array}{l}
         \mbox{\tt<body>}                                            \\
                                                                     \\ \\
{\tt~~~~~~}
\cleft{\begin{array}{l}
         \mbox{\tt<div>}                                             \\
                                                                     \\ \\
{\tt~~~~~~}
\cleft{\begin{array}{l}
         \mbox{\tt<p>}                                               \\
                                                                     \\ \\
           {\tt~~~~~~}\underbrace{\mbox{\tt<a>......</a>}}           \\
 \mynote[4cm]{All tags should be in ``lowercase''!}                  \\ \\
           {\tt~~~~~~}\underbrace{\mbox{\tt<span>......</span>}}     \\
                                                                     \\ \\
         \mbox{\tt</p>}                                              \\
       \end{array}}                                                  \\
                                                                     \\ \\
         \mbox{\tt</div>}                                            \\
       \end{array}}                                                  \\
                                                                     \\ \\
         \mbox{\tt</body>}                                           \\
       \end{array}}                                                  \\
                                                                     \\ \\
         \mbox{\tt</html>}                                           \\
       \end{array}}
\)                                                                   \\
\\
\tabb
%
%
 See Sec.~\ref{sec:basic_htmldom_nodes}
 for the definitions of a ``node'' and the ``node tree''
 in HTML DOM.

\Xdeft{Empty elements (without closing tags)}
\verb+<meta +%
  \redtt{/}%
  \verb+>, <link +%
  \redtt{/}%
  \verb+>,+          \\
\verb+<br +%
  \redtt{/}%
  \verb+>, <hr +%
  \redtt{/}%
  \verb+>, <img +%
  \redtt{/}%
  \verb+>, <input +%
  \redtt{/}%
  \verb+>,+          \\
\verb+<area +%
  \redtt{/}%
  \verb+>, <col +%
  \redtt{/}%
  \verb+>, <frame +%
  \redtt{/}%
  \verb+>+           \\
\tabb
\newpage
\subsection{Attributes}
\label{sec:basic_html_attributes}
%
%
\deft
\verb+<+%
  \redtt{tag}%
  \verb+ type="+%
  \redtt{type choice of this tag}%
  \verb+"+                                \\
\verb++ \mynote[6cm]{Both single (`...') or double (``...'') quotes can be used.} \\
\verb+     id="+%
  \redtt{id of this tag}%
  \verb+"+                                \\
\verb+     name="+%
  \redtt{name of this tag}%
  \verb+"+                                \\
\verb++ \mynote[6cm]{All attributes should be in ``lowercase''!} \\
\verb+     value="+%
  \redtt{value given for this tag}%
  \verb+"+                                \\
\verb++                                   \\
\verb+     class="+%
  \redtt{specified class (for this tag)}%
  \verb+"+ \mynote{See Sec.~\ref{sec:basic_css} for CSS} \\
\verb+     style="+%
  \redtt{specified style for this tag}%
  \verb+"+                                \\
\verb++                                   \\
\verb+     +%
  \redtt{(/)}%
  \verb+ >+                               \\
\tabb
 See Secs.~\ref{sec:basic_html_a},
 \ref{sec:basic_html_img},
 and \ref{sec:basic_html_input}
 for some specified attributes combined with
 the \verb+<a>+,
 the \verb+<img>+,
 and the \verb+<input type="radio">+,
 the \verb+<input type="checkbox">+
 as well as with the
 \verb+<option>+ tags.
%
%
%
\newpage
\subsection{Headings}
\label{sec:basic_html_headings}
%
%
\begin{templateenumerate}
\myitem
\label{item:basic_html_headings.html}
 {\cmbfsfx basic\_html\_headings.html}
 (cf.~Template \ref{item:basic_css_headings.html})
\deft
\verb+<body>+                              \\
\verb++                                    \\
\verb+  <h1> +%
  \redtt{Main Title of the Website}%
  \verb+ </h1>+                            \\
\verb++                                    \\
\verb+  <h2> +%
  \redtt{Title of the web page}%
  \verb+ </h2>+                            \\
\verb++                                    \\
\verb+  <hr +%
  \redtt{/}%
  \verb+>+                                 \\
\verb++                                    \\
\verb+  <h3> +%
  \redtt{Main title of the category}%
  \verb+ </h3>+                            \\
\verb+  <h4> +%
  \redtt{Title of blocks in the category}%
  \verb+ </h4>+                            \\
\verb++                                    \\
\verb+  <hr +%
  \redtt{/}%
  \verb+>+                                 \\
\verb++                                    \\
\verb+  <h5> +%
  \redtt{Title of the section}%
  \verb+ </h5>+                            \\
\verb++                                    \\
\verb+  <div>+                             \\
\verb++                                    \\
\verb+    <h6> +%
  \redtt{Title of the paragraph}%
  \verb+ </h6>+                            \\
\verb++                                    \\
\verb+    <p>+                             \\
\verb+      +%
  \redtt{Some sentences ...}%
  \verb+   <br +%
  \redtt{/}%
  \verb+>+                                 \\
\verb+        +%
  \redtt{Some sentences ...}%
  \verb+ <br +%
  \redtt{/}%
  \verb+>+                                 \\
\verb++                                    \\
\verb+      <span> +%
  \redtt{Some more sentences ...}%
  \verb+ </span>+                          \\
\verb+    </p>+                            \\
\verb++                                    \\
\verb+    <hr +%
  \redtt{/}%
  \verb+>+                                 \\
\verb++                                    \\
\verb+  </div>+                            \\
\verb++                                    \\
\verb+</body>+                             \\
\tabb
\newpage
\picin{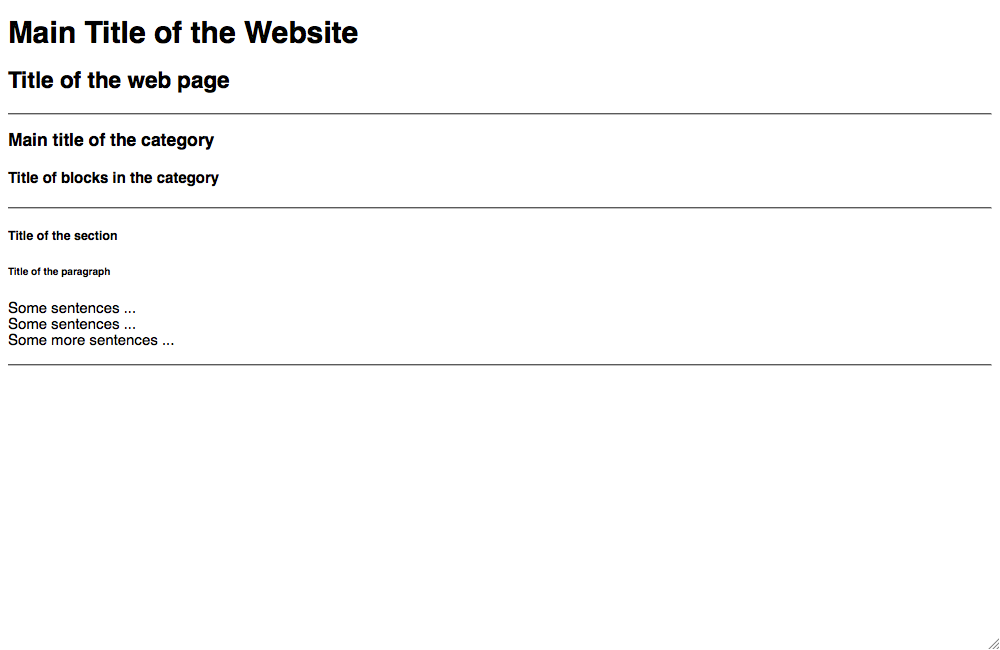}
{basic\_html\_headings}
{Headings.}
\end{templateenumerate}
\newpage
\subsection{Text formatting}
\label{sec:basic_html_text_formatting}
%
%
\Xdeft{Bold face}
\verb+<b> +%
  \redtt{Bold face}%
  \verb+ </b>+      \\
\verb+<strong> +%
  \redtt{Bold face}%
  \verb+ </strong>+ \\
\tabb
\Xdeft{Italic}
\verb+<i> +%
  \redtt{Italic}%
  \verb+ </i>+    \\
\verb+<em> +%
  \redtt{Italic}%
  \verb+ </em>+   \\
\tabb
\Xdeft{Resize}
\verb+<big> +%
  \redtt{Text with a bigger font size}%
  \verb+ </big>+   \\
\verb+<small> +%
  \redtt{Text with a smaller font size}%
  \verb+ </small>+ \\
\tabb
\Xdeft{Super-/Subscript}
\verb+x<sup>+%
  \redtt{n}%
  \verb+</sup>+
  \hspace{2.875cm} $\lTo$ \hspace{1cm}
  x$^{\rm n}$                            \\
\verb+x<sub>+%
  \redtt{n}%
  \verb+</sub>+
  \hspace{2.875cm} $\lTo$ \hspace{1cm}
  x$_{\rm n}$                            \\
\tabb
\Xdeft{Inserted and deleted texts}
\verb+<ins> +%
  \redtt{XXXXXX}%
  \verb+ </ins>+
  \hspace{2cm} $\lTo$ \hspace{1cm}
  \underline{XXXXXX}                              \\
\verb+<del> +%
  \redtt{XXXXXX}%
  \verb+ </del>+
  \hspace{2cm} $\lTo$ \hspace{1cm}
  X\hspace{-2.05ex}\raisebox{0.25ex}{$-$}%
  X\hspace{-2.56ex}\raisebox{0.25ex}{$-\!\!\!-$}%
  X\hspace{-2.56ex}\raisebox{0.25ex}{$-\!\!\!-$}%
  X\hspace{-2.56ex}\raisebox{0.25ex}{$-\!\!\!-$}%
  X\hspace{-2.56ex}\raisebox{0.25ex}{$-\!\!\!-$}%
  X\hspace{-2.46ex}\raisebox{0.25ex}{$-\!\!\!-$}  \\
\tabb
\Xdeft{Acronym and abbreviations}
\verb+<acronym title="+%
  \redtt{complete words for the acronym}%
  \verb+">      +%
  \redtt{An acronym}%
  \verb+      </acronym>+                      \\
\verb+<abbr    title="+%
  \redtt{complete words for the abbreviation}%
  \verb+"> +%
  \redtt{An abbreviation}%
  \verb+ </abbr>+                              \\
\tabb
\Xdeft{Preformatted text}
\verb+<pre>+                                          \\
\verb+  +%
  \redtt{Some script ...}%
  {\tt ~~~~~~~~~~~~~~~~~~~~~~~~~~~}
  \hspace{0.5cm} \white{$\lTo$}\hspace{0.5cm}
  \verb+  Some script ...+                            \\
\verb+    +%
  \redtt{Some script with a larger indent...}%
  {\tt ~~~~~}
  \hspace{0.5cm} $\lTo$\hspace{0.5cm}
  \verb+    Some script with a larger indent...+      \\
\verb+      +%
  \redtt{Some script with the largest indent...}%
  \hspace{0.5cm} \white{$\lTo$}\hspace{0.5cm}
  \verb+      Some script with the largest indent...+ \\
\verb+</pre>+                                         \\
\tabb
\newpage
\subsection{Links and the anchor {\sf a}}
\label{sec:basic_html_a}
%
%
\deft
\verb+<a href="+%
  \redtt{http://www.xxxxxx.xxx/xxxxxx/xxx}%
  \verb+.html"+                             \\
\verb+   href="+%
  \redtt{../xxxxxx/xxx}%
  \verb+.html"+                             \\
\verb++                                     \\
\verb+   href="+%
  \redtt{http://www.xxxxxx.xxx/xxxxxx/xxx}%
  \verb+.html+%
  \redtt{\#xxxxxx}%
  \verb+"+                                  \\
\verb+   href="+%
  \redtt{../xxxxxx/xxx}%
  \verb+.html+%
  \redtt{\#xxxxxx}%
  \verb+"+                                  \\
\verb+   href="+%
  \redtt{\#xxxxxx}%
  \verb+"+                                  \\
\verb++                                     \\
\verb+   target="+%
  \redtt{\_top}%
  \verb+/+%
  \redtt{\_blank}%
  \verb+">+                                 \\
\verb+   +%
  \redtt{Description of this link ......}%
  \verb+</a>+                               \\
\tabb
\Xdeft{Link for sending e-mails}
\verb+<a href="mailto:+%
  \redtt{xxxxxx@xxxxxx.xxx.xx}%
  \verb+?+                            \\
\verb+         cc=+%
  \redtt{yyyyyy1@yyyyyy1.yyy1.yy1}%
  \verb+,+                            \\
\verb+            +%
  \redtt{yyyyyy2@yyyyyy2.yyy2.yy2}%
  \verb+&+                            \\
\verb+         bcc=+%
  \redtt{zzzzzz@zzzzzz.zzz.zz}%
  \verb+&+                            \\
\verb+         subject=+%
  \redtt{......\%20......\%20......}%
  \verb+&+                            \\
\verb+         body=+%
  \redtt{text of the e-mail}%
  \verb+">+                           \\
\verb+   +%
  \redtt{Send e-mail to some body}%
  \verb+</a>+                         \\
\tabb
 See Sec.~\ref{sec:basic_php_email_sending}
 for e-mail sending in PHP.
\newpage
\subsection{\sf img}
\label{sec:basic_html_img}
%
%
\deft
\verb+<img src="+%
  \redtt{xxxxxx.jpg}%
  \verb+"+                                  \\
\verb+     alt="+%
  \redtt{description of the image}%
  \verb+"+                                  \\
\verb++                                     \\
\verb+     height="+%
  \redtt{height of the image (in units of `px')}%
  \verb+"+ \mynote[-0.1cm]{$\lGetsto$ {\tt style="height: ...; width: ..."}} \\
\verb+     width="+%
  \redtt{width of the image (in units of `px')}%
  \verb+"+                                  \\
\verb++                                     \\
\verb+     border="+%
  \redtt{width of the border of the image}%
  \verb+"+                                  \\
\verb+     align="+%
  \redtt{top}%
  \verb+/+%
  \redtt{middle}%
  \verb+/+%
  \redtt{bottom}%
  \verb+"+                                  \\
\verb++                                     \\
\verb+     />+                              \\
\tabb
\deft
\verb+<img src="+%
  \redtt{xxxxxx.jpg}%
  \verb+"+                      \\
\verb+      .+                  \\
\verb+      .+                  \\
\verb+      .+                  \\
\verb+     usemap="#+%
  \redtt{name of the used map}%
  \verb+" />+                    \\
\verb++                         \\
\verb+<map name="+%
  \redtt{name of the used map}%
  \verb+"> +                    \\
\verb++                         \\
\verb+  <area shape="+%
  \redtt{rect}%
  \verb+/+%
  \redtt{circle}%
  \verb+"+                      \\
\verb+        coords="+%
  \redtt{x\_1,y\_1,x\_2,y\_2}%
  \verb+/+%
  \redtt{x\_0,y\_0,r}%
  \verb+"+                      \\
\verb+        href="+%
  \redtt{......}%
  \verb+"+                      \\
\verb+        />+               \\
\verb++                         \\
\verb+</map>+                   \\
\tabb
\Xdeft{Link with an image}
\verb+<a href="+%
  \redtt{../xxxxxx/xxxxxx.jpg}%
  \verb+"+                      \\
\verb+   target="+%
  \bluett{\_blank}%
  \verb+">+                     \\
\verb+   <img src="+%
  \redtt{../xxxxxx/xxxxxx.jpg}%
  \verb+" /><a>+                \\
\tabb
\Xdeft{Link for sending e-mails with an envelope icon}
\verb+<a href="mailto:+%
  \redtt{xxxxxx@xxxxxx.xxx.xx}%
  \verb+?+                            \\
\verb+         cc=+%
  \redtt{yyyyyy1@yyyyyy1.yyy1.yy1}%
  \verb+,+                            \\
\verb+            +%
  \redtt{yyyyyy2@yyyyyy2.yyy2.yy2}%
  \verb+&+                            \\
\verb+         bcc=+%
  \redtt{zzzzzz@zzzzzz.zzz.zz}%
  \verb+&+                            \\
\verb+         subject=+%
  \redtt{......\%20......\%20......}%
  \verb+&+                            \\
\verb+         body=+%
  \redtt{text of the e-mail}%
  \verb+">+                           \\
\verb+   <img src="+%
  \bluett{icons/envelope.gif}%
  \verb+"+                            \\
\verb+        alt="+%
  \bluett{Send E-Mail}%
  \verb+"+                            \\
\verb+        style="+%
  \bluett{border:~0; width:~32px; height:~32px}%
  \verb+" /></a>+                         \\
\tabb
\newpage
\subsection{Lists {\sf ol}, {\sf ul} and {\sf li}}
\label{sec:basic_html_lists}
%
%
\Xdeft{Ordered list}
\verb+<ol type="+%
  \redtt{1}%
  \verb+/+%
  \redtt{A}%
  \verb+/+%
  \redtt{a}%
  \verb+/+%
  \redtt{I}%
  \verb+/+%
  \redtt{i}%
  \verb+">+         \\
\verb++             \\
\verb+  <li> +%
  \redtt{......}%
  \verb++           \\
\verb+       </li>+ \\
\verb++             \\
\verb+  <li> +%
  \redtt{......}%
  \verb++           \\
\verb+       </li>+ \\
\verb++             \\
\verb+</ol>+        \\
\tabb
\Xdeft{Unordered list}
\verb+<ul type="+%
  \redtt{disc}%
  \verb+/+%
  \redtt{circle}%
  \verb+/+%
  \redtt{square}%
  \verb+">+         \\
\verb++             \\
\verb+  <li> +%
  \redtt{......}%
  \verb++           \\
\verb+       </li>+ \\
\verb++             \\
\verb+  <li> +%
  \redtt{......}%
  \verb++           \\
\verb+       </li>+ \\
\verb++             \\
\verb+</ul>+        \\
\tabb
\newpage
\subsection{{\sf form} and {\sf fieldset}}
\label{sec:basic_html_form}
%
%
\deft
\verb+<form>+         \\
\verb++               \\
\verb+  <fieldset>+   \\
\verb+    <legend> +%
  \redtt{Title of this input field}%
  \verb+ </legend>+   \\
\verb+       .+       \\
\verb+       .+       \\
\verb+       .+       \\
\verb+  </fieldset>+  \\
\verb++               \\
\verb++               \\
\verb+  <fieldset>+   \\
\verb+       .+       \\
\verb+       .+       \\
\verb+       .+       \\
\verb+  </fieldset>+  \\
\verb++               \\
\verb+</form>+        \\
\tabb
 See Sec.~\ref{sec:basic_php_form}
 for the complete script for \verb+<form>+ tag. 
\newpage
\subsection{{\sf input}, {\sf select}, and {\sf textarea}}
\label{sec:basic_html_input}
%
%
\deft
\verb+<input type="text"     +%
  \redtt{value="the (default) text/number"}%
  \verb++                              \\
\verb+       type="password" +%
  \redtt{value="the (default) password"}%
  \verb++                              \\
\verb++                                \\
\verb+       type="radio"    +%
  \redtt{(checked="checked") (disabled="disabled")}%
  \verb++                              \\
\verb+       type="chechbox" +%
  \redtt{(checked="checked") (disabled="disabled")}%
  \verb++                              \\
\verb++                                \\
\verb+       type="file"+
  \mynote{See Secs.~\ref{sec:basic_php_file_handling}
          and \ref{sec:basic_php_file_uploading}
          for file handling and uploading} \\
\verb++                                \\
\verb+       type="button"   value="+%
  \redtt{Yes!}%
  \verb+"+                             \\
\verb+       type="image"    src="+%
  \redtt{../xxxxxx/xxxxxx.jpg}%
  \verb+"+ \mynote{See Sec.~\ref{sec:basic_html_img}} \\
\verb++                                \\
\verb+       type="submit"   value="+%
  \bluett{Submit}%
  \verb+"+                             \\
\verb+       type="reset"    value="+%
  \bluett{Reset}%
  \verb+"+                             \\
\verb++                                \\
\verb+       type="hidden"   value="+%
  \redtt{the transported text/password/number}%
  \verb+"+                             \\
\verb++                                \\
\verb+       name="+%
  \redtt{the name of the input/transported variable}%
  \verb+"+                             \\
\verb+       />+                       \\
\tabb
\deft
\verb+<select name="+%
  \redtt{the name of this seletion}%
  \verb+">+                          \\
\verb++                              \\
\verb+  <optgroup label="+%
  \redtt{label for this category}%
  \verb+"+                           \\
\verb++                              \\
\verb+    <option value="+%
  \redtt{the value of this option}%
  \verb+" +%
  \redtt{(selected="selected")}%
  \verb+>+                           \\
\verb+          +%
  \redtt{the value of this option}%
  \verb++                            \\
\verb+    </option>+                 \\
\verb++                              \\
\verb+    <option value="+%
  \redtt{the value of this option}%
  \verb+">+                          \\
\verb+          +%
  \redtt{the value of this option}%
  \verb++                            \\
\verb+    </option>+                 \\
\verb++                              \\
\verb+  </optgroup>+                 \\
\verb++                              \\
\verb+  <optgroup label="+%
  \redtt{label for this category}%
  \verb+"+                           \\
\verb++                              \\
\verb+    <option value="+%
  \redtt{the value of this option}%
  \verb+" +%
  \redtt{(selected="selected")}%
  \verb+>+                           \\
\verb+          +%
  \redtt{the value of this option}%
  \verb++                            \\
\verb+    </option>+                 \\
\verb++                              \\
\verb+  </optgroup>+                 \\
\verb++                              \\
\verb+</select>+                     \\
\tabb
\deft
\verb+<textarea>+        \\
\verb+  +%
  \redtt{Some text ...}%
  \verb++                \\
\verb+      .+           \\
\verb+      .+           \\
\verb+      .+           \\
\verb+</textarea>+       \\
\tabb
\deft
\verb+<label for="+%
  \redtt{the id of the specified item}%
  \verb+">+                                            \\
\verb+  +%
  \redtt{Some text/Description of the specified item}%
  \verb+ +                                             \\
\verb+</label>+                                        \\
\verb++                                                \\
\verb+<input id="+%
  \redtt{an id for this item}%
  \verb+" /> +                                         \\
\tabb
\newpage
\subsection{\sf table}
\label{sec:basic_html_table}
%
%
\begin{templateenumerate}
\myitem
\label{item:basic_html_table.html}
 {\cmbfsfx basic\_html\_table.html}
 (cf.~Template \ref{item:basic_css_table.html})
\deft
\verb+<table border="+%
  \redtt{width of the border of the tabel}%
  \verb+"+                          \\
\verb+       cellspacing="+%
  \redtt{distance between two cells (in units of `px')}%
  \verb+"+                          \\
\verb+       cellpadding="+%
  \redtt{distance between the inner border of the cell and the text (in units of `px')}%
  \verb+"+                          \\
\verb++                             \\
\verb+       frame="+%
  \redtt{border}%
  \verb+/+%
  \redtt{box}%
  \verb+"+                          \\
\verb+             "+%
  \redtt{void}%
  \verb+"+                          \\
\verb+             "+%
  \redtt{above}%
  \verb+/+%
  \redtt{below}%
  \verb+"+                          \\
\verb+             "+%
  \redtt{hsides}%
  \verb+/+%
  \redtt{vsides}%
  \verb+"+                          \\
\verb+             "+%
  \redtt{lhs}%
  \verb+/+%
  \redtt{rhs}%
  \verb+"+                          \\
\verb+       >+                     \\
\verb++                             \\
\verb+  <caption> +%
  \redtt{Title of this table}%
  \verb+ </caption>+                \\
\verb++                             \\
\verb+  <tr>+                       \\
\verb++                             \\
\verb+    <th>+                     \\
\verb+        +%
  \redtt{Table header 1}%
  \verb++                           \\
\verb+    </th>+                    \\
\verb++                             \\
\verb+    <th colspan="+%
  \redtt{number of columns}%
  \verb+">+                         \\
\verb+      +%
  \redtt{Table header 2+3}%
  \verb++                           \\
\verb+    </th>+                    \\
\verb++                             \\
\verb+  </tr>+                      \\
\verb++                             \\
\verb+  <tr>+                       \\
\verb++                             \\
\verb+    <td rowspan="+%
  \redtt{number of rows}%
  \verb+">+                         \\
\verb+      +%
  \redtt{Table data (2+3)-1}%
  \verb++                           \\
\verb+    </td>+                    \\
\verb++                             \\
\verb+    <td>+                     \\
\verb+        +%
  \redtt{Table data 2-2}%
  \verb++                           \\
\verb+    </td>+                    \\
\verb++                             \\
\verb+    <td>+                     \\
\verb+        +%
  \redtt{Table data 2-3}%
  \verb++                           \\
\verb+    </td>+                    \\
\verb++                             \\
\verb+  </tr>+                      \\
\verb++                             \\
\verb+  <tr>+                       \\
\verb++                             \\
\verb+    <td>+                     \\
\verb+        +%
  \redtt{Table data 3-2}%
  \verb++                           \\
\verb+    </td>+                    \\
\verb++                             \\
\verb+    <td>+                     \\
\verb+        +%
  \redtt{Table data 3-3}%
  \verb++                           \\
\verb+    </td>+                    \\
\verb++                             \\
\verb+  </tr>+                      \\
\verb++                             \\
\verb+</table>+                     \\
\tabb
\newpage
\picin{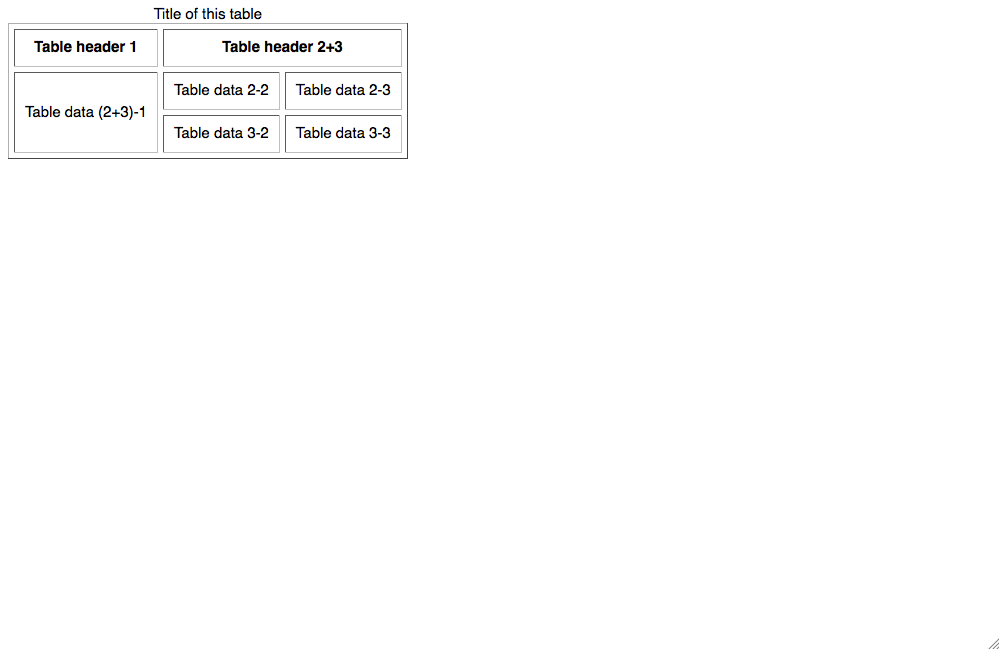}
{basic\_html\_table}
{A sample table with two combined cells.}
\newpage
\myitem
\label{item:basic_html_table2.html}
 {\bf Style declarations for grouped one or more rows}
 (see Template \ref{item:basic_css_table2.html})
%
\deft
\verb+<head>+                    \\
\verb++                          \\
\verb+  <style type="text/css">+ \\
\verb++                          \\
\verb+    thead      {+%
  \redtt{style declarations}%
  \verb+}+                       \\
\verb+    tbody      {+%
  \redtt{style declarations}%
  \verb+}+                       \\
\verb+    tfoot      {+%
  \redtt{style declarations}%
  \verb+}+                       \\
\verb++                          \\
\verb+  </style>+                \\
\verb++                          \\
\verb+</head>+                   \\
\verb++                          \\
\verb+<body>+                    \\
\verb++                          \\
\verb+  <table>+                 \\
\verb++                          \\
\verb+    <thead>+               \\
\verb++                          \\
\verb+      <tr>+                \\
\verb+        .+                 \\
\verb+        .+                 \\
\verb+        .+                 \\
\verb+      </tr>+               \\
\verb++                          \\
\verb+    </thead>+              \\
\verb++                          \\
\verb+    <tbody>+               \\
\verb++                          \\
\verb+      <tr>+                \\
\verb+        .+                 \\
\verb+        .+                 \\
\verb+        .+                 \\
\verb+      </tr>+               \\
\verb++                          \\
\verb+    </tbody>+              \\
\verb++                          \\
\verb+    <tfoot>+               \\
\verb++                          \\
\verb+      <tr>+                \\
\verb+        .+                 \\
\verb+        .+                 \\
\verb+        .+                 \\
\verb+      </tr>+               \\
\verb++                          \\
\verb+    </tfoot>+              \\
\verb++                          \\
\verb+  </table>+                \\
\verb++                          \\
\verb+</body>+                   \\
\tabb
\newpage
\Xdeft{Defining attribute values for one or more columns}
\verb+<col style="+%
  \redtt{style declarations}%
  \verb+" +%
  \redtt{/}%
  \verb+>+                   \\
\verb++                      \\
\verb+<colgroup span="+%
  \redtt{number of columns}%
  \verb+" style="+%
  \redtt{style declarations}%
  \verb+">+                  \\
\verb+</colgroup>+           \\
\tabb
\end{templateenumerate}
\newpage
\subsection{Frames and {\sf frameset}}
\label{sec:basic_html_frameset}
%
%
\deft
\verb+<!DOCTYPE html+                              \\
\verb+PUBLIC "-//W3C//DTD XHTML 1.0 +%
  \redtt{Frameset}%
  \verb+//EN"+                                     \\
\verb+"http://www.w3.org/TR/xhtml1/DTD/xhtml1-+%
  \redtt{frameset}%
  \verb+.dtd">+                                    \\
\tabb
\deft
\verb+<frameset cols="+%
  \redtt{...}%
  \verb+%, +%
  \redtt{...}%
  \verb+%, +%
  \redtt{...}%
  \verb+%"+              \\
\verb+          rows="+%
  \redtt{...}%
  \verb+%, +%
  \redtt{...}%
  \verb+%, +%
  \redtt{...}%
  \verb+%"+ \\
\verb+          >+       \\
\verb++                  \\
\verb+  <frame src="+%
  \redtt{xxxxxx}%
  \verb+.html" +%
  \redtt{/}%
  \verb+>+               \\
\verb+  <frame src="+%
  \redtt{xxxxxx}%
  \verb+.html" +%
  \redtt{/}%
  \verb+>+               \\
\verb+  <frame src="+%
  \redtt{../xxxxxx/xxxxxx}%
  \verb+.html"+          \\
\verb+         noresize +%
  \redtt{/}%
  \verb+>+               \\
\verb++                  \\
\verb+  <noframes>+      \\
\verb+    +%
  \redtt{Sorry, your browser does not handle frames!}%
  \verb++                \\
\verb+  </noframes>+     \\
\verb++                  \\
\verb+</frameset>+       \\
\tabb
\Xdeft{Inline frame}
\verb+<iframe src ="+%
  \redtt{xxxxxx.html}%
  \verb+"+                                             \\
\verb+        height="+%
  \redtt{the height of this inline frame}%
  \verb+"+                                             \\
\verb+        width="+%
  \redtt{the width of this inline frame}%
  \verb+"+                                             \\
\verb+        >+                                       \\
\verb++                                                \\
\verb+  <p> +%
  \redtt{Sorry, your browser does not handle iframe!}%
  \verb+ </p>+                                         \\
\verb++                                                \\
\verb+</iframe>+                                       \\
\tabb
\newpage
\subsection{Characters and symbols}
\label{sec:basic_html_characters_symbols}
%
%
\tabt{Reserved characters in HTML} {1.15} {|c p{2cm} p{11.16cm}|}
\makebox[1cm][c]{\tt \symbol{38}} & \verb+&amp;+  & (ampersand)                       \\
\verb+<+                          & \verb+&lt;+   & (less than)                       \\
\verb+>+                          & \verb+&gt;+   & (greater than)                    \\
\verb+'+                          & \verb+&apos;+ & (apostrophe, does not work in IE) \\
\verb+"+                          & \verb+&quot;+ & (quotation mark)                  \\
\tabb
\tabt{ISO 8859-1 symbols} {1.15} {|c p{2cm} p{11.16cm}|}
\makebox[1cm][c]{~}  & \verb+&nbsp;+  & (non-breaking space)        \\
\tabb
 Detailed lists of the ISO 8859-1 symbols,
 the ISO 8859-1 characters
 as well as the math symbols, Greek letters
 and other entities supported by HTML
 can be found in Ref.~\cite{W3Schools}.

\newpage
\section{CSS (Cascading Style Sheets)}
\label{sec:basic_css}
%
%
%
\subsection{Comments}
\label{sec:basic_css_comments}
%
%
\deft
\verb+/*   +%
  \redtt{Comment line}%
  \verb+   */+               \\
\verb++                      \\
\verb+/*+                    \\
\verb+  +%
  \redtt{Some comments ...}%
  \verb++                    \\
\verb+      .+               \\
\verb+      .+               \\
\verb+      .+               \\
\verb+*/+                    \\
\tabb
 See Secs.~\ref{sec:basic_html_comments},
 \ref{sec:basic_js_comments},
 and \ref{sec:basic_php_comments}
 for comments in HTML, JavaScript, and PHP,
 respectively.
\newpage
\subsection{Three ways for inserting CSS}
\label{sec:basic_css_three_ways}
%
%
\Xdeft{External style sheet}
\verb+<head>+                                     \\
\verb++                                           \\
\verb+  <link rel="stylesheet" type="text/css"+   \\
\verb+        href="+%
  \redtt{folder of the css file}%
  \verb+/+%
  \redtt{name of the css file}%
  \verb+.css" />+
  \mynote[0.5cm]{See Secs.~\ref{sec:basic_css_four_forms}
                 and \ref{sec:basic_css_headings}} \\
\verb++                                           \\
\verb+</head>+                                    \\
\tabb
\Xdeft{Internal style sheet}
\verb+<head>+                    \\
\verb++                          \\
\verb+  <style type="text/css">+ \\
\verb+       .+                  \\
\verb+       .+ \mynote{See Secs.~\ref{sec:basic_css_four_forms}
                        and \ref{sec:basic_css_headings}} \\
\verb+       .+                  \\
\verb+  </style>+                \\
\verb++                          \\
\verb+</head>+                   \\
\tabb
\Xdeft{Inline style}
\verb+<body>+               \\
\verb++                     \\
\verb+  <+%
  \redtt{tag}%
  \redtt{ (class="the style class")}%
  \verb++                   \\
\verb+       style="+%
  \redtt{(extra) style declarations}%
  \verb+" +%
  \redtt{(/)}%
  \verb+>+                  \\
\verb+       +%
  \redtt{Some text ...}%
  \verb++                   \\
\verb+  +%
  \redtt{(</tag>)}%
  \verb++                   \\
\verb++ \mynote[6.5cm]{See Secs.~\ref{sec:basic_css_four_forms}
                             and \ref{sec:basic_css_headings}} \\ 
\verb+  <span+%
  \redtt{ (class="the style class")}%
  \verb++                   \\
\verb+        style="+%
  \redtt{style declarations}%
  \verb+">+                 \\
\verb+        +%
  \redtt{Some text ...}%
  \verb++                   \\
\verb+  </span>+            \\
\verb++                     \\
\verb+</body>+              \\
\tabb
\newpage
\subsection{Four forms for defining styles}
\label{sec:basic_css_four_forms}
%
%
\Xdeft{Forms for style declaration (class/id/attribute selectors)}
\verb+<style type="text/css">+ \\
\verb++                        \\
\verb+     .+%
  \redtt{general class}%
  \verb+                    {+%
  \redtt{style declaration}%
  \verb+}+                     \\
\verb++                        \\
\verb+  +%
  \redtt{tag}%
  \verb+(, +%
  \redtt{tag2}%
  \verb+, +%
  \redtt{tag3}%
  \verb+, ...)               {+%
  \redtt{style declaration}%
  \verb+}+                     \\
\verb+  +%
  \redtt{tag}%
  \verb+.+%
  \redtt{specified class}%
  \verb+                  {+%
  \redtt{style declaration}%
  \verb+}+                     \\
\verb++ \mynote[10cm]{See Sec.~\ref{sec:basic_css_headings}} \\ 
\verb+  #+%
  \redtt{specified id}%
  \verb+                        {+%
  \redtt{style declaration}%
  \verb+}+                     \\
\verb+  #+%
  \redtt{specified id}%
  \verb+   +%
  \redtt{tag}%
  \verb+                  {+%
  \redtt{style declaration}%
  \verb+}+                     \\
\verb+  #+%
  \redtt{specified id}%
  \verb+  .+%
  \redtt{specified class}%
  \verb+      {+%
  \redtt{style declaration}%
  \verb+}+                     \\
\verb++                        \\
\verb+  [+%
  \redtt{(user-defined) attribute}%
  \verb+]           {+%
  \redtt{style declaration}%
  \verb+}+                     \\
\verb+  +%
  \redtt{tag}%
  \verb+ [+%
  \redtt{attribute}%
  \verb+]                      {+%
  \redtt{style declaration}%
  \verb+}+                     \\
\verb++                        \\
\verb+</style>+                \\
\tabb
\Xdeft{Examples}
\verb+<style type="text/css">+ \\
\verb++                        \\
\verb+  +%
  \redtt{h1}%
  \verb+                      {+%
  \redtt{style declaration for the "h1" tag}%
  \verb+}+                     \\
\verb+  +%
  \redtt{h2}%
  \verb+, +%
  \redtt{h3}%
  \verb+                  {+%
  \redtt{style declaration for both the "h2" and "h3" tags}%
  \verb+}+                     \\
\verb++                        \\
\verb+       .+%
  \redtt{en}%
  \verb+                {+%
  \redtt{style declaration for the "en" class}%
  \verb+}+                     \\
\verb+  +%
  \redtt{input}%
  \verb+.+%
  \redtt{data}%
  \verb+              {+%
  \redtt{style declaration for the "data" class of the "input" tag}%
  \verb+}+                     \\
\verb++                        \\
\verb+  #+%
  \redtt{main\_title}%
  \verb+             {+%
  \redtt{style declaration for the tag with the id of "main\_title"}%
  \verb+}+                     \\
\verb++                        \\
\verb+  [+%
  \redtt{id}%
  \verb+]                    {+%
  \redtt{style declaration for all tags with the "id" attribute}%
  \verb+}+                     \\
\verb+  +%
  \redtt{input}%
  \verb+[+%
  \redtt{type}%
  \verb+="+%
  \redtt{text}%
  \verb+"]      {+%
  \redtt{style declaration for the "input" tag with the type of "text"}%
  \verb+}+                     \\
\verb+  +%
  \redtt{img}%
  \verb+[+%
  \redtt{class}%
  \verb+|=+%
  \redtt{result}%
  \verb+]      {+%
  \redtt{style declaration for the "img" tag with the class of "result"}%
  \verb+}+                     \\
\verb++                        \\
\verb+</style>+                \\
\tabb
\Xdeft{Using defined style classes}
\verb+<body>+            \\
\verb++                  \\
\verb+  <+%
  \redtt{tag}%
  \verb+ class="+%
  \redtt{the unique choice of the general/specified class}%
  \verb+" +%
  \redtt{(/)}%
  \verb+>+               \\
\verb+       +%
  \redtt{Some text ...}%
  \verb++                \\
\verb+  +%
  \redtt{(</tag>)}%
  \verb++                \\
\verb++                  \\
\verb+  <+%
  \redtt{tag}%
  \verb+ id="+%
  \redtt{the specified id of this tag}%
  \verb+" +%
  \redtt{(/)}%
  \verb+>+               \\
\verb+       +%
  \redtt{Some text ...}%
  \verb++                \\
\verb+  +%
  \redtt{(</tag>)}%
  \verb++                \\
\verb++                  \\
\verb+</body>+           \\
\tabb
\newpage
\subsection{Syntax for style declaration}
\label{sec:basic_css_headings}
%
%
\begin{templateenumerate}
\myitem
\label{item:basic_css_headings.css}
 {\cmbfsfx basic\_css\_headings.css}
 (Template \ref{item:css_headings} included)
\Xdeft{Example:
       An external css file}
\verb+  +%
  \redtt{h1}%
  \verb+          {font-style:  +%
  \redtt{italic}%
  \verb+}+                                 \\
\verb+  +%
  \redtt{h2}%
  \verb+, +%
  \redtt{h4}%
  \verb+      {font-weight: +%
  \redtt{normal}%
  \verb+}+                         \\
\verb++                            \\
\verb+    .+%
  \redtt{en}%
  \verb+       {font-family: +%
  \redtt{"Times New Roman"}%
  \verb+}+                         \\
\verb+    .+%
  \redtt{en2}%
  \verb+      {font-family: +%
  \redtt{sans-serif}%
  \verb+}+                         \\
\verb++                            \\
\verb+  +%
  \redtt{h3.en}%
  \verb+       {font-family: +%
  \redtt{sans-serif}%
  \verb+;        color: +%
  \redtt{red}%
  \verb+}+                         \\
\verb+  +%
  \redtt{h3.en2}%
  \verb+      {font-family: +%
  \redtt{"Times New Roman"}%
  \verb+; color: +%
  \redtt{black}%
  \verb+}+                         \\
\tabb
\myitem
\label{item:basic_css_headings.html}
 {\cmbfsfx basic\_css\_headings.html}
\Xdeft{Example:
       Internal style declarations}
\verb+<style type="text/css">+     \\
\verb++                            \\
\verb+  +%
  \redtt{h2}%
  \verb+          {color: +%
  \redtt{green}%
  \verb+}+                         \\
\verb+  +%
  \redtt{h6}%
  \verb+          {text-align: +%
  \redtt{left}%
  \verb+; margin-left: +%
  \redtt{25pt}%
  \verb+}+                         \\
\verb++                            \\
\verb+</style>+                    \\
\tabb
\Xdeft{Example:
       Using the above style declarations
       {\rm (cf.~Template \ref{item:basic_html_headings.html})}}
\verb+<body +%
  \bluett{class="}%
  \verb++%
  \redtt{en}%
  \verb++%
  \bluett{"}%
  \verb+>+ \\
\verb++                                    \\
\verb+  <h1> +%
  \redtt{Main Title of the Website}%
  \verb+ </h1>+                            \\
\verb++                                    \\
\verb+  <h2> +%
  \redtt{Title of the web page}%
  \verb+ </h2>+                            \\
\verb++                                    \\
\verb+  <hr +%
  \redtt{/}%
  \verb+>+                                 \\
\verb++                                    \\
\verb+  <h3 +%
  \bluett{class="}%
  \verb++%
  \redtt{en}%
  \verb++%
  \bluett{"}%
  \verb+> +%
  \redtt{Main title of the category}%
  \verb+ </h3>+                            \\
\verb+  <h4> +%
  \redtt{Title of blocks in the category}%
  \verb+ </h4>+                            \\
\verb++                                    \\
\verb+  <hr +%
  \redtt{/}%
  \verb+>+                                 \\
\verb++                                    \\
\verb+  <h5> +%
  \redtt{Title of the section}%
  \verb+ </h5>+                            \\
\verb++                                    \\
\verb+  <div +%
  \bluett{style="margin-left:}%
  \verb+ +%
  \redtt{50pt}%
  \verb++%
  \bluett{"}%
  \verb+>+                                 \\
\verb++                                    \\
\verb+    <h6> +%
  \redtt{Title of the paragraph}%
  \verb+ </h6>+                            \\
\verb++                                    \\
\verb+    <p>+                             \\
\verb+      +%
  \redtt{Some sentences ...}%
  \verb+   <br +%
  \redtt{/}%
  \verb+>+                                 \\
\verb+        +%
  \redtt{Some sentences ...}%
  \verb+ <br +%
  \redtt{/}%
  \verb+>+                                 \\
\verb++                                    \\
\verb+      <span +%
  \bluett{class="}%
  \verb++%
  \redtt{en2}%
  \verb++%
  \bluett{"}%
  \verb+> +%
  \redtt{Some sentences ...}%
  \verb+ </span>+                          \\
\verb+    </p>+                            \\
\verb++                                    \\
\verb+    <hr +%
  \redtt{/}%
  \verb+>+                                 \\
\verb++                                    \\
\verb+  </div>+                            \\
\verb++                                    \\
\verb+</body>+ \\
\tabb
\newpage
\picin{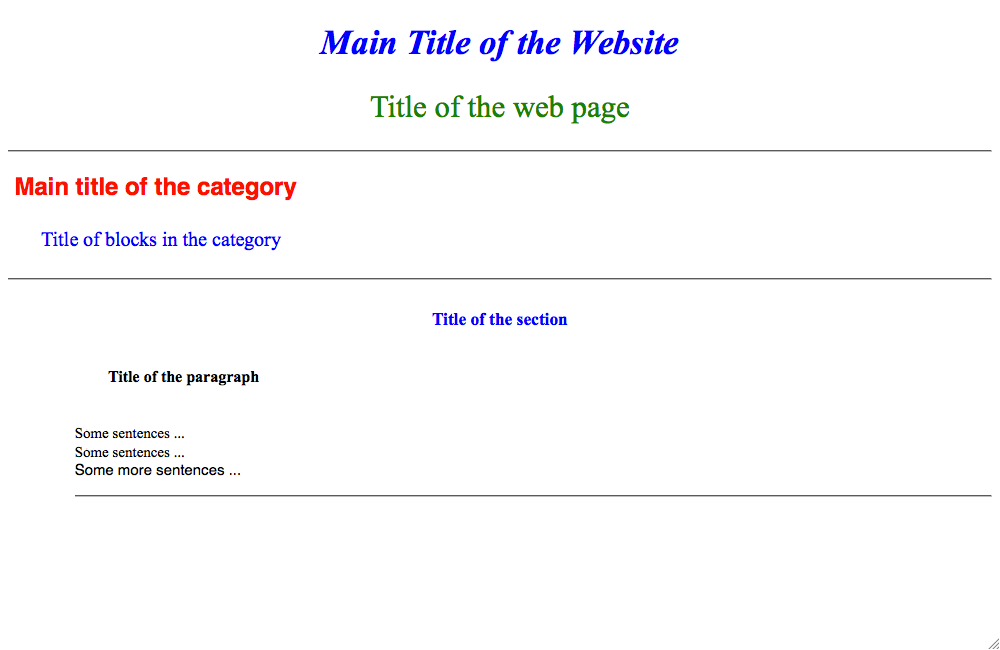}
{basic\_css\_headings}
{Headings given in Template \ref{item:basic_html_headings.html}
 with style declarations
 (cf.~Figure \ref{fig:basic_html_headings}).}
\newpage
\myitem
\label{item:basic_css_table.html}
 {\cmbfsfx basic\_css\_table.html}
 (cf.~Templates \ref{item:basic_html_table.html}
            and \ref{item:basic_js-htmldom_table.html})
%
\deft
\verb+<style type="text/css">+               \\
\verb+  table        {line-height:  +%
  \redtt{36pt}%
  \verb+;   margin-left:  +%
  \redtt{24pt}%
  \verb+;+                                   \\
\verb+                border-style: +%
  \redtt{groove}%
  \verb+; border-color:     ;+               \\
\verb+                border-top-width:   +%
  \redtt{5pt}%
  \verb+; border-bottom-width: +%
  \redtt{0pt}%
  \verb+;+                                   \\
\verb+                border-left-width: +%
  \redtt{10pt}%
  \verb+; border-right-width:  +%
  \redtt{5pt}%
  \verb+}+                                   \\
\verb++                                      \\
\verb+  caption      {font-family: +%
  \redtt{"Times New Roman"}%
  \verb+; font-size: +%
  \redtt{20pt}%
  \verb+;+                                   \\
\verb+                font-style:  +%
  \redtt{italic}%
  \verb+; font-weight: +%
  \redtt{bold}%
  \verb+;+                                   \\
\verb+                color: +%
  \redtt{\#00008B}%
  \verb+}+                                   \\
\verb++                                      \\
\verb+  tr           {font-size:  +%
  \redtt{16pt}%
  \verb+;   color: +%
  \redtt{\#006400}%
  \verb+;+                                   \\
\verb+                text-align: +%
  \redtt{center}%
  \verb+; width: +%
  \redtt{100pt}%
  \verb+}+                                   \\
\verb++                                      \\
\verb+  th           {font-family: +%
  \redtt{"Times New Roman"}%
  \verb+; color: +%
  \redtt{blue}%
  \verb+;+                                   \\
\verb+                text-align:  +%
  \redtt{center}%
  \verb+; width: +%
  \redtt{150pt}%
  \verb+;+                                   \\
\verb+                background-color: +%
  \redtt{\#FDE4D0}%
  \verb+}+                                   \\
\verb+</style>+                              \\
\defc
\verb++                                      \\
\verb+<table border="+%
  \redtt{  1}%
  \verb+"+                          \\
\verb+       cellspacing="+%
  \redtt{ 5px}%
  \verb+"+                          \\
\verb+       cellpadding="+%
  \redtt{10}%
  \verb+"+                          \\
\verb++                             \\
\verb+       frame="+%
  \redtt{border}%
  \verb+">+                         \\
\verb++                             \\
\verb+  <caption> +%
  \redtt{Title of this}%
  \verb+ +%
  \bluett{<span style="color:}%
  \verb++%
  \redtt{ red}%
  \verb++%
  \bluett{; background-color:}%
  \verb++%
  \redtt{ \#FFF8DC}%
  \verb++%
  \bluett{">}%
  \verb++%
  \redtt{table}%
  \verb++%
  \bluett{</span>}%
  \verb+ </caption>+                \\
\verb++                             \\
\verb+  <tr>+                       \\
\verb++                             \\
\verb+    <th>+                     \\
\verb+        +%
  \redtt{Table header 1}%
  \verb++                           \\
\verb+    </th>+                    \\
\verb++                             \\
\verb+    <th colspan="+%
  \redtt{2}%
  \verb+">+                         \\
\verb+      +%
  \redtt{Table header 2+3}%
  \verb++                           \\
\verb+    </th>+                    \\
\verb++                             \\
\verb+  </tr>+                      \\
\verb++                             \\
\verb+  <tr +%
  \bluett{style="background-color:}%
  \verb++%
  \redtt{ yellow}%
  \verb+"+%
  \bluett{>}%
  \verb++                           \\
\verb++                             \\
\verb+    <td rowspan="+%
  \redtt{2}%
  \verb+">+                         \\
\verb+      +%
  \redtt{Table data (2+3)-1}%
  \verb++                           \\
\verb+    </td>+                    \\
\verb++                             \\
\verb+    <td>+                     \\
\verb+        +%
  \redtt{Table data 2-2}%
  \verb++                           \\
\verb+    </td>+                    \\
\verb++                             \\
\verb+    <td>+                     \\
\verb+        +%
  \redtt{Table data 2-3}%
  \verb++                           \\
\verb+    </td>+                    \\
\verb++                             \\
\verb+  </tr>+                      \\
\verb++                             \\
\verb+  <tr>+                       \\
\verb++                             \\
\verb+    <td>+                     \\
\verb+        +%
  \redtt{Table data 3-2}%
  \verb++                           \\
\verb+    </td>+                    \\
\verb++                             \\
\verb+    <td>+                     \\
\verb+        +%
  \redtt{Table }%
  \verb++%
  \bluett{<span style="font-family:}%
  \verb++%
  \redtt{ 'Times New Roman'}%
  \verb++%
  \bluett{; color:}%
  \verb++%
  \redtt{ red}%
  \verb++%
  \bluett{">}%
  \verb++%
  \redtt{data 3-3}%
  \verb++%
  \bluett{</span>}%
  \verb++                           \\
\verb+    </td>+                    \\
\verb++                             \\
\verb+  </tr>+                      \\
\verb++                             \\
\verb+</table>+                     \\
\tabb
\newpage
\picin{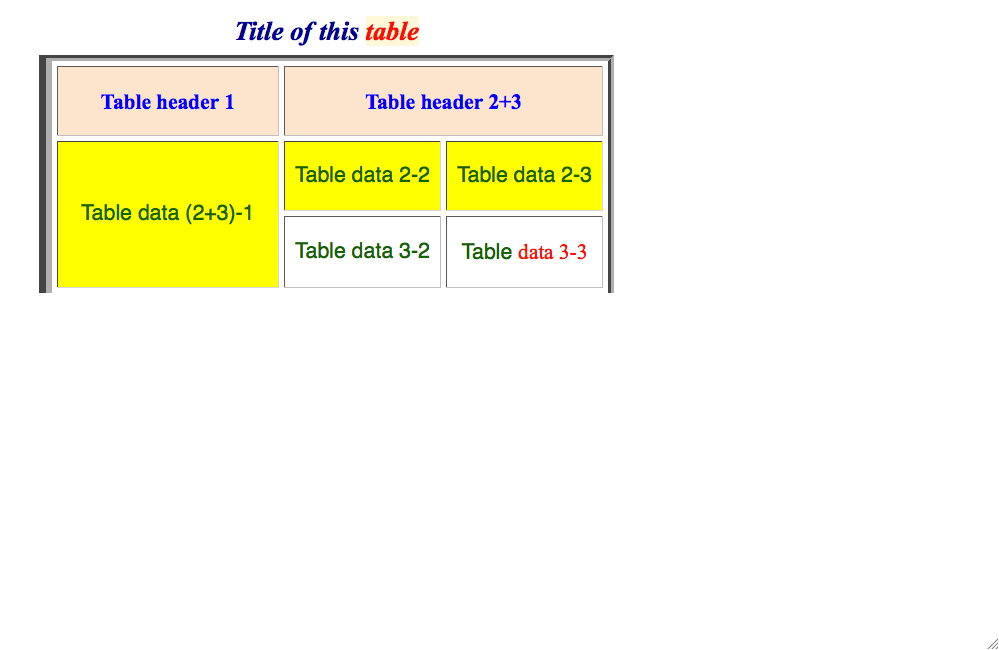}
{basic\_css\_table}
{The sample table given in Template \ref{item:basic_html_table.html}
 with style declarations
 (cf.~Figure \ref{fig:basic_html_table}).}
\newpage
\myitem
\label{item:basic_css_table2.html}
 {\cmbfsfx basic\_css\_table2.html}
 (cf.~Templates \ref{item:basic_html_table2.html}
            and \ref{item:basic_js-htmldom_table2.html})
\Xdeft{Example:
       Style declarations for grouped one or more rows}
\verb+<style type="text/css">+               \\
\verb+  table      {line-height:  +%
  \redtt{24pt}%
  \verb+;  margin-left:  +%
  \redtt{36pt}%
  \verb+;+                                   \\
\verb+              border-style: +%
  \redtt{solid}%
  \verb+; border-color: +%
  \redtt{blue}%
  \verb+;+                                   \\
\verb+              border-top-width:  +%
  \redtt{5pt}%
  \verb+; border-bottom-width: +%
  \redtt{4pt}%
  \verb+;+                                   \\
\verb+              border-left-width: +%
  \redtt{0pt}%
  \verb+; border-right-width:  +%
  \redtt{0pt}%
  \verb+}+                                   \\
\verb++                                      \\
\verb+  tr         {color: +%
  \redtt{red}%
  \verb+;   text-align: +%
  \redtt{center}%
  \verb+; width: +%
  \redtt{100pt}%
  \verb+}+                                   \\
\verb+  td         {color: +%
  \redtt{green}%
  \verb+; text-align: +%
  \redtt{center}%
  \verb+; width: +%
  \redtt{100pt}%
  \verb+}+                                   \\
\verb++                                      \\
\verb+  thead      {font-size: +%
  \redtt{13pt}%
  \verb+;   font-weight: +%
  \redtt{normal}%
  \verb+;+                                   \\
\verb+              color: +%
  \redtt{blue}%
  \verb+;+                                   \\
\verb+              line-height: +%
  \redtt{18pt}%
  \verb+; background-color: +%
  \redtt{\#AFEEEE}%
  \verb+}+                                   \\
\verb++                                      \\
\verb+  tbody      {font-family: +%
  \redtt{"Times New Roman"}%
  \verb+; font-style: +%
  \redtt{italic}%
  \verb+;+                                   \\
\verb+              background-color: +%
  \redtt{\#ADD8E6}%
  \verb+}+                                   \\
\verb++                                      \\
\verb+  tfoot      {font-family: +%
  \redtt{sans-serif}%
  \verb+;+                                   \\
\verb+              text-align: +%
  \redtt{left}%
  \verb+}+                                   \\
\verb+</style>+                              \\
\defc
\verb++                                      \\
\verb+<table>+                               \\
\verb++                                      \\
\verb+  <thead>+                             \\
\verb++                                      \\
\verb+    <tr>+                              \\
\verb++                                      \\
\verb+      <th>+                            \\
\verb+        +%
  \redtt{Table head 1}%
  \verb++                                    \\
\verb+      </th>+                           \\
\verb++                                      \\
\verb+      <td>+                            \\
\verb+        +%
  \redtt{Table head 2}%
  \verb++                                    \\
\verb+      </td>+                           \\
\verb++                                      \\
\verb+    </tr>+                             \\
\verb++                                      \\
\verb+  </thead>+                            \\
\verb++                                      \\
\verb+  <tbody>+                             \\
\verb++                                      \\
\verb+    <tr +%
  \bluett{style="color:}%
  \verb++%
  \redtt{ yellow}%
  \verb++%
  \bluett{"}%
  \verb+>+                                   \\
\verb++                                      \\
\verb+      <th>+                            \\
\verb+        +%
  \redtt{Table body 1}%
  \verb++                                    \\
\verb+      </th>+                           \\
\verb++                                      \\
\verb+      <td>+                            \\
\verb+        +%
  \redtt{Table body 2}%
  \verb++                                    \\
\verb+      </td>+                           \\
\verb++                                      \\
\verb+    </tr>+                             \\
\verb++                                      \\
\verb+  </tbody>+                            \\
\verb++                                      \\
\verb+  <tfoot>+                             \\
\verb++                                      \\
\verb+    <tr>+                              \\
\verb++                                      \\
\verb+      <th>+                            \\
\verb+        +%
  \redtt{Table foot 1}%
  \verb++                                    \\
\verb+      </th>+                           \\
\verb++                                      \\
\verb+      <td +%
  \bluett{style="font-size:}%
  \verb++%
  \redtt{ 16}%
  \verb++%
  \bluett{pt; color:}%
  \verb++%
  \redtt{ black}%
  \verb++%
  \bluett{"}%
  \verb+>+                                   \\
\verb+        +%
  \redtt{Table foot 2}%
  \verb++                                    \\
\verb+      </td>+                           \\
\verb++                                      \\
\verb+    </tr>+                             \\
\verb++                                      \\
\verb+  </tfoot>+                            \\
\verb++                                      \\
\verb+</table>+                              \\
\tabb
 More examples for style declarations
 can be found in Sec.~\ref{sec:main.css}.
\newpage
\picin{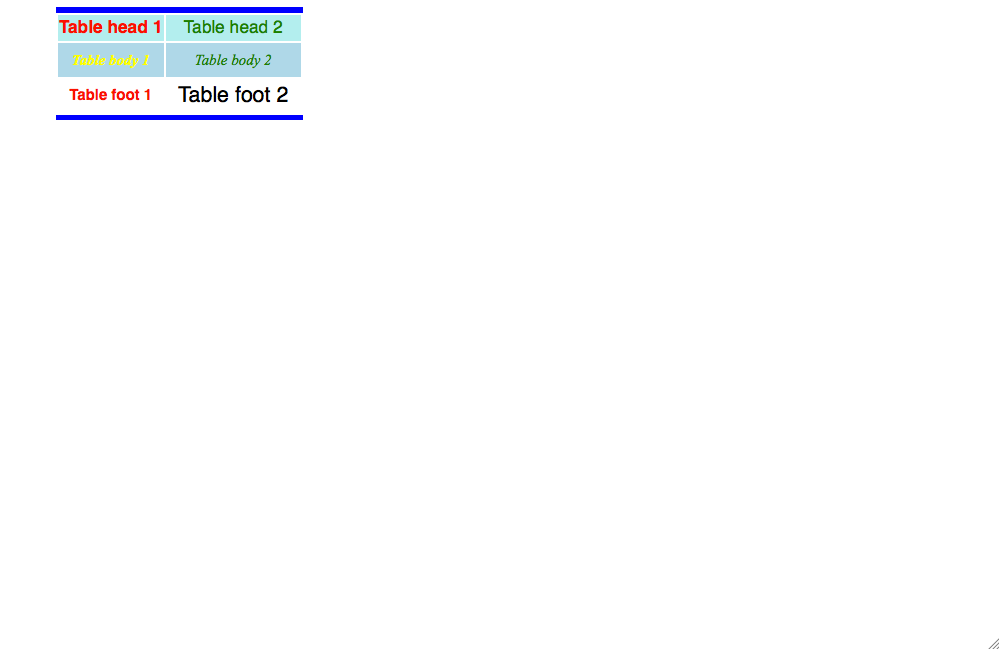}
{basic\_css\_table2}
{Another sample table
 with style declarations.}
\end{templateenumerate}
\newpage
\subsection{Image handling}
\label{sec:basic_css_image_handling}
%
%
\deft
\verb+<style type="text/css">+               \\
\verb++                                      \\
\verb+  +%
  \redtt{img}%
  \verb+      {position: +%
  \redtt{left}%
  \verb+/+%
  \redtt{right}%
  \verb+/+%
  \redtt{top}%
  \verb+/+%
  \redtt{bottom}%
  \verb+;+                                   \\
\verb+            z-index: +%
  \redtt{1}%
  \verb+/+%
  \redtt{-1}%
  \verb+;+ \mynote[3cm]{1: frontground/--1: background} \\
\verb++                                      \\
\verb+            visibility: +%
  \redtt{visible}%
  \verb+/+%
  \redtt{hidden}%
  \verb+;+                                   \\

\verb++                                      \\
\verb+            opacity: +%
  \redtt{0.5}%
  \verb+;+ \mynote[3.15cm]{for Firefox}      \\
\verb+            filter: alpha(opacity = +%
  \redtt{50}%
  \verb+)}+ \mynote[0.91cm]{for Internet Explorer} \\
\verb++                                      \\
\verb+</style>+                              \\
\tabb
\newpage
\subsection{Frequently used style properties and values}
\label{sec:basic_css_properties_values}
%
%
\Xdeft{Background}
\verb+  background-color+                    \\
\verb+  background-image: url('+%
  \redtt{../xxxxxx/xxxxxx.jpg}%
  \verb+')+                                  \\
\verb+  background-position: +%
\verb+left, right, top, bottom, center+      \\
\verb+  background-repeat: +%
\verb+repeat, repeat-x, repeat-y, no-repeat+ \\
\tabb
\Xdeft{Font}
\verb+  font-family: +%
\verb+"Times New Roman", Times, serif, sans-serif+ \\
\verb+  font-weight: bold+                         \\
\verb+  font-style: italic+                        \\
\verb+  font-size+                                 \\
\tabb
\Xdeft{Text}
\verb+  color+                               \\
\verb+  height+                              \\
\verb+  width+                               \\
\verb+  line-height+                         \\
\verb+  text-align: left, right, center+     \\
\verb+  vertical-align: top, bottom, middle+ \\
\verb+  text-indent+                         \\
\verb+  text-decoration: +%
\verb+underline, overline, line-through+     \\
\verb++                                      \\
\verb+  p:first-letter+                      \\
\tabb
\Xdeft{Margin}
\verb+  margin+        \\
\verb+  margin-left+   \\
\verb+  margin-right+  \\
\verb+  margin-top+    \\
\verb+  margin-bottom+ \\
\tabb
\Xdeft{Links}
\verb+  a:link+    \\
\verb+  a:visited+ \\
\verb+  a:hover+   \\
\verb+  a:active+  \\
\tabb
\Xdeft{Lists}
\verb+  list-style-type: decimal, disc+ \\
\tabb
\Xdeft{Border}
\verb+  border+                         \\
\verb+  border-left+                    \\
\verb+  border-right+                   \\
\verb+  border-top+                     \\
\verb+  border-bottom+                  \\
\verb++                                 \\
\verb+  border-style: +%
\verb+solid, dashed, dotted, double, +%
\verb+groove, ridge, inset, outset, +%
\verb+hidden, none+                     \\
\verb+  border-color+                   \\
\verb+  border-width+                   \\
\tabb
\Xdeft{Padding}
\verb+  padding+        \\
\verb+  padding-left+   \\
\verb+  padding-right+  \\
\verb+  padding-top+    \\
\verb+  padding-bottom+ \\
\tabb
\Xdeft{Cursor}
\verb+  cursor: +%
\verb+default, pointer, progress, wait, +%
\verb+text, help, move, crosshair,+                        \\
\verb+          e-resize, w-resize, n-resize, s-resize, +%
\verb+ne-resize, nw-resize, se-resize, sw-resize+          \\
\tabb
 See Sec.~\ref{sec:basic_htmldom_style_properties}
 for frequently used HTML DOM properties for style declaration.

\newpage
\section{JavaScript}
\label{sec:basic_js}
%
%
\subsection{Start with JavaScript}
\label{sec:basic_js_start}
%
%
\deft
\verb+<script type="text/javascript">+ \\
\verb++                                \\
\verb+  document.write(+%
  \bluett{"}%
  \verb++%
  \redtt{some text}%
  \verb++%
  \bluett{"}%
  \verb+);+                            \\
\verb++                                \\
\verb+  document.write(+%
  {\color{blue}%
  \verb-"-%
  \redtt{some text }%
  \verb-" + "-%
  {\color{red}\verb+\'+}%
  \redtt{single quote}%
  {\color{red}\verb+\'+}%
  \verb-" + ' -%
  {\color{red}\verb+\&+}%
  \verb- ' + "-%
  {\color{red}\verb+<br />+}%
  \verb-" + '-%
  {\color{red}\verb+\"+}%
  \redtt{double quote}%
  {\color{red}\verb+\"+}%
  \verb-'-}%
  \verb+);+                            \\
\verb++                                \\
\verb+</script>+                       \\
\tabb
 See Sec.~\ref{sec:basic_php_start}
 for basic PHP syntax.
\newpage
\subsection{Comments}
\label{sec:basic_js_comments}
%
%
\deft
\verb+<script type="text/javascript">+ \\
\verb+//   +%
  \redtt{Comment line}%
  \verb++                              \\
\verb++                                \\
\verb+/*   +%
  \redtt{Comment line}%
  \verb+   */+                         \\
\verb++                                \\
\verb+/*+                              \\
\verb+  +%
  \redtt{Some comments ...}%
  \verb++                              \\
\verb+      .+                         \\
\verb+      .+                         \\
\verb+      .+                         \\
\verb+      .+                         \\
\verb+*/+                              \\
\verb+</script>+                       \\
\tabb
 See Secs.~\ref{sec:basic_html_comments},
 \ref{sec:basic_css_comments},
 and \ref{sec:basic_php_comments}
 for comments in HTML, CSS, and PHP,
 respectively.
\newpage
\subsection{Variables}
\label{sec:basic_js_variables}
%
%
\deft
\verb+<script type="text/javascript">+ \\
\verb+  var +%
  \redtt{xxx}%
  \verb+;+                             \\
\verb+  var +%
  \redtt{xxx\_xxx}%
  \verb+;+                             \\
\verb++                                \\
\verb+  var +%
  \redtt{xxx}%
  \verb+     = +%
  \redtt{number}%
  \verb+;+                             \\
\verb+  var +%
  \redtt{xxx\_xxx}%
  \verb+ = +%
  \redtt{"string"}%
  \verb+;+                             \\
\verb++                                \\
\verb+  var +%
  \redtt{yyy}%
  \verb+     = +%
  \redtt{number}%
  \verb-   + -%
  \redtt{xxx}%
  \verb+;+                             \\
\verb+  var +%
  \redtt{yyy\_yyy}%
  \verb+ = +%
  \redtt{"string"}%
  \verb- + -%
  \redtt{xxx\_xxx}%
  \verb+;+                             \\
\verb++                                \\
\verb+  var +%
  \redtt{yyy\_yyy}%
  \verb+ = +%
  \redtt{"string"}%
  \verb- + -%
  \redtt{xxx}%
  \verb+;+                             \\
\verb+</script>+                       \\
\tabb
 See Sec.~\ref{sec:basic_php_variables}
 for variables in PHP.
\newpage
\subsection{Arrays}
\label{sec:basic_js_arrays}
%
%
\deft
\verb+<script type="text/javascript">+ \\
\verb+  var +%
  \redtt{xxx}%
  \verb+     = Array();+                   \\
\verb+  var +%
  \redtt{xxx\_xxx}%
  \verb+ = Array();+                   \\
\verb++                                \\
\verb+  for (+%
  \redtt{n\_xxx}%
  \verb+ = +%
  \bluett{1}%
  \verb+; +%
  \redtt{n\_xxx}%
  \verb+ <= +%
  \redtt{n\_xxx\_max}%
  \verb+; +%
  \redtt{n\_xxx}%
  \verb+ +%
  \bluett{++}%
  \verb+)+                             \\
\verb+  {+                             \\
\verb+    +%
  \redtt{xxx}%
  \verb+[+%
  \redtt{n\_xxx}%
  \verb+] = Array();+                  \\
 \verb+  }+                            \\
\verb++                                \\
\verb+  +%
  \redtt{xxx}%
  \verb+[+%
  \redtt{n\_xxx}%
  \verb+][+%
  \redtt{m\_xxx}%
  \verb+]  = +%
  \redtt{number}%
  \verb+;+                             \\
\verb+  +%
  \redtt{xxx\_xxx}%
  \verb+[+%
  \redtt{n\_xxx\_xxx}%
  \verb+] = +%
  \redtt{"string"}%
  \verb+;+                             \\
\verb+</script>+                       \\
\tabb
 See Sec.~\ref{sec:basic_php_arrays}
 for arrays in PHP.
\newpage
\subsection{Function defining}
\label{sec:basic_js_function_defining}
%
%
\deft
\verb+<script type="text/javascript">+ \\
\verb+  function +%
  \redtt{function\_without\_parameter}%
  \verb+()+                            \\
\verb+  {+                             \\
\verb+      .+                         \\
\verb+      .+                         \\
\verb+      .+                         \\
\verb+  }+                             \\
\verb++                                \\
\verb+  function +%
  \redtt{function\_with\_parameters}%
  \verb+(+%
  \redtt{xxx}%
  \verb+, +%
  \redtt{yyy}%
  \verb+, +%
  \redtt{...}%
  \verb+)+                             \\
\verb+  {+                             \\
\verb+      .+                         \\
\verb+      .+                         \\
\verb+      .+                         \\
\verb+  }+                             \\
\verb+</script>+                       \\
\tabb
 Examples can be found in Sec.~\ref{sec:basic_htmldom_examples}.
\newpage
\subsection{{\sf if}}
\label{sec:basic_js_if}
%
%
\deft
\verb+<script type="text/javascript">+ \\
\verb+  if ((+%
  \redtt{xxx}%
  \verb+     +%
  \bluett{==}%
  \verb+ +%
  \redtt{yyy}%
  \verb+     +%
  {\color{blue}\verb+&&+}%
  \verb++                              \\
\verb+       +%
  \redtt{xxx\_xxx}%
  \verb+ +%
  \bluett{!=}%
  \verb+ +%
  \redtt{yyy\_yyy}%
  \verb+   ) +%
  \bluett{||}%
  \verb++                              \\
\verb+      (+%
  \redtt{condition}%
  \verb+            ) +%
  \bluett{||}%
  \verb++                              \\
\verb+      (+%
  \bluett{!}%
  \verb+(+%
  \redtt{condition}%
  \verb+)         )   )+               \\
\verb+  {+                             \\
\verb+      .+                         \\
\verb+      .+                         \\
\verb+      .+                         \\
\verb+  }+                             \\
\verb++                                \\
\verb+  +%
  \bluett{else}%
  \verb++ \mynote[2.5cm]{$\lGetsto$ \bluett{elseif} in PHP (Sec.~\ref{sec:basic_php_if})} \\
\verb+  +%
  \bluett{if}%
  \verb+ (+%
  \redtt{conditions}%
  \verb+)+                             \\
\verb+  {+                             \\
\verb+      .+                         \\
\verb+      .+                         \\
\verb+      .+                         \\
\verb+  }+                             \\
\verb++                                \\
\verb+  else+                          \\
\verb+  {+                             \\
\verb+      .+                         \\
\verb+      .+                         \\
\verb+      .+                         \\
\verb+  }+                             \\
\verb+</script>+                       \\
\tabb
 See Sec.~\ref{sec:basic_php_if}
 for \verb+if+, \verb+else if+ and \verb+else+ comments in PHP.

\deft
\verb+<script type="text/javascript">+ \\
\verb+ +%
  \redtt{xxx}%
  \verb+ = (+%
  \redtt{condition}%
  \verb+) +%
  \bluett{?}%
  \verb+ +%
  \redtt{yyy}%
  \verb+ +%
  \bluett{:}%
  \verb+ +%
  \redtt{zzz}%
  \verb+;+     \\
\verb+</script>+                       \\
\tabb
\newpage
\subsection{{\sf switch}}
\label{sec:basic_js_switch}
%
%
\deft
\verb+<script type="text/javascript">+ \\
\verb+  switch (+%
  \redtt{xxx}%
  \verb+)+                             \\
\verb+  {+                             \\
\verb+    case '+%
  \redtt{x}%
  \verb+':+                            \\
\verb+      +%
  \redtt{......}%
  \verb+;+                             \\
\verb+      break;+                    \\
\verb++                                \\
\verb+    case '+%
  \redtt{y}%
  \verb+':+                            \\
\verb+      +%
  \redtt{......}%
  \verb+;+                             \\
\verb+      break;+                    \\
\verb++                                \\
\verb+    case '+%
  \redtt{z}%
  \verb+':+                            \\
\verb+      +%
  \redtt{......}%
  \verb+;+                             \\
\verb+      break;+                    \\
\verb++                                \\
\verb+    default:+                    \\
\verb+      +%
  \redtt{......}%
  \verb+;+                             \\
\verb+  }+                             \\
\verb+</script>+                       \\
\tabb
 See Sec.~\ref{sec:basic_php_switch}
 for \verb+switch+ comment in PHP.
\newpage
\subsection{{\sf for}}
\label{sec:basic_js_for}
%
%
\deft
\verb+<script type="text/javascript">+ \\
\verb+  for (+%
  \redtt{n\_xxx}%
  \verb+ = +%
  \bluett{1}%
  \verb+; +%
  \redtt{n\_xxx}%
  \verb+ <= +%
  \redtt{n\_xxx\_max}%
  \verb+; +%
  \redtt{n\_xxx}%
  \verb+ +%
  \bluett{++}%
  \verb+)+                             \\
\verb+  {+                             \\
\verb+     .+                          \\
\verb+     .+                          \\
\verb+     .+                          \\
\verb+  }+                             \\
\verb+</script>+                       \\
\tabb
\deft
\verb+<script type="text/javascript">+ \\
\verb+  var +%
  \redtt{yyy}%
  \verb+ = Array();+                   \\
\verb++                                \\
\verb+  +%
  \redtt{yyy}%
  \verb+[0] = +%
  \redtt{"zzz\_0"}%
  \verb+;+                             \\
\verb+  +%
  \redtt{yyy}%
  \verb+[1] = +%
  \redtt{"zzz\_1"}%
  \verb+;+                             \\
\verb+     .+                          \\
\verb+     .+                          \\
\verb+     .+                          \\
\verb++                                \\
\verb+  for (+%
  \redtt{xxx}%
  \verb+ in +%
  \redtt{yyy}%
  \verb+)+                             \\
\verb+  {+                             \\
\verb+     .+                          \\
\verb+     .+                          \\
\verb+     .+                          \\
\verb+  }+                             \\
\verb+</script>+                       \\
\tabb
 See Sec.~\ref{sec:basic_php_for}
 for \verb+for+ and \verb+foreach as+ comments in PHP.
\newpage
\subsection{{\sf while}}
\label{sec:basic_js_while}
%
%
\deft
\verb+<script type="text/javascript">+ \\
\verb+  while (+%
  \redtt{conditions}%
  \verb+)+                             \\
\verb+  {+                             \\
\verb+     .+                          \\
\verb+     .+                          \\
\verb+     .+                          \\
\verb+  }+                             \\
\verb+</script>+                       \\
\tabb
\deft
\verb+<script type="text/javascript">+ \\
\verb+  do+                            \\
\verb+  {+                             \\
\verb+     .+                          \\
\verb+     .+                          \\
\verb+     .+                          \\
\verb+  }+                             \\
\verb+  while (+%
  \redtt{conditions}%
  \verb+)+%
  \bluett{;}%
  \verb++                              \\
\verb+</script>+                       \\
\tabb
 See Sec.~\ref{sec:basic_php_while}
 for \verb+while+ and \verb+do while+ comments in PHP.
\newpage
\subsection{Events}
\label{sec:basic_js_events}
%
%
\Xdeft{Frequently used events}
\verb+<+%
  \redtt{tag}%
  \verb+ onload =      "+%
  \redtt{......}%
  \verb+"+                 \\
\verb+     onunload =    "+%
  \redtt{......}%
  \verb+"+                 \\
\verb++                    \\
\verb+     onkeydown =   "+%
  \redtt{......}%
  \verb+"+                 \\
\verb+     onkeypress =  "+%
  \redtt{......}%
  \verb+"+                 \\
\verb+     onkeyup =     "+%
  \redtt{......}%
  \verb+"+                 \\
\verb++                    \\
\verb+     onclick =     "+%
  \redtt{......}%
  \verb+"+                 \\
\verb+     ondblclick =  "+%
  \redtt{......}%
  \verb+"+                 \\
\verb++                    \\
\verb+     onmousedown = "+%
  \redtt{......}%
  \verb+"+                 \\
\verb+     onmouseup =   "+%
  \redtt{......}%
  \verb+"+                 \\
\verb++                    \\
{\tt~~~$\;$}%
\(
       {\begin{array}{l}
          \mbox{\tt onmouseover~=~"\red{......}"} \\
          \mbox{\tt onmouseout~=~~"\red{......}"}
        \end{array}}
        \cBigr{~}
\) \mynote[0.25cm]{Together!} \\
%
\verb+     onmousemove = "+%
  \redtt{......}%
  \verb+"+                 \\
\verb++                    \\
\verb+     +%
  \redtt{(/)}%
  \verb+ >+                \\
\tabb
 Examples can be found in Sec.~\ref{sec:basic_htmldom_examples}.
\newpage
\subsection{Frequently used JavaScript methods}
\label{sec:basic_js_methods}
%
%
\Xdeft{{\tt new Date()} object}
\verb+<script type="text/javascript">+                        \\
\verb+  document.write(new Date().getFullYear());      +%
  A four digit representation of a year                       \\
\verb++                                                       \\
\verb+  document.write(new Date().getMonth());         +%
  A numeric representation of a month 
 (\red{0} to \red{11})                                        \\
\verb++                                                       \\
\verb+  document.write(new Date().getDate());          +%
  The day of the month 
 (1 to 31)                                                    \\
\verb+  document.write(new Date().getDay());           +%
  A numeric representation of a day of the week (\red{0} to \red{6}) \\
\verb++                                                       \\
\verb+  document.write(new Date().getHours());         +%
  24-hour format of an hour (0 to 23)                         \\
\verb+  document.write(new Date().getUTCHours());      +%
  24-hour format of an hour, according to universal time      \\
\verb++                                                       \\
\verb+  document.write(new Date().getMinutes());       +%
  Minutes without leading zeros (0 to 59)                     \\
\verb+  document.write(new Date().getSeconds());       +%
  Seconds without leading zeros (0 to 59)                     \\
\verb+</script>+                                              \\
\tabb
\Xdeft{{\tt Math} object}
\verb+Math.abs(), Math.ceil(), Math.floor(), Math.round(), Math.max(), Math.min(),+     \\
\verb+Math.sin(), Math.cos(), Math.tan(),+       \\
\verb+Math.asin(), Math.acos(), Math.atan(),+ \\
\verb+Math.sqrt(), Math.pow(), Math.exp(), Math.log(),+              \\
\verb+Math.random()+                                             \\
\tabb
 See Sec.~\ref{sec:basic_php_other_functions}
 for frequently used PHP functions.

\newpage
\section{HTML DOM (Document Object Model)}
\label{sec:basic_htmldom}
%
%
%
\subsection{Nodes and node tree}
\label{sec:basic_htmldom_nodes}
%
%
\Xdeft{Nodes}
\\
\(
%
{\rm Document~node -\!\!\!-\!\!\!- Root~node}
\cleft{\begin{array}{l}
         \mbox{\tt<html>}                                            \\
                                                                     \\ \\
%
{\rm Node}
\cleft{\begin{array}{l}
         \mbox{\tt<head>}                                            \\
                                                                     \\ \\
           {\tt~~~~~~}\stackrel{\underbrace{\mbox{\tt<title> ...... </title>}}}{\rm Node} \\
                                                                     \\ \\
         \mbox{\tt</head>}                                           \\
       \end{array}}                                                  \\
                                                                     \\ \\
%
{\rm Node}
\cleft{\begin{array}{l}
         \mbox{\tt<body>}                                            \\
                                                                     \\ \\
{\tt~~~~~~}
\cleft{\begin{array}{l}
         \mbox{\tt<div>}                                             \\
                                                                     \\ \\
{\tt~~~~~~}
\cleft{\begin{array}{l}
         \mbox{\tt<p>}                                               \\
                                                                     \\ \\
           {\tt~~~~~~}\stackrel{\underbrace{\mbox{\tt<!-- ...... -->}}}{\rm Comment~node} \\
                                                                     \\ \\ \\ \\
           {\tt~~~~~~}\underbrace{\mbox{\tt<a href="...">......</a>}} \\
                                                                     \\ \\ \\ \\
           {\tt~~~~~~}\underbrace{\mbox{\tt<span>......</span>}}     \\
                                                                     \\ \\
         \mbox{\tt</p>}                                              \\
       \end{array}}                                                  \\
                                                                     \\ \\
         \mbox{\tt</div>}                                            \\
       \end{array}}                                                  \\
                                                                     \\ \\
         \mbox{\tt</body>}                                           \\
       \end{array}}                                                  \\
                                                                     \\ \\
         \mbox{\tt</html>}                                           \\
       \end{array}}
\)                                                                   \\
\\
\tabb
 See Sec.~\ref{sec:basic_html_tags}
 for the basic structure of an HTML document.

\vspace{-18.4cm}
\hspace{5.93cm}
\begin{picture}(6  , 12  )
\thicklines
\put(3.5 ,  0.5 ){\line(1, 0){1.5}}
\put(3.5 ,  0   ){\line(0, 1){0.5}}
\put(5.1 ,  0.2 ){\makebox(1.6, 0.6){Text node}}
\put(1.65, 11.2 ){\line(1, 0){1.5}}
\put(1.65, 10.7 ){\line(0, 1){0.5}}
\put(3.25, 10.9 ){\makebox(1.6, 0.6){Text node}}
\put(3.35,  4.95){\line(1, 0){1.5}}
\put(3.35,  4.45){\line(0, 1){0.5}}
\put(4.95,  4.65){\makebox(1.6, 0.6){Text node}}
\put(3   ,  2.6 ){\line(1, 0){1.5}}
\put(3   ,  2.1 ){\line(0, 1){0.5}}
\put(4.6 ,  2.3 ){\makebox(2.4, 0.6){Attribute node}}
\thinlines
\end{picture}
\newpage
\subsection{Methods}
\label{sec:basic_htmldom_methods}
%
%
\deft
\verb+<script type="text/javascript">+ \\
\verb+  +%
  \redtt{node\_object}%
  \verb+.getElementById("+%
  \redtt{specified\_id}%
  \verb+").+%
  \redtt{property}%
  \verb++                              \\
\verb+  +%
  \redtt{node\_object}%
  \verb+.getElementsByName("+%
  \redtt{specified\_name}%
  \verb+")[+%
  \redtt{number}%
  \verb+].+%
  \redtt{property}%
  \verb++                              \\
\verb+  +%
  \redtt{node\_object}%
  \verb+.getElementsByTagName("+%
  \redtt{specified\_tag}%
  \verb+")[+%
  \redtt{number}%
  \verb+].+%
  \redtt{property}%
  \verb++                              \\
\verb+</script>+                       \\
\tabb
 See Sec.~\ref{sec:basic_htmldom_properties}
 for HTML DOM properties.
\newpage
\subsection{Properties}
\label{sec:basic_htmldom_properties}
%
%
\deft
\verb+<script type="text/javascript">+ \\
\verb+  +%
  \redtt{node\_object}%
  \verb+.parentNode.+%
  \redtt{property}%
  \verb++                              \\
\verb+  +%
  \redtt{node\_object}%
  \verb+.childNodes[+%
  \redtt{number}%
  \verb+].+%
  \redtt{property}%
  \verb++                              \\
\verb++                                \\
\verb+  +%
  \redtt{node\_object}%
  \verb+.firstChild.+%
  \redtt{property}%
  \verb++                              \\
\verb+  +%
  \redtt{node\_object}%
  \verb+.lastChild.+%
  \redtt{property}%
  \verb++                              \\
\verb++                                \\
\verb+  +%
  \redtt{node\_object}%
  \verb+.previousSibling.+%
  \redtt{property}%
  \verb++                              \\
\verb+  +%
  \redtt{node\_object}%
  \verb+.nextSibling.+%
  \redtt{property}%
  \verb++                              \\
\verb++                                \\
\verb+  +%
  \redtt{node\_object}%
  \verb+.attributes[+%
  \redtt{number}%
  \verb+].+%
  \redtt{property}%
  \verb++                              \\
\verb++                                \\
\verb+  +%
  \redtt{node\_object}%
  \verb+.nodeName  = "+%
  \redtt{......}%
  \verb+";+                            \\
\verb+  +%
  \redtt{node\_object}%
  \verb+.nodeType  = "+%
  \redtt{......}%
  \verb+";+                            \\
\verb+  +%
  \redtt{node\_object}%
  \verb+.nodeValue = "+%
  \redtt{......}%
  \verb+";+                            \\
\verb++                                \\
\verb+  +%
  \redtt{node\_object}%
  \verb+.name  = "+%
  \redtt{......}%
  \verb+";+                            \\
\verb+  +%
  \redtt{node\_object}%
  \verb+.type  = "+%
  \redtt{......}%
  \verb+";+                            \\
\verb+  +%
  \redtt{node\_object}%
  \verb+.value = "+%
  \redtt{......}%
  \verb+";+                            \\
\verb++                                \\
\verb+  +%
  \redtt{node\_object}%
  \verb+.innerHTML = "+%
  \redtt{......}%
  \verb+";+                            \\
\verb+</script>+                       \\
\tabb
\Xdeft{Links}
\verb+  +%
  \redtt{anchor\_node\_object}%
  \verb+.href+ \\
\verb+  +%
  \redtt{anchor\_node\_object}%
  \verb+.src+  \\
\tabb
\Xdeft{\tt input}
\verb+  +%
  \redtt{input\_node\_object}%
  \verb+.checked  = true, false+ \\
\verb+  +%
  \redtt{input\_node\_object}%
  \verb+.disabled = true, false+ \\
\tabb
\newpage
\subsection{Examples}
\label{sec:basic_htmldom_examples}
%
%
\begin{templateenumerate}
%
%
\myitem
\label{item:modification2_form-main_functions.php}
 {\bf Modification 2 of {\cmbfsfx form-main\_functions.php}}
 (cf.~Template \ref{item:modification_form-main_functions.php})
%
\deft
\verb+<input type="radio"+                                                    \\
\verb+       onmouseover = "this.nextSibling.nextSibling.style.color      = '+%
  \bluett{red}%
  \verb+';+                                                                   \\
\verb+                      this.nextSibling.nextSibling.style.fontWeight = '+%
  \bluett{bold}%
  \verb+'"+                                                                   \\
\verb+       onmouseout =  "this.nextSibling.nextSibling.style.color      = '+%
  \bluett{black}%
  \verb+';+                                                                   \\
\verb+                      this.nextSibling.nextSibling.style.fontWeight = '+%
  \bluett{normal}%
  \verb+'"+                                                                   \\
\verb+       onclick =     "this.checked = false"+                            \\
\verb+       />+                                                              \\
\verb+       <span> +%
  \redtt{Some text ...}%
  \verb+ </span>+                                                             \\
\tabb
\newpage
\picin{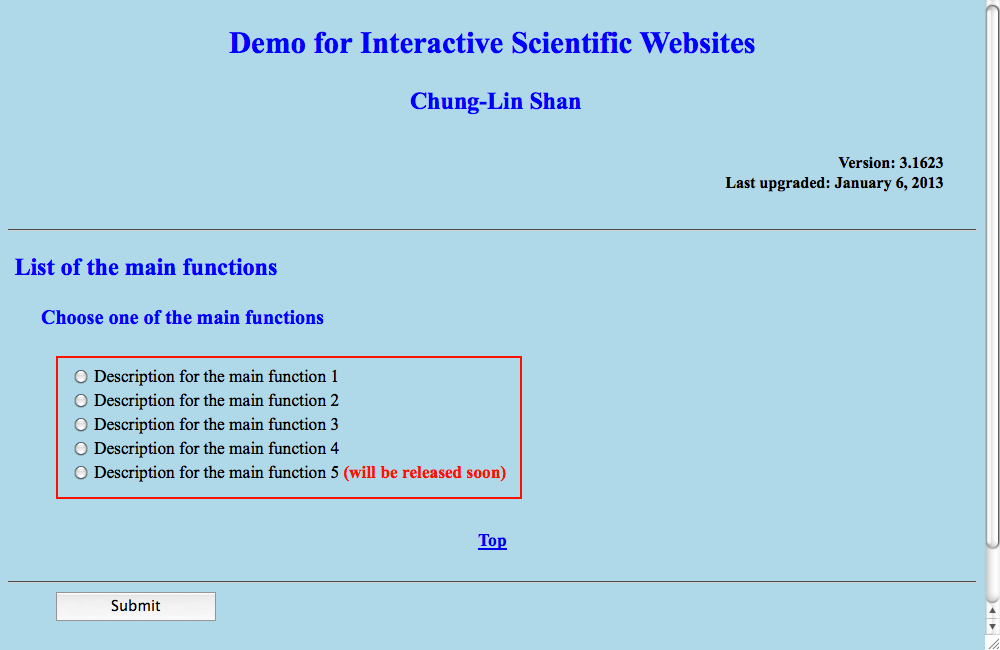}
{modification2\_main\_function\_choosing}
{Emphasizing the ``blocked'' option 5 for the main function
 (after clicking the option,
  cf.~Figure \ref{fig:modification_main_function_choosing-2}).}
\newpage
\myitem
\label{item:basic_js-htmldom_table3.html}
 {\cmbfsfx basic\_js-htmldom\_table3.html}
%
\deft
\verb+<script type="text/javascript">+                             \\
\verb+  function +%
  \bluett{change\_color}%
  \verb+(+%
  \bluett{row\_number}%
  \verb+, +%
  \bluett{item\_number}%
  \verb+, +%
  \bluett{color\_user}%
  \verb+)+                             \\
\verb+  {+                             \\
\verb+    if (navigator.appName == "Microsoft Internet Explorer")+ \\
\verb+    {+                                                       \\
\verb+      +%
  \bluett{item\_number}%
  \verb+ = +%
  \bluett{item\_number}%
  \verb+;+                                                         \\
\verb+    }+                                                       \\
\verb++                                                            \\
\verb+    else+                                                    \\
\verb+//  if (navigator.userAgent.search("Chrome")    != -1 ||+    \\
\verb+//      navigator.userAgent.search("Firefox")   != -1 ||+    \\
\verb+//      navigator.userAgent.search("Opera")     != -1 ||+    \\
\verb+//      navigator.userAgent.search("Safari")    != -1 ||+    \\
\verb+//      navigator.userAgent.search("Navigator") != -1   )+   \\
\verb+    {+                                                       \\
\verb+      +%
  \bluett{item\_number}%
  \verb+ = 2 * +%
  \bluett{item\_number}%
  \verb- + 1;-                                                     \\
\verb+    }+                                                       \\
\verb++                                                            \\
\verb+      document.getElementById(+%
  \bluett{row\_number}%
  \verb+)+                             \\
\verb+              .childNodes[+%
  \bluett{item\_number}%
  \verb+]+                             \\
\verb+              .style.color+      \\
\verb+    = +%
  \bluett{color\_user}%
  \verb+;+                             \\
\verb+  }+                             \\
\verb++                                \\
\verb+  function +%
  \bluett{recover\_color}%
  \verb+(+%
  \bluett{row\_number}%
  \verb+, +%
  \bluett{item\_number}%
  \verb+)+                             \\
\verb+  {+                             \\
\verb+    if (navigator.appName == "Microsoft Internet Explorer")+ \\
\verb+    {+                                                       \\
\verb+      +%
  \bluett{item\_number}%
  \verb+ = +%
  \bluett{item\_number}%
  \verb+;+                                                         \\
\verb+    }+                                                       \\
\verb++                                                            \\
\verb+    else+                                                    \\
\verb+//  if (navigator.userAgent.search("Chrome")    != -1 ||+    \\
\verb+//      navigator.userAgent.search("Firefox")   != -1 ||+    \\
\verb+//      navigator.userAgent.search("Opera")     != -1 ||+    \\
\verb+//      navigator.userAgent.search("Safari")    != -1 ||+    \\
\verb+//      navigator.userAgent.search("Navigator") != -1   )+   \\
\verb+    {+                                                       \\
\verb+      +%
  \bluett{item\_number}%
  \verb+ = 2 * +%
  \bluett{item\_number}%
  \verb- + 1;-                                                     \\
\verb+    }+                                                       \\
\verb++                                                            \\
\verb+      document.getElementById(+%
  \bluett{row\_number}%
  \verb+)+                             \\
\verb+              .childNodes[+%
  \bluett{item\_number}%
  \verb+]+                             \\
\verb+              .style.color+      \\
\verb+    = '+%
  \bluett{black}%
  \verb+';+                            \\
\verb+  }+                             \\
\verb+</script>+                       \\
\defc
\verb++                                \\
\verb+<table border="+%
  \bluett{1}%
  \verb+">+                            \\
\verb++                                \\
\verb+  <tr id="+%
  \redtt{row\_1}%
  \verb+">+                            \\
\verb++                                \\
\verb+    <th onmouseover="+%
  \bluett{change\_color}%
  \verb+('+%
  \redtt{row\_1}%
  \verb+', +%
  \bluett{0}%
  \verb+, '+%
  \redtt{red}%
  \verb+')"+                            \\
\verb+        onmouseout="+%
  \bluett{recover\_color}%
  \verb+('+%
  \redtt{row\_1}%
  \verb+', +%
  \bluett{0}%
  \verb+)">+                           \\
\verb+        +%
  \redtt{cell(1,1)}%
  \verb++                              \\
\verb+    </th>+                       \\
\verb++                                \\
\verb+    <th onmouseover="+%
  \bluett{change\_color}%
  \verb+('+%
  \redtt{row\_1}%
  \verb+', +%
  \bluett{1}%
  \verb+, '+%
  \redtt{blue}%
  \verb+')"+                            \\
\verb+        onmouseout="+%
  \bluett{recover\_color}%
  \verb+('+%
  \redtt{row\_1}%
  \verb+', +%
  \bluett{1}%
  \verb+)">+                           \\
\verb+        +%
  \redtt{cell(1,2)}%
  \verb++                              \\
\verb+    </th>+                       \\
\verb++                                \\
\verb+  </tr>+                         \\
\verb++                                \\
\verb+  <tr id="+%
  \redtt{row\_2}%
  \verb+">+                            \\
\verb++                                \\
\verb+    <th onmouseover="+%
  \bluett{change\_color}%
  \verb+('+%
  \redtt{row\_2}%
  \verb+', +%
  \bluett{0}%
  \verb+, '+%
  \redtt{lightgreen}%
  \verb+')"+                            \\
\verb+        onmouseout="+%
  \bluett{recover\_color}%
  \verb+('+%
  \redtt{row\_2}%
  \verb+', +%
  \bluett{0}%
  \verb+)">+                           \\
\verb+        +%
  \redtt{cell(2,1)}%
  \verb++                              \\
\verb+    </th>+                       \\
\verb++                                \\
\verb+    <th onmouseover="+%
  \bluett{change\_color}%
  \verb+('+%
  \redtt{row\_2}%
  \verb+', +%
  \bluett{1}%
  \verb+, '+%
  \redtt{gold}%
  \verb+')"+                            \\
\verb+        onmouseout="+%
  \bluett{recover\_color}%
  \verb+('+%
  \redtt{row\_2}%
  \verb+', +%
  \bluett{1}%
  \verb+)">+                           \\
\verb+        +%
  \redtt{cell(2,2)}%
  \verb++                              \\
\verb+    </th>+                       \\
\verb++                                \\
\verb+  </tr>+                         \\
\verb++                                \\
\verb+</table>+                        \\
\tabb
\newpage
\picqin{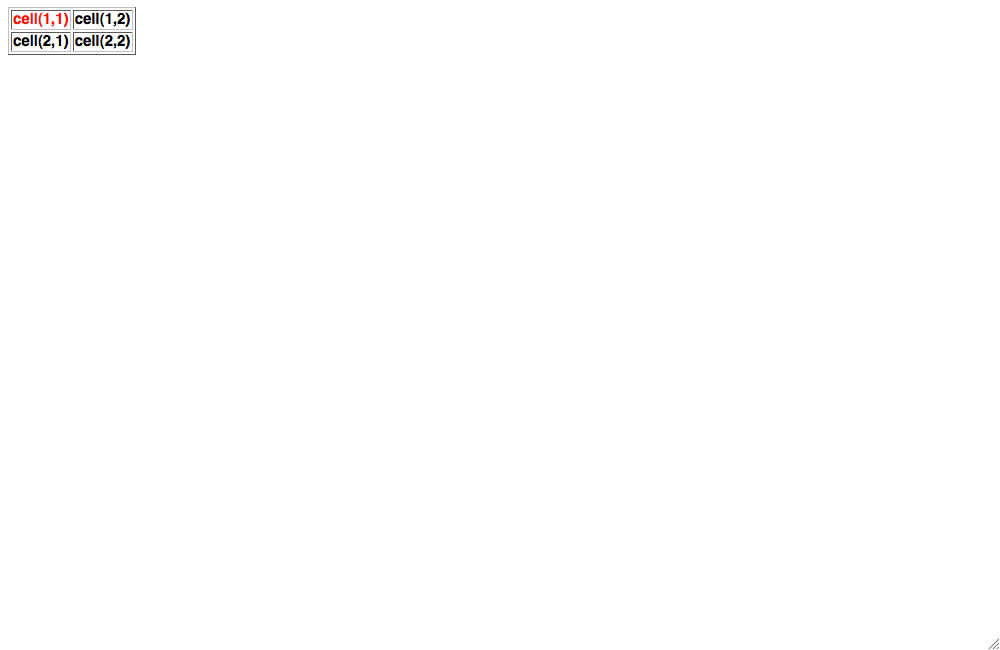}
{basic\_js-htmldom\_table3}
{Changing the text color of the cell element (1,1)
 by moving the cursor over the cell.}
{Changing the text color of the cell element (1,2)
 by moving the cursor over the cell.}
{Changing the text color of the cell element (2,1)
 by moving the cursor over the cell.}
{Changing the text color of the cell element (2,1)
 by moving the cursor over the cell.}
\newpage
\myitem
\label{item:basic_js-htmldom_table.html}
 {\bf Modification of {\cmbfsfx basic\_css\_table.html}}
 (cf.~Template \ref{item:basic_css_table.html})
%
\deft
\verb+<style type="text/css">+               \\
\verb+  table        {line-height:  +%
  \redtt{36pt}%
  \verb+;   margin-left:  +%
  \redtt{24pt}%
  \verb+;+                                   \\
\verb+                border-style: +%
  \redtt{groove}%
  \verb+; border-color:     ;+               \\
\verb+                border-top-width:   +%
  \redtt{5pt}%
  \verb+; border-bottom-width: +%
  \redtt{0pt}%
  \verb+;+                                   \\
\verb+                border-left-width: +%
  \redtt{10pt}%
  \verb+; border-right-width:  +%
  \redtt{5pt}%
  \verb+}+                                   \\
\verb++                                      \\
\verb+  caption      {font-family: +%
  \redtt{"Times New Roman"}%
  \verb+; font-size: +%
  \redtt{20pt}%
  \verb+;+                                   \\
\verb+                font-style:  +%
  \redtt{italic}%
  \verb+; font-weight: +%
  \redtt{bold}%
  \verb+;+                                   \\
\verb+                color: +%
  \redtt{\#00008B}%
  \verb+}+                                   \\
\verb++                                      \\
\verb+  tr           {font-size:  +%
  \redtt{16pt}%
  \verb+;   color: +%
  \redtt{\#006400}%
  \verb+;+                                   \\
\verb+                text-align: +%
  \redtt{center}%
  \verb+; width: +%
  \redtt{100pt}%
  \verb+}+                                   \\
\verb++                                      \\
\verb+  th           {font-family: +%
  \redtt{"Times New Roman"}%
  \verb+; color: +%
  \redtt{blue}%
  \verb+;+                                   \\
\verb+                text-align:  +%
  \redtt{center}%
  \verb+; width: +%
  \redtt{150pt}%
  \verb+;+                                   \\
\verb+                background-color: +%
  \redtt{\#FDE4D0}%
  \verb+}+                                   \\
\verb+</style>+                              \\
\defc
\verb++                                                          \\
\verb+<style type="text/javascript">+                            \\
\verb+  if (navigator.appName == "Microsoft Internet Explorer")+ \\
\verb+  {+                                                       \\
\verb+    +%
  \bluett{childNodes\_No}%
  \verb+ = 0;+                                                   \\
\verb+  }+                                                       \\
\verb++                                                          \\
\verb+  else+                                                    \\
\verb+//if (navigator.userAgent.search("Chrome")    != -1 ||+    \\
\verb+//    navigator.userAgent.search("Firefox")   != -1 ||+    \\
\verb+//    navigator.userAgent.search("Opera")     != -1 ||+    \\
\verb+//    navigator.userAgent.search("Safari")    != -1 ||+    \\
\verb+//    navigator.userAgent.search("Navigator") != -1   )+   \\
\verb+  {+                                                       \\
\verb+    +%
  \bluett{childNodes\_No}%
  \verb- = 2 * 0 + 1;-                                           \\
\verb+  }+                                                       \\
\verb++                                                          \\
\verb+  function +%
  \bluett{erase\_text}%
  \verb+()+                                                      \\
\verb+  {+                                                       \\
\verb+      document.getElementById("+%
  \redtt{ex\_1}%
  \verb+")+                                                      \\
\verb+              .childNodes[+%
  \bluett{childNodes\_No}%
  \verb+]+                                                       \\
\verb+              .innerHTML+                                  \\
\verb+    = "";+                                                 \\
\verb+  }+                                                       \\
\verb++                                                          \\
\verb+  function +%
  \bluett{change\_font}%
  \verb+()+                                                      \\
\verb+  {+                                                       \\
\verb+      document.getElementById("+%
  \redtt{ex\_2}%
  \verb+")+                                                      \\
\verb+              .nextSibling.nextSibling+                    \\
\verb+              .style.fontSize+                             \\
\verb+    = "+%
  \redtt{30pt}%
  \verb+";+                                                      \\
\verb++                                                          \\
\verb+      document.getElementById("+%
  \redtt{ex\_2}%
  \verb+")+                                                      \\
\verb+              .nextSibling.nextSibling+                    \\
\verb+              .style.color+                                \\
\verb+    = "+%
  \redtt{blue}%
  \verb+";+                                                      \\
\verb+  }+                                                       \\
\verb++                                                          \\
\verb+  function +%
  \bluett{recover\_font}%
  \verb+()+                                                      \\
\verb+  {+                                                       \\
\verb+      document.getElementById("+%
  \redtt{ex\_2}%
  \verb+")+                                                      \\
\verb+              .nextSibling.nextSibling+                    \\
\verb+              .style.color+                                \\
\verb+    = "+%
  \redtt{red}%
  \verb+";+                                                      \\
\verb+  }+                                                       \\
\verb++                                                          \\
\verb+  function +%
  \bluett{change\_bgcolor}%
  \verb+()+                                                      \\
\verb+  {+                                                       \\
\verb+      var +%
  \bluett{target\_cell}%
  \verb++                                                        \\
\verb+    = document.getElementById("+%
  \redtt{ex\_2}%
  \verb+")+                                                      \\
\verb+              .parentNode+                                 \\
\verb+              .nextSibling.nextSibling+                    \\
\verb+              .childNodes[+%
  \bluett{childNodes\_No}%
  \verb+];+                                                      \\
\verb++                                                          \\
\verb+      +%
  \bluett{target\_cell}%
  \verb+.style.backgroundColor+                                  \\
\verb+    = "+%
  \redtt{red}%
  \verb+";+                                                      \\
\verb++                                                          \\
\verb+      +%
  \bluett{target\_cell}%
  \verb+.nextSibling.nextSibling+                                \\
\verb+                 .style.backgroundColor+                   \\
\verb+    = "+%
  \redtt{blue}%
  \verb+";+                                                      \\
\verb+  }+                                                       \\
\verb+</style>+                                                  \\
\defc
\verb++                             \\
\verb+<table border="+%
  \bluett{1}%
  \verb+"+                          \\
\verb+       cellspacing="+%
  \redtt{ 5px}%
  \verb+"+                          \\
\verb+       cellpadding="+%
  \redtt{10}%
  \verb+"+                          \\
\verb++                             \\
\verb+       frame="+%
  \redtt{border}%
  \verb+"+                          \\
\verb++                             \\
\verb+       +%
  \bluett{onclick="erase\_text()"}%
  \verb+>+                          \\
\verb++                             \\
\verb+  <caption> +%
  \redtt{Title of this}%
  \verb+ +%
  \bluett{<span style="color:}%
  \verb++%
  \redtt{ red}%
  \verb++%
  \bluett{; background-color:}%
  \verb++%
  \redtt{ \#FFF8DC}%
  \verb++%
  \bluett{">}%
  \verb++%
  \redtt{table}%
  \verb++%
  \bluett{</span>}%
  \verb+ </caption>+                \\
\verb++                             \\
\verb+  <tr +%
  \bluett{id="}%
  \verb++%
  \redtt{ex\_1}%
  \verb++%
  \bluett{"}%
  \verb+>+                          \\
\verb++                             \\
\verb+    <th +%
  \bluett{onmouseover="this.style.color='}%
  \verb++%
  \redtt{white}%
  \verb++%
  \bluett{'"}%
  \verb+>+                          \\
\verb+      +%
  \redtt{Table header 1}%
  \verb++                           \\
\verb+    </th>+                    \\
\verb++                             \\
\verb+    <th colspan="+%
  \redtt{2}%
  \verb+"+                          \\
\verb+        +%
  \bluett{onmouseover="this.style.color='}%
  \verb++%
  \redtt{white}%
  \verb++%
  \bluett{'"}%
  \verb++                           \\
\verb+        +%
  \bluett{onmouseout= "this.style.color='}%
  \verb++%
  \redtt{green}%
  \verb++%
  \bluett{'"}%
  \verb+>+                          \\
\verb+      +%
  \redtt{Table header 2+3}%
  \verb++                           \\
\verb+    </th>+                    \\
\verb++                             \\
\verb+  </tr>+                      \\
\verb++                             \\
\verb+  <tr +%
  \bluett{style="background-color:}%
  \verb++%
  \redtt{ yellow}%
  \verb+"+%
  \bluett{>}%
  \verb++                           \\
\verb++                             \\
\verb+    <td +%
  \bluett{id="}%
  \verb++%
  \redtt{ex\_2}%
  \verb++%
  \bluett{"}%
  \verb+ rowspan="+%
  \redtt{2}%
  \verb+">+                         \\
\verb+      +%
  \redtt{Table data (2+3)-1}%
  \verb++                           \\
\verb+    </td>+                    \\
\verb++                             \\
\verb+    <td>+                     \\
\verb+      +%
  \bluett{<span onclick=~~~"change\_font()"}%
  \verb++                           \\
\verb+            +%
  \bluett{ondblclick="recover\_font()">}%
  \verb++                           \\
\verb+        +%
  \redtt{Table data 2-2}%
  \verb++                           \\
\verb+      +%
  \bluett{</span>}%
  \verb++                           \\
\verb+    </td>+                    \\
\verb++                             \\
\verb+    <td>+                     \\
\verb+        +%
  \redtt{Table data 2-3}%
  \verb++                           \\
\verb+    </td>+                    \\
\verb++                             \\
\verb+  </tr>+                      \\
\verb++                             \\
\verb+  <tr +%
  \bluett{onmouseover="change\_bgcolor()"}%
  \verb+>+                          \\
\verb++                             \\
\verb+    <td>+                     \\
\verb+        +%
  \redtt{Table data 3-2}%
  \verb++                           \\
\verb+    </td>+                    \\
\verb++                             \\
\verb+    <td>+                     \\
\verb+        +%
  \redtt{Table }%
  \verb++%
  \bluett{<span style="font-family:}%
  \verb++%
  \redtt{ 'Times New Roman'}%
  \verb++%
  \bluett{; color:}%
  \verb++%
  \redtt{ red}%
  \verb++%
  \bluett{">}%
  \verb++%
  \redtt{data 3-3}%
  \verb++%
  \bluett{</span>}%
  \verb++                           \\
\verb+    </td>+                    \\
\verb++                             \\
\verb+  </tr>+                      \\
\verb++                             \\
\verb+</table>+                     \\
\tabb
\newpage
\picsin{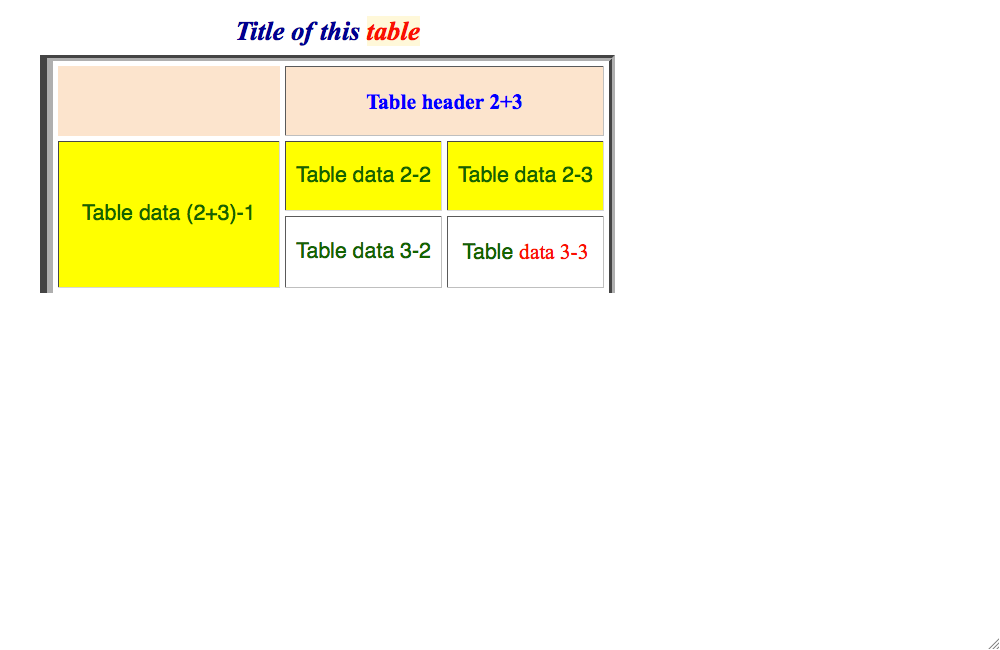}
{basic\_js-htmldom\_table}
{Changing the table style
 given by Template \ref{item:basic_css_table.html}
 (after clicking ``Title of this table'',
  cf.~Figure \ref{fig:basic_css_table}).}
{Changing the table style
 given by Template \ref{item:basic_css_table.html}
 (after clicking the cell of ``Table header 1'',
  cf.~Figure \ref{fig:basic_css_table}).}
{Changing the table style
 given by Template \ref{item:basic_css_table.html}
 (after moving the cursor onto the cell of ``Table header 2+3'',
  cf.~Figure \ref{fig:basic_css_table}).}
{Changing the table style
 given by Template \ref{item:basic_css_table.html}
 (after moving the cursor out the cell of ``Table header 2+3'',
  cf.~Figure \ref{fig:basic_js-htmldom_table-3}).}
{Changing the table style
 given by Template \ref{item:basic_css_table.html}
 (after clicking the words ``Table data 2-2'',
  cf.~Figure \ref{fig:basic_css_table}).}
{Changing the table style
 given by Template \ref{item:basic_css_table.html}
 (after double--clicking the words ``Table data 2-2'',
  cf.~Figure \ref{fig:basic_js-htmldom_table-5}).}
{Changing the table style
 given by Template \ref{item:basic_css_table.html}
 (after moving the cursor onto the cell of ``Table data 3-2'' or ``Table data 3-3'',
  cf.~Figure \ref{fig:basic_css_table}).}
\newpage
\myitem
\label{item:basic_js-htmldom_table2.html}
 {\bf Modification of {\cmbfsfx basic\_css\_table2.html}}
 (cf.~Template \ref{item:basic_css_table2.html})
%
\deft
\verb+<style type="text/css">+               \\
\verb+  table      {line-height:  +%
  \redtt{24pt}%
  \verb+;  margin-left:  +%
  \redtt{36pt}%
  \verb+;+                                   \\
\verb+              border-style: +%
  \redtt{solid}%
  \verb+; border-color: +%
  \redtt{blue}%
  \verb+;+                                   \\
\verb+              border-top-width:  +%
  \redtt{5pt}%
  \verb+; border-bottom-width: +%
  \redtt{4pt}%
  \verb+;+                                   \\
\verb+              border-left-width: +%
  \redtt{0pt}%
  \verb+; border-right-width:  +%
  \redtt{0pt}%
  \verb+}+                                   \\
\verb++                                      \\
\verb+  tr         {color: +%
  \redtt{red}%
  \verb+;   text-align: +%
  \redtt{center}%
  \verb+; width: +%
  \redtt{100pt}%
  \verb+}+                                   \\
\verb+  td         {color: +%
  \redtt{green}%
  \verb+; text-align: +%
  \redtt{center}%
  \verb+; width: +%
  \redtt{100pt}%
  \verb+}+                                   \\
\verb++                                      \\
\verb+  thead      {font-size: +%
  \redtt{13pt}%
  \verb+;   font-weight: +%
  \redtt{normal}%
  \verb+;+                                   \\
\verb+              color: +%
  \redtt{blue}%
  \verb+;+                                   \\
\verb+              line-height: +%
  \redtt{18pt}%
  \verb+; background-color: +%
  \redtt{\#AFEEEE}%
  \verb+}+                                   \\
\verb++                                      \\
\verb+  tbody      {font-family: +%
  \redtt{"Times New Roman"}%
  \verb+; font-style: +%
  \redtt{italic}%
  \verb+;+                                   \\
\verb+              background-color: +%
  \redtt{\#ADD8E6}%
  \verb+}+                                   \\
\verb++                                      \\
\verb+  tfoot      {font-family: +%
  \redtt{sans-serif}%
  \verb+;+                                   \\
\verb+              text-align: +%
  \redtt{left}%
  \verb+}+                                   \\
\verb+</style>+                              \\
%
%
\verb++                                                          \\
\verb+<style type="text/javascript">+                            \\
\verb+  if (navigator.appName == "Microsoft Internet Explorer")+ \\
\verb+  {+                                                       \\
\verb+    +%
  \bluett{childNodes\_No}%
  \verb+ = 0;+                                                   \\
\verb+  }+                                                       \\
\verb++                                                          \\
\verb+  else+                                                    \\
\verb+//if (navigator.userAgent.search("Chrome")    != -1 ||+    \\
\verb+//    navigator.userAgent.search("Firefox")   != -1 ||+    \\
\verb+//    navigator.userAgent.search("Opera")     != -1 ||+    \\
\verb+//    navigator.userAgent.search("Safari")    != -1 ||+    \\
\verb+//    navigator.userAgent.search("Navigator") != -1   )+   \\
\verb+  {+                                                       \\
\verb+    +%
  \bluett{childNodes\_No}%
  \verb- = 2 * 0 + 1;-                                           \\
\verb+  }+                                                       \\
\defc
\verb++                                                          \\
\verb+  function +%
  \bluett{change\_text}%
  \verb+(+%
  \redtt{replaced\_text}%
  \verb+)+                                                       \\
\verb+  {+                                                       \\
\verb+      document.getElementById("+%
  \redtt{ex\_1}%
  \verb+")+                                                      \\
\verb+              .childNodes[+%
  \bluett{childNodes\_No}%
  \verb+]+                                                       \\
\verb+              .innerHTML+                                  \\
\verb+    = +%
  \redtt{replaced\_text}%
  \verb+;+                                                       \\
\verb++                                                          \\
\verb+      document.getElementsByTagName("+%
  \redtt{tfoot}%
  \verb+")[+%
  \bluett{0}%
  \verb+]+ \\
\verb+              .childNodes[+%
  \bluett{childNodes\_No}%
  \verb+].childNodes[+%
  \bluett{childNodes\_No}%
  \verb+]+ \\
\verb+              .innerHTML+ \\
\verb-    =  new Date().getUTCFullYear()   + "/"   +-  \\
\verb+      (new Date().getUTCMonth() +%
  \redtt{+ 1}%
  \verb-) + "/"   +- \\
\verb-       new Date().getUTCDate()       + "   " +- \\
\verb-       new Date().getUTCHours()      + ":"   +- \\
\verb-       new Date().getUTCMinutes()    + ":"   +- \\
\verb+       new Date().getUTCSeconds();+ \\
\verb++ \\
\verb+    if (new Date().getUTCHours() <= +%
  \redtt{16}%
  \verb+)+ \\
\verb+    {+ \\
\verb+        document.getElementsByTagName("+%
  \redtt{tfoot}%
  \verb+")[+%
  \bluett{0}%
  \verb+]+ \\
\verb+                .childNodes[+%
  \bluett{childNodes\_No}%
  \verb+].childNodes[+%
  \bluett{childNodes\_No}%
  \verb+]+ \\
\verb+                .innerHTML+ \\
\verb+      = document.getElementsByTagName("+%
  \redtt{tfoot}%
  \verb+")[+%
  \bluett{0}%
  \verb+]+ \\
\verb+                .childNodes[+%
  \bluett{childNodes\_No}%
  \verb+].childNodes[+%
  \bluett{childNodes\_No}%
  \verb+]+ \\
\verb-                .innerHTML              + '<br />' +- \\
\verb-         new Date().getUTCFullYear()    + "/"      +- \\
\verb+        (new Date().getUTCMonth() +%
  \redtt{+~~1}%
  \verb-) + "/"      +- \\
\verb-         new Date().getUTCDate()        + "   "    +- \\
\verb+        (new Date().getUTCHours() +%
  \redtt{+~~8}%
  \verb-) + ":"      +- \\
\verb-         new Date().getUTCMinutes()     + ":"      +- \\
\verb+         new Date().getUTCSeconds();+ \\
\verb+    }+ \\
\verb++ \\
\verb+    else+ \\
\verb+    if (new Date().getUTCHours() > +%
  \redtt{16}%
  \verb+)+ \\
\verb+    {+ \\
\verb+        document.getElementsByTagName("+%
  \redtt{tfoot}%
  \verb+")[+%
  \bluett{0}%
  \verb+]+ \\
\verb+                .childNodes[+%
  \bluett{childNodes\_No}%
  \verb+].childNodes[+%
  \bluett{childNodes\_No}%
  \verb+]+ \\
\verb+                .innerHTML+ \\
\verb+      = document.getElementsByTagName("+%
  \redtt{tfoot}%
  \verb+")[+%
  \bluett{0}%
  \verb+]+ \\
\verb+                .childNodes[+%
  \bluett{childNodes\_No}%
  \verb+].childNodes[+%
  \bluett{childNodes\_No}%
  \verb+]+ \\
\verb-                .innerHTML              + '<br />' +- \\
\verb-         new Date().getUTCFullYear()    + "/"      +- \\
\verb+        (new Date().getUTCMonth() +%
  \redtt{+~~1}%
  \verb-) + "/"      +- \\
\verb+        (new Date().getUTCDate()  +%
  \redtt{+~~1}%
  \verb-) + "   "    +- \\
\verb+        (new Date().getUTCHours() +%
  \redtt{- 16}%
  \verb-) + ":"      +- \\
\verb-         new Date().getUTCMinutes()     + ":"      +- \\
\verb+         new Date().getUTCSeconds();+ \\
\verb+    }+ \\
\verb+  }+                                                       \\
\defc
\verb++                                                          \\
\verb+  function +%
  \bluett{change\_font}%
  \verb+(+%
  \bluett{new\_size}%
  \verb+, +%
  \bluett{new\_color}%
  \verb+)+                                                       \\
\verb+  {+                                                       \\
\verb+      document.getElementById("+%
  \redtt{ex\_2}%
  \verb+")+                                                      \\
\verb+              .nextSibling.nextSibling+                    \\
\verb+              .style.fontSize+                             \\
\verb+    = +%
  \bluett{new\_size}%
  \verb+;+                                                       \\
\verb++                                                          \\
\verb+      document.getElementById("+%
  \redtt{ex\_2}%
  \verb+")+                                                      \\
\verb+              .nextSibling.nextSibling+                    \\
\verb+              .style.color+                                \\
\verb+    = +%
  \bluett{new\_color}%
  \verb+;+                                                       \\
\verb+  }+                                                       \\
\verb++                                                          \\
\verb+  function +%
  \bluett{recover\_font}%
  \verb+(+%
  \bluett{old\_color}%
  \verb+)+                                                      \\
\verb+  {+                                                       \\
\verb+      document.getElementById("+%
  \redtt{ex\_2}%
  \verb+")+                                                      \\
\verb+              .nextSibling.nextSibling+                    \\
\verb+              .style.color+                                \\
\verb+    = +%
  \bluett{old\_color}%
  \verb+;+                                                       \\
\verb+  }+                                                       \\
\verb++                                                          \\
\verb+  function +%
  \bluett{change\_bgcolor}%
  \verb+(+%
  \bluett{bgcolor\_1}%
  \verb+, +%
  \bluett{bgcolor\_2}%
  \verb+)+                                                      \\
\verb+  {+                                                       \\
\verb+      var +%
  \bluett{target\_cell}%
  \verb++                                                        \\
\verb+    = document.getElementById("+%
  \redtt{ex\_2}%
  \verb+")+                                                      \\
\verb+              .parentNode.parentNode+                      \\
\verb+              .nextSibling.nextSibling+                    \\
\verb+              .childNodes[+%
  \bluett{childNodes\_No}%
  \verb+].childNodes[+%
  \bluett{childNodes\_No}%
  \verb+];+                                                      \\
\verb++                                                          \\
\verb+      +%
  \bluett{target\_cell}%
  \verb+.style.backgroundColor+                                  \\
\verb+    = +%
  \bluett{bgcolor\_1}%
  \verb+;+                                                       \\
\verb++                                                          \\
\verb+      +%
  \bluett{target\_cell}%
  \verb+.nextSibling.nextSibling+                                \\
\verb+                 .style.backgroundColor+                   \\
\verb+    = +%
  \bluett{bgcolor\_1}%
  \verb+;+                                                       \\
\verb+  }+                                                       \\
\verb+</style>+                                                  \\
\defc
\verb++                                      \\
\verb+<table +%
  \bluett{onclick="change\_text('}%
  \verb++%
  \redtt{Ba ba ba...~Ba ba ba...~Ba ba ba...}%
  \verb++%
  \bluett{')"}%
  \verb+>+                                   \\
\verb++                                      \\
\verb+  <thead>+                             \\
\verb++                                      \\
\verb+    <tr +%
  \bluett{id="}%
  \verb++%
  \redtt{ex\_1}%
  \verb++%
  \bluett{"}%
  \verb+>+                                   \\
\verb++                                      \\
\verb+      <th +%
  \bluett{onmouseover="this.style.color='}%
  \verb++%
  \redtt{white}%
  \verb++%
  \bluett{'"}%
  \verb+>+                                   \\
\verb+        +%
  \redtt{Table head 1}%
  \verb++                                    \\
\verb+      </th>+                           \\
\verb++                                      \\
\verb+      <td +%
  \bluett{onmouseover="this.style.color='}%
  \verb++%
  \redtt{white}%
  \verb++%
  \bluett{'"}%
  \verb++                           \\
\verb+          +%
  \bluett{onmouseout= "this.style.color='}%
  \verb++%
  \redtt{green}%
  \verb++%
  \bluett{'"}%
  \verb+>+                          \\
\verb+        +%
  \redtt{Table head 2}%
  \verb++                                    \\
\verb+      </td>+                           \\
\verb++                                      \\
\verb+    </tr>+                             \\
\verb++                                      \\
\verb+  </thead>+                            \\
\verb++                                      \\
\verb+  <tbody>+                             \\
\verb++                                      \\
\verb+    <tr +%
  \bluett{style="color:}%
  \verb++%
  \redtt{ yellow}%
  \verb++%
  \bluett{"}%
  \verb+>+                                   \\
\verb++                                      \\
\verb+      <th +%
  \bluett{id="}%
  \verb++%
  \redtt{ex\_2}%
  \verb++%
  \bluett{"}%
  \verb+>+                                   \\
\verb+        +%
  \redtt{Table body 1}%
  \verb++                                    \\
\verb+      </th>+                           \\
\verb++                                      \\
\verb+    <td>+                              \\
\verb+      +%
  \bluett{<span onclick=~~~"change\_font('}%
  \verb++%
  \redtt{30pt}%
  \verb++%
  \bluett{', '}%
  \verb++%
  \redtt{blue}%
  \verb++%
  \bluett{')"}%
  \verb++                                    \\
\verb+            +%
  \bluett{ondblclick="recover\_font('}%
  \verb++%
  \redtt{red}%
  \verb++%
  \bluett{')">}%
  \verb++                                    \\
\verb+        +%
  \redtt{Table data 2-2}%
  \verb++                                    \\
\verb+      +%
  \bluett{</span>}%
  \verb++                                    \\
\verb+    </td>+                             \\
\verb++                                      \\
\verb+    </tr>+                             \\
\verb++                                      \\
\verb+  </tbody>+                            \\
\verb++                                      \\
\verb+  <tfoot +%
  \bluett{onmouseover="change\_bgcolor('}%
  \verb++%
  \redtt{yellow}%
  \verb++%
  \bluett{', '}%
  \verb++%
  \redtt{red}%
  \verb++%
  \bluett{')"}%
  \verb+>+                                   \\
\verb++                                      \\
\verb+    <tr>+                              \\
\verb++                                      \\
\verb+      <th>+                            \\
\verb+        +%
  \redtt{Table foot 1}%
  \verb++                                    \\
\verb+      </th>+                           \\
\verb++                                      \\
\verb+      <td +%
  \bluett{style="font-size:}%
  \verb++%
  \redtt{ 16}%
  \verb++%
  \bluett{pt; color:}%
  \verb++%
  \redtt{ black}%
  \verb++%
  \bluett{"}%
  \verb+>+                                   \\
\verb+        +%
  \redtt{Table foot 2}%
  \verb++                                    \\
\verb+      </td>+                           \\
\verb++                                      \\
\verb+    </tr>+                             \\
\verb++                                      \\
\verb+  </tfoot>+                            \\
\verb++                                      \\
\verb+</table>+                              \\
\tabb
 More examples can be found in Chap.~\ref{chap:special_effects}.
\newpage
\picsin{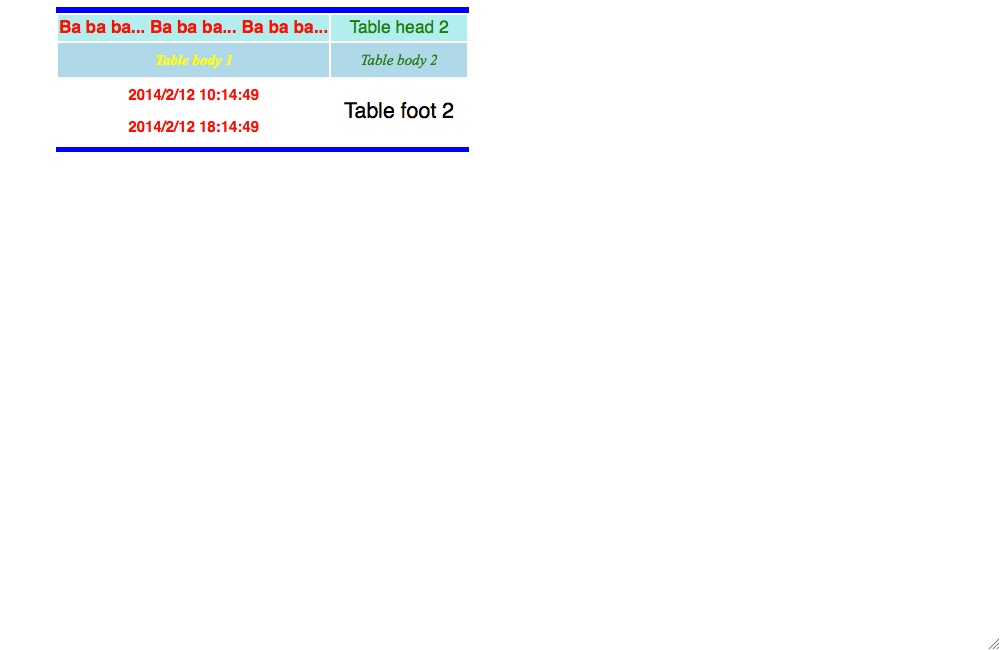}
{basic\_js-htmldom\_table2}
{Changing the table style
 given by Template \ref{item:basic_css_table2.html}
 (after clicking the ``thick--blue top or bottom line'',
  cf.~Figure \ref{fig:basic_css_table2}).}
{Changing the table style
 given by Template \ref{item:basic_css_table2.html}
 (after clicking the cell of ``Table head 1'',,
  cf.~Figure \ref{fig:basic_css_table2}).}
{Changing the table style
 given by Template \ref{item:basic_css_table2.html}
 (after moving the cursor onto the cell of ``Table head 2'',
  cf.~Figure \ref{fig:basic_css_table2}).}
{Changing the table style
 given by Template \ref{item:basic_css_table2.html}
 (after moving the cursor out the cell of ``Table head 2'',
  cf.~Figure \ref{fig:basic_js-htmldom_table2-3}).}
{Changing the table style
 given by Template \ref{item:basic_css_table2.html}
 (after clicking the words ``Table body 2'',
  cf.~Figure \ref{fig:basic_css_table2}).}
{Changing the table style
 given by Template \ref{item:basic_css_table2.html}
 (after double--clicking the words ``Table body 2'',
  cf.~Figure \ref{fig:basic_js-htmldom_table2-5}).}
{Changing the table style
 given by Template \ref{item:basic_css_table2.html}
 (after moving the cursor onto the cell of ``Table foot 1'' or ``Table foot 2'',
  cf.~Figure \ref{fig:basic_css_table2}).}
\end{templateenumerate}
\newpage
\subsection{Frequently used HTML DOM properties for style declaration}
\label{sec:basic_htmldom_style_properties}
%
%
\Xdeft{Background}
\verb+  style.backgroundColor+                       \\
\verb+  style.backgroundImage = "url('+%
  \redtt{../xxxxxx/xxxxxx.jpg}%
  \verb+')"+                                         \\
\verb+  style.backgroundPosition = +%
\verb+"left", "right", "top", "bottom", "center"+    \\
\verb+  style.backgroundRepeat = +%
\verb+"repeat", "repeat-x", "repeat-y", "no-repeat"+ \\
\tabb
\Xdeft{Font}
\verb+  style.fontFamily = +%
\verb+"Times New Roman", "Times", "serif", "sans-serif"+ \\
\verb+  style.fontWeight = "bold"+                       \\
\verb+  style.fontStyle = "italic"+                      \\
\verb+  style.fontSize+                                  \\
\tabb
\Xdeft{Text}
\verb+  style.color+                                     \\
\verb+  style.height+                                    \\
\verb+  style.width+                                     \\
\verb+  style.lineHeight+                                \\
\verb+  style.textAlign = "left", "right", "center"+     \\
\verb+  style.verticalAlign = "top", "bottom", "middle"+ \\
\verb+  style.textIndent+                                \\
\verb+  style.textDecoration = +%
\verb+"underline", "overline", "line-through"+           \\
\tabb
\Xdeft{Margin}
\verb+  style.margin+       \\
\verb+  style.marginLeft+   \\
\verb+  style.marginRight+  \\
\verb+  style.marginTop+    \\
\verb+  style.marginBottom+ \\
\tabb
%

%
%

%
\Xdeft{Image}
\verb+  style.visibility = "visible", "hidden"+ \\
\verb+  style.zIndex = "1", "-1"+               \\
\tabb
\Xdeft{Lists}
\verb+  style.listStyleType = "decimal", "disc"+ \\
\tabb
\Xdeft{Border}
\verb+  style.border+                          \\
\verb+  style.borderLeft+                      \\
\verb+  style.borderRight+                     \\
\verb+  style.borderTop+                       \\
\verb+  style.borderBottom+                    \\
\verb++                                        \\
\verb+  style.borderStyle = +%
\verb+"solid", "dashed", "dotted", "double"+   \\
\verb+                      +%
\verb+"groove", "ridge", "inset", "outset", +%
\verb+"hidden", "none"+                        \\
\verb+  style.borderColor+                     \\
\verb+  style.borderWidth+                     \\
\tabb
\Xdeft{Padding}
\verb+  style.padding+       \\
\verb+  style.paddingLeft+   \\
\verb+  style.paddingRight+  \\
\verb+  style.paddingTop+    \\
\verb+  style.paddingBottom+ \\
\tabb
\Xdeft{Cursor}
\verb+  style.cursor = +%
\verb+"default", "pointer", "progress", "wait", +%
\verb+"text", "help", "move", "crosshair",+                                \\
\verb+                 "e-resize", "w-resize", "n-resize", "s-resize", +   \\
\verb+                 "ne-resize", "nw-resize", "se-resize", "sw-resize"+ \\
\tabb
 See Sec.~\ref{sec:basic_css_properties_values}
 for frequently used style properties and values in CSS.
\newpage
\subsection{Navigator object}
\label{sec:basic_htmldom_navigator}
%
%
\deft
\verb+<script type="text/javascript">+                           \\
\verb+  if (navigator.appName == "Microsoft Internet Explorer")+ \\
\verb+  {+                                                       \\
\verb+     .+                                                    \\
\verb+     .+                                                    \\
\verb+     .+                                                    \\
\verb+  }+                                                       \\
\verb++                                                          \\
\verb+  else+                                                    \\
\verb+//if (navigator.userAgent.search("Chrome")    != -1 ||+    \\
\verb+//    navigator.userAgent.search("Firefox")   != -1 ||+    \\
\verb+//    navigator.userAgent.search("Opera")     != -1 ||+    \\
\verb+//    navigator.userAgent.search("Safari")    != -1 ||+    \\
\verb+//    navigator.userAgent.search("Navigator") != -1   )+   \\
\verb+  {+                                                       \\
\verb+     .+                                                    \\
\verb+     .+                                                    \\
\verb+     .+                                                    \\
\verb+  }+                                                       \\
\verb+</script>+                                                 \\
\tabb
\newpage
\subsection{Location object}
\label{sec:basic_htmldom_location}
%
%
\deft
\verb+<script type="text/javascript">+                              \\
\verb+  if (navigator.appName == "Microsoft Internet Explorer")+    \\
\verb+  {                                                      +    \\
\verb+    window.location = "#+%
  \redtt{specified\_id}%
  \verb+";+ \mynote{Before defining the \redtt{specified\_id}!}     \\
\verb+  }+                                                          \\
\verb+</script>+                                                    \\
\verb++                                                             \\
\verb+<h3 id="+%
  \redtt{specified\_id}%
  \verb+"> +%
  \redtt{Main title of the catagory}%
  \verb+ </h3>+                                                     \\
\verb++                                                             \\
\verb+<script type="text/javascript">+                              \\
\verb+  if (navigator.userAgent.search("Chrome")     != -1     ||+  \\
\verb+      (navigator.userAgent.search("Firefox")   != -1 &&+      \\
\verb+       navigator.userAgent.search("Navigator") == -1   ) ||+  \\
\verb+      navigator.userAgent.search("Opera")      != -1     ||+  \\
\verb+      (navigator.userAgent.search("Chrome")    == -1 &&+      \\
\verb+       navigator.userAgent.search("Safari")    != -1   )   )+ \\
\verb+  {+                                                          \\
\verb+    window.location = "#+%
  \redtt{specified\_id}%
  \verb+";+ \mynote{After defining the \redtt{specified\_id}!}      \\
\verb+  }+                                                          \\
\verb+</script>+                                                    \\
\verb++                                                             \\
\verb+      .+                                                      \\
\verb+      .+                                                      \\
\verb+      .+                                                      \\
\verb++                                                             \\
\verb+<script type="text/javascript">+                              \\
\verb+  if (navigator.userAgent.search("Navigator") != -1)+         \\
\verb+  {+                                                          \\
\verb+    window.location = "#+%
  \redtt{specified\_id}%
  \verb+";+ \mynote{At the end of this paragraph!}                  \\
\verb+  }+                                                          \\
\verb+</script>+                                                    \\
\tabb
\Xdeft{Methods}
\verb+reload(), replace("+%
  \redtt{new\_url}%
  \verb+")+                 \\
\tabb

\newpage
\section{PHP (PHP: Hypertext Preprocessor)}
\label{sec:basic_php}
%
%
\subsection{Start with PHP}
\label{sec:basic_php_start}
%
%
\deft
\verb+<?php+              \\
\verb++                   \\
\verb+  echo +%
  \bluett{"}%
  \verb++%
  \redtt{some text}%
  \verb++%
  \bluett{";}%
  \verb++                 \\
\verb++                   \\
\verb+  echo +%
  {\color{blue}%
  \verb-"-%
  \redtt{some text }%
  \verb-" . "-%
  \redtt{'single quote'}%
  \verb-" . ' -%
  {\color{red}\verb+&+}%
  \verb- ' . "-%
  {\color{red}\verb+<br />+}%
  \verb-" . '-%
  \redtt{"double quote"}%
  \verb-' . "\r\n";-}%
  \verb++                 \\
\verb++                   \\
\verb+?>+                 \\
\tabb
 See Sec.~\ref{sec:basic_js_start}
 for basic JavaScript syntax.
\newpage
\subsection{Comments}
\label{sec:basic_php_comments}
%
%
\deft
\verb+<?php+                 \\
\verb+//   +%
  \redtt{Comment line}%
  \verb++                    \\
\verb++                      \\
\verb+/*   +%
  \redtt{Comment line}%
  \verb+   */+               \\
\verb++                      \\
\verb+/*+                    \\
\verb+  +%
  \redtt{Some comments ...}%
  \verb++                    \\
\verb+      .+               \\
\verb+      .+               \\
\verb+      .+               \\
\verb+      .+               \\
\verb+*/+                    \\
\verb+?>+                    \\
\tabb
 See Secs.~\ref{sec:basic_html_comments},
 \ref{sec:basic_css_comments},
 and \ref{sec:basic_js_comments}
 for comments in HTML, CSS, and JavaScript,
 respectively.
\newpage
\subsection{Variables}
\label{sec:basic_php_variables}
%
%
\deft
\verb+<?php+        \\
\verb+  +%
  \bluett{\$}%
  \verb++%
  \redtt{xxx}%
  \verb+     = +%
  \redtt{number}%
  \verb+;+          \\
\verb+  +%
  \bluett{\$}%
  \verb++%
  \redtt{xxx\_xxx}%
  \verb+ = +%
  \redtt{"string"}%
  \verb+;+          \\
\verb++             \\
\verb+  +%
  \bluett{\$}%
  \verb++%
  \redtt{yyy}%
  \verb+     = +%
  \redtt{number}%
  \verb-   + -%
  \bluett{\$}%
  \verb++%
  \redtt{xxx}%
  \verb+;+          \\
\verb+  +%
  \bluett{\$}%
  \verb++%
  \redtt{yyy\_yyy}%
  \verb+ = +%
  \redtt{"string"}%
  \verb+ . +%
  \bluett{\$}%
  \verb++%
  \redtt{xxx\_xxx}%
  \verb+;+          \\
\verb++             \\
\verb+  +%
  \bluett{\$}%
  \verb++%
  \redtt{yyy\_yyy}%
  \verb+ = +%
  \redtt{"string"}%
  \verb+ . +%
  \bluett{\$}%
  \verb++%
  \redtt{xxx}%
  \verb+;+          \\
\verb+?>+           \\
\tabb
 See Sec.~\ref{sec:basic_js_variables}
 for variables in JavaScript.
\newpage
\subsection{Arrays}
\label{sec:basic_php_arrays}
%
%
\deft
\verb+<?php+        \\
\verb+  +%
  \bluett{\$}%
  \verb++%
  \redtt{xxx}%
  \verb+[+%
  \bluett{\$}%
  \verb++%
  \redtt{n\_xxx}%
  \verb+]     = +%
  \redtt{number}%
  \verb+;+          \\
\verb+  +%
  \bluett{\$}%
  \verb++%
  \redtt{xxx\_xxx}%
  \verb+[+%
  \bluett{\$}%
  \verb++%
  \redtt{n\_xxx}%
  \verb+] = +%
  \redtt{"string"}%
  \verb+;+          \\
\verb++             \\
\verb+  +%
  \bluett{\$}%
  \verb++%
  \redtt{xxx}%
  \verb+[+%
  \bluett{\$}%
  \verb++%
  \redtt{n\_xxx}%
  \verb+][+%
  \bluett{\$}%
  \verb++%
  \redtt{m\_xxx}%
  \verb+] = +%
  \redtt{number}%
  \verb+;+          \\
\verb++             \\
\verb+  +%
  \bluett{\$}%
  \verb++%
  \redtt{variable\_name}%
  \verb+[+%
  \bluett{\$}%
  \verb++%
  \redtt{n\_item}%
  \verb+] = +%
  \bluett{\$}%
  \verb+_POST[+%
  \bluett{"}%
  \verb++%
  \redtt{variable\_name}%
  \verb++%
  \bluett{"}%
  \verb+][+%
  \bluett{\$}%
  \verb++%
  \redtt{n\_item}%
  \verb+];+         \\
\verb+  +%
  \bluett{\$}%
  \verb++%
  \redtt{variable\_name}%
  \verb+[+%
  \bluett{\$}%
  \verb++%
  \redtt{n\_item}%
  \verb+] = +%
  \bluett{\$}%
  \verb+_GET[+%
  \bluett{"}%
  \verb++%
  \redtt{variable\_name}%
  \verb++%
  \bluett{"}%
  \verb+][+%
  \bluett{\$}%
  \verb++%
  \redtt{n\_item}%
  \verb+];+         \\
\verb++             \\

\verb+  +%
  \bluett{\$}%
  \verb++%
  \redtt{personal\_file}%
  \verb+[+%
  \bluett{\$}%
  \verb++%
  \redtt{n\_file}%
  \verb+] = +%
  \bluett{\$}%
  \verb+_FILES["+%
  \redtt{personal\_file}%
  \verb+"]["name"][+%
  \bluett{\$}%
  \verb++%
  \redtt{n\_file}%
  \verb+];+                                    \\

\verb+?>+           \\
\tabb
 See Sec.~\ref{sec:basic_js_arrays}
 for arrays in JavaScript.
\newpage
\subsection{{\sf if}}
\label{sec:basic_php_if}
%
%
\deft
\verb+<?php+               \\
\verb+  if (($+%
  \redtt{xxx}%
  \verb+     +%
  \bluett{==}%
  \verb+ $+%
  \redtt{yyy}%
  \verb+     +%
  {\color{blue}\verb+&&+}%
  \verb++                  \\
\verb+       $+%
  \redtt{xxx\_xxx}%
  \verb+ +%
  \bluett{!=}%
  \verb+ $+%
  \redtt{yyy\_yyy}%
  \verb+   ) +%
  \bluett{||}%
  \verb++                  \\
\verb+      (+%
  \redtt{condition}%
  \verb+              ) +%
  \bluett{||}%
  \verb++                  \\
\verb+      (+%
  \bluett{!}%
  \verb+(+%
  \redtt{condition}%
  \verb+)           )   )+ \\
\verb+  {+                 \\
\verb+      .+             \\
\verb+      .+             \\
\verb+      .+             \\
\verb+  }+                 \\
\verb++                    \\
\verb+  +%
  \bluett{elseif}%
  \verb+ (+%
  \redtt{conditions}%
  \verb+)+ \mynote{$\lGetsto$ \bluett{else if} in JaveScript (Sec.~\ref{sec:basic_js_if})} \\
\verb+  {+                 \\
\verb+      .+             \\
\verb+      .+             \\
\verb+      .+             \\
\verb+  }+                 \\
\verb++                    \\
\verb+  else+              \\
\verb+  {+                 \\
\verb+      .+             \\
\verb+      .+             \\
\verb+      .+             \\
\verb+  }+                 \\
\verb+?>+                  \\
\tabb
 See Sec.~\ref{sec:basic_js_if}
 for \verb+if+, \verb+else if+ and \verb+else+ comments in JaveScript.
\newpage
\subsection{{\sf switch}}
\label{sec:basic_php_switch}
%
%
\deft
\verb+<?php+             \\
\verb+  switch ($+%
  \redtt{xxx}%
  \verb+)+               \\
\verb+  {+               \\
\verb+    case +%
  \redtt{x}%
  \verb+:+               \\
\verb+      +%
  \redtt{......}%
  \verb+;+               \\
\verb+      break;+      \\
\verb++                  \\
\verb+    case +%
  \redtt{y}%
  \verb+:+               \\
\verb+      +%
  \redtt{......}%
  \verb+;+               \\
\verb+      break;+      \\
\verb++                  \\
\verb+    case +%
  \redtt{z}%
  \verb+:+               \\
\verb+      +%
  \redtt{......}%
  \verb+;+               \\
\verb+      break;+      \\
\verb++                  \\
\verb+    default:+      \\
\verb+      +%
  \redtt{......}%
  \verb+;+               \\
\verb+  }+               \\
\verb+?>+                \\
\tabb
 See Sec.~\ref{sec:basic_js_switch}
 for \verb+switch+ comment in JaveScript.
\newpage
\subsection{{\sf for}}
\label{sec:basic_php_for}
%
%
\deft
\verb+<?php+        \\
\verb+  for ($+%
  \redtt{n\_xxx}%
  \verb+ = +%
  \bluett{1}%
  \verb+; $+%
  \redtt{n\_xxx}%
  \verb+ <= $+%
  \redtt{n\_xxx\_max}%
  \verb+; $+%
  \redtt{n\_xxx}%
  \verb+ +%
  \bluett{++}%
  \verb+)+          \\
\verb+  {+          \\
\verb+     .+       \\
\verb+     .+       \\
\verb+     .+       \\
\verb+  }+          \\
\verb+?>+           \\
\tabb
\deft
\verb+<?php+         \\
\verb+  $+%
  \redtt{yyy}%
  \verb+ = array(+%
  \redtt{"zzz\_1"}%
  \verb+, +%
  \redtt{"zzz\_2"}%
  \verb+, +%
  \redtt{...}%
  \verb+);+          \\
\verb++              \\
\verb+  foreach ($+%
  \redtt{yyy}%
  \verb+ as $+%
  \bluett{value}%
  \verb+)+           \\
\verb+  {+           \\
\verb+     .+        \\
\verb+     .+        \\
\verb+     .+        \\
\verb+  }+           \\
\verb+?>+            \\
\tabb
 See Sec.~\ref{sec:basic_js_for}
 for \verb+for+ and \verb+for in+ comments in JaveScript.
\newpage
\subsection{{\sf while}}
\label{sec:basic_php_while}
%
%
\deft
\verb+<?php+     \\
\verb+  while (+%
  \redtt{conditions}%
  \verb+)+                             \\
\verb+  {+                             \\
\verb+     .+                          \\
\verb+     .+                          \\
\verb+     .+                          \\
\verb+  }+                             \\
\verb+?>+        \\
\tabb
\deft
\verb+<?php+     \\
\verb+  do+                            \\
\verb+  {+                             \\
\verb+     .+                          \\
\verb+     .+                          \\
\verb+     .+                          \\
\verb+  }+                             \\
\verb+  while (+%
  \redtt{conditions}%
  \verb+)+%
  \bluett{;}%
  \verb++                             \\
\verb+?>+        \\
\tabb
 See Sec.~\ref{sec:basic_js_while}
 for \verb+while+ and \verb+do while+ comments in JaveScript.
\newpage
\subsection{{\sf form}}
\label{sec:basic_php_form}
%
%
\deft
\verb+<form action="+%
  \redtt{login\_check}%
  \verb+.php"+                              \\
\verb+      method="+%
  \redtt{post}%
  \verb+/+%
  \redtt{get}%
  \verb+"+                                  \\
\verb+      enctype="multipart/form-data">+ \\
\verb++                    \\
\verb+  <fieldset>+        \\
\verb++                    \\
\verb+    <legend> +%
  \bluett{Login}%
  \verb+ </legend>+        \\
\verb++                    \\
\verb+    +%
  \bluett{User name}%
  \verb+:+                 \\
\verb+    <input type="+%
  \bluett{text}%
  \verb+"     name="+%
  \bluett{user\_name}%
  \verb+"     /> <br/ >+   \\
\verb++                    \\
\verb+    +%
  \bluett{Password}%
  \verb+:+                 \\
\verb+    <input type="+%
  \bluett{password}%
  \verb+" name="+%
  \bluett{user\_password}%
  \verb+" /> <br/ >+       \\
\verb++                    \\
\verb+  </fieldset>+       \\
\verb++                    \\
\verb+       .+            \\
\verb+       .+            \\
\verb+       .+            \\
\verb++                    \\
\verb+  <input type="submit" name="+%
  \bluett{submit}%
  \verb+" value="+%
  \bluett{Submit}%
  \verb+" />+         \\
\verb+  <br />+       \\
\verb++               \\
\verb+</form>+        \\
\tabb
\Xdeft{login\_check.php}
\verb+<?php+                \\
\verb+  $+%
  \redtt{user\_name}%
  \verb+     = +%
  \bluett{\$\_POST}%
  \verb+["+%
  \redtt{user\_name}%
  \verb+"];+                \\
\verb+  $+%
  \redtt{user\_password}%
  \verb+ = +%
  \bluett{\$\_POST}%
  \verb+["+%
  \redtt{user\_password}%
  \verb+"];+                \\
\verb+?>+                   \\
\tabb
\modificationdeft
\verb+<?php+       \\
\verb+  $+%
  \redtt{xxx}%
  \verb+ = +%
  \bluett{\$\_GET}%
  \verb+["+%
  \redtt{xxx}%
  \verb+"];+       \\
\verb+  $+%
  \redtt{yyy}%
  \verb+ = +%
  \bluett{\$\_GET}%
  \verb+["+%
  \redtt{yyy}%
  \verb+"];+       \\
\verb+?>+          \\
\tabb
\newpage
\subsection{File handling}
\label{sec:basic_php_file_handling}
%
%
\deft
\verb+<?php+             \\
\verb+  $+%
  \redtt{data\_input}%
  \verb+ = fopen("+%
  \redtt{data.txt}%
  \verb+", "r");+        \\
\verb++                  \\
\verb+  do+              \\
\verb+  {+               \\
\verb+     .+            \\
\verb+     .+            \\
\verb+     .+            \\
\verb+  }+               \\
\verb+  while (!feof($+%
  \redtt{data\_input}%
  \verb+));+             \\
\verb++                  \\
\verb+  fclose($+%
  \redtt{data\_input}%
  \verb+);+              \\
\verb++                  \\
\verb+     .+            \\
\verb+     .+            \\
\verb+     .+            \\
\verb++                  \\
\verb+  $+%
  \redtt{result\_output}%
  \verb+ = fopen("+%
  \redtt{result.txt}%
  \verb+", "+%
  \redtt{w}%
  \verb+/+%
  \redtt{a}%
  \verb+");+             \\
\verb+     .+            \\
\verb+     .+            \\
\verb+     .+            \\
\verb+  fclose($+%
  \redtt{result\_output}%
  \verb+);+              \\
\verb+?>+                \\
\tabb
\Xdeft{Frequently used functions for file handling}
\verb+fopen(), fclose(),+        \\
\verb+fputs(), fscanf(),+        \\
\verb+file_exists(), feof(),+    \\
\verb+link(), unlink(), copy(),+ \\
\verb+move_uploaded_file(),+     \\
\verb+mkdir(), rmdir()+          \\
\tabb
 See Sec.~\ref{sec:basic_php_file_uploading}
 for more details about the \verb+move_uploaded_file()+ function.
\newpage
\subsection{File uploading}
\label{sec:basic_php_file_uploading}
%
%
\deft
\verb+<form action="+%
  \redtt{get\_file}%
  \verb+.php"+                              \\
\verb+      method="post"+                  \\
\verb+      enctype="multipart/form-data">+ \\
\verb++               \\
\verb+  <fieldset>+   \\
\verb++               \\
\verb+    <legend> +%
  \redtt{File uploading}%
  \verb+ </legend>+   \\
\verb++               \\
\verb+    <input type="file" name="+%
  \redtt{uploaded\_file}%
  \verb+" /> <br />+  \\
\verb++               \\
\verb+  </fieldset>+  \\
\verb++               \\
\verb+       .+       \\
\verb+       .+       \\
\verb+       .+       \\
\verb++               \\
\verb+  <input type="submit" name="+%
  \bluett{submit}%
  \verb+" value="+%
  \bluett{Submit}%
  \verb+" />+         \\
\verb+  <br />+       \\
\verb++               \\
\verb+</form>+        \\
\tabb
\Xdeft{get\_file.php}
\verb+<?php+                           \\
\verb+  move_uploaded_file($_FILES["+%
  \redtt{uploaded\_file}%
  \verb+"]["tmp_name"],+               \\
\verb+                     "+%
  \redtt{folder\_name}%
  \verb+/" . $_FILES["+%
  \redtt{uploaded\_file}%
  \verb+"]["name"]);+                  \\
\verb++                                \\
\verb+  if (file_exists("+%
  \redtt{folder\_name}%
  \verb+/" . $_FILES["+%
  \redtt{uploaded\_file}%
  \verb+"]["name"]))+                  \\
\verb+  {+                             \\
\verb+    if ($_FILES["+%
  \redtt{uploaded\_file}%
  \verb+"]["type"] = "+%
  \bluett{image/jpeg}%
  \verb+")+                            \\
\verb+    {+                           \\
\verb+       .+                        \\
\verb+       .+                        \\
\verb+       .+                        \\
\verb+    }+                           \\
\verb++                                \\
\verb+    if ($_FILES["+%
  \redtt{uploaded\_file}%
  \verb+"]["size"] > +%
  \redtt{20000}%
  \verb+)+                             \\
\verb+    {+                           \\
\verb+       .+                        \\
\verb+       .+                        \\
\verb+       .+                        \\
\verb+    }+                           \\
\verb++                                \\
\verb+     .+                          \\
\verb+     .+                          \\
\verb+     .+                          \\
\verb+  }+                             \\
\verb+?>+                              \\
\tabb
 More examples for file uploading and handling
 can be found in Chap.~\ref{chap:file_handling}.
\newpage
\subsection{E-mail sending}
\label{sec:basic_php_email_sending}
%
%
\deft
\verb+<?php+                                  \\
\verb+  mail(+%
  \redtt{e-mail address of a receiver}%
  \verb+,+                                    \\
\verb+       +%
  \redtt{subject of this e-mail}%
  \verb+,+                                    \\
\verb+       +%
  \redtt{message of this e-mail}%
  \verb+,+                                    \\
\verb+       'From: +%
  \redtt{xxxxxx@xxxxxx.xxx.xx}%
  \verb+'     . "\r\n" .+                     \\
\verb+       'Cc:   +%
  \redtt{yyyyyy1@yyyyyy1.yyy1.yy1}%
  \verb+' . ', '   .+                         \\
\verb+             '+%
  \redtt{yyyyyy2@yyyyyy2.yyy2.yy2}%
  \verb+' . "\r\n" .+                         \\
\verb+       'Bcc:  +%
  \redtt{zzzzzz@zzzzzz.zzz.zz}%
  \verb+',+                                   \\
\verb+       +%
  \redtt{an additional parameter (optional)}%
  \verb+);+                                   \\ 
\verb+?>+                                     \\
\tabb
 See Sec.~\ref{sec:basic_html_a}
 for e-mail sending with an \verb+<a>+ tag in HTML.
\Xdeft{Example}
\verb+<?php+                                                                \\
\verb+  if (+%
  \redtt{\$\_POST["confirm\_email"] == "yes"}%
  \verb+)+                                                                  \\
\verb+  {+                                                                  \\
\verb+    mail($_POST["+%
  \redtt{new\_participant\_email\_address}%
  \verb+"],+                                                                \\
\verb++                                                                     \\
\verb+         '[Auto submission] Registration confirmation',+              \\
\verb++                                                                     \\
\verb+         'Dear '. $_POST["+%
  \redtt{new\_participant\_title}%
  \verb+"]   . ' ' .+                                                       \\
\verb+                  $_POST["+%
  \redtt{new\_participant\_surname}%
  \verb+"] . ',' . "\r\n" .+                                                \\
\verb+                                                            "\r\n" .+ \\
\verb++                                                                     \\
\verb+         'Herewith we would be glad to confirm '          .+          \\
\verb+         'your registration by +%
  \redtt{......}%
  \verb+ '                   .+                                             \\
\verb+         'with the following personal information:'       . "\r\n" .+ \\
\verb+                                                            "\r\n" .+ \\
\verb++                                                                     \\
\verb+         'Title:   ' . $_POST["+%
  \redtt{new\_participant\_title}%
  \verb+"]    . "\r\n" .+                                                   \\
\verb+         'Surname: ' . $_POST["+%
  \redtt{new\_participant\_surname}%
  \verb+"]  . "\r\n" .+                                                     \\
\verb+         'Name:    ' . $_POST["+%
  \redtt{new\_participant\_name}%
  \verb+"]     . "\r\n" .+                                                  \\
\verb++                                                                     \\
\verb+             .+                                                       \\
\verb+             .+                                                       \\
\verb+             .+                                                       \\
\verb++                                                                     \\
\verb+         'Thank you very much for your registration. '    .+          \\
\verb+         'Should you have any questions, '                .+          \\
\verb+         'please don\'t hesitate to contact us.'          . "\r\n" .+ \\
\verb+                                                            "\r\n" .+ \\
\verb++                                                                     \\
\verb+         'We are looking forward to seeing you.'          . "\r\n" .+ \\
\verb+                                                            "\r\n" .+ \\
\verb++                                                                     \\
\verb+         'Sincerely yours,'                               . "\r\n" .+ \\
\verb+                                                            "\r\n" .+ \\
\verb+                                                            "\r\n" .+ \\
\verb++                                                                     \\
\verb+         'The organization committee'                     . "\r\n" .+ \\
\verb+                                                            "\r\n" ,+ \\
\verb++                                                                     \\
\verb+         'From: +%
  \redtt{xxxxxx@xxxxxx.xxx.xx}%
  \verb+'                     . "\r\n" .+                                   \\
\verb+         'Cc:   +%
  \redtt{yyyyyy1@yyyyyy1.yyy1.yy1}%
  \verb+'                 . ', '   .+                                       \\
\verb+               '+%
  \redtt{yyyyyy2@yyyyyy2.yyy2.yy2}%
  \verb+'                 . "\r\n" .+                                       \\
\verb+         'Bcc:  +%
  \redtt{zzzzzz@zzzzzz.zzz.zz}%
  \verb+'                               );+                                 \\
\verb+  }+                                                                  \\
\verb+?>+                                                                   \\
\tabb
\newpage
\subsection{Frequently used functions}
\label{sec:basic_php_other_functions}
%
%
%
\begin{templateenumerate}
%
%
\myitem
\label{item:php_date_time.php}
 {\bf Date and time}
%
\deft
\verb+<?php+                                                            \\
\verb+  echo gmdate(Y) . '<br /> ' . "\r\n";      // +%
  A four digit representation of a year                                 \\
\verb+  echo gmdate(y) . '<br /> ' . "\r\n";      // +%
  A two digit representation of a year                                  \\
\verb++                                                                 \\
\verb+  echo gmdate(m) . '<br /> ' . "\r\n";      // +%
  A numeric representation of a month (01 to 12)                        \\
\verb+  echo gmdate(n) . '<br /> ' . "\r\n";      // +%
  A numeric representation of a month                                   \\
\verb+                                            // +%
  without leading zeros (1 to 12)                                       \\
\verb+  echo gmdate(F) . '<br /> ' . "\r\n";      // +%
  A full textual representation of a month                              \\
\verb+  echo gmdate(M) . '<br /> ' . "\r\n";      // +%
  A short textual representation of a month (three letters)             \\
\verb++                                                                 \\
\verb+  echo gmdate(z) . '<br /> ' . "\r\n";      // +%
  The day of the year (0 to 365)                                        \\
\verb+  echo gmdate(d) . '<br /> ' . "\r\n";      // +%
  The day of the month (01 to 31)                                       \\
\verb+  echo gmdate(j) . '<br /> ' . "\r\n";      // +%
  The day of the month without leading zeros (1 to 31)                  \\
\verb+  echo gmdate(l) . '<br /> ' . "\r\n";      // +%
  A full textual representation of a day  of the week                   \\
\verb+                                            // +%
  (lowercase ``L'') \\
\verb+  echo gmdate(D) . '<br /> ' . "\r\n";      // +%
  A textual representation of a day  of the week (three letters)        \\
\verb++                                                                 \\
\verb+  echo gmdate(A) . '<br /> ' . "\r\n";      // +%
  Uppercase AM or PM                                                    \\
\verb+  echo gmdate(a) . '<br /> ' . "\r\n";      // +%
  Lowercase am or pm                                                    \\
\verb++                                                                 \\
\verb+  echo gmdate(H) . '<br /> ' . "\r\n";      // +%
  24-hour format of an hour (00 to 23)                                  \\
\verb+  echo gmdate(G) . '<br /> ' . "\r\n";      // +%
  24-hour format of an hour (0 to 23)                                   \\
\verb+  echo gmdate(h) . '<br /> ' . "\r\n";      // +%
  12-hour format of an hour (01 to 12)                                  \\
\verb+  echo gmdate(g) . '<br /> ' . "\r\n";      // +%
  12-hour format of an hour (1 to 12)                                   \\
\verb++                                                                 \\
\verb+  echo gmdate(i) . '<br /> ' . "\r\n";      // +%
  Minutes with leading zeros (00 to 59)                                 \\
\verb++                                                                 \\
\verb+  echo gmdate(s) . '<br /> ' . "\r\n";      // +%
  Seconds with leading zeros (00 to 59)                                 \\
\verb+?>+                                                               \\
\tabb
\newpage
\picin{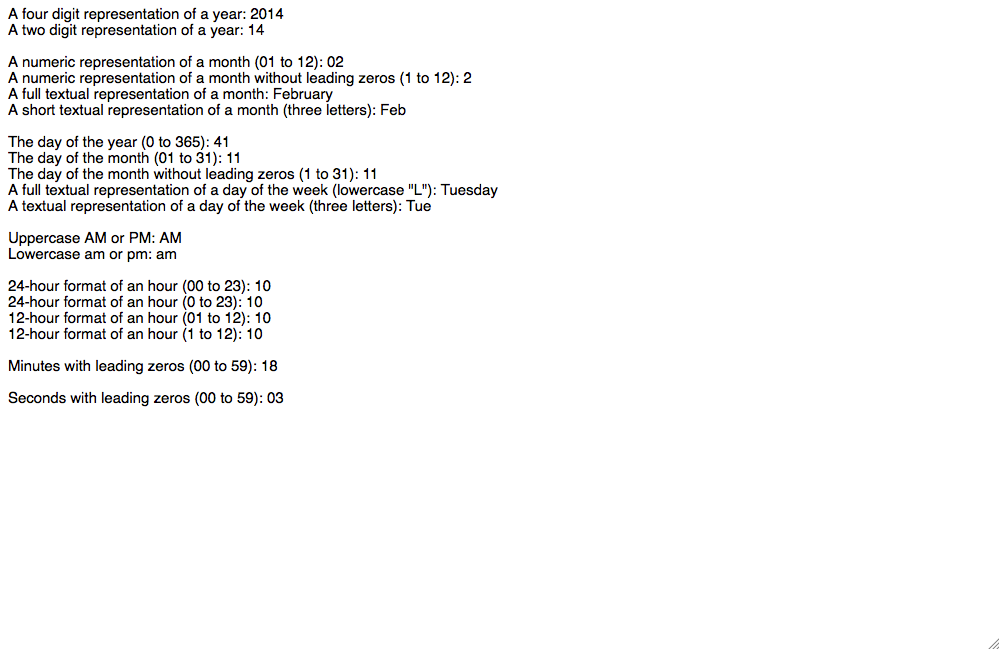}
{basic\_php\_date\_time}
{A collection of presentations for date and time
 defined in PHP.}
\newpage
\Xdeft{Math}
\verb+abs(), ceil(), floor(), round(), max(), min(),+     \\
\verb+sin(), cos(), tan(), sinh(), cosh(), tanh(),+       \\
\verb+asin(), acos(), atan(), asinh(), acosh(), atanh(),+ \\
\verb+sqrt(), pow(), exp(), log(), log10(),+              \\
\verb+rand()+                                             \\
\tabb
\Xdeft{String}
\verb+printf(), fprintf(), sprintf()+ \\
\tabb
\Xdeft{File including}
\verb+<?php+            \\
\verb+  include("+%
  \redtt{folder\_name}%
  \verb+/+%
  \redtt{full\_file\_name\_with\_the\_extension}%
  \verb+");+            \\
\verb+?>+               \\
\tabb
\Xdeft{Shell commands}
\verb+<?php+               \\
\verb+  shell_exec("+%
  \redtt{a shell command}%
  \verb+");+               \\
\verb+?>+                  \\
\tabb
 See Sec.~\ref{sec:basic_js_methods}
 for frequently used JavaScript methods.
\end{templateenumerate}
\newpage
\subsection{Basic combination of HTML with PHP}
\label{sec:combination_html_php}
%
%
%
\begin{templateenumerate}
%
%
\myitem
\label{item:php_input-1.php}
 {\bf PHP inserted into HTML}
\deft
\verb+  <input type="+%
  \bluett{text}%
  \verb+"+                \\
\verb+         class="+%
  \bluett{table}%
  \verb+"+                \\
\verb+         name="+%
  \redtt{integer\_1}%
  \verb+"+                \\
\verb+         value="+%
  \bluett{<?php echo }%
  \verb++%
  \redtt{\$integer\_1}%
  \verb++%
  \bluett{; ?>}%
  \verb+" />+             \\
\tabb
\newpage
\myitem
\label{item:php_input-2.php}
 {\bf PHP includes HTML}
\deft
\verb++%
  \bluett{<?php}%
  \verb++                  \\
\verb+  +%
  \bluett{echo '}%
  \verb+  <input type="+%
  \bluett{text}%
  \verb+"+%
  {\color{blue}\verb+'          . "\r\n";+}%
  \verb++                  \\
\verb+  +%
  \bluett{echo '}%
  \verb+         class="+%
  \bluett{table}%
  \verb+"+%
  {\color{blue}\verb+'        . "\r\n";+}%
  \verb++                  \\
\verb+  +%
  \bluett{echo '}%
  \verb+         name="+%
  \redtt{integer\_1}%
  \verb+"+%
  {\color{blue}\verb+'     . "\r\n";+}%
  \verb++                  \\
\verb+  +%
  \bluett{echo '}%
  \verb+         value="+%
  \bluett{' .}%
  \verb+ +%
  \redtt{\$integer\_1}%
  \verb+ +%
  \bluett{.~'}%
  \verb+" />+%
  \bluett{';}%
  \verb++                  \\
\verb++%
  \bluett{?>}%
  \verb++                  \\
\tabb
\newpage
\myitem
\label{item:php_if-1.php}
 {\bf PHP with {\tt if/elseif/else} comments inserted into HTML}
\deft
\verb+  <input type="+%
  \bluett{text}%
  \verb+"+                 \\
\verb+         class="+%
  \bluett{table}%
  \verb+"+                 \\
\verb+         name="+%
  \redtt{integer\_1}%
  \verb+"+                 \\
\verb+         value="+%
  \bluett{<?php}%
  \verb++                  \\
\verb+                  +%
  \bluett{if (}%
  \verb++%
  \redtt{......}%
  \verb++%
  \bluett{)}%
  \verb++                  \\
\verb+                  +%
  {\color{blue}\verb+{+}%
  \verb++                  \\
\verb+                  +%
  \bluett{  echo }%
  \verb++%
  \redtt{\$integer\_1\_1}%
  \verb++%
  \bluett{;}%
  \verb++                  \\
\verb+                  +%
  {\color{blue}\verb+}+}%
  \verb++                  \\
\verb++                    \\
\verb+                  +%
  \bluett{elseif (}%
  \verb++%
  \redtt{......}%
  \verb++%
  \bluett{)}%
  \verb++                  \\
\verb+                  +%
  {\color{blue}\verb+{+}%
  \verb++                  \\
\verb+                  +%
  \bluett{  echo }%
  \verb++%
  \redtt{\$integer\_1\_2}%
  \verb++%
  \bluett{;}%
  \verb++                  \\
\verb+                  +%
  {\color{blue}\verb+}+}%
  \verb++                  \\
\verb++                    \\
\verb+                  +%
  \bluett{else}%
  \verb++                  \\
\verb+                  +%
  {\color{blue}\verb+{+}%
  \verb++                  \\
\verb+                  +%
  \bluett{  echo }%
  \verb++%
  \redtt{\$integer\_1\_3}%
  \verb++%
  \bluett{;}%
  \verb++                  \\
\verb+                  +%
  {\color{blue}\verb+}+}%
  \verb++                  \\
\verb+                +%
  \bluett{?>}%
  \verb+" />+              \\
\tabb
\newpage
\myitem
\label{item:php_if-2.php}
 {\bf PHP with {\tt if/elseif/else} comments includes HTML}
\deft
\verb++%
  \bluett{<?php}%
  \verb++                  \\
\verb+  +%
  \bluett{echo '}%
  \verb+  <input type="+%
  \bluett{text}%
  \verb+"+%
  {\color{blue}\verb+'              . "\r\n";+}%
  \verb++                  \\
\verb+  +%
  \bluett{echo '}%
  \verb+         class="+%
  \bluett{table}%
  \verb+"+%
  {\color{blue}\verb+'            . "\r\n";+}%
  \verb++                  \\
\verb+  +%
  \bluett{echo '}%
  \verb+         name="+%
  \redtt{integer\_1}%
  \verb+"+%
  {\color{blue}\verb+'         . "\r\n";+}%
  \verb++                  \\
\verb++                    \\
\verb+  +%
  \bluett{if (}%
  \verb++%
  \redtt{......}%
  \verb++%
  \bluett{)}%
  \verb++                  \\
\verb+  +%
  {\color{blue}\verb+{+}%
  \verb++                  \\
\verb+    +%
  \bluett{echo '}%
  \verb+         value="+%
  \bluett{' .}%
  \verb+ +%
  \redtt{\$integer\_1\_1}%
  \verb+ +%
  \bluett{.~'}%
  \verb+" />+%
  \bluett{';}%
  \verb++                  \\
\verb+  +%
  {\color{blue}\verb+}+}%
  \verb++                  \\
\verb++                    \\
\verb+  +%
  \bluett{elseif (}%
  \verb++%
  \redtt{......}%
  \verb++%
  \bluett{)}%
  \verb++                  \\
\verb+  +%
  {\color{blue}\verb+{+}%
  \verb++                  \\
\verb+    +%
  \bluett{echo '}%
  \verb+         value="+%
  \bluett{' .}%
  \verb+ +%
  \redtt{\$integer\_1\_2}%
  \verb+ +%
  \bluett{.~'}%
  \verb+" />+%
  \bluett{';}%
  \verb++                  \\
\verb+  +%
  {\color{blue}\verb+}+}%
  \verb++                  \\
\verb++                    \\
\verb+  +%
  \bluett{else}%
  \verb++                  \\
\verb+  +%
  {\color{blue}\verb+{+}%
  \verb++                  \\
\verb+    +%
  \bluett{echo '}%
  \verb+         value="+%
  \bluett{' .}%
  \verb+ +%
  \redtt{\$integer\_1\_3}%
  \verb+ +%
  \bluett{.~'}%
  \verb+" />+%
  \bluett{';}%
  \verb++                  \\
\verb+  +%
  {\color{blue}\verb+}+}%
  \verb++                  \\
\verb++%
  \bluett{?>}%
  \verb++                  \\
\tabb
\newpage
\myitem
\label{item:php_if-3.php}
 {\bf Modification of PHP with {\tt if/elseif/else} comments includes HTML}
\deft
\verb++%
  \bluett{<?php}%
  \verb++                  \\
\verb+  +%
  \bluett{echo '}%
  \verb+  <input type="+%
  \bluett{text}%
  \verb+"+%
  {\color{blue}\verb+'              . "\r\n";+}%
  \verb++                  \\
\verb+  +%
  \bluett{echo '}%
  \verb+         class="+%
  \bluett{table}%
  \verb+"+%
  {\color{blue}\verb+'            . "\r\n";+}%
  \verb++                  \\
\verb++                    \\
\verb+  +%
  \bluett{if (}%
  \verb++%
  \redtt{......}%
  \verb++%
  \bluett{)}%
  \verb++                  \\
\verb+  +%
  {\color{blue}\verb+{+}%
  \verb++                  \\
\verb+    +%
  \bluett{echo '}%
  \verb+         name="+%
  \redtt{integer\_1\_1}%
  \verb+"+%
  {\color{blue}\verb+'     . "\r\n";+}%
  \verb++                  \\
\verb+    +%
  \bluett{echo '}%
  \verb+         value="+%
  \bluett{' .}%
  \verb+ +%
  \redtt{\$integer\_1\_1}%
  \verb+ +%
  \bluett{.~'}%
  \verb+" />+%
  \bluett{';}%
  \verb++                  \\
\verb+  +%
  {\color{blue}\verb+}+}%
  \verb++                  \\
\verb++                    \\
\verb+  +%
  \bluett{elseif (}%
  \verb++%
  \redtt{......}%
  \verb++%
  \bluett{)}%
  \verb++                  \\
\verb+  +%
  {\color{blue}\verb+{+}%
  \verb++                  \\
\verb+    +%
  \bluett{echo '}%
  \verb+         name="+%
  \redtt{integer\_1\_2}%
  \verb+"+%
  {\color{blue}\verb+'     . "\r\n";+}%
  \verb++                  \\
\verb+    +%
  \bluett{echo '}%
  \verb+         value="+%
  \bluett{' .}%
  \verb+ +%
  \redtt{\$integer\_1\_2}%
  \verb+ +%
  \bluett{.~'}%
  \verb+" />+%
  \bluett{';}%
  \verb++                  \\
\verb+  +%
  {\color{blue}\verb+}+}%
  \verb++                  \\
\verb++                    \\
\verb+  +%
  \bluett{else}%
  \verb++                  \\
\verb+  +%
  {\color{blue}\verb+{+}%
  \verb++                  \\
\verb+    +%
  \bluett{echo '}%
  \verb+         name="+%
  \redtt{integer\_1\_3}%
  \verb+"+%
  {\color{blue}\verb+'     . "\r\n";+}%
  \verb++                  \\
\verb+    +%
  \bluett{echo '}%
  \verb+         value="+%
  \bluett{' .}%
  \verb+ +%
  \redtt{\$integer\_1\_3}%
  \verb+ +%
  \bluett{.~'}%
  \verb+" />+%
  \bluett{';}%
  \verb++                  \\
\verb+  +%
  {\color{blue}\verb+}+}%
  \verb++                  \\
\verb++%
  \bluett{?>}%
  \verb++                  \\
\tabb
\newpage
\myitem
\label{item:php_if-4.php}
 {\bf Modification 2 of PHP with {\tt if/elseif/else} comments includes HTML}
\deft
\verb++%
  \bluett{<?php}%
  \verb++                  \\
\verb+  +%
  \bluett{if (}%
  \verb++%
  \redtt{......}%
  \verb++%
  \bluett{)}%
  \verb++                  \\
\verb+  +%
  {\color{blue}\verb+{+}%
  \verb++                  \\
\verb+    +%
  \bluett{echo '}%
  \verb+  <input type="+%
  \bluett{text}%
  \verb+"+%
  {\color{blue}\verb+'             . "\r\n";+}%
  \verb++                  \\
\verb+    +%
  \bluett{echo '}%
  \verb+         class="+%
  \bluett{table}%
  \verb+"+%
  {\color{blue}\verb+'           . "\r\n";+}%
  \verb++                  \\
\verb+    +%
  \bluett{echo '}%
  \verb+         name="+%
  \redtt{integer\_1}%
  \verb+"+%
  {\color{blue}\verb+'        . "\r\n";+}%
  \verb++                  \\
\verb+    +%
  \bluett{echo '}%
  \verb+         value="+%
  \bluett{' .}%
  \verb+ +%
  \redtt{\$integer\_1}%
  \verb+    +%
  \bluett{.~'}%
  \verb+" />+%
  \bluett{';}%
  \verb++                  \\
\verb+  +%
  {\color{blue}\verb+}+}%
  \verb++                  \\
\verb++                    \\
\verb+  +%
  \bluett{else}%
  \verb++                  \\
\verb+  +%
  {\color{blue}\verb+{+}%
  \verb++                  \\
\verb+    +%
  \bluett{echo '}%
  \verb+  <+%
  \bluett{select}%
  \verb+ name="+%
  \redtt{integer\_2}%
  \verb+">+%
  {\color{blue}\verb+'      . "\r\n";+}%
  \verb++                  \\
\verb++                    \\
\verb+    +%
  \bluett{echo '}%
  \verb+    <+%
  \bluett{option}%
  \verb+ value="+%
  \redtt{integer\_2\_1}%
  \verb+">+%
  {\color{blue}\verb+' . "\r\n";+}%
  \verb++                  \\
\verb+    +%
  \bluett{echo}%
  \verb+         +%
  \redtt{\$integer\_2\_1}%
  \verb++%
  {\color{blue}\verb+               . "\r\n";+}%
  \verb++                  \\
\verb+    +%
  \bluett{echo '}%
  \verb+    <+%
  \bluett{/option}%
  \verb+>+%
  {\color{blue}\verb+'                    . "\r\n";+}%
  \verb++                  \\
\verb++                    \\\verb+    +%
  \bluett{echo '}%
  \verb+    <+%
  \bluett{option}%
  \verb+ value="+%
  \redtt{integer\_2\_2}%
  \verb+">+%
  {\color{blue}\verb+' . "\r\n";+}%
  \verb++                  \\
\verb+    +%
  \bluett{echo}%
  \verb+         +%
  \redtt{\$integer\_2\_2}%
  \verb++%
  {\color{blue}\verb+               . "\r\n";+}%
  \verb++                  \\
\verb+    +%
  \bluett{echo '}%
  \verb+    <+%
  \bluett{/option}%
  \verb+>+%
  {\color{blue}\verb+'                    . "\r\n";+}%
  \verb++                  \\
\verb++                    \\
\verb++                    \\
\verb+                  .+ \\
\verb+                  .+ \\
\verb+                  .+ \\
\verb++                    \\
\verb++                    \\
\verb+    +%
  \bluett{echo '}%
  \verb+  <+%
  \bluett{/select}%
  \verb+>+%
  {\color{blue}\verb+'                      . "\r\n";+}%
  \verb++                  \\
\verb+  +%
  {\color{blue}\verb+}+}%
  \verb++                  \\
\verb++%
  \bluett{?>}%
  \verb++                  \\
\tabb
\end{templateenumerate}

\newpage
\section{Colors}
\label{sec:basic_colors}
 In this section
 I list some most frequently used colors (in my opinion)
 for building a web page.
 Note that
 the W3C HTML and CSS standards have listed only 16 valid color names:
 aqua, black, blue, fuchsia, gray, green,
 lime, maroon, navy, olive, purple, red,
 silver, teal, white, and yellow;
 other colors have to be defined with their HEX values.
%
%
\subsection{Red series}
\label{sec:basic_colors_red}
%
%
\colordefst
 & & & & \\
\colordeftitle
 & & & & \\
\hline
 & & & & \\
\colordef{red}            {1~~~,~0~~~,~0~~~}   {\#FF0000}
\colordef{brilliantred}   {1~~~,~0.28,~0.28}   {\#FF4848}
\colordef{darkred}        {0.55,~0~~~,~0~~~}   {\#8B0000}
\hline
 & & & & \\
\colordef{orange}         {1~~~,~0.65,~0~~~}   {\#FFA500}
\colordef{darkorange}     {1~~~,~0.55,~0~~~}   {\#FF8C00}
\colordef{orangered}      {1~~~,~0.27,~0~~~}   {\#FF4500}
\hline
 & & & & \\
\colordef{salmon}         {0.98,~0.5~,~0.45}   {\#FA8072}
\colordef{coral}          {1~~~,~0.5~,~0.31}   {\#FF7F50}
\colordef{tomato}         {1~~~,~0.39,~0.28}   {\#FF6347}
\hline
 & & & & \\
\colordef{palerose}       {1~~~,~0.76,~0.88}   {\#FFC2E0}
\colordef{pink}           {1~~~,~0.75,~0.80}   {\#FFC0CB}
\colordef{hotpink}        {1~~~,~0.41,~0.71}   {\#FF69B4}
\colordef{deeppink}       {1~~~,~0.08,~0.58}   {\#FF1493}
\colordef{crimson}        {0.86,~0.08,~0.24}   {\#DC143C}
\hline
 & & & & \\
\colordef{chocolate}      {0.82,~0.41,~0.12}   {\#D2691E}
\colordef{firebrick}      {0.7~,~0.13,~0.13}   {\#B22222}
\colordef{brown}          {0.65,~0.16,~0.16}   {\#A52A2A}
\tabb
\newpage
\subsection{Green series}
\label{sec:basic_colors_green}
%
%
\colordefst
 & & & & \\
\colordeftitle
 & & & & \\
\hline
 & & & & \\
\colordef{green}                {0~~~,~0.5~,~0~~~}   {\#008000}
\colordef{darkgreen}            {0~~~,~0.39,~0~~~}   {\#006400}
\hline
 & & & & \\
\colordef{lime}                 {0~~~,~1~~~,~0~~~}   {\#00FF00}
\colordef{limegreen}            {0.2~,~0.8~,~0.2~}   {\#32CD32}
\colordef{forestgreen}          {0.13,~0.55,~0.13}   {\#228B22}
\hline
 & & & & \\
\colordef{lightgreen}           {0.56,~0.93,~0.56}   {\#90EE90}
\colordef{springgreen}          {0~~~,~1~~~,~0.5~}   {\#00FF7F}
\hline
 & & & & \\
\colordef{palelimegreen}        {0.92,~0.98,~0.74}   {\#EBFABD}
\colordef{greenyellow}          {0.68,~1~~~,~0.18}   {\#ADFF2F}
\colordef{lawngreen}            {0.49,~0.99,~0~~~}   {\#7CFC00}
\hline
 & & & & \\
\colordef{brilliantspringbud}   {0.57,~0.92,~0.04}   {\#92EB0A}
\colordef{yellowgreen}          {0.6~,~0.8~,~0.2~}   {\#9ACD32}
\colordef{olive}                {0.5~,~0.5~,~0~~~}   {\#808000}
\tabb
\newpage
\subsection{Blue series}
\label{sec:basic_colors_blue}
%
%
\colordefst
 & & & & \\
\colordeftitle
 & & & & \\
\hline
 & & & & \\
\colordef{blue}               {0~~~,~0~~~,~1~~~}   {\#0000FF}
\colordef{darkblue}           {0~~~,~0~~~,~0.55}   {\#00008B}
\hline
 & & & & \\
\colordef{cyan}               {0~~~,~1~~~,~1~~~}   {\#00FFFF}
\colordef{darkcyan}           {0~~~,~0.55,~0.55}   {\#008B8B}
\hline
 & & & & \\
\colordef{lightgrayishcyan}   {0.75,~0.95,~0.95}   {\#C0F2F2}
\colordef{lightgreyishcyan}   {0.75,~0.95,~0.95}   {\#C0F2F2}
\colordef{paleturquoise}      {0.69,~0.93,~0.93}   {\#AFEEEE}
\colordef{lightblue}          {0.68,~0.85,~0.9 }   {\#ADD8E6}
\colordef{skyblue}            {0.53,~0.81,~0.92}   {\#87CEEB}
\hline
 & & & & \\
\colordef{deepskyblue}        {0~~~,~0.75,~1~~~}   {\#00BFFF}
\colordef{dodgerblue}         {0.12,~0.56,~1~~~}   {\#1E90FF}
\colordef{royalblue}          {0.25,~0.41,~0.88}   {\#4169E1}
\colordef{slateblue}          {0.42,~0.35,~0.8~}   {\#6A5ACD}
\tabb
\newpage
\subsection{Violet series}
\label{sec:basic_colors_violet}
%
%
\colordefst
 & & & & \\
\colordeftitle
 & & & & \\
\hline
 & & & & \\
\colordef{plum}           {0.87,~0.63,~0.87}   {\#DDA0DD}
\colordef{violet}         {0.93,~0.51,~0.93}   {\#EE82EE}
\colordef{magenta}        {1~~~,~0~~~,~1~~~}   {\#FF00FF}
\hline
 & & & & \\
\colordef{mediumpurple}   {0.58,~0.44,~0.85}   {\#9370D8}
\colordef{blueviolet}     {0.54,~0.17,~0.89}   {\#8A2BE2}
\colordef{darkviolet}     {0.58,~0~~~,~0.83}   {\#9400D3}
\colordef{purple}         {0.5 ,~0~~~,~0.5 }   {\#800080}
\tabb
\newpage
\subsection{Yellow series}
\label{sec:basic_colors_yellow}
%
%
\colordefst
 & & & & \\
\colordeftitle
 & & & & \\
\hline
 & & & & \\
\colordef{verypaleyellow}   {1~~~,~1~~~,~0.8~}   {\#FFFFCC}
\colordef{yellow}           {1~~~,~1~~~,~0~~~}   {\#FFFF00}
\hline
 & & & & \\
\colordef{gold}             {1~~~,~0.84,~0~~~}   {\#FFD700}
\colordef{goldenrod}        {0.85,~0.65,~0.13}   {\#DAA520}
\colordef{darkgoldenrod}    {0.72,~0.53,~0.04}   {\#B8860B}
\tabb
\newpage
\subsection{White-black series}
\label{sec:basic_colors_white_black}
%
%
\colordefst
 & & & & \\
\colordeftitle
 & & & & \\
\hline
 & & & & \\
\colordef{white}       {1~~~,~1~~~,~1~~~}   {\#FFFFFF}
\colordef{gainsboro}   {0.86,~0.86,~0.86}   {\#DCDCDC}
\colordef{silver}      {0.75,~0.75,~0.75}   {\#C0C0C0}
\hline
 & & & & \\
\colordef{gray}        {0.5~,~0.5~,~0.5~}   {\#808080}
\colordef{grey}        {0.5~,~0.5~,~0.5~}   {\#808080}
\hline
 & & & & \\
\colordef{dimgray}     {0.41,~0.41,~0.41}   {\#696969}
\colordef{dimgrey}     {0.41,~0.41,~0.41}   {\#696969}
\hline
 & & & & \\
\colordef{darkgray}    {0.25,~0.25,~0.25}   {\#404040}
\colordef{darkgrey}    {0.25,~0.25,~0.25}   {\#404040}
\hline
 & & & & \\
\colordef{black}       {0~~~,~0~~~,~0~~~}   {\#000000}
\tabb
\newpage
\subsection{Non-white backgrounds}
\label{sec:basic_colors_non_white}
%
%
\colordefst
 & & & & \\
\colordeftitle
 & & & & \\
\hline
 & & & & \\
\colordef{cornsilk}            {1~~~,~0.97,~0.86}   {\#FFF8DC}
\colordef{navajowhite}         {1~~~,~0.87,~0.68}   {\#FFDEAD}
\colordef{wheat}               {0.96,~0.87,~0.70}   {\#F5DEB3}
\hline
 & & & & \\
\colordef{lightbrownishgray}   {0.99,~0.89,~0.82}   {\#FDE4D0}
\colordef{lightbrownishgrey}   {0.99,~0.89,~0.82}   {\#FDE4D0}
\colordef{peachpuff}           {1~~~,~0.85,~0.73}   {\#FFDAB9}
\tabb

\addemptypage
\end{appendix}
\newpage
\begin{center}
{\LARGE\bf Acknowledgments}
\end{center}
\vspace{0.75cm}
 The author would like to thank
 James Yi-Yu Liu
 for useful discussion about the basic idea
 of including user--defined functions
 into the source code.

 The author would also like to thank
 the friendly hospitality of
 the Institut f\"ur Physik,
 Humboldt-Universit\"at zu Berlin,
 the College of Physics Science and Technology,
 Xinjiang University,
 the Graduate School of Science and Engineering for Research,
 University of Toyama,
 the Department of Astronomy and the Institute of Theoretical Physics and Astrophysics,
 Xiamen University,
 the Institute of Modern Physics,
 Chinese Academy of Sciences
 and
 the Center for High Energy Physics,
 Peking University,
 where part of this work was completed.

 This work
 was partially supported
 by the National Science Council of R.O.C.~%
 under contracts no.~NSC-98-2811-M-006-044 and
 no.~NSC-99-2811-M-006-031
 as well as
 by the National Center of Theoretical Sciences, R.O.C.
 and
 the Department of Physics,
 National Cheng Kung University, Taiwan.
\addemptypage
\newpage
\begin{center}
{\LARGE\bf References}
\end{center}
\vspace{0.75cm}
\begin{enumerate}
\settowidth{\labelwidth}{~~~}
\setlength{\leftmargin}{\labelwidth}
\renewcommand{\makelabel}{}
\bibitem{W3Schools}
 {W3Schools online web tutorials,
  {\tt \url{http://www.w3schools.com/}}.}
\bibitem{validators}
 {Online validators:
  {\tt \url{http://validator.w3.org/}}                             for (X)HTML,
  {\tt \url{http://jigsaw.w3.org/css-validator/validator.html.en}} for CSS.}
\end{enumerate}
\end{document}